\documentclass[journal]{IEEEtran}
\usepackage{amsmath,latexsym,amssymb,stmaryrd,pifont}
\usepackage{setspace,stackrel,tikz,graphicx}
\usepackage{exscale,relsize,subfig,textcomp,stackrel,setspace,float}
\usepackage{mdframed,pifont}
\usepackage{multicol,xfrac}
\usepackage{multirow, booktabs}
\usepackage{mathtools}
\usepackage{tabularx}
\usepackage[input-decimal-markers=.]{siunitx}
\usepackage{longtable}
\usepackage[T1]{fontenc}
\usepackage[font=footnotesize]{caption}
\usepackage{lipsum}
\usepackage{multicol}
\usepackage{graphicx}
\usepackage[utf8]{inputenc}
\pdfoutput=1
\usepackage{amsthm}
\newcommand{\xmark}{\ding{55}}%

\begin{document}

\title{\vspace{0cm}\LARGE \textcolor{black}{Chirp Spread Spectrum-based Waveform Design and Detection Mechanisms for LPWAN-based IoT -- A Survey}\vspace{0em}}
\makeatletter
\patchcmd{\@maketitle}
  {\addvspace{0\baselineskip}\egroup}
  {\addvspace{0\baselineskip}\egroup}
  {}
  {}
\makeatother
\author{Ali Waqar Azim,  Ahmad Bazzi, Raed Shubair, Marwa Chafii
\thanks{Ali Waqar Azim is with Department of Telecommunication Engineering,  University of Engineering and Technology,  Taxila,  Pakistan (email: aliwaqarazim@gmail.com).}
\thanks{Ahmad Bazzi and Raed Shubair is with with Engineering Division, New York University (NYU) Abu Dhabi, 129188, UAE (email: \{ahmad.bazzi,raed.shubair\}@nyu.edu).}
\thanks{Marwa Chafii is with Engineering Division, New York University (NYU) Abu Dhabi, 129188, UAE and NYU WIRELESS, NYU Tandon School of Engineering, Brooklyn, 11201, NY, USA (email: marwa.chafii@nyu.edu).}}
\maketitle
\begin{abstract}
\textcolor{black}{The Long Range protocol, commonly referred to as LoRa, is a widely adopted and highly regarded method of utilizing chirp spread spectrum (CSS) techniques at the physical (PHY) layer to facilitate low-power wide-area network (LPWAN) connectivity. By tailoring the spreading factors, LoRa can achieve a diverse array of spectral and energy efficiency (EE) levels, making it amenable to a plethora of Internet-of-Things (IoT) applications that rely on LPWAN infrastructure. However, a primary drawback of LoRa is its relatively low data transfer rate. Despite this, there has been a dearth of research dedicated to enhancing the data transfer capabilities of LoRa until recently, when a plethora of CSS-based PHY layer alternatives to LoRa for LPWANs was proposed. This survey, for the first time, presents a comprehensive examination of the waveform design of these CSS-based PHY layer alternatives, which have been proposed between \(2019\) and \(2022\). A total of fifteen alternatives to LoRa are analyzed and compared. Other surveys on LoRa focus on topics such as LoRa networking, the deployment of LoRa in massive IoT networks, and LoRa architectural considerations. In contrast, this study delves deeply into the waveform design of alternatives to LoRa. The CSS schemes studied in this study are classified into three categories: single chirp, multiple chirps, and multiple chirps with index modulation, based on the number of activated frequency shifts activated for un-chirped symbols. The transceiver architecture of these schemes is thoroughly explicated. Additionally, we propose coherent/non-coherent detection mechanisms for specific schemes that have not been previously documented in the literature. We also provide some key insights and recommendations based on the performance of the schemes. The performance of the schemes is evaluated based on metrics such as EE, spectral efficiency, the bit-error-rate (BER) in additive white Gaussian noise, and BER in the presence of phase and frequency offsets. Finally, we highlight some open research issues and future research directions in this field.}
\end{abstract}
\begin{IEEEkeywords}
Waveform design, LoRa, chirp spread spectrum, IoT.
\end{IEEEkeywords}
\IEEEpeerreviewmaketitle

\section{Introduction}
\textcolor{black}{\IEEEPARstart{T}{he} advent of Low-Power Wide-Area Networks (LPWANs) represents a significant technological advancement for the Internet-of-Things (IoT). These networks serve as a conduit for vast quantities of intelligent devices may gain access to the centralized infrastructure of the internet. Through the utilization of LPWANs, it becomes possible to achieve ubiquitous connectivity between an extensive array of sensors, thereby facilitating data collection and promoting cooperation among them. This, in turn, enables the proliferation of various intelligent IoT application development, paving the way for the realization of an innovative way of living characterized by the interconnectedness and remote control of everyday devices via the internet. As such, it is anticipated that IoT devices will become an ever-present reality in the imminent future \cite{zanella2014internet}.}

\textcolor{black}{The design of waveforms represents a fundamental aspect of the realization of pervasive connectivity in IoT. This is because the implementation of an IoT network necessitates the connection of a vast number of devices, and the achievement of low-power consumption through the optimization of waveform design is of paramount importance. This is particularly relevant in light of the fact that the nodes within an IoT network are typically battery-powered terminals, rendering regular maintenance of each node a challenging and unwieldy task. Based on the transmission frequency, the battery life of these nodes is typically targeted to be five to ten years. To this end, various LPWAN standards have been proposed, utilizing both licensed and unlicensed frequency bands, such as the industrial, scientific, and medical (ISM) bands. One notable example of a standard that utilizes a licensed band is Narrowband-IoT (NB-IoT), as proposed by the 3rd Generation Partnership Project (3GPP) \cite{3gpp}. An alternative license-free standard is LoRa Wide-Area Network (LoRaWAN), which has gained widespread deployment compared to other contenders such as Sigfox and Ingenu \cite{goursaud2015dedicated}. This study focuses specifically on LoRaWAN due to its prevalent implementation \cite{pasolini2021lora}.}

\textcolor{black}{LoRaWAN protocol employs LoRa for modulation and demodulation. LoRa, a proprietary technique developed by Semtech, has emerged as a highly promising physical (PHY) layer method for achieving both low-power consumption and wide-scale connectivity over extended distances, as evidenced by various studies such as \cite{lorawan_spec,de2017lorawan, lora, petajajarvi2017performance,raza2017low}. Although the specifics of LoRa's design remain largely undisclosed, it is widely acknowledged that its waveform design is a derivative of the chirp spread spectrum (CSS) technique. Recent efforts have been made to theoretically interpret the properties of LoRa, with notable examples including \cite{lora,chiani2019lora, pasolini2021lora}. In particular, Vangelista in \cite{lora} describes LoRa as a frequency-shift chirp modulation, in which there is a frequency-shift keying (FSK) component that is the un-chirped symbol and a spreading component that spreads the bandwidth of the un-chirped symbol in the entire bandwidth. The frequency spreading of LoRa makes it resilient to Doppler shift. It is also inherently robust against doubly dispersive channels while offering high sensitivity in detecting potentially weak signals while consuming extremely low power, as demonstrated in studies such as \cite{liu2021deeplora,lora_mod_basics}. Furthermore, LoRa's constant envelope property allows for the use of low-cost devices such as power-efficient non-linear amplifiers, as highlighted in studies such as \cite{sundaram2019survey, augustin2016study,zhou2019design}.}

\textcolor{black}{LoRa operates within the ISM band and utilizes three different bandwidths; \(125\) kHz, \(250\) kHz, and \(500\) kHz. The bit rate of LoRa is determined by the parameter known as the spreading factor (SF), which varies between \(6\) and \(12\). By utilizing different SFs and code rates, a wide range of throughputs can be achieved with LoRa, making it a versatile option for a diverse array of applications, ranging from smart cities to agriculture, as observed in studies such as \cite{sundaram2019survey}. However, the achievable bit rate/spectral efficiency (SE) of LoRa is relatively low, which may present a significant limitation. To overcome the limitation of low achievable SE, LoRa modems can also be equipped with an FSK transceiver to double the throughput. However, this added feature can be costly and may increase power consumption due to the requirement for additional computational resources.}

\textcolor{black}{To alleviate the limitations inherent in LoRa, various PHY layer variants of CSS have been proposed in the literature as direct competitors to LoRa. The foremost objectives of these schemes are to either enhance spectral efficiency (SE) or to improve energy efficiency (EE). These improvements are directly linked to the waveform design of these schemes. It is noteworthy that the waveform design and properties of CSS alternatives to LoRa vary significantly; for example, certain schemes incorporate a phase-shift (PS) to the chirp (as exemplified in the work of PSK-LoRa \cite{psk_lora}), while other designs employ varying chirp rates (CRs) (as demonstrated in SSK-LoRa \cite{ssk_lora} and DCRK-LoRa \cite{dcrk_css}), as well as the index modulation (IM) paradigm (as represented by FSCSS-IM \cite{fscss_im} and IQCIM \cite{iqcim}), among others. While a limited number of CSS schemes for LPWAN-based IoT have been proposed in the literature in recent years, to the best of our knowledge, a study consolidating the waveform design of these CSS PHY layer alternatives to LoRa has yet to be made available in the literature. This survey comprehensively provides the waveform design of several CSS alternatives to LoRa proposed in the literature from \(2019\) to \(2022\). Table \ref{tab1} provides a list of acronyms that are used in this study.}
\begingroup
\setlength{\tabcolsep}{6pt} 
\renewcommand{\arraystretch}{0.96} 
\begin{flushleft}
\begin{table}[tbh]
\caption{List of acronyms used in this study.}
\centering
\begin{tabular}{llcc}
\hline
\hline
\textbf{Acronym} & \textbf{Meaning} \\
\hline
ADR & Adaptive Data Rate\\
AWGN&Additive White Gaussian Noise\\
BER&Bit-Error Rate\\
\textcolor{black}{CR} & \textcolor{black}{Chirp Rate}\\
CSI & Channel State Information\\
CSS&Chirp Spread Spectrum\\
DCRK-LoRaS& Discrete CR Keying LoRa\\
\textcolor{black}{DFRC} & \textcolor{black}{Dual Function Radar Communications}\\
DFT& Discrete Fourier Transform\\
DM-CSS & Dual-Mode CSS\\
DO-CSS & Dual Orthogonal CSS\\
EE&Energy Efficiency\\
E-LoRa & Extended-LoRa\\
ePSK-CSS & Enhanced Phase-Shift Keying CSS\\
FAP& Frequency Activation Pattern\\
FF & Fundamental Frequency\\
FS& Frequency-Shift\\
FSCM & Frequency-Shift Chirp Modulation\\
FSCSS-IM & Frequency-Shift CSS with IM\\
FSK & Frequency-Shift Keying\\
GCSS& Group-based CSS\\
IM & Index Modulation\\
ICS-LORa& Interleaved Chirp Spreading-LoRa\\
IoT & Internet-of-Things\\
IQ-CIM& In-phase and Quadrature Chirp IM\\
IQ-CSS& In-phase and Quadrature CSS\\
ISM & Industrial, Scientific and Medical\\
IQ-TDM-CSS& In-phase and Quadrature\\
{}& Time Domain Multiplexed CSS\\
I/Q & In-phase/Quadrature\\
LoRa&Long Range\\
LoRaWAN& LoRa Wide-Area Network\\
LPWAN & Low-Power Wide-Area Networks\\
MAC & Media Access Control\\
ML & Maximum Likelihood\\
NB-IoT& Narrowband-IoT\\
\textcolor{black}{OCDM} & \textcolor{black}{Orthogonal Chirp-Division}\\
{}& \textcolor{black}{Multiplexing}\\
PAPR & Peak-to-average Power Ratio\\
PHY layer& Physical layer\\
PS & Phase-Shift\\
PSK-LoRa& Phase-Shift Keying LoRa\\
QoS& Quality of Service\\
SE & Spectral efficiency\\
SF & Spreading Factor\\
SNR & Signal-to-noise Ratio\\
SSK-LoRa&Slope-Shift Keying LoRa\\
SSK-ICS-LoRa& Slope-Shift Keying Interleaved \\
{}&Chirp Spreading LoRa\\
TDM-CSS & Time Domain Multiplexed CSS\\
UAV & Unmanned Ariel Vehicle\\
\hline 
\hline
\end{tabular}
\label{tab1}
\end{table}
\end{flushleft}
\endgroup
\vspace{-5mm}
\subsection{LoRa Surveys}
\textcolor{black}{Amid a plethora of literature extant pertaining to LoRa, various surveys have been conducted to assess a multitude of facets of the technology, including but not limited to: the PHY and data link layer performance, the various MAC protocols employed, deployment considerations, implementation strategies, simulation testbeds, and field performance evaluations, etc. Additionally, certain studies have sought to examine the potential applications of LoRa within the domain of smart cities and agriculture, etc. Even with the abundance of literature pertaining to LoRA, a dearth of surveys exists that delves explicitly into the waveform design of PHY layer CSS-based alternatives to LoRa. As such, we have established a set of inclusion and exclusion criteria, as detailed in Table \ref{inclusion_exclusion_tab}, to guide our examination of the relevant literature in the area.}
\begingroup
\setlength{\tabcolsep}{6pt} 
\renewcommand{\arraystretch}{0.99} 
\begin{table}[tbh]
\caption{\textcolor{black}{Inclusion exclusion criteria for LoRa-based surveys.}}
\centering
\color{black}\begin{tabular}{llcc}
\hline
\hline
\textbf{Inclusion Criteria} & \textbf{Exclusion Criteria}  \\
\hline
1. Written in English&1. Written in language\\
language& other than English. \\
2. Must be published & 2. Published before \(2016\).\\
between \(2016\) and \(2022\).\\
3. Keywords: LPWANs,  & 3. Presents an incomplete\\
 LoRa or LoRaWAN & study.\\
appear in title or abstract. & \\
4. Published in journals & \\
conference proceedings, &\\
symposiums or workshops. & \\
5. Preprints in arXiv.\\
6. Provides a complete & \\
study on the topic.&\\
\hline 
\hline
\end{tabular}
\label{inclusion_exclusion_tab}
\end{table}
\endgroup
\begin{table*}[h]
  \caption{\textcolor{black}{A comparison between this study and relevant surveys on LoRa available in the literature.}}
  \label{tab_surveys}
  \centering
  \color{black}\begin{tabular}{*{6}{c}}
    \hline
    \hline
    \textbf{Year} & \textbf{Survey} &  \textbf{PHY Waveform}  & \textbf{Detection Mechanisms} & \textbf{Performance} & \textbf{Recommendations}  \\
     &  & \textbf{Design}  & \textbf{Principles}  & \textbf{Evaluation}  & \textbf{on Waveform Design}  \\
    \hline
    \(2016\) & \cite{centenaro2016long}& \xmark & \xmark &\xmark & \xmark \\
    \(2016\) &\cite{augustin2016study}& \xmark & \xmark & \xmark & \xmark \\
    \(2017\) &\cite{marais2017lora}&\xmark & \xmark & \xmark & \xmark \\
    \(2017\) &\cite{sinha2017survey}&\xmark & \xmark & \xmark & \xmark \\
    \(2017\) &\cite{raza2017low}&\xmark & \xmark & \xmark & \xmark \\
    \(2017\) &\cite{lavric2017internet}&\xmark & \xmark & \checkmark (only LoRa) & \xmark \\
    \(2017\) &\cite{wang2017survey}&\xmark & \xmark & \xmark & \xmark \\
    \(2017\) &\cite{migabo2017comparative}&\xmark & \xmark & \xmark & \xmark \\
    \(2018\) &\cite{haxhibeqiri2018survey}& \checkmark (only LoRa)  &\xmark &\checkmark (only LoRa) &\xmark\\
    \(2018\) &\cite{saari2018lora}&\xmark & \xmark & \xmark & \xmark \\
    \(2019\) &\cite{sundaram2019survey}&\xmark & \xmark & \xmark & \xmark \\
    \(2019\) &\cite{sarker2019survey}&\xmark & \xmark & \xmark & \xmark \\
    \(2019\) &\cite{qin2019low}&\xmark & \xmark & \xmark & \xmark \\
    \(2019\) &\cite{erturk2019survey}&\xmark & \xmark & \xmark & \xmark \\
    \(2019\) &\cite{andrade2019comprehensive}&\xmark & \xmark & \xmark & \xmark \\
    \(2020\) &\cite{cotrim2020lorawan}&\xmark & \xmark & \xmark & \xmark \\
    \(2020\) &\cite{kufakunesu2020survey}&\xmark & \xmark & \xmark & \xmark \\
    \(2020\) &\cite{noura2020lorawan}&\xmark & \xmark & \xmark & \xmark \\
    \(2020\) &\cite{alenezi2020ultra}&\xmark & \xmark & \xmark & \xmark \\
    \(2020\) &\cite{staikopoulos2020image}&\xmark & \xmark & \xmark & \xmark \\
    \(2020\) &\cite{marais2020survey}&\xmark & \xmark & \xmark & \xmark \\
    \(2020\) &\cite{raychowdhury2020survey}&\xmark & \xmark & \xmark & \xmark \\
    \(2021\) &\cite{da2021survey}&\xmark & \xmark & \xmark & \xmark \\
    \(2021\) &\cite{gkotsiopoulos2021performance}& \checkmark (only LoRa)  &\xmark &\xmark &\xmark\\
    \(2021\) &\cite{benkahla2021review}&\xmark & \xmark & \xmark & \xmark \\
    \(2021\) &\cite{ghazali2021systematic}&\xmark & \xmark & \xmark & \xmark \\
    \(2022\) &\cite{almuhaya2022survey}&\xmark & \xmark & \xmark & \xmark \\
    \(2022\) &\cite{cheikh2022multi}&\xmark & \xmark & \xmark & \xmark \\
    \(2022\) &\cite{li2022lora}&\xmark & \xmark & \xmark & \xmark \\
    \(2022\) &\cite{sun2022recent}&\xmark & \xmark & \xmark & \xmark \\
    \(2022\) &\cite{jouhari2022survey}&\xmark & \xmark & \xmark & \xmark \\
    \(2022\) &\cite{idris2022survey}&\xmark & \xmark & \xmark & \xmark \\
    \(2022\) &\cite{pagano2022survey}&\xmark & \xmark & \xmark & \xmark \\
    \(2022\) &\cite{milarokostas2022comprehensive}&\xmark & \xmark & \xmark & \xmark \\
    \(2022\) &\cite{banti2022lorawan}&\xmark & \xmark & \xmark & \xmark \\
    \(2023\) &This study&\checkmark & \checkmark & \checkmark & \checkmark \\
    \hline
    \hline
  \end{tabular}
\end{table*}

\textcolor{black}{The present study aims to comprehensively examine the PHY layer waveform design of alternatives to LoRa based on CSS and proposed specifically for LPWANs. In consonance with the parameters detailed in Table \ref{inclusion_exclusion_tab}, Table \ref{tab_surveys}  enumerates a selection of surveys chosen for comparative analysis. The primary objective of this study is to expound upon the various mechanisms of detection, contrast the performance of these alternatives across a plethora of scenarios, and proffer suggestions for improvement as well as potential areas for further research. The comparison of these chosen surveys with the present study, as evaluated against the aforementioned four criteria, is exhibited in Table \ref{tab_surveys}. It is worth noting that, with a few exceptions, most other surveys do not delve into the intricacies of the PHY layer waveform design of CSS-based alternatives to LoRa for LPWANs in great depth and the ones which do; they only consider the state-of-the-art LoRa.}
\begin{table*}[t]
  \caption{Summary of related surveys and comparative works.}
  \label{tab_surveys2}
  \centering
  \begin{tabular}{*{2}{llc}}
    \hline
    \hline
    \textbf{Survey} &  \textbf{Short Description}  \\
    \hline
   \cite{centenaro2016long}&Provides an overview of LPWANs, and discusses its advantages for applications related to smart cities.\\
   \cite{augustin2016study} & Provides an overview of LoRa, and evaluates the PHY and data link layer performance.\\
   \cite{marais2017lora} &Compares the strength and limitations of different LoRa/LoRaWAN test beds.\\
    \cite{sinha2017survey}&Compares NB-IoT and LoRa considering different performance metrics.\\
   \cite{raza2017low}&Surveys the design and techniques employed in LPWANs offering wide-area coverage.\\
   \cite{lavric2017internet} &Evaluates the performance of LoRa considering IoT requirements.\\
    \cite{wang2017survey}& Surveys different LPWAN solutions in context of machine-to-machine communications.\\
   \cite{migabo2017comparative} &Compares research and industrial states of NB-IoT and LoRa considering different metrics.\\
   \cite{haxhibeqiri2018survey}&Consolidates LoRa and LoRaWAN research considering PHY and network layer aspects.\\
   \cite{saari2018lora}& Identify LoRa applications, research trends, and practical applications of LoRa in IoT deployment.\\
    \cite{sundaram2019survey} &Surveys technical challenges in LoRa networks and their possible solutions.\\
   \cite{sarker2019survey}&Explores LoRa application fields and integration of edge computation capability in IoT.\\
   \cite{qin2019low}& Investigates NB-IoT and LoRa in the context of challenges preventing LPWANs from widespread adoption.\\
   \cite{erturk2019survey}& Provides an overview of LoRaWAN, including MAC protocols and its architecture.\\
  \cite{andrade2019comprehensive}& Studies the use of LoRa in developing smart cities.\\
  \cite{cotrim2020lorawan}&Provides a review of the multi-hop proposals for LoRaWAN.\\
    \cite{kufakunesu2020survey}&Surveys ADR algorithms for LoRaWAN. \\
    \cite{noura2020lorawan}&Reviews LoRaWAN architecture, applications, security concerns, and possible countermeasures.\\
    \cite{alenezi2020ultra} &Reviews LoRa and LoRaWAN, and discusses applying ultra-dense network concept on LPWAN.\\
    \cite{staikopoulos2020image}&Reviews the methods for image transfer via LoRa infrastructure. \\
    \cite{marais2020survey}&Reviews use cases requiring confirmed traffic and investigate LoRaWAN confirmed traffic. \\
 \cite{raychowdhury2020survey}& Surveys LoRa performance under different scenarios and difficulties in its implementation. \\
   \cite{da2021survey}&Surveys the available tools for simulating LoRa networks in NS-3 network simulator.\\
    \cite{gkotsiopoulos2021performance}&Presents an overview of LoRa PHY layer, MAC protocols, deployment, and end device transmission settings.\\
    \cite{benkahla2021review}& Provides experimental performance evaluation of ADR enhancements in a mobile node scenario.\\
   \cite{ghazali2021systematic}&Studies the real-time deployments of UAV-based LoRa network.\\
   \cite{almuhaya2022survey} &Compares simulation tools for investigating and analyzing LoRa/LoRaWAN network performance.\\
   \cite{cheikh2022multi}&Provides an overview on LoRa, insights for building energy-efficient IoT infrastructures, and IoT devices.\\
   \cite{li2022lora}&Categorizes and compares LoRa networking techniques.\\
 \cite{sun2022recent}&Surveys LoRa from a systematic perspective. \\
    \cite{jouhari2022survey}&Surveys the scalability challenges/solutions to assist LoRaWAN deployment in massive IoT networks.\\
   \cite{idris2022survey}&Provides an overview of LoRaWAN, discusses open-source simulation tools, and evaluates their performance.\\
   \cite{pagano2022survey}& Surveys LoRa-based solutions in smart agriculture.\\
    \cite{milarokostas2022comprehensive} & Provides an overview of simulators supported by cloud infrastructure and discusses data-related operations.\\
  \cite{banti2022lorawan} &Surveys LoRaWAN protocols considering EE perspective.\\
    \hline
    \hline
  \end{tabular}
\end{table*}

\textcolor{black}{The lack of literature on the PHY layer waveform design of CSS-based protocols for LPWANs is demonstrated by Table \ref{tab_surveys}. To furnish a more exhaustive understanding of the target areas of the selected surveys per the criteria outlined in Table \ref{inclusion_exclusion_tab}, Table \ref{tab_surveys2} elaborates on this subject matter by providing a succinct synopsis for the reader's edification. }

\textcolor{black}{The authors in \cite{centenaro2016long} provide an overview of sub-GHz LPWANs and explore various techniques for providing connectivity in IoT environments, highlighting the advantages of such approaches over extant paradigms in terms of efficiency, effectiveness, and architectural design, specifically in the context of smart cities applications. The authors in \cite{augustin2016study} presents a general overview of LoRa and undertake an in-depth analysis of its functional components, utilizing both field tests and simulations to evaluate the performance of the PHY and data link layers, subsequently proposing potential solutions to enhance performance. In \cite{marais2017lora}, the authors conduct a retrospective analysis of prior work related to LoRa and LoRaWAN testbeds, identifying flaws and strengths by comparing the testbeds and further recognizing the limitations of LoRaWANs. }

\textcolor{black}{The authors in \cite{sinha2017survey} present a comprehensive survey of NB-IoT and LoRa as efficient solutions for connecting devices in LPWANs and ascertaining that LoRa has advantages in terms of battery lifetime, capacity, and cost, while licensed NB-IoT offers QoS, latency, reliability, and range benefits. In \cite{raza2017low}, the authors review the design goals and techniques that LPWANs employ to offer wide-area coverage to low-power devices at the expense of low data rates and further analyze several emerging LPWA technologies, the standardization activities carried out by different standards development organizations and the industrial consortia built around individual LPWA technologies. In \cite{lavric2017internet}, a thorough evaluation of LoRa is conducted by examining the obstacles impeding IoT development, identifying the challenges associated with such obstacles, and subsequently elucidation the need for solutions to address these problems. Additionally, the authors of this study also perform a comprehensive analysis of the LoRaWAN communication protocol architecture requirements and LoRa modulation performance. }

\textcolor{black}{Similarly, in \cite{wang2017survey}, the authors survey the benefits and drawbacks of LPWAN machine-to-machine solutions, as well as the corresponding deployment strategies, and identify areas of knowledge deficiency through the presentation of a summary of research directions aimed at improving the performance of surveyed low power and long-range machine-to-machine communication technologies. Like \cite{sinha2017survey}, \cite{migabo2017comparative} focuses on providing a comprehensive and comparative survey study of the current research and industrial states of NB-IoT and LoRa, with a particular emphasis on analyzing their power efficiency, capacity, QoS, reliability, and range of coverage. \cite{haxhibeqiri2018survey} offers an overview of research work on LoRa and LoRaWAN, published between \(2015\) and September \(2018\), as retrievable through both the Google Scholar and IEEE Explore databases. This work evaluates the performance of LoRa and LoRaWAN, considering security and reliability mechanisms, PHY layer aspects, and network layer aspects. Furthermore, the authors of this study also proffer possible improvements and extensions to the standard.}

\textcolor{black}{In \cite{saari2018lora}, Saari \textit{et al.} survey the most recent trends in research and practical applications of LoRa and also provide recommendations to fully exploit the advantages of LoRa-based technologies in the development of IoT systems and solutions. \cite{sundaram2019survey} surveys technical challenges in LoRa deployment and inspires work on improving the performance of LoRa networks and enabling more practical deployments. In  \cite{sarker2019survey}, the authors explore and analyze different application fields which use LoRa and investigate potential improvement opportunities and considerations. Furthermore, the authors also propose a generic architecture to integrate edge computation capability in IoT-based applications for enhanced performance. \cite{qin2019low} compares competing candidates of LPWANs, such as NB-IoT and LoRa, in terms of technical fundamentals and large-scale deployment potential. The authors also identify several challenges that prevent LPWAN technologies from moving from theory to widespread practice.}
 
\textcolor{black}{\cite{erturk2019survey} provides an overview of LoRaWAN, including MAC protocols and its architecture. In \cite{andrade2019comprehensive}, Andrade and Yoo identify the technological barriers that hamper the development of solutions, find possible future trends that could exist in the context of smart cities and IoT, and understand how the industry and academy could exploit them. \cite{cotrim2020lorawan} reviews the state-of-the-art multi-hop proposals for LoRaWAN. by providing comparative analysis and classification, considering technical characteristics, intermediate devices function, and network topologies. Moreover, the authors also discuss open issues and future directions to realize the full potential of multi-hop networking for LPWANs. \cite{kufakunesu2020survey} survey ADR solutions as an optimization approach to improve throughput, EE and scalability in LoRaWANs. }

\textcolor{black}{In \cite{noura2020lorawan}, Noura \textit{et al.} provide a comprehensive analysis of the LoRaWAN architecture, including its various applications and associated security concerns. They also outline countermeasures to address existing vulnerabilities and prevent potential attacks on the LoRaWAN protocol. Similarly, \cite{alenezi2020ultra} conducts a literature review on LoRa and advances in the LoRaWAN protocol, highlighting the challenges faced by the technology and exploring the potential for utilizing ultra-dense network concepts in the context of low-power wide area networks. \cite{staikopoulos2020image} evaluates the various methods for image transfer via LoRa infrastructure, noting the limitations of each approach and identifying key challenges that must be addressed to achieve reliable image transfer over a LoRa network. In \cite{marais2020survey}, Marais \textit{et al.} examine the use cases that necessitate confirmed traffic in LoRaWAN, surveying recent literature on the topic and discussing the mechanisms by which confirmed traffic is implemented. They also identify and propose solutions to open research challenges that must be overcome to enhance the feasibility of confirmed traffic.}

\textcolor{black}{\cite{raychowdhury2020survey} provides an overview of the performance of LoRa in different scenarios, as well as an examination of the implementation hindrances of this technology. Additionally, \cite{da2021survey} presents a survey of the available tools for simulating LoRa networks in the NS-3 network simulator, detailing their main features and limitations. Finally, in reference \cite{gkotsiopoulos2021performance}, Gkotsiopoulos \textit{et al.} present an overview of LoRa-based networks, analyzing their behavior and categorizing them based on their technological breakthroughs, in terms of various performance determinants, including PHY layer characteristics, deployment and hardware features, end device transmission settings, LoRa MAC protocols, and application requirements. The literature reviewed in \cite{benkahla2021review} evaluates the advancements in ADR and presents the experimental results of their implementation in a mobile node scenario. }

\textcolor{black}{The study conducted in \cite{ghazali2021systematic} critically examines the practical deployment of UAV-based LoRa communication networks, explicitly emphasizing the communication setup. In \cite{almuhaya2022survey}, the authors present a comprehensive comparative analysis of various LPWAN technologies, including NB-IoT, SigFox, Telensa, Ingenu, and LoRa/LoRaWAN, providing a technical overview of LoRa/LoRaWAN by highlighting its key features, opportunities, open issues, and simulation tools for evaluating network performance. \cite{cheikh2022multi} offers a tutorial on the LoRa standard and examines existing solutions, hot topics, and future insights for the construction of energy-efficient IoT infrastructures and devices, with a particular focus on LoRa/LoRaWAN EE across the PHY, MAC, and network layers. \cite{li2022lora} compares and categorizes state-of-the-art LoRa networking techniques. \cite{sun2022recent} summarizes research focusing on the performance analysis and security vulnerabilities of LoRa networks and potential countermeasures. }

\textcolor{black}{\cite{jouhari2022survey} investigates scalability challenges in LoRaWAN and the current state-of-the-art solutions in the PHY and MAC layers, summarizing existing literature solutions and proposing potential enhancements for future research. \cite{idris2022survey} provides a chronological survey of available IoT simulation tools using a systemic approach and presents an overview of LoRa/LoRaWAN technology, including its architecture, transmission parameters, device classes, and simulation tools, and evaluates LoRaWAN performance in terms of Packet Delivery Ratio, memory usage, execution time, and the number of collisions. \cite{pagano2022survey} surveys LoRa-based solutions for irrigation systems, plantation and crop monitoring, tree monitoring, and livestock monitoring, analyzing their scalability, interoperability, network architecture, and energy efficiency. \cite{milarokostas2022comprehensive} offers a comparative overview of cloud-based simulators for LPWANs. In \cite{banti2022lorawan}, the authors identify and discuss critical aspects and research challenges in designing a LoRaWAN communication protocol from an EE perspective and propose a GreenLoRaWAN communication protocol that prioritizes EE, robustness, and scalability.}
\subsection{Motivation and Contributions}
\textcolor{black}{It is evident that there exists a lack of surveys that consolidates the various PHY layer CSS-based alternatives to LoRa for LPWANs (cf. Table \ref{tab_surveys2}). As such, our primary objective is to thoroughly explicate a comparative analysis of the CSS-based waveform designs proposed in the literature between \(2019\) and \(2022\) as a viable alternative to LoRa. This study is distinct from other surveys on the topic in that it diverges from the conventional approach of providing a general overview of LoRa technology, focusing on networking, deployment, architecture, and applications.}

\textcolor{black}{The contributions of this comparative study are summarized as follows:}
\begin{itemize}
\item \textcolor{black}{We classify all the CSS schemes between \(2019\) and \(2022\) into three distinct taxonomies based on the number of FSs activated for the un-chirped symbol. These three taxonomies are (i) single chirp (SC); (ii) multiple chirp (MC); and (iii) MC with IM (MC-IM). This method of characterization has yet to be previously done in the literature. Each taxonomy has unique advantages and disadvantages, which shall be elucidated in the subsequent analysis.}
\item \textcolor{black}{Within the purview of SC, MC, and MC-IM taxonomies, we expound upon the various design methodologies utilized in implementing CSS-based schemes for IoT based on LPWANs. These design methodologies are the fundamental tenets of various CSS modulations. Furthermore, these methodologies furnish guidance for enhancing the efficacy of the protocols in terms of various performance metrics.}
\item\textcolor{black}{ We present a comprehensive analysis of coherent and non-coherent detection principles from a waveform design standpoint. To the best of our knowledge, these principles have yet to be explicitly expounded upon in the literature.}
\item \textcolor{black}{Thereafter, we furnish a comprehensive analysis of the PHY waveform design of each CSS-based scheme for LPWAN, in which the time domain symbol structure, as well as the coherent and non-coherent detection mechanisms (where applicable), are also expounded upon. Furthermore, for the first time, we also present the coherent detection mechanisms for four of the fifteen schemes included in the study, which have yet to be documented in the extant literature. Additionally, for the sake of clarity and ease of implementation, we also provide the block diagram of the transceivers of these schemes. We have recreated the transceiver block diagrams to conform with the notations utilized throughout this study.}
\item \textcolor{black}{Based on the waveform design of each scheme, we proffer several critical insights concerning the ease of implementation, potential performance limitations, and issues inherent in the waveform design, among other considerations.}
\item \textcolor{black}{Following a systematic examination of the examined CSS schemes for LPWANs in terms of SE, a categorical designation of the schemes into six distinct groups is established based on the maximum attainable SE. Subsequently, an investigation is conducted to assess the performance of the CSS-based schemes within each grouping concerning the SE and EE trade-off and bit error rate (BER) performance in an additive white Gaussian noise (AWGN) channel. Furthermore, an analysis of the BER performance is performed, accounting for phase and frequency offsets inherent in low-cost devices utilized in LPWANs.}
\item \textcolor{black}{In this study, certain open research issues about the waveform design of CSS schemes were brought to attention. We also propose a framework to improve the performance of the schemes. The framework proposed in this study will likely catalyze the emergence of new CSS schemes. Furthermore, we intend this work to stimulate further exploration and investigation into this study area.}
\item \textcolor{black}{In conclusion, a thorough examination of the various CSS schemes' pros and cons was conducted, utilizing the previously established performance metrics. A comparative analysis is performed to identify the most advantageous CSS schemes within each grouping. Specific attention is given to those CSS schemes that may serve as viable alternatives to LoRa in different environmental and operational scenarios. Note that the evaluation of the CSS schemes' relative merits and shortcomings is conducted by comparing and contrasting them based on previously discussed performance metrics.}
\end{itemize}

\textcolor{black}{To encourage reproducible research and to facilitate the replication of the results presented in this study, public versions of the MATLAB codes for all the schemes discussed in this study will be made available.}
\subsection{Paper Organization}
\begin{figure*}[h]
  \includegraphics[width=\textwidth,height=11cm]{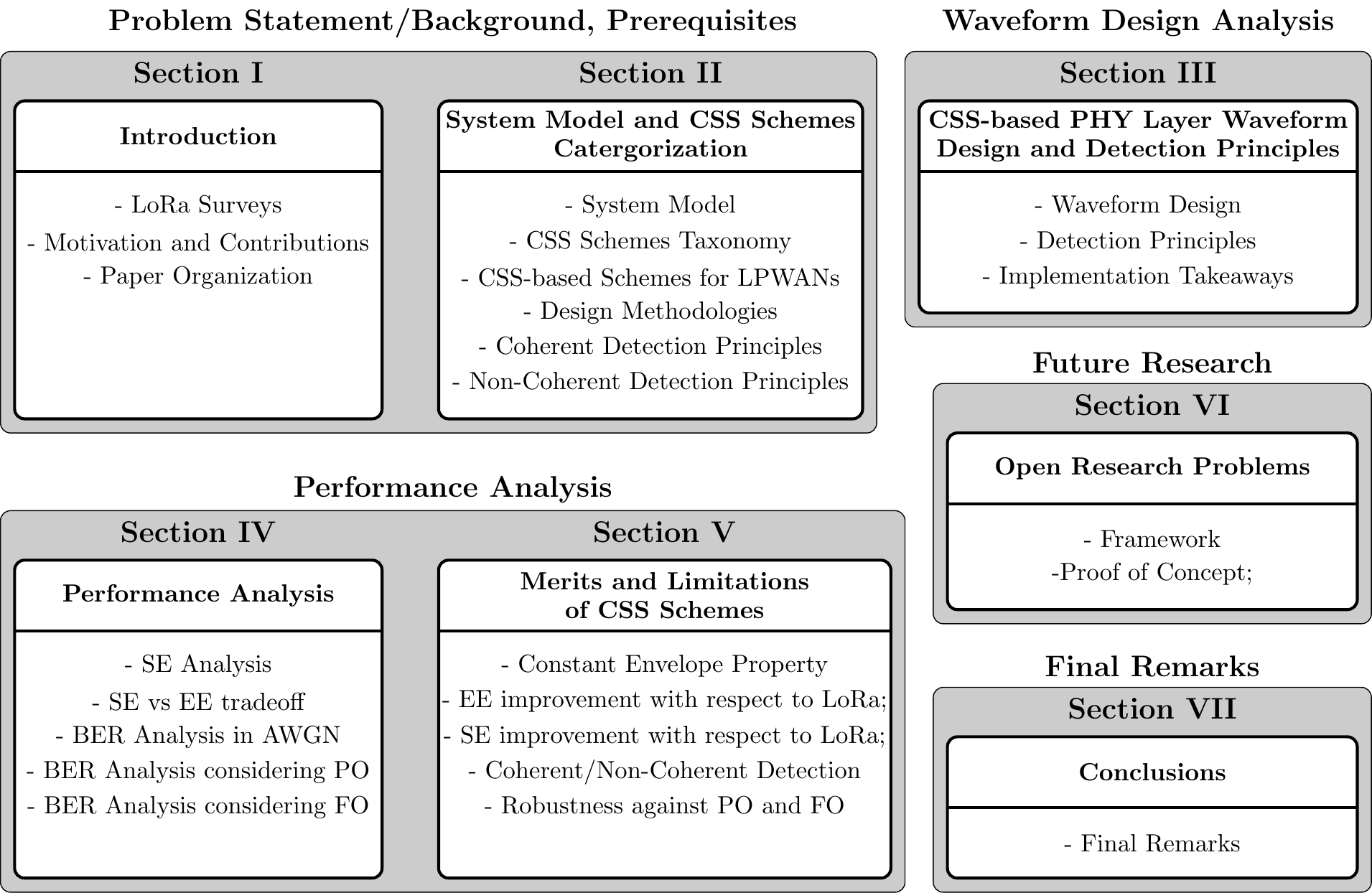}
  \caption{\textcolor{black}{Section names and the main components of the conducted study.}}
  \label{intro_fig}
\end{figure*}

\textcolor{black}{The overall structure of this survey is illustrated in Fig. \ref{intro_fig}. The rest of the manuscript is organized in the following manner: In Section \ref{sec2}, we commence by presenting the system model adopted in this  study, wherein we subsequently categorize the schemes in SC, MC and MC with IM and offer some salient observations from these categories concerning waveform design. Within the context of SC, MC, and MC-IM, various design methodologies are elucidated with regards to their specific characteristics and constraints. Furthermore, Section \ref{sec2} also endeavors to provide a comprehensive explication of the coherent and non-coherent CSS detection principles. The waveform design and detection mechanisms of the CSS schemes for IoT based on LPWANs are expounded in Section \ref{sec3}, along with some pertinent observations regarding different design aspects. In Section \ref{sec4}, we undertake a performance comparison of the studied CSS schemes in terms of SE, SE/EE trade-off, BER performance in the AWGN channel, and BER performance taking into account phase and frequency offsets. In Section \ref{sec5}, we delve into the individual merits and limitations of the investigated CSS schemes and propose alternative CSS schemes to LoRa, which may exhibit superior performance under certain conditions. In Section \ref{framework}, we provide open research issues and future research directions. Finally, in Section \ref{sec6}, we present the conclusions drawn from this study.}
%
%
\section{\textcolor{black}{System Model and CSS Schemes Categorization}}\label{sec2}
\textcolor{black}{This section presents a generalized system model as the theoretical foundation for our analysis. Subsequently, we classify the CSS schemes under examination into three distinct categories based on the number of activated FSs for the un-chirped symbol, with these categories being SC, MC and MC-IM.}
\subsection{System Model}
\begin{figure}[h]\centering
\includegraphics[trim={2 0 0 0},clip,scale=0.99]{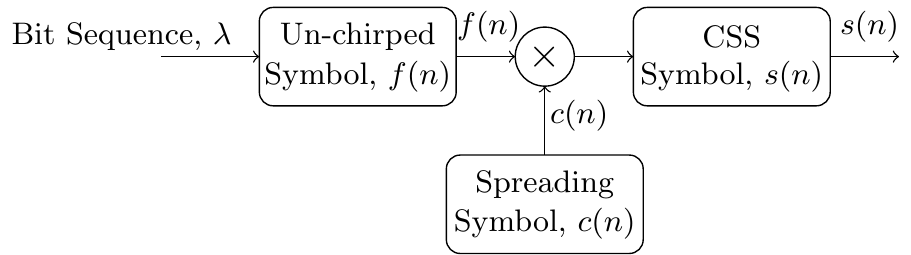}
  \caption{\textcolor{black}{Generalized system model adopted in this study.}}
\label{sys_model}
\end{figure}
\textcolor{black}{In this study, we adopt a generalized system model as in \cite{lora} and \cite{chiani2019lora} without loss of generality as illustrated in Fig. \ref{sys_model}. Specifically, we consider a discrete-time chirp symbol, \(s(n)\), for \(n \in \llbracket 0, M-1\rrbracket\), where \(M\) is the number of samples of duration \(T_\mathrm{c}\). The sample rate, \(T_\mathrm{c}\) corresponds to a sampling frequency of \(F_\mathrm{c} = \sfrac{M}{T_\mathrm{c}} = B\); thus, each discrete-time chirp symbol is fully represented by \(M\) samples taken at the rate \(F_\mathrm{c}\). By definition, \(s(n)\) is the product of two components, namely (i) an un-chirped symbol, \(f(n)\), and (ii) a spreading/chirp symbol, \(c(n)\); therefore, \(s(n) = f(n)c(n)\), where \(c(n)\) serves to spread the information in the bandwidth, \(B = \sfrac{M}{T_\mathrm{s}}\), where \(T_\mathrm{s}\) is the symbol duration.  Moreover, when \(f(n)\) is chirped using \(c(n)\), the FS(s) has an injective mapping to a unique cyclic time-shift(s) of the chirped signal, \(s(n)\).}

\textcolor{black}{A total of \(M\) FSs are available for activation in the un-chirped symbol, where one or multiple FSs can be activated. It is assumed that the FS is the information-bearing element at this point. The FS(s) are chosen from the set \([0, M-1]\). It is worth noting that, in addition to the FS(s), other components of the transmit symbol structure, such as the PS, the use of multiple CRs, and interleaving on the chirped symbol, etc. can also be utilized to convey information. These parameters provide added flexibility in incorporating more bits into the CSS symbols. As such, in the subsequent sections of this study, we shall undertake a comprehensive examination of different design aspects related to CSS-based schemes for LPWANs.}

\textcolor{black}{In LPWANs, the CSS symbols are characterized by a limited frequency bandwidth of \(500\) kHz or less. As such, it is appropriate to posit that the channel through which these symbols propagate exhibits flat fading behavior, wherein the attenuation experienced by the signal is constant across the entire bandwidth, \(B\). The discrete-time baseband received symbol is mathematically represented as:}
\begin{equation}\label{channel}
y(n) = s(n) + w(n),
\end{equation}
\textcolor{black}{for \(n = \llbracket 0, M-1\rrbracket\), where w(n) represents the samples of AWGN, which possess a single-sided noise power spectral density of \(N_0\) and a noise variance of \(\sigma_n^2 = N_0B\). In this formulation, the channel attenuation is assumed to be unity and, as such, is not explicitly present in eq. (\ref{channel}). This is due to the assumption that a priori knowledge of the CSI is available; thus, the channel does not need to be explicitly represented.}
\subsection{\textcolor{black}{CSS Schemes Taxonomy}}
\textcolor{black}{In the context of LPWAN-based IoT, the utilization of CSS schemes necessitates the presence of one or multiple FSs within the un-chirped symbol. Consequently, the initial characterization of CSS schemes is predicated on the number and configuration of activated FSs, categorizing three distinct taxonomies: SC, MC, and MC-IM, as previously alluded to. It is imperative to reiterate that the scope of the present analysis is restricted to those CSS schemes specifically devised for LPWANs, wherein not all available FSs are activated for the un-chirped symbols, as in OCDM, etc.}

\textcolor{black}{The previously enumerated three taxonomies exhibit distinctions in the methodology employed to attain an un-chirped symbol. We consider that a total of \(M\) available FSs for activation}.
\subsubsection{\textcolor{black}{SC CSS Schemes}}
\textcolor{black}{In the SC schemes, only a single FS is activated for the un-chirped symbol. Then, the un-chirped symbol in SC schemes is mathematically represented as follows:}
\begin{equation}\label{sc}
\textcolor{black}{f(n) = \exp\left\{j\frac{2\pi}{M}kn\right\},}
\end{equation}
\textcolor{black}{where \(k \in \left[0,M-1\right]\) denotes the a activated FS and \(n=\llbracket 0, M-1\rrbracket\). It is underlined that \(k\) is obtained by converting the input bit sequence from its binary representation to its equivalent decimal form.}

\textcolor{black}{From a waveform design standpoint, the SC schemes boast a plethora of advantageous characteristics. These include: (i) high EE due to the activation of a single FS; (ii) constant envelope for the chirped symbol, which is beneficial for the utilization of low-cost components within LPWANs; (iii) simple waveform design, implementation, forthright detection mechanisms, and maintenance; (iv) improved robustness to interference and phase and frequency offsets; (v) lower latency resulting from reduced complexity in terms of signal processing; (vi) more cost-effective as they require fewer components and less complex signal processing.}

\textcolor{black}{Despite the plethora of advantageous characteristics enumerated above, one of the most substantial drawbacks of these SC systems is their relatively low achievable SE, which precludes the widespread utilization of these schemes, particularly for high data-rate IoT applications.}
\subsubsection{\textcolor{black}{MC CSS Schemes}}
\textcolor{black}{In MC schemes, multiple FSs are activated in the un-chirped symbol rather than activating a single FS. Assuming that a total of \(q>1\) FSs are activated, the un-chirped symbol, \(f(n)\), for MC schemes can be represented as:}
\begin{equation}\label{mc}
\textcolor{black}{f(n) = \sum_{q}\exp\left\{j\frac{2\pi}{M}k_q n\right\},}
\end{equation}
\textcolor{black}{where \(k_q \in \left[0,M-1\right]\) are the activated FS, such that \(k_q \neq k_p\) for \(q \neq p\). From the standpoint of waveform design, it is typical to select a value of \(q\) that is relatively diminutive, such as \(2\) or maybe \(3\). This is because augmenting the value of \(q\) results in a decrease in the EE metric (at the cost of an increase in SE), rendering it inapplicable for LPWANs. The implementation of activating \(q\) FSs can be accomplished through various techniques. A potential strategy includes partitioning the total number of FSs, \(M\), into \(q\) subsets, with each subset comprising \(\sfrac{M}{q}\) FSs. Subsequently, one FS from each designated subset is activated in correspondence to the specific input bit sequence.}

\textcolor{black}{MC schemes also offer numerous advantages, such as (i) higher SE relative to SC CSS schemes; (ii) negligible decrease in EE due to activation of additional FS(s); (iii) multiple FSs can be used to implement diversity; and (iv) acceptable robustness against the phase and frequency offsets.}

\textcolor{black}{The utilization of MC schemes has its limitations. One such limitation is the potential for non-constant envelope symbols resulting from the activation of multiple FSs, necessitating the utilization of costly equipment. Additionally, MC schemes often exhibit a higher PAPR than SC schemes, thereby exacerbating the non-linearities of different system components. Furthermore, the design of transceivers for MC systems is often complicated by the need for complex signal-processing algorithms to facilitate the detection of multiple active FSs. Finally, MC schemes may not possess the same robustness against phase and frequency offsets as SC modulations.}
\subsubsection{\textcolor{black}{MC-IM CSS Schemes}}
\textcolor{black}{In MC-IM, multiple FSs are simultaneously activated following a predefined frequency activation pattern (FAP). The information bits are encoded within the FAP rather than in the individual FSs activated within the un-chirped symbol. This method typically utilizes two to three FSs, as activating a higher number may negatively impact EE and increase system complexity. The set of available FSs is represented as $\boldsymbol{\Omega} = \left\{0,1,\cdots, M-1\right\}$, where $\varsigma$ FSs are to be activated. The FAP, denoted as $\boldsymbol{\sigma}= \{\sigma_1, \sigma_2,\cdots, \sigma_\varsigma\}\in \boldsymbol{\Omega}$, is determined through binary-to-decimal (bi2de) conversion on \(\lfloor \log_2 \binom{M}{\varsigma}\rfloor\) bits. The resulting un-chirped symbol in MC-IM is then formulated as a function of the FAP as:}
\begin{equation}\label{mc-im}
\textcolor{black}{f(n) = \sum_{k\in \boldsymbol{\sigma}}\exp\left\{j\frac{2\pi}{M}k n\right\}.}
\end{equation}

\textcolor{black}{The utilization of MC-IM schemes not only confers the benefits inherent to MC schemes but also potentially yields superior SE via incorporating additional bits in the FAPs. However, the employment of MC-IM schemes also incurs added complexity, as detecting FAPs necessitates sophisticated signal processing algorithms, thereby exacerbating latency and power consumption.}
\subsubsection{\textcolor{black}{Takeaways}}
\textcolor{black}{It is unequivocally apparent that each taxonomic classification of modulation schemes possesses its own set of inherent limitations. Specifically, SC schemes are optimal from a waveform design perspective as they afford the utilization of low-cost equipment, are comparatively simple to implement and maintain and possess minimal signal processing requirements. However, such schemes' maximum achievable SE is limited, rendering them ill-suited for applications requiring high data rates. Conversely, MC schemes exhibit an improvement in SE relative to SC schemes but are accompanied by the added complexity of signal processing and other associated issues. Furthermore, MC-IM schemes, being a relatively nascent concept, necessitates a comprehensive analysis to optimize various parameters to enhance their overall performance.}

\textcolor{black}{In the practical implementation of LPWANs for IoT applications, LoRa is commonly employed, as it belongs to the class of SC schemes and is one of the first CSS-based schemes proposed for LPWANS. However, with the advent of CSS-based schemes from other taxonomies, it is possible that schemes other than LoRa, belonging to different taxonomies, exhibit superior performance under varying circumstances. Consequently, a comprehensive analysis from a practical implementation standpoint is imperative. From a theoretical perspective, SC schemes can be employed in scenarios where low-cost equipment is compulsory, different offsets are anticipated, and SE is not the paramount concern. Conversely, MC and MC-IM schemes can be implemented in scenarios where a higher SE is required, and the channel is relatively clean. At the same time, equipment cost should not be a critical factor.}
\subsection{\textcolor{black}{CSS-based Schemes for LPWANs}}
\textcolor{black}{A plethora of modulation schemes predicated on using CSS has been put forth in the literature. However, the focus of this inquiry is precisely on those CSS modulation schemes proposed explicitly for implementation within the framework of LPWANs to facilitate IoT applications. To ensure clarity and concision, we outline the parameters for the inclusion and exclusion of the various modulation schemes in Table \ref{inclusion_exclusion_css} that will be the subject of study in the context of this comparative study.}
\begingroup
\setlength{\tabcolsep}{6pt} 
\renewcommand{\arraystretch}{0.99} 
\begin{table}[tbh]
\caption{\textcolor{black}{Inclusion exclusion criteria for the schemes compared in this study.}}
\centering
\color{black}\begin{tabular}{llcc}
\hline
\hline
\textbf{Inclusion Criteria} & \textbf{Exclusion Criteria}  \\
\hline
1. Written in English&1. Written in language\\
language& other than English. \\
2. Must be published & 2. Published before \(2016\)\\
between \(2016\) and \(2022\).& and after \(2022\).\\
3. Proposed for low-data  & 3. Proposed for other\\
rate LPWANs for IoT. & applications requiring \\
applications.& high data rate, such as\\
{}& DFRC, etc.\\
4. Published in journals & 4. Preprints in arXiv.\\
conference proceedings, &\\
symposiums or workshops. & \\
\hline 
\hline
\end{tabular}
\label{inclusion_exclusion_css}
\end{table}
\endgroup
\begin{table*}[h]
  \caption{\textcolor{black}{CSS schemes fulfilling the inclusion criteria to be discussed in this study.}}
   \label{tab_schemes}
  \centering
  \color{black}\begin{tabular}{*{4}{c}}
    \hline
    \hline
   \bfseries{Ref.}  & \bfseries{Name}    & \bfseries {Time} & \bfseries {Taxonomy}   \\
    \hline
    \hline
    \cite{lora_mod_basics, lora} & LoRa/FSCM & \(2015/2017\) & SC\\
     \cite{ics_lora} & ICS-LoRa & \(2019\) & SC\\
     \cite{e_lora} & E-LoRa & \(2019\) & SC \\
     \cite{psk_lora} & PSK-LoRa & \(2019\) & SC\\
     \cite{do_css} & DO-CSS & \(2019\) & MC\\
      \cite{ssk_lora} & SSK-LoRa & \(2020\) & SC\\
      \cite{iqcss} & IQ-CSS & \(2020\) & MC\\
       \cite{dcrk_css} & DCRK-LoRa & \(2021\) & SC\\
       \cite{fscss_im} & FSCSS-IM & \(2021\) & MC-IM\\
       \cite{iqcim} & IQ-CIM & \(2021\) & MC-IM\\
        \cite{ssk_ics_lora} & SSK-ICS-LoRa & \(2022\) & SC\\
         \cite{epsk_lora} & ePSK-CSS & \(2022\) & MC\\
          \cite{gcss} & GCSS & \(2022\) & MC \\
           \cite{tdm_lora} & TDM-CSS & \(2022\) & MC\\        
            \cite{tdm_lora} & IQ-TDM-CSS & \(2022\) & MC\\
            \cite{dm_css} & DM-CSS & \(2022\) & MC\\         
    \hline
    \hline
  \end{tabular}
\end{table*}
\textcolor{black}{Following the inclusion criteria established for this study, a chronologically ordered listing of schemes and their respective categorization can be observed in Table \ref{tab_schemes}. It is worth noting that various CSS-based schemes, such as those detailed in references \cite{ouyang2016orthogonal,li2017ofdm,ouyang2017chirp,omar2019designing,omar2020spectrum,csahin2020dft,hoque2020index,omar2021performance,csahin2021index, azim2022dual}, exist within the literature. However, these schemes do not adhere to the inclusion criteria established for the current analysis. It is noteworthy that protocols belonging to the same taxonomical classification may exhibit different design methodologies from the perspective of waveform design. Thus, a thorough comprehension of these disparate methodologies is essential, as expounded upon in the subsequent sections of this study.}

\begin{figure*}[t]
  \includegraphics[width=\textwidth,height=11cm]{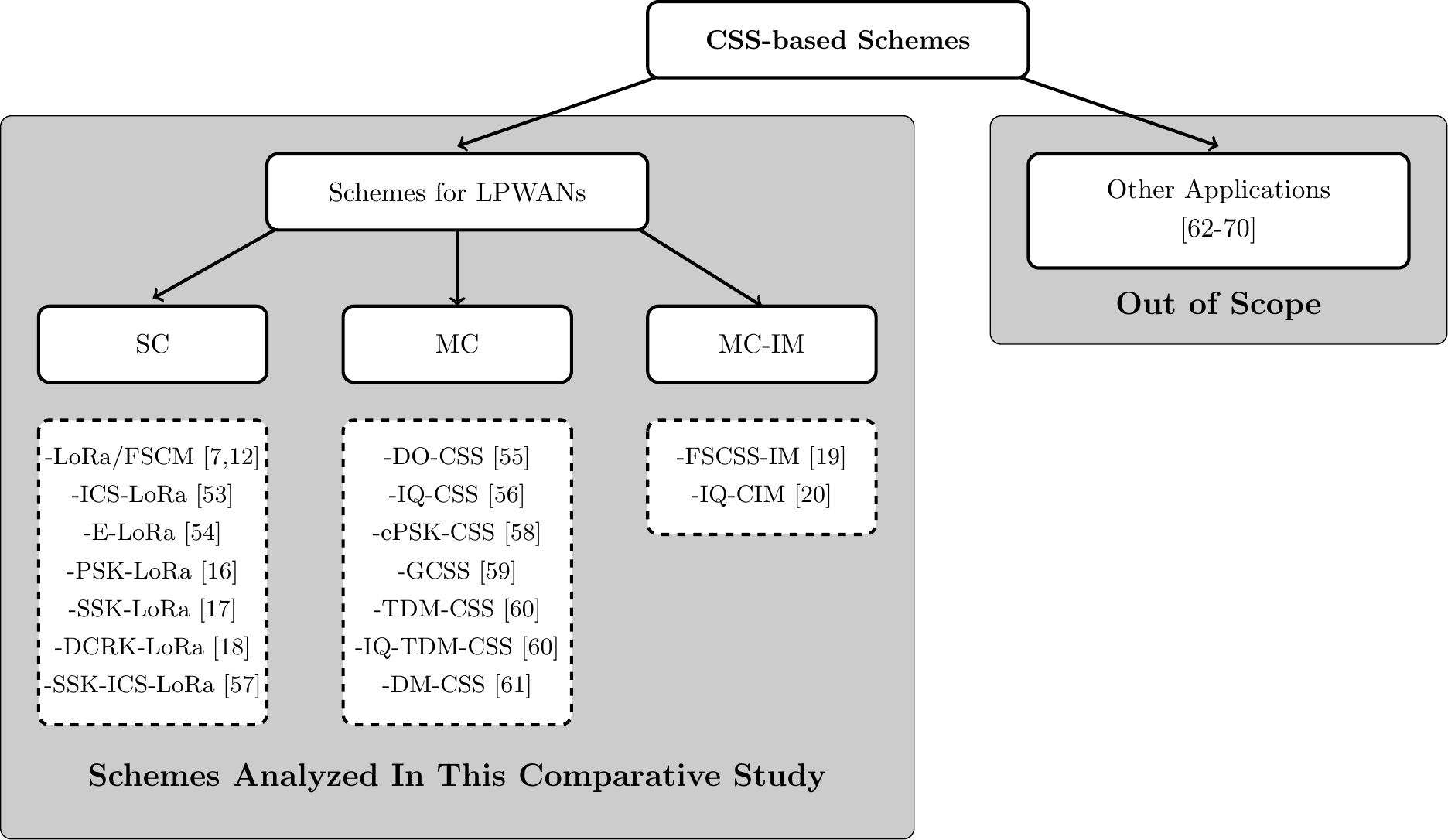}
  \caption{\textcolor{black}{CSS schemes analyzed in this study and their categorization based on the taxonomy.}}
  \label{flowchart}
\end{figure*}

\textcolor{black}{Following the established parameters of inclusion and exclusion criteria and the taxonomic categorizations of the CSS schemes, Fig. \ref{flowchart} depicts a visual representation of the schemes within the purview of our comparative investigation, as well as those schemes which fall outside the scope of our examination}.
\subsection{\textcolor{black}{Design Methodologies}}
\begin{figure*}[t]
  \includegraphics[width=\textwidth,height=1.75cm]{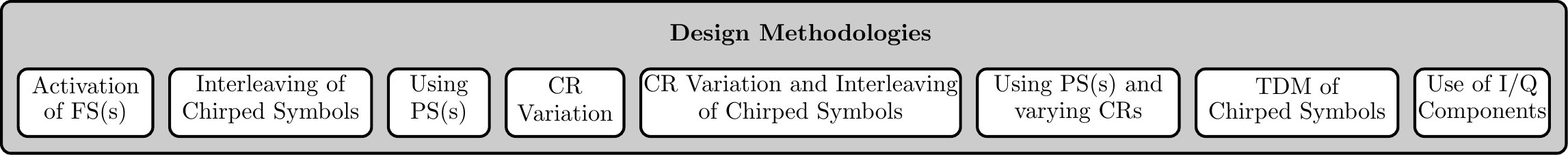}
  \caption{\textcolor{black}{Design methodologies for the CSS schemes to attain different waveform designs.}}
  \label{des_method}
\end{figure*}
\begin{figure}[t]\centering
\includegraphics[trim={0 0 0 0},clip,scale=0.99]{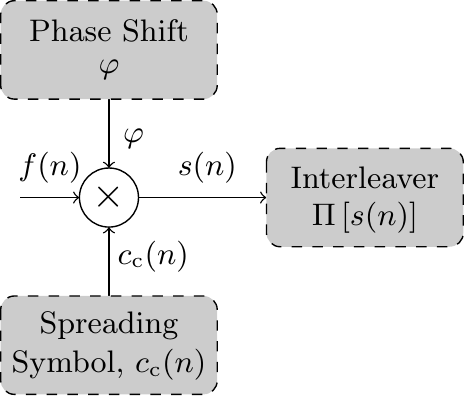}
  \caption{\textcolor{black}{System Model for different design methodologies.}}
\label{design_methods}
\end{figure}
\textcolor{black}{This section provides a comprehensive exposition regarding the various design methodologies that can be implemented for the waveform design pertaining to LPWANs. It is to be noted that these design methodologies pertain exclusively to the un-chirped symbols, \(f(n)\) unless explicitly indicated otherwise. We shall categorically specify the methods in question concerned with the chirped symbols, \(s(n)\). These methodologies can be explained with the help of Fig. \ref{design_methods}, where dashed and shaded blocks depict different design methodologies that can be employed. Moreover, different design methodologies that can be applied to attain different waveform designs are illustrated in Fig. \ref{des_method}.}
\subsubsection{\textcolor{black}{Activation of Frequency Shift(s)}}
\textcolor{black}{A rudimentary design methodology commonly involves the deployment of a single FS (as depicted in eq. (\ref{sc}) ) or multiple FSs (as depicted in eqs. (\ref{mc}) and (\ref{mc-im})) for the un-chirped symbol. Schemes adhering to other design methodologies also active either a single FS or multiple FSs. Thus, activating FS(s) is fundamental for all the CSS schemes. Of the CSS schemes analyzed in this comparative examination, LoRa activates a single FS, DO-CSS and GCSS adhere to activating multiple FSs. Furthermore, FCSS-IM activates a multiple of FSs and falls under the categorization of MC-IM taxonomy.}
\subsubsection{\textcolor{black}{Interleaving the Chirped Symbol}}
\textcolor{black}{Another means to enhance the design of the LoRa waveform is to perform interleaving of the chirped symbol, \(s(n)\) denoted as \(\Pi\left[s(n)\right]\). It is imperative to acknowledge that the un-chirped symbol is similar to those utilized in systems classified under the SC or MC taxonomies. In a rudimentary manner, the chirped symbol is either interleaved or not, resulting in the ICS-LoRa scheme, which conveys one extra bit compared to conventional LoRa. However, the diversification of interleaving patterns can be explored, as evidenced in \cite{8923337}, resulting in a further increase in the transmitted bits. Nevertheless, it is crucial to note that an increase in the interleaving patterns shall lead to an augmentation in low correlation patterns, thereby yielding a detrimental effect on the scheme's performance. Thus, a reasonable selection of interleave patterns is imperative, with preference given to patterns exhibiting high correlation values.}
\subsubsection{\textcolor{black}{Using Phase Shift(s)}}
\textcolor{black}{An alternate strategy for the waveform design of CSS systems is the integration of PS(s) into the un-chirped symbol, in addition to FS(s). This creates a composite symbol, \(f(n)\varphi\), which can then be chirped utilizing a spreading symbol, \(c(n)\). This design approach showcases a multitude of advantages and several drawbacks. On the one hand, it can enhance the SE and EE of the CSS scheme considerably. On the other hand, its performance can be significantly degraded in PO, a common occurrence in low-cost devices utilized in LPWANs. Furthermore, the utilization of PS mandates a coherent detection mechanism, as opposed to the de-chirped received symbol's FS, which may be detected through non-coherent means. Nevertheless, if only two PSs are used, non-coherent detection is also possible. However, deploying a limited number of PSs would constrict the enhancement in SE to a lower level. The SC scheme which uses PS is PSK-LoRa, while the MC approach using PSs is ePSK-LoRa.}
\subsubsection{\textcolor{black}{Chirp Rates Variation}}
\textcolor{black}{An alternate design paradigm involves varying CRs within the spreading symbol, which was previously denoted as \(c(n)\). The spreading symbol can be represented in a more generalized manner as follows:}
\begin{equation}\label{cr}
\textcolor{black}{c_\mathrm{c}(n)=\exp\left\{j\frac{\pi}{M}\gamma_\mathrm{c}n^2\right\}, }
\end{equation}
\textcolor{black}{where \(\gamma_\mathrm{c}\) is the CR, \(\mathrm{c}\) is the indexing variable whose values are \(\mathrm{c} = \llbracket 0, M_\mathrm{c}-1\rrbracket\), and \(M_\mathrm{c}\) is the number of different CRs that can used. In the sequel of the article, if \(\gamma_\mathrm{c} =1\), we will refer to the spreading symbol as an up-chirp symbol, \(c_\mathrm{u}(n)\).}

\textcolor{black}{In this methodology, the chirped symbol is obtained by multiplying the un-chirped symbol with a spreading symbol with varying chirp rates, i.e., \(f(n)c_\mathrm{c}(n)\). In this particular design paradigm, the methodology necessitates only an alteration of the CR. At the same time, the un-chirped symbol is the same as in SC or MC schemes. The utilization of varying CRs affords a higher EE and SE than methodologies that rely solely on using a single CR. However, we must note that the detection complexity of these schemes increases linearly with the number of CRs utilized. The SC schemes that adopt this methodology include DCRK-LoRa and SSK-LoRa; however, it is worth mentioning that this methodology can also be easily adapted for implementation in MC schemes.}
\subsubsection{\textcolor{black}{Chirp Rates Variation and Interleaving the Chirped Symbol}}
\textcolor{black}{The design methodology under consideration involves the integration of two previous methodologies, namely the utilization of varying CRs and the interleaving of the resulting chirped symbol. The chirped symbol, \(s(n)\), is obtained via the multiplication of the un-chirped symbol, \(f(n)\) and the spreading symbol, \(c_\mathrm{c}\), i.e., \(s(n)=f(n)c_\mathrm{c}(n)\), following which it is subjected to interleaving, i.e., \(\Pi\left[s(n)\right]\). This design paradigm elevates the SE by incorporating additional information bits into the CRs and interleaved symbols. The scheme belonging to the SC taxonomy of the CSS schemes, which adheres to this methodology, is SSK-ICS-LoRa.}
\subsubsection{\textcolor{black}{Using Phase Shift(s) and Varying Chirp Rates}}
\textcolor{black}{This methodology synergistically incorporates FS(s) with integrating PS(s) and varying CRs; thus, the chirped symbol is obtained as \(s(n)=f(n)\varphi c_\mathrm{c}(n)\). This design methodology affords a substantial degree of operational latitude in waveform design, as it allows to alter PS(s) and vary the CRs. The eventual outcome of this approach is a noteworthy improvement in SE and an enhancement in EE. Nevertheless, the utilization of diverse design methodologies within this method heightens the complexity of detection. Furthermore, it is only possible to employ coherent detection when the number of PS(s) is substantial. The CSS scheme employing this methodology is DM-CSS, which falls within the MC taxonomy. However, it is important to note that in DM-CSS, only two PS(s) are utilized; thus, non-coherent detection remains a viable option.}
\subsubsection{\textcolor{black}{Time Domain Multiplexing of Chirped Symbols}}
\textcolor{black}{TDM design methodology results in a MC CSS scheme as it multiplexed different chirped symbols. Let's consider two distinct chirped signals, denoted as \(s_1(n)\) and \(s_2(n)\); then, through the implementation of TDM, the resulting transmitted chirp signal is \(s(n)=s_1(n)+s_2(n)\). It is imperative to note that the underlying un-chirped signals, \(f_1(n)\) and \(f_2(n)\), within \(s_1(n)\) and \(s_2(n)\), respectively, can be obtained via a single FS or multiple FSs. The spreading symbols, \(c_1(n)\) and \(c_2(n)\), use varying CRs to facilitate a straightforward detection process. This design approach dramatically enhances the scheme's SE. However, it is essential to acknowledge that the scheme's complexity increases proportionally with the number of multiplexed symbols. Additionally, it is worth mentioning that TDM-based schemes do not possess a constant envelope characteristic. In the existing literature, TDM-CSS and IQ-TDM-CSS are among the MC schemes incorporating TDM design methodology.}
\subsubsection{\textcolor{black}{Use of In-phase and Quadrature Components}}
\textcolor{black}{This methodology exploits the complex nature inherent within the un-chirped signal, \(f(n)\), by inducing simultaneous activation of FS in both in-phase and quadrature components. As a result, a substantial enhancement in SE is observed for schemes that adopt this methodology. Despite its advantages, we must note that schemes that employ this methodology are highly vulnerable to phase and frequency shifts, thus rendering its practicality relatively limited. Furthermore, in most scenarios, the non-coherent detection mechanism of these schemes is not possible or is notably complex. We can implement this methodology across CSS schemes that belong to all taxonomies, with E-LoRa exemplifying its utilization in the SC category through the utilization of I/Q components. Other CSS schemes that belong to the MC and MC-IM taxonomies, such as IQ-CSS and IQ-CIM, respectively, also adopt this methodology. IQ-TDM-CSS also exploits the I/Q components; however, in this study, we classify it as a TDM approach.}
\subsubsection{\textcolor{black}{Takeaways}}
\textcolor{black}{Each methodology used for the waveform design has distinct advantages and drawbacks. As a general trend, such methodologies do not exhibit any discernible impact on the inherent attributes of the SC, MC, and MC-IM taxonomies. For instance, they do not alter envelope characteristics; SC-based approaches maintain a constant envelope, while MC and MC-IM schemes exhibit a non-constant envelope. In addition, these design methodologies are not predisposed to create any complications regarding coherent or non-coherent detection, except for utilizing PS(s) and I/Q components that mandate the utilization of a coherent detection mechanism. Note that these design methodologies serve as the means by which efficient waveforms can be generated for LPWANs, with the SE metric commonly being associated with the taxonomy and not the waveform design methodology.}

\textcolor{black}{Utilizing specific design methodologies can result in a substantial effect on the operational efficiency of resulting waveform designs. For instance, the utilization of PS(s) and I/Q components renders the configurations vulnerable to phase and frequency offsets, thus rendering these design methodologies less viable from a practicality standpoint.}
\begin{table*}[h]
  \caption{\textcolor{black}{Differentiation of SC CSS schemes based on design methodologies.}}
   \label{sc_methodology}
  \centering
  \color{black}\begin{tabular}{*{8}{c}}
    \hline
    \hline
   \bfseries{FS}  & \bfseries{Interleaving}    & \bfseries {PS} & \bfseries {CR}& \bfseries {CR Variation}  & \bfseries {PS and CR} &  \bfseries {TDM} & \bfseries {I/Q}    \\
     {}  & {}    &  {} & \bfseries {Variation}& \bfseries {and Interleaving}  & \bfseries {Variation} &  {} & \bfseries{Components}     \\
    \hline
    \hline
   LoRa & ICS-LoRa & PSK-LoRa & SSK-LoRa & SSK-ICS-LoRa & - &  Not Applicable & E-LoRa\\
     &  &  & DCRK-LoRa & &  &  &\\
    \hline
    \hline
  \end{tabular}
\end{table*}
\begin{table*}[h]
  \caption{\textcolor{black}{Differentiation of MC CSS schemes based on design methodologies.}}
   \label{mc_methodology}
  \centering
  \color{black}\begin{tabular}{*{8}{c}}
    \hline
    \hline
   \bfseries{FS}  & \bfseries{Interleaving}    & \bfseries {PS} & \bfseries {CR}& \bfseries {CR Variation}  & \bfseries {PS and CR} &  \bfseries {TDM} & \bfseries {I/Q}    \\
     {}  & {}    &  {} & \bfseries {Variation}& \bfseries {and Interleaving}  & \bfseries {Variation} &  {} & \bfseries{Components}    \\
    \hline
    \hline
   DO-CSS & - & ePSK-CSS & - & - & DM-CSS &  TDM-CSS & IQ-CSS\\
    GCSS &  &  &  & &  & IQ-TDM-CSS&\\
    \hline
    \hline
  \end{tabular}
\end{table*}
\begin{table*}[h]
  \caption{\textcolor{black}{Differentiation of MC-IM CSS schemes based on design methodologies.}}
   \label{mc_im_methodology}
  \centering
  \color{black}\begin{tabular}{*{8}{c}}
    \hline
    \hline
   \bfseries{FS}  & \bfseries{Interleaving}    & \bfseries {PS} & \bfseries {CR}& \bfseries {CR Variation}  & \bfseries {PS and CR} &  \bfseries {TDM} & \bfseries {I/Q}    \\
     {}  & {}    &  {} & \bfseries {Variation}& \bfseries {and Interleaving}  & \bfseries {Variation} &  {} & \bfseries{Components}    \\
    \hline
    \hline
   FSCSS-IM & - & - & - & - & - &  -& IQ-CIM\\
   
    \hline
    \hline
  \end{tabular}
\end{table*}
\begin{figure}[tb]\centering
\includegraphics[trim={0 0 0 0},clip,scale=0.8]{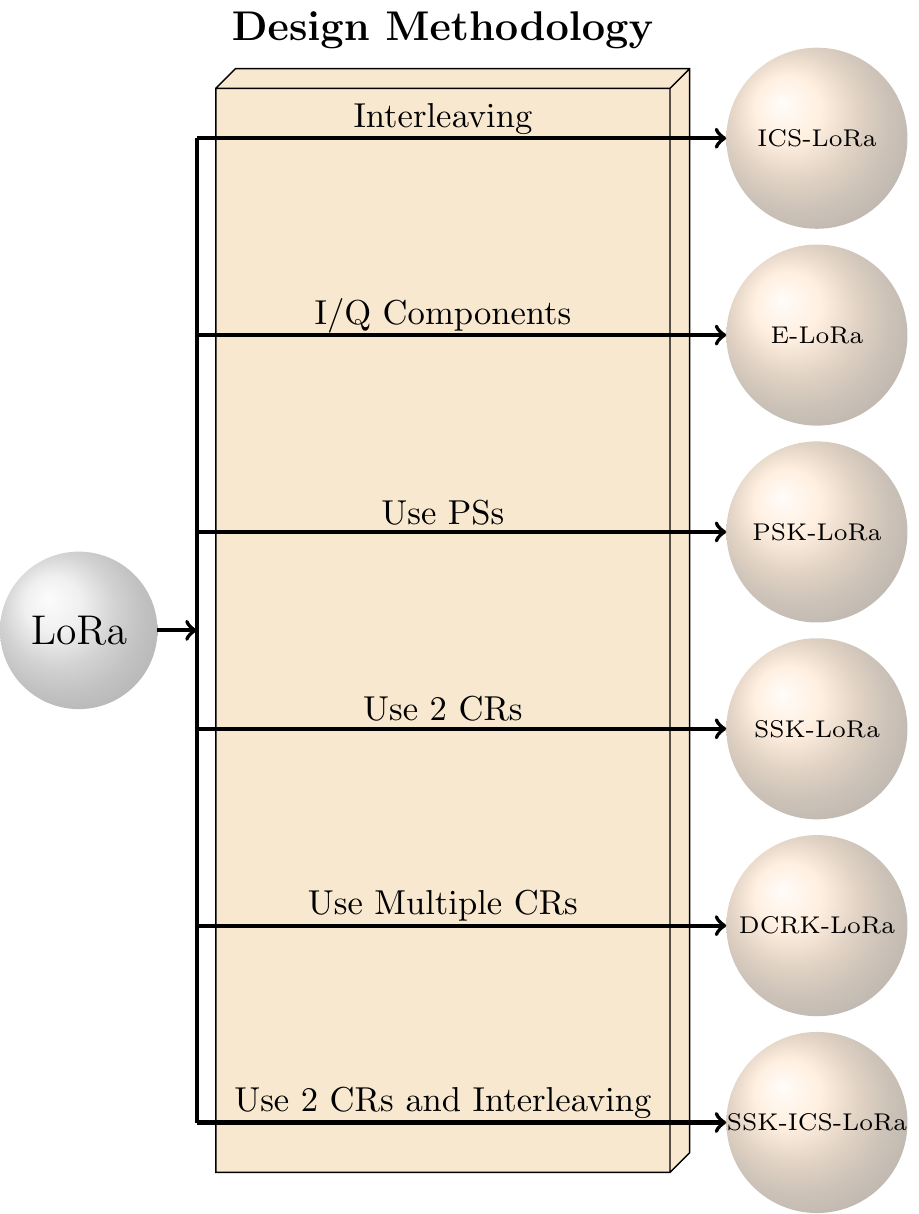}
  \caption{\textcolor{black}{Graphical representation of how different SC schemes are derived from LoRa using different design methodologies.}}
\label{sc_schemes_fig}
\end{figure}
\begin{figure}[tb]\centering
\includegraphics[trim={0 0 0 0},clip,scale=0.8]{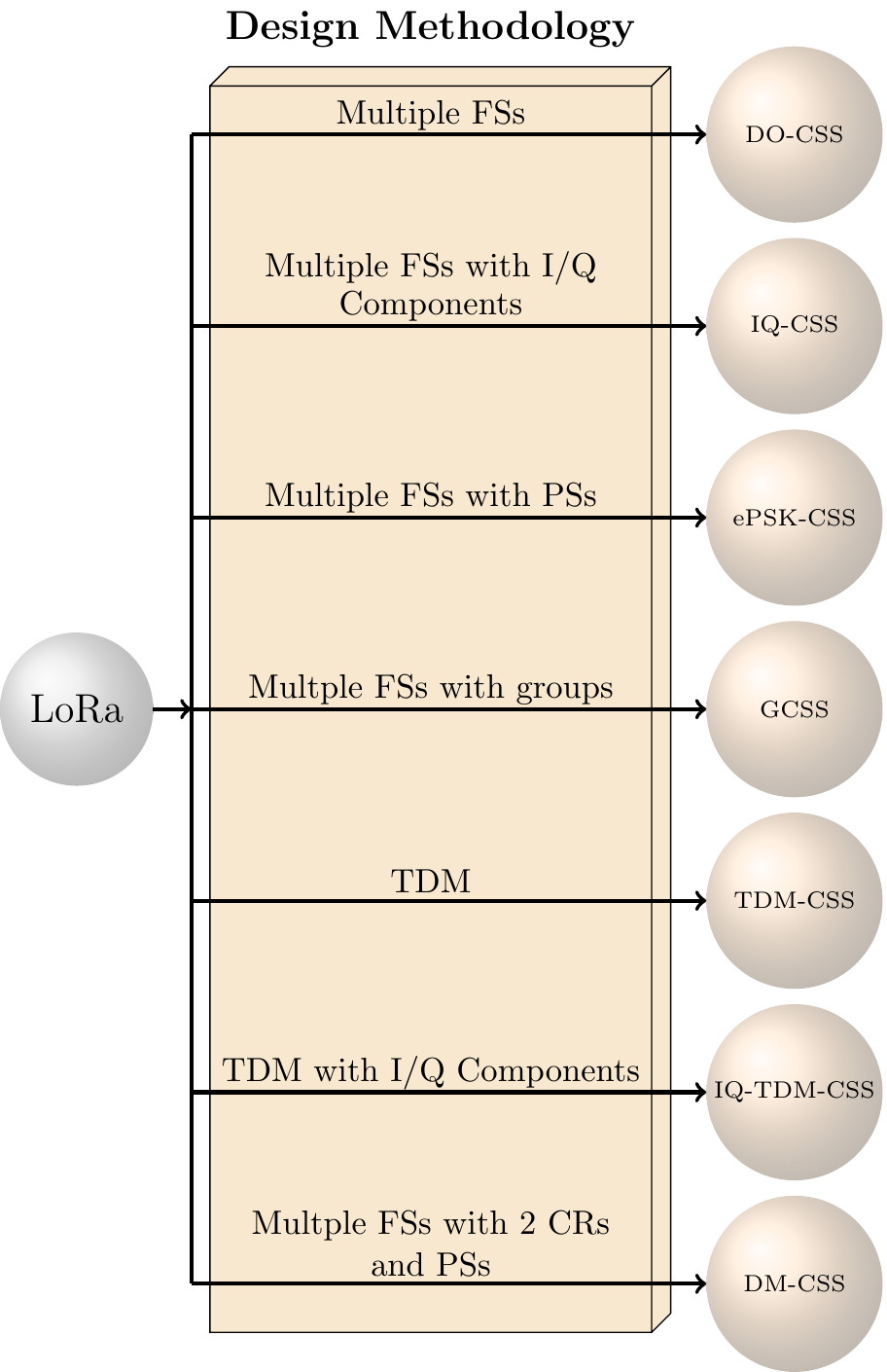}
  \caption{\textcolor{black}{Graphical representation of how different MC schemes are derived from LoRa using different design methodologies.}}
\label{mc_schemes_fig}
\end{figure}

\textcolor{black}{The design methodologies applicable to SC, MC, and MC-IM taxonomies are presented in Tables \ref{sc_methodology}, \ref{mc_methodology}, and \ref{mc_im_methodology}, respectively. Given that both SC and MC schemes are derived from classical LoRa, the graphical representation of their relationship with LoRa based on different design methodologies is illustrated in Figs. \ref{sc_schemes_fig} and \ref{mc_schemes_fig}, respectively. A cursory examination reveals that certain design methodologies are exclusively pertinent to a specific taxonomy, as demonstrated by the fact that we can utilize all design methodologies in SC schemes except the methodology that incorporates both PS and CRs variation. Additionally, literature has yet to be documented exploring MC schemes incorporating interleaving, CRs variation, and a combination of interleaving and CR variation. The existing MC-IM schemes in the scholarly corpus either employ FSs or harness I/Q components. Thus, the differentiation serves as a foundation for future research by illuminating the design methodologies that remain unexplored in this field.}
\subsection{\textcolor{black}{Detection Principles}}
\textcolor{black}{Within this purview, we delve into the different detection mechanisms we can leverage for CSS schemes. Coherent detection is implemented when there exists a priori information regarding the CSI; in contrast, non-coherent detection is exercised when CSI information is not accessible. Additionally, when the magnitude of the PS vector exceeds a cardinality of \(2\) and when the I/Q components are utilized, then coherent detection becomes an imperative requirement. It is to be noted that, despite its efficacy, coherent detection bears a relatively elevated level of complexity, rendering it less practical to execute in cost-sensitive LPWANs.}
\subsubsection{\textcolor{black}{Coherent Detection Principles}}
\textcolor{black}{The fundamental principle of coherent detection rests on the assumption that all the information-bearing elements are to be estimated. Thus, in its most rudimentary form, coherent detection primarily focuses on estimating the FS, i.e., \(k\) (cf. eq. (\ref{sc})). Nevertheless, in other CSS schemes, it may be necessary to extend the estimation procedure to detect other information-bearing elements, such as PS(s) and the like. Nonetheless, the coherent detection process can effectively estimate all these parameters by concurrently optimizing the Likelihood Function across all the viable orthonormal bases.}

The coherent detection process dictates to maximize the probability of receiving \(\boldsymbol{y} = \left[y(0),y(1), \cdots, y(M-1)\right]^\mathrm{T}\) given \(\boldsymbol{s} = \left[s(0),s(1), \cdots, s(M-1)\right]^\mathrm{T}\), i.e., \(\mathrm{prob}\left(\boldsymbol{y}\vert \boldsymbol{s}\right)\). Considering that all the possible symbols are equiprobable, the likelihood function, \(\mathrm{prob}\left(\boldsymbol{y}\vert \boldsymbol{s}\right)\) \textcolor{black}{adheres to the Gaussian probability density function as follows}:
\begin{equation}\label{cd1}
\begin{split}
\mathrm{prob}\left(\boldsymbol{y}\vert \boldsymbol{s}\right) &= \left(\frac{1}{2\pi \sigma_n^2}\right)^M\exp\left\{\frac{\Vert\boldsymbol{y}-\boldsymbol{s}\Vert^2}{2\sigma_n^2}\right\}\\
& = \mathrm{C_{st}}\exp\left\{\frac{\Re\left\{\langle \boldsymbol{y},\boldsymbol{s}\rangle\right\}}{\sigma_n^2}\right\},
\end{split}
\end{equation}
where \(\Vert\cdot\Vert^2\) evaluates Euclidean norm, \textcolor{black}{the real component of a complex argument is determined through \(\Re\{\cdot\}\), the inner product is evaluated via \(\langle \cdot, \cdot \rangle\), and the constant term is defined as} \(\mathrm{C_{st}}= ({1}/{2\pi \sigma_n^2})^M\exp\left\{-(\Vert \boldsymbol{y}\Vert^2 + \Vert \boldsymbol{s}\Vert^2)/2\sigma_n^2\right\}\). \textcolor{black}{Consequently, the coherent detection problem is reduced to the following simplified form:}
\begin{equation}\label{cd2}
\hat{k}_\mathrm{coh}=\mathrm{arg}\max_{k} ~\mathrm{prob}\left(\boldsymbol{y}\vert \boldsymbol{s}\right)=\mathrm{arg}\max_{k} ~\Re\left\{\langle \boldsymbol{y},\boldsymbol{s}\rangle\right\}.
\end{equation}

\textcolor{black}{Eq. (\ref{cd2}) conveys that with the utilization of coherent detection, the calculation of the inner product between the received signal vector, \(\boldsymbol{y}\), and all feasible transmit symbols must be performed. This computation is computationally intensive. However, because the transmit symbol is represented as \(s(n)= f(n)c_\mathrm{u}(n)\), the simplification of \(\Re\left\{\langle \boldsymbol{y},\boldsymbol{s}\rangle\right\}\) can be achieved as follows:}
\begin{equation}\label{cd3}
\begin{split}
\Re\left\{\langle \boldsymbol{y},\boldsymbol{s}\rangle\right\}&=\Re\left\{\sum_{n=0}^{M-1}y(n) \overline{s}(n)\right\}\\
&=\Re\left\{\sum_{n=0}^{M-1}y(n) \overline{f}(n) c_\mathrm{d}(n)\right\}\\
&=\Re\left\{\sum_{n=0}^{M-1}r(n) \overline{f}(n) \right\}\\
&=\Re\left\{\sum_{n=0}^{M-1}r(n) \exp\left\{-j\frac{2\pi}{M}nk\right\} \right\}\\
&=\Re\left\{R(k) \right\},
\end{split}
\end{equation}
where \(\overline{(\cdot)}\) is the conjugate operator, \(r(n) = y(n)c_\mathrm{d}(n)\), and \(R(k)\) is the DFT of \(r(n)\) evaluated at \(k\)th index. Here, \(c_\mathrm{d}(n) = \exp\left\{-j\frac{\pi}{M}n^2\right\}\) is the down-chirp which is the conjugate of \(c_\mathrm{u}(n)\). Now, the detection problem \textcolor{black}{eq.} (\ref{cd3}) in its simplified form becomes:
\begin{equation}\label{cd4}
\hat{k}_\mathrm{coh} =\mathrm{arg}\max_{k}~\Re\left\{R(k) \right\}.
\end{equation}

\textcolor{black}{In practice, the process commences with the correlation of the received signal, \(y(n)\) with \(c_\mathrm{d}(n)\), yielding the \(r(n)\). Subsequently, the DFT of \(r(n)\) is computed, resulting in \(R(k)\). The estimation of the transmitted FS, represented as \(\hat{k}\), is acquired by taking the real argument of \(R(k)\). It is noteworthy that for coherent detection, only the real component of \(R(k)\) is deemed pertinent, leading to the neglect of all noise present in the imaginary components.}

\textcolor{black}{It is imperative to emphasize that eq. (\ref{cd4}) is only applicable to specific CSS designs, such as LoRa, and may not hold for other CSS configurations. The inner product of \(\langle \boldsymbol{y},\boldsymbol{s}\rangle\) may diverge greatly based on the waveform design of the transmitted symbol. Hence, in subsequent discussions on applying coherent detection to CSS schemes, we will deduce the evaluation of  \(\langle \boldsymbol{y},\boldsymbol{s}\rangle\) based on the specific signal structure of the transmitted symbol of each CSS scheme considered in this study. }
\subsubsection{\textcolor{black}{Non-Coherent Detection Principles}}
\textcolor{black}{According to previous discourse, in the absence of a priori knowledge of CSI, non-coherent detection may be leveraged to ascertain the information-bearing element, assumed to be the transmit FS, \(k\). This non-coherent detection methodology is more feasible due to its comparatively reduced computational demands. Thus, it is a viable option for implementation in LPWANs where constraints such as limited computational power and the cost-effectiveness of components are paramount considerations. Hence, through the utilization of non-coherent detection, we can determine the transmit FS, \(k\), as:} 
\begin{equation}\label{ncd1}
\begin{split}
\hat{k}_\mathrm{non-coh}&=\mathrm{arg}\max_{k} ~\left\vert \sum_{n=0}^{M-1}y(n) \overline{s}(n)\right \vert \\
&=\mathrm{arg}\max_{k} ~\left\vert R(k) \right \vert,
\end{split}
\end{equation}
where \(\vert \cdot \vert\) is the absolute operator, and the second equality holds owing to eq. (\ref{cd3}).

\textcolor{black}{In the context of non-coherent detection-based estimation of the transmit FS, we commence by performing a DFT of the de-chirped received signal, \(r(n)\), to yield \(R(k)\). Subsequently, the argument of the absolute maximum of \(R(k)\) is computed, resulting in the index corresponding to the highest peak. Notably, in contrast to coherent detection, non-coherent detection incorporates the effect of both real and imaginary noise components. Consequently, it is anticipated that in an AWGN channel, the performance of coherent detection would surpass that of non-coherent detection. It is also noteworthy that while a coherent detector enables the correction of phase rotation induced by the channel, non-coherent detection cannot mitigate the random phase rotation imposed by the channel.}
\subsubsection{\textcolor{black}{Takeaways}}
\textcolor{black}{Coherent detection is a formidable technique from a waveform design perspective, exhibiting high performance in ideal operating conditions. However, its applicability in low-power and low-cost applications is questionable given the requirement of prior knowledge of the CSI, increased computational complexity, and heightened sensitivity to phase and frequency offsets. In contrast, non-coherent detection constitutes a feasible and computationally tractable mechanism. Yet, it may fall short of providing the performance and robustness attainable through coherent detection in scenarios with more stringent requirements.}
\section{CSS-based PHY Layer Waveform Design and Detection Principles}\label{sec3}
\textcolor{black}{In the sequel, the purview of our examination is limited to the waveform design and detection mechanisms pertaining to uncoded variants of CSS-based PHY layer to maintain simplicity and clarity. Investigating the performance of these schemes incorporating appropriate coding is a matter of future exploration. It should be noted that the CSS schemes analyzed in this study, as enumerated in Table \ref{tab_schemes}, are depicted based on specific taxonomical classifications.}
\subsection{\textcolor{black}{Single Carrier CSS Schemes}}
\subsubsection{LoRa}
\textcolor{black}{LoRa is a proprietary SC CSS scheme developed by the Semtech corporation, which allows for the trade-off of sensitivity and data rates for fixed channel bandwidths. Although the Semtech corporation has not publicly disclosed the specifics of the LoRa modulation scheme, \cite{lora}, and \cite{chiani2019lora} have provided a comprehensive theoretical overview of the LoRa modulation scheme, including an optimal low-complexity detection process based on DFT. A scalable parameter, SF, characterizes the LoRa modulation scheme, denoted by \(\lambda\), and is defined in the interval \(\lambda = \llbracket 6,12\rrbracket\). In actuality, \(\lambda\) equals the number of bits transmitted by a LoRa symbol of duration \(T_\mathrm{s}\). Based on \(\lambda\), a total of \(M\) distinct orthogonal symbols can be realized for LoRa, defined through \(M\) distinct cyclic time shifts of the chirp signal. These cyclic time shifts correspond to FSs of the complex conjugate of the chirp signal, i.e., the down-chirp signal. It is imperative to note that LoRa operates at the PHY layer and is agnostic to higher-layer implementations, allowing for its coexistence with other network architectures.}
\paragraph{Transmission}
\begin{figure}[tb]\centering
\includegraphics[trim={0 0 0 0},clip,scale=1]{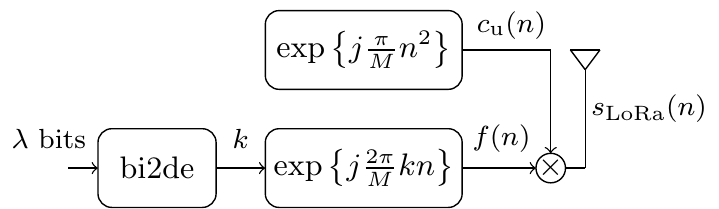}
  \caption{LoRa transmitter architecture. }
\label{fig1tx}
\end{figure}
\textcolor{black}{The transmitter architecture of LoRa is depicted in Fig. \ref{fig1tx}. Consider that there are \(M\) FSs within the bandwidth, \(B\). The activated FS, denoted by \(k\), is determined through the bi2de (binary-to-decimal) conversion of incoming \(\lambda\) bits. Utilizing \(k\), a discrete-time un-chirped symbol, \(f(n)\), is generated as outlined in eq. (\ref{sc}). This un-chirped symbol is then subjected to spreading through the application of an up-chirp symbol, \(c_\mathrm{u}(n)\), resulting in the generation of the LoRa signal, \(s_\mathrm{LoRa}(n)\). The discrete-time chirped signal, \(s_\mathrm{LoRa}(n)\), can be represented as:}
\begin{equation}\label{eq3}
s_\mathrm{LoRa}(n)= f(n)c_\mathrm{u}(n) = \exp\left\{j\frac{\pi}{M}n(2k+n)\right\}.
\end{equation}

\textcolor{black}{It is imperative to note that the normalization factors have been omitted for notational simplicity. The energy of a LoRa symbol} is \(E_\mathrm{s} = \sfrac{1}{M}\sum_{n=0}^{M-1}\vert s_\mathrm{LoRa}(n) \vert^2 = 1\). 
\paragraph{Detection}
\begin{figure}[tb]\centering
\includegraphics[trim={0 0 0 0},clip,scale=1]{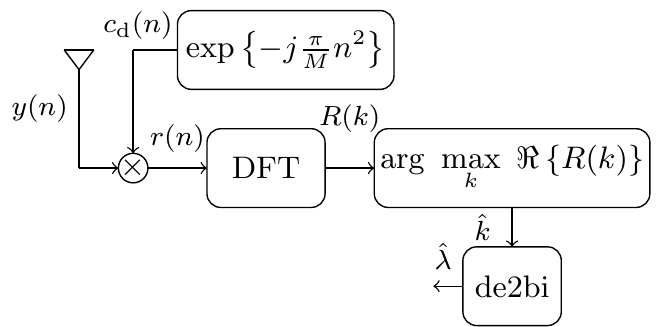}
  \caption{Coherent detector architecture for LoRa.}
\label{fig1rx}
\end{figure}
\textcolor{black}{The received LoRa symbol can be detected by the coherent or non-coherent method. The implementation of a coherent detector for the LoRa symbol is depicted in  Fig. \ref{fig1rx}. Coherent detection entails computing the inner product of the received symbol vector, represented as $\boldsymbol{y}$, and the pre-defined transmit symbol vector, $\boldsymbol{s}\mathrm{LoRa} = \left[{s}\mathrm{LoRa}(0), {s}\mathrm{LoRa}(1), \cdots, {s}\mathrm{LoRa}(M-1)\right]^\mathrm{T}$, i.e., the dot product $\langle \boldsymbol{y},\boldsymbol{s}_\mathrm{LoRa}\rangle$ is computed as:}
\begin{equation}\label{in_pd_lora}
\langle \boldsymbol{y},\boldsymbol{s}_\mathrm{LoRa}\rangle = \sum_{n=0}^{M-1}y(n)\overline{s}_\mathrm{LoRa}=R(k),
\end{equation}
\textcolor{black}{where the realization of equation (\ref{in_pd_lora}) is achieved by adhering to the procedure delineated in eq. (\ref{cd3}). Subsequently, by employing the principle of coherent detection as outlined in eq. (\ref{cd4}), the estimation of the received symbol's FS, $\hat{k}$, can be ascertained as follows:}
\begin{equation}\label{eqr1}
\hat{k} = \mathrm{arg} \max_{k} ~\Re\left\{R(k)\right\}.
\end{equation}

\textcolor{black}{Therefore, for the implementation of coherent detection, the process begins by performing a de-chirping of the received symbol, $y(n)$, using a down-chirp, $c_\mathrm{d}(n)$, resulting in $r(n)= y(n)c_\mathrm{d}(n)$ which serves as the received equivalent of $f(n)$. Thereafter, the output of the de-chirping process, $r(n)$, is subjected to a DFT evaluation, which yields $R(k)$, for $k \in \llbracket 0, M-1 \rrbracket$. It is worth highlighting that the detection problem for LoRa is equivalent to that presented in eq. (\ref{cd4}).}
\begin{figure}[tb]\centering
\includegraphics[trim={0 0 0 0},clip,scale=1]{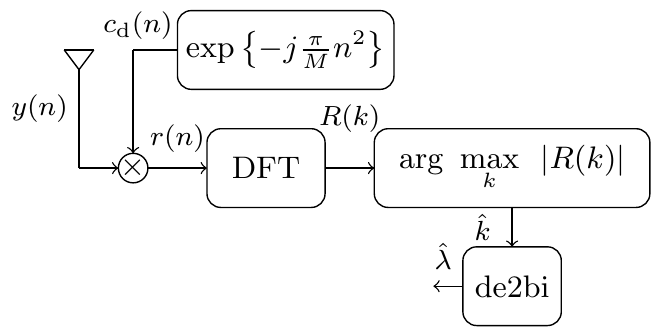}
  \caption{Non-coherent detector architecture for LoRa.}
\label{fig2rx}
\end{figure}

\textcolor{black}{The non-coherent detector utilized in the LoRa is depicted in Fig. \ref{fig2rx}. After the evaluation of \(R(k)\) in a manner congruent to the evaluation procedure utilized in the coherent detection method, the estimation of the FS, denoted as \(\hat{k}\), is determined through the implementation of the non-coherent detection principle, as outlined in equation (\ref{ncd1}) as:}
\begin{equation}\label{eqr2}
\hat{k} = \mathrm{arg} \max_{k} ~\left \vert R(k)\right\vert.
\end{equation}

\textcolor{black}{After the utilization of the decimal-to-binary conversion function (de2bi) to obtain the binary form of $\hat{k}$, the bit sequence, $\hat{\lambda}$, comprising the estimate of the transmitted bit sequence, is determined.}
\paragraph{\textcolor{black}{Takeaways}}
\textcolor{black}{LoRa has a simplistic and versatile waveform design, in which SFs may be adjusted to comply with specific application requisites, such as the intended coverage range, data transfer rate, and energy expenditure. Moreover, the SC nature of LoRa fortifies its advantageous features in the context of low-cost implementation in LPWANs. LoRa's simple waveform architecture is conducive to sustaining satisfactory performance levels under varying channel conditions. LoRa detection mechanisms are notably facile to realize, with non-coherent detection manifesting comparatively lower complexity vis-à-vis coherent detection. Coherent detection yields superior performance outcomes by virtue of its exclusion of noise on the imaginary constituents of the received frequency-domain un-chirped symbol. Conversely, non-coherent detection takes both the real and imaginary noise components of the received frequency-domain un-chirped symbol into account, culminating in comparatively inferior performance. Nonetheless, a small number of bits per symbol transmitted by LoRa precludes its attractiveness as an adaptive waveform choice that can cater to a wide range of applications requiring higher data rates. Furthermore, in LoRa's waveform design, not all the parameters that can improve the performance have been targeted; therefore, there is considerable room for enhancement in the waveform design.}
\subsubsection{Interleaved Chirp-Spreading (ICS)-LoRa}
\textcolor{black}{The first PHY layer-based approach aimed at elevating SE relative to the LoRa is ICS-LoRa, as outlined in the seminal work presented in \cite{ics_lora}. This scheme belongs to the SC taxonomy and leverages interleaved chirped symbols as a waveform design methodology. The coexistence of LoRa and ICS-LoRa was studied in \cite{ics_lora2}, concluding that a LoRa network can support both protocols simultaneously. \cite{ics_lora3} explored the utilization of ICS-LoRa as a parallel logical network to augment the capacity of a LoRa network. The number of transmitted bits in ICS-LoRa is augmented by including both the interleaved version of the LoRa symbol and the conventional LoRa symbol. However, this increases one bit per transmitted symbol of duration \(T_\mathrm{s}\), thereby yielding \(\lambda + 1\) bits transmitted per ICS-LoRa symbol.}
\begin{figure}[tb]\centering
\includegraphics[trim={0 0 0 0},clip,scale=1]{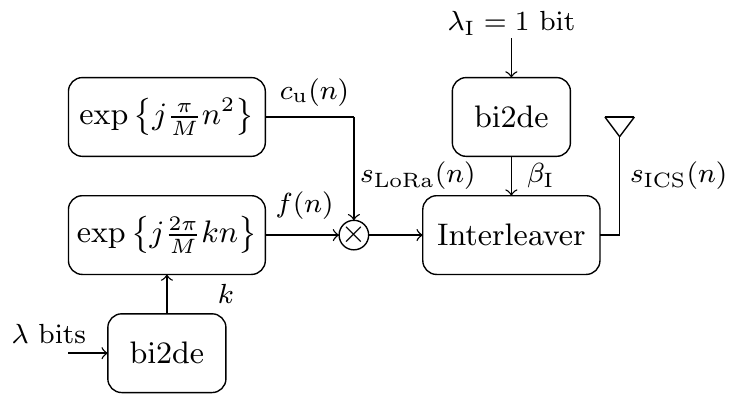}
  \caption{ICS-LoRa transmitter architecture. }
\label{fig2tx}
\end{figure}
\paragraph{Transmission}
\textcolor{black}{The implementation architecture of the ICS-LoRa transmitter is graphically represented in Fig. \ref{fig2tx}. Adhering to the LoRa protocol, the process commences by attain the FS, \(k\) by converting $\lambda$ bits using a bi2de converter. Subsequently, the output, $f(n)$, is generated using \(k\) following the eq. (\ref{sc}). The derived function, $f(n)$, is then up-chirped through using $c_\mathrm{u}(n)$, resulting in the LoRa transmit symbol, $s_\mathrm{LoRa}(n)$, as per eq. (\ref{eq3}). The binary input, $\lambda_\mathrm{I}=1$, is utilized to determine the value of $\beta_\mathrm{I}$ which decides whether to transmit the conventional LoRa symbol, $s_\mathrm{LoRa}(n)$, or its interleaved equivalent, i.e., $\Pi \left[s_\mathrm{LoRa}(n)\right]$, where $\Pi[\cdot]$ represents the interleaving operation. The value of $\beta_\mathrm{I}$ is obtained post-conversion via the bi2de converter and is restricted to the binary set $\left\{0,1\right\}$, as indicated by $\beta_\mathrm{I} = \llbracket 0,2^{\lambda_\mathrm{I}}-1\rrbracket$. The discrete-time ICS-LoRa transmit symbol is given by:}
\begin{equation}\label{eq05}
s_\mathrm{ICS}(n)= \begin{dcases}
    s_\mathrm{LoRa}(n) & \quad \beta_\mathrm{I}  = 0\\
   \Pi\left[s_\mathrm{LoRa}(n)\right] & \quad \beta_\mathrm{I}  = 1
   \end{dcases}.
\end{equation}

\textcolor{black}{The interleave function, as shown in Fig. 2, performs a shuffling operation on the \(s_\mathrm{LoRa}(n)\) of length \(M\), whereby the 2nd and 3rd quartered segments are rearranged in a permutated manner, as outlined below:}
\begin{equation}\label{eq06}
\Pi \left[s_\mathrm{LoRa}(n)\right]= 
  \begin{dcases}
    s_\mathrm{LoRa}(n) & \quad n= \llbracket 0,\sfrac{M}{4}\rrbracket \\
   s_\mathrm{LoRa}(n+\sfrac{M}{4}) & \quad n= \llbracket \sfrac{M}{4},\sfrac{M}{2}\rrbracket  \\
   s_\mathrm{LoRa}(n-\sfrac{M}{4}) & \quad n= \llbracket \sfrac{M}{2},\sfrac{3M}{4}\rrbracket  \\
   s_\mathrm{LoRa}(n) & \quad n= \llbracket \sfrac{3M}{4},M\rrbracket  \\
  \end{dcases}.
\end{equation}

The symbol energy of a \textcolor{black}{ICS-LoRa} symbol is \(E_\mathrm{s} = \sfrac{1}{M}\sum_{n=0}^{M-1}\vert s_\mathrm{ICS}(n) \vert^2 = 1\). 
\paragraph{Detection}
\begin{figure}[tb]\centering
\includegraphics[trim={0 0 0 0},clip,scale=1]{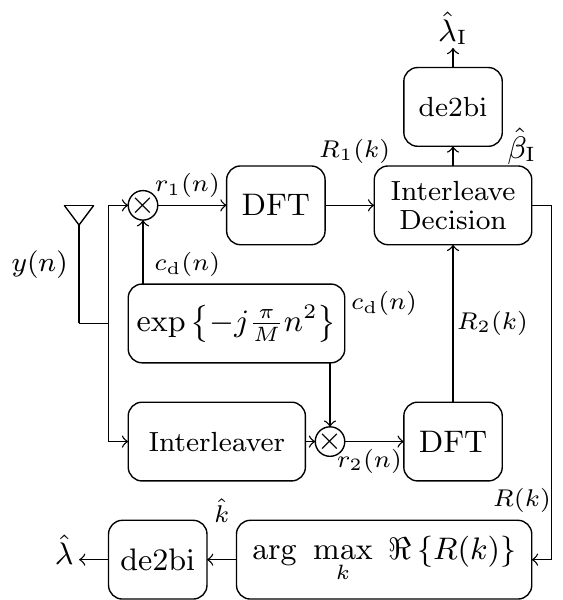}
  \caption{Coherent detector architecture for ICS-LoRa.}
\label{fig3rx}
\end{figure}
\textcolor{black}{In \cite{ics_lora}, the authors proffered only a non-coherent method for detecting the ICS-LoRa symbol. However, it is also feasible to employ a coherent detection approach for symbol detection in ICS-LoRa. In this work, we extend the scope of detection in ICS-LoRa by presenting both coherent and non-coherent detection mechanisms.}

\textcolor{black}{Fig. \ref{fig3rx} depicts the coherent detection architecture for ICS-LoRa. As specified by eq. (\ref{eq05}), it is evident that LoRa is a particular instance of ICS-LoRa. Consequently, the inner product calculation of the received symbol vector, represented by $\boldsymbol{y}$, and the transmit symbol vector, $\boldsymbol{s}_\mathrm{ICS} = \left[s_\mathrm{ICS}(0), s_\mathrm{ICS}(1), \cdots, s_\mathrm{ICS}(M-1)\right]^\mathrm{T}$, i.e., $\langle \boldsymbol{y}, \boldsymbol{s}_\mathrm{ICS}\rangle$, must be computed for two distinct scenarios, namely, (i) when the conventional LoRa symbol is transmitted, i.e., $\beta_\mathrm{I} = 0$, and (ii) the interleaved LoRa symbol is transmitted, i.e., $\beta_\mathrm{I} = 1$. For the former scenario (i), the inner product term $\langle \boldsymbol{y}, \boldsymbol{s}_\mathrm{ICS}\rangle$ evaluates to:}
\begin{equation}\label{in_pd_ics_lora}
\begin{split}
\langle \boldsymbol{y}, \boldsymbol{s}_\mathrm{ICS}\rangle & =  \sum_{n=0}^{M-1}y(n)\overline{s}_\mathrm{ICS}\\
&= \sum_{n=0}^{M-1}y(n)\overline{f}(n)c_\mathrm{d}(n)\\
&= \sum_{n=0}^{M-1}r_1(n)\overline{f}(n)\\
&=R_1(k),
\end{split}
\end{equation}
where \(r_1(n) = y(n)c_\mathrm{d}(n)\). Conversely, when the interleaved LoRa is transmitted, then \(\langle \boldsymbol{y}, \boldsymbol{s}_\mathrm{ICS}\rangle\) yields:
\begin{equation}\label{in_pd_ics_lora2}
\begin{split}
\langle \boldsymbol{y}, \boldsymbol{s}_\mathrm{ICS}\rangle & =  \sum_{n=0}^{M-1}y(n)\overline{s}_\mathrm{ICS}\\
&= \sum_{n=0}^{M-1}\Pi\left[y(n)\right]\overline{f}(n)c_\mathrm{d}(n)\\
&= \sum_{n=0}^{M-1}r_2(n)\overline{f}(n)\\
&=R_2(k),
\end{split}
\end{equation}
where \(r_2(n) = \Pi\left[y(n)\right]c_\mathrm{d}(n)\).

\textcolor{black}{As can be deduced from eqs. (\ref{in_pd_ics_lora}) and (\ref{in_pd_ics_lora2}), the received symbol, $y(n)$ and its interleaved counterpart, $\Pi\left[y(n)\right]$, undergo de-chirping utilizing $c_\mathrm{d}(n)$ leading to $r_1(n) = y(n)c_\mathrm{d}(n)$ and $r_2(n)= \Pi\left[y(n)\right]c_\mathrm{d}(n)$. Subsequently, the DFT is computed for both $r_1(n)$ and $r_2(n)$, yielding $R_1(k)$ and $R_2(k)$, respectively. The subsequent step is to identify whether a conventional LoRa symbol or an interleaved LoRa symbol was transmitted. This is determined through the calculation of two parameters; namely, $\kappa_1^\mathrm{coh} = \max~\Re\left\{R_1(k)\right\}$ and $\kappa_2^\mathrm{coh} = \max~\Re\left\{R_2(k)\right\}$ in the interleave decision block. These parameters then facilitate the determination of $\hat{\beta}_\mathrm{I}$ as:}
\begin{equation}\label{eqr3}
\hat{\beta}_\mathrm{I}= 
  \begin{dcases}
    0 & \quad \kappa_1^\mathrm{coh}>\kappa_2^\mathrm{coh}\\
   1 & \quad \kappa_1^\mathrm{coh}<\kappa_2^\mathrm{coh} \\
  \end{dcases},
\end{equation}
\textcolor{black}{which facilitates the discernment of the bit sequence, $\hat{\lambda}_\mathrm{I}$, signifying the interleaved information in the transmit symbol post transformation by de2bi.}

\textcolor{black}{Subsequently, the determination of the utilization of either $R_1(k)$ or $R_2(k)$ in the identification of $\hat{k}$ is made using $\kappa_1^\mathrm{coh}$ and $\kappa_2^\mathrm{coh}$, as follows:}
\begin{equation}\label{eqr4}
R(k)= 
  \begin{dcases}
    R_1(k) & \quad \kappa_1^\mathrm{coh}>\kappa_2^\mathrm{coh}\\
   R_2(k) & \quad \kappa_1^\mathrm{coh}<\kappa_2^\mathrm{coh} \\
  \end{dcases}.
\end{equation}

Finally, \textcolor{black}{using the coherent detection, the FS \(\hat{k}\)} is evaluated as:
\begin{equation}\label{eqr5}
\hat{k} = \mathrm{arg} \max_{k} ~\Re\left\{R(k)\right\},
\end{equation}
that \textcolor{black}{determines} \(\hat{\lambda}\) after de2bi conversion. 
\begin{figure}[tb]\centering
\includegraphics[trim={0 0 0 0},clip,scale=1]{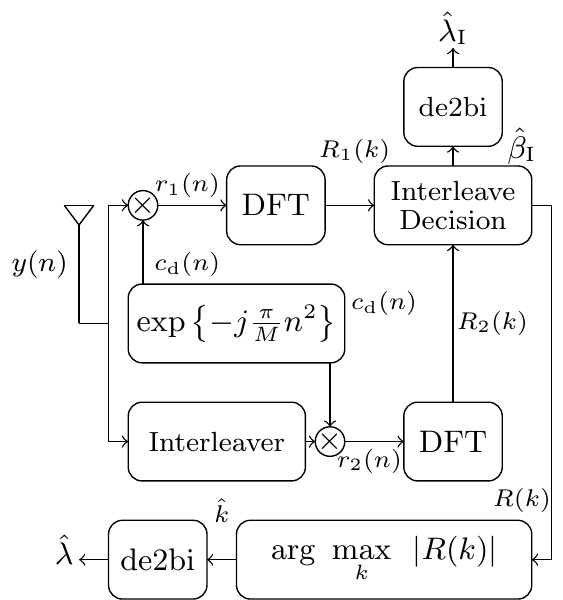}
  \caption{Non-coherent detector architecture for ICS-LoRa.}
\label{fig4rx}
\end{figure}

\textcolor{black}{The non-coherent detector configuration for ICS-LoRa is depicted in Fig. \ref{fig4rx}. A distinction between non-coherent detection and coherent detection is evident in two key aspects. The first differentiation arises in the interleave decision block, where, instead of utilizing $\kappa_1^\mathrm{coh}$ and $\kappa_2^\mathrm{coh}$, the utilization of $\kappa_1^\mathrm{non-coh} = \max~\vert R_1(k) \vert$, and $\kappa_2^\mathrm{non-coh} = \max~\vert R_2(k)\vert$ is employed in the determination of $\hat{\lambda}_\mathrm{I}$ and $R(k)$. Secondly, a deviation is noted in identifying the FS, executed using the following criterion:}
\begin{equation}\label{eqr6}
\hat{k} = \mathrm{arg} \max_{k} ~\left\vert R(k)\right\vert.
\end{equation}

\textcolor{black}{After bi2de conversion of \(\hat{k}\), the transmitted bit sequence, \(\hat{\lambda}\) is attained.}
\paragraph{\textcolor{black}{Takeaways}}
\textcolor{black}{ICS-LoRa, a technological advancement over conventional LoRa, offers the capability of transmitting one additional bit, albeit the enhancement in the number of transmitted bits is marginal. ICS-LoRa's inherent superiority lies in its improved EE, necessitating less energy per bit to correctly detect bits for the target BER. Like LoRa, ICS-LoRa is endowed with a constant envelop, facilitating low-cost implementation and augmenting the robustness of waveform design. Conversely, the detection mechanisms of ICS-LoRa are relatively intricate compared to LoRa, given the compulsion to discern interleaved versions of LoRa symbols. Additionally, ICS-LoRa symbols lack orthogonality with LoRa symbols because of the use of interleaved symbols, thereby engendering high correlation peaks approximately around the correct DFT index, which may contribute to BER degradation. Although the waveform design allows for a larger number of interleaving patterns, augmenting them diminishes the cross-correlation, which may ultimately lead to performance deterioration.}
\subsubsection{Extended (E)-LoRa}
\textcolor{black}{The E-LoRa technique represents a PHY layer strategy that endeavors to augment the number of bits transmitted per symbol compared to conventional LoRa, through the utilization of I/Q components. Unlike the standard LoRa scenario in which the symbol \(f(n)\), as depicted in eq. (\ref{sc}), restricts the information transmission solely to the in-phase components, E-LoRa harnesses the benefits of both I/Q components, thereby enabling the encoding of an additional bit per symbol. Hence, the total number of bits transmitted per symbol of duration \(T_{\mathrm{s}}\) in E-LoRa is \(\lambda + 1\).}
\paragraph{Transmission}
\begin{figure}[tb]\centering
\includegraphics[trim={0 0 0 0},clip,scale=1]{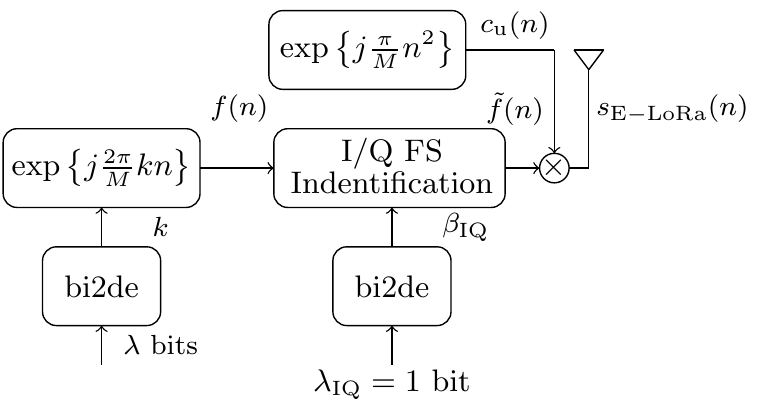}
  \caption{E-LoRa transmitter architecture. }
\label{fig3tx}
\end{figure}
\textcolor{black}{The architecture of the E-LoRa transmitter is depicted in  Fig. \ref{fig3tx}. The transmitter utilizes a classical LoRa modulation technique, wherein the FS, \(k\), is ascertained based on the bi2de conversion of \(\lambda\) bits. \(k\) is then employed to determine the discrete time signal \(f(n)\) as expressed in equation (\ref{sc}). Subsequently, the I/Q component encoding of the un-chirped symbol, represented as \(\tilde{f}(n)\), is achieved using \(\lambda_{\mathrm{IQ}}\), which yields \(\beta_\mathrm{IQ}\) that determines the encoding of the information into either the in-phase or the quadrature component as follows:}

\begin{equation}\label{eq08}
\tilde{f}(n)= \begin{dcases}
    f(n) & \quad \beta_\mathrm{IQ} = 0\\
   jf(n) & \quad \beta_\mathrm{IQ} = 1
   \end{dcases}.
\end{equation}

\textcolor{black}{Afterwards,} \(\tilde{f}(n)\) is chirped using an up-chirp, \(c_\mathrm{u}(n)\) \textcolor{black}{yielding} \(s_\mathrm{E-LoRa}(n)=\tilde{f}(n)c_\mathrm{u}(n)\), that is mathematically \textcolor{black}{given as:} 

\begin{equation}\label{eq09}
s_\mathrm{E-LoRa}(n)=  \begin{dcases}
    \exp\left\{j\frac{\pi}{M}n(2k+n)\right\} &  \beta_\mathrm{IQ} = 0\\
   j\exp\left\{j\frac{\pi}{M}n(2k+n)\right\}&  \beta_\mathrm{IQ} = 1
\end{dcases}.
\end{equation}

The symbol energy of E-LoRa is \(E_\mathrm{s} = \sfrac{1}{M}\sum_{n=0}^{M-1}\vert s_\mathrm{E-LoRa}(n) \vert^2 = 1\).
\paragraph{Detection}
\begin{figure}[tb]\centering
\includegraphics[trim={0 0 0 0},clip,scale=1]{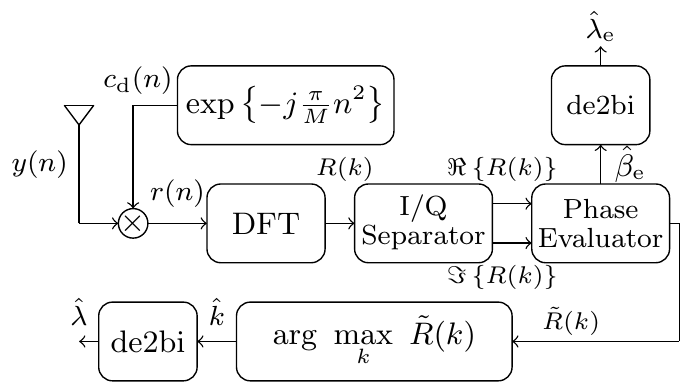}
  \caption{Coherent detector architecture for E-LoRa.}
\label{fig5rx}
\end{figure}
\textcolor{black}{The analysis of the E-LoRa symbol structure reveals that the information is conveyed through the phase of the un-chirped symbol, thus necessitating the use of coherent detection as the only viable option for retrieving said transmitted information. The architecture of the coherent detector in E-LoRa is illustrated in Fig. \ref{fig5rx}. It is noteworthy that the utilization of either the in-phase or quadrature un-chirped symbols in E-LoRa results in varying inner product calculations between the received symbol vector, represented by $\boldsymbol{y}$, and the transmit symbol vector, represented by $\boldsymbol{s}_\mathrm{E-LoRa} = \left[{s}_\mathrm{E-LoRa}(0),{s}_\mathrm{E-LoRa}(1), \cdots, {s}_\mathrm{E-LoRa}(M-1)\right]^\mathrm{T}$, as quantified by $\langle \boldsymbol{y}, \boldsymbol{s}_\mathrm{E-LoRa}\rangle$. Hence, a comprehensive evaluation of $\langle \boldsymbol{y}, \boldsymbol{s}_\mathrm{E-LoRa}\rangle$ is performed to consider both cases. It is worth mentioning that when the bit encoding is performed through the in-phase components, the inner product calculation, $\langle \boldsymbol{y}, \boldsymbol{s}_\mathrm{E-LoRa}\rangle$, is similar to  LoRa; thus we have:}
\begin{equation}\label{in_pd_elora}
\begin{split}
\textcolor{black}{\langle \boldsymbol{y}, \boldsymbol{s}_\mathrm{E-LoRa}\rangle = R(k).}
\end{split}
\end{equation}

\textcolor{black}{In contrast, when the quadrature component of the un-chirped symbol is used, the result of the inner product, $\langle \boldsymbol{y}, \boldsymbol{s}_\mathrm{E-LoRa}\rangle$, is as follows:}
\begin{equation}\label{in_pd_elora2}
\begin{split}
\langle \boldsymbol{y}, \boldsymbol{s}_\mathrm{E-LoRa}\rangle & = \sum_{n=0}^{M-1}y(n)\overline{s}_\mathrm{E-LoRa}(n)\\
& = -j\sum_{n=0}^{M-1}r(n)\textcolor{black}{\overline{\tilde{f}}(n)}\\
& = -jR(k).
\end{split}
\end{equation}

When the real argument of eqs. (\ref{in_pd_elora}) and (\ref{in_pd_elora2}) is evaluated, we attain the following:
\begin{equation}\label{eqr8}
\Re\left\{\langle \boldsymbol{y},\boldsymbol{s}_\mathrm{E-LoRa}\rangle\right\}=  \begin{dcases}
    \Re\left\{R(k)\right\} &  \beta_\mathrm{IQ}=0\\
   \Im\left\{R(k)\right\}&  \beta_\mathrm{IQ}=1
\end{dcases}.
\end{equation}

\textcolor{black}{Therefore, in the context of coherent detection, the received symbol $y(n)$ undergoes de-chirping using $c_\mathrm{d}(n)$ to yield $r(n)$. Subsequently, the DFT of $r(n)$ results in $R(k)$. As per eq. (\ref{eq08}), the information is transmitted through either the in-phase component or the quadrature component of the un-chirped symbol. Moreover, eq. (\ref{eqr8}) dictates that the separation of the I/Q components of $R(k)$ is required, which is performed by the I/Q separator block, resulting in $\Re\left\{R(k)\right\}$ and $\Im\left\{R(k)\right\}$. The subsequent evaluation of $\kappa_1 = \max~\Re\left\{R(k)\right\}$ and $\kappa_2 = \max~\Im\left\{R(k)\right\}$ in the phase evaluator block then determines $\hat{\beta}_\mathrm{IQ}$ based on the following criterion:}
\begin{equation}\label{eqr8}
\hat{\beta}_\mathrm{IQ}=  \begin{dcases}
    0 &  \kappa_1>\kappa_2\\
   1&  \kappa_1<\kappa_2
\end{dcases}.
\end{equation}

\textcolor{black}{The parameter $\hat{\lambda}\mathrm{IQ}$ is determined via the de2bi conversion of \(\hat{\beta}_\mathrm{IQ}\). The phase evaluator block plays a crucial role in this process, as it facilitates determining the information regarding the presence or absence of a $\sfrac{\pi}{2}$ rotation introduced by the transmitter based on the value of $\beta_\mathrm{IQ}$ .}

Moreover, based on the values of \(\kappa_1\) and \(\kappa_2\), \(\tilde{R}(k)\) which is required for further evaluation \textcolor{black}{is obtained} as:
\begin{equation}\label{eqr9}
\tilde{R}(k)=  \begin{dcases}
    \Re\left\{R(k)\right\} &  \kappa_1>\kappa_2\\
   \Im\left\{R(k)\right\}&  \kappa_1<\kappa_2
\end{dcases}.
\end{equation}

Subsequently, the FS,  \(\hat{k}\) is identified as:
\begin{equation}\label{eqr10}
\hat{k} = \mathrm{arg} \max_{k} ~ \tilde{R}(k).
\end{equation}

Then \(\hat{\lambda}\) is determined after de2bi conversion of \(\hat{k}\).
\paragraph{\textcolor{black}{Takeaways}}
\textcolor{black}{E-LoRa represents a pioneering technique that leverages I/Q components as a design methodology for waveform design leading to a SC scheme with a constant envelop. This methodology offers an incremental addition of one extra bit per transmitted symbol compared to LoRa, because of the use of the in-phase or quadrature component. Notwithstanding, deploying the I/Q component as a design methodology is not conducive to non-coherent detection. Consequently, only more intricate coherent detection may be employed, reducing its practicability. Furthermore, utilizing I/Q components precipitates vulnerability for the waveform design resilience against potential offsets that may manifest in low-cost devices utilized in LPWANs.}
\subsubsection{Phase-Shift Keying (PSK)-LoRa}
\textcolor{black}{PSK-LoRa waveform design employs the PSs methodology to encode more bits in the transmit symbol. It has been established that all the previously explored schemes can transmit one additional bit per symbol with respect to the LoRa protocol. However, it is to be noted that PSK-LoRa stands distinct in its ability to transmit a higher number of additional bits compared to its LoRa counterpart. The information transmission in LoRa occurs via the FS of the un-chirped symbol. Conversely, the PSK-LoRa uses PSK alphabets to transmit additional information in the un-chirped signal's PS \cite{psk_lora}. PSK-LoRa is an SC scheme, as only one FS is activated, akin to other SC schemes. The amount of additional bits transmitted by the PSK-LoRa symbol is contingent upon the cardinality of the PSK alphabets, \(M_\varphi\). As a result, in addition to the \(\lambda\) bits transmitted via the FS, the FS imparts an additional \(\log_2(M_\varphi)\) bits, yielding a total of \(\lambda + \log_2(M_\varphi)\) bits per symbol of duration \(T_\mathrm{s}\).}
\begin{figure}[tb]\centering
\includegraphics[trim={0 0 0 0},clip,scale=1]{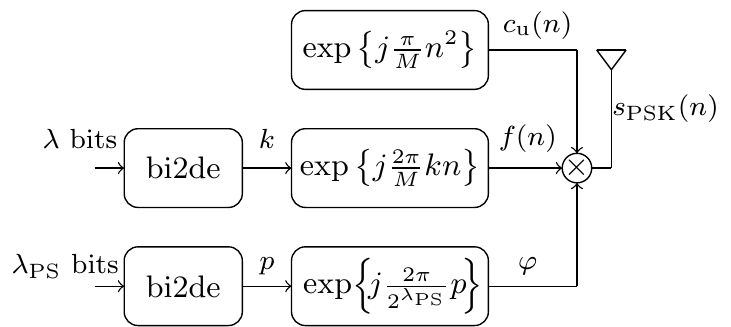}
  \caption{PSK-LoRa transmitter architecture. }
\label{fig4tx}
\end{figure}
\paragraph{Transmission}
\textcolor{black}{The transmitter configuration of the PSK-LoRa is depicted in Fig. \ref{fig4tx}. Within the PSK-LoRa framework, the FS, \(k\), is determined by \(\lambda\) bits, which enables obtaining the un-chirped symbol, \(f(n)\) (cf. eq. (\ref{sc})). Furthermore, an additional \(\lambda_\mathrm{PS} = \log_2(M_\varphi)\) bits are utilized to determine the PS parameter, \(p= \llbracket 0,2^{\lambda_\mathrm{PS}}-1 \rrbracket\), as indicated by the following equation:} 
\begin{equation}\label{eq11}
\varphi =\exp\left\{ \frac{j2\pi}{2^{\lambda_\mathrm{PS}}}p\right\}.
\end{equation}

The resulting discrete time symbol chirped PSK-LoRa symbol is given as:
\begin{equation}\label{eq12}
\begin{split}
s_\mathrm{PSK}(n) &= f(n)\varphi c_\mathrm{u}(n) \\
&=\exp\left\{j\pi \left(\frac{2}{M}kn +\frac{2p}{2^{\lambda_\mathrm{PS}}}+\frac{n^2}{M}\right)\right\}.
\end{split}
\end{equation}

\textcolor{black}{The symbol} energy of PSK-LoRa is \(E_\mathrm{s} = \sfrac{1}{M}\sum_{n=0}^{M-1}\vert s_\mathrm{PSK}(n) \vert^2 = 1\). 
\paragraph{Detection}
\textcolor{black}{Like E-LoRa, PSK-LoRa encodes information within the symbol phase, thereby rendering it infeasible to establish a non-coherent detector as it would result in the loss of the information in the PSs. Consequently, only two alternatives exist, that are, coherent detection and the co-called semi-coherent detection. In the case of coherent detection, the ML criterion is utilized to determine both the FS and the PS. On the other hand, in semi-coherent detection, the FS is determined non-coherently, while the PS is estimated using the ML criterion.}
\begin{figure}[tb]\centering
\includegraphics[trim={0 0 0 0},clip,scale=1]{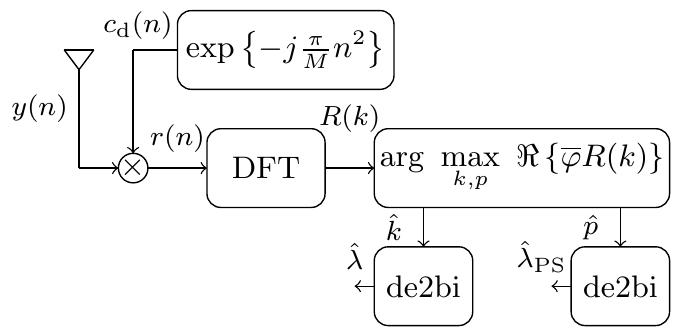}
  \caption{Coherent detector architecture for PSK-LoRa.}
\label{fig6rx}
\end{figure}

\textcolor{black}{The coherent detector of PSK-LoRa is depicted in Fig. \ref{fig6rx}. The process of coherent detection entails the calculation of the inner product between the received symbol vector, represented by \(\boldsymbol{y}\), and the transmit symbol vector, denoted by \(\boldsymbol{s}_\mathrm{PSK}\). The latter can be expressed in vectorial form as \(\boldsymbol{s}_\mathrm{PSK} = \left[{s}_\mathrm{PSK}(0), {s}_\mathrm{PSK}(1), \cdots, {s}_\mathrm{PSK}(M-1)\right]^\mathrm{T}\). The evaluation of the inner product between the two vectors, \(\boldsymbol{y}\) and \(\boldsymbol{s}_\mathrm{PSK}\) yields:}
\begin{equation}\label{eqr11}
\begin{split}
\langle \boldsymbol{y}, \boldsymbol{s}_\mathrm{PSK} \rangle &= \sum_{n=0}^{N-1} y(n)s_\mathrm{PSK}(n)\\
& = \overline{\varphi}\sum_{n=0}^{N-1} y(n)\overline{f}(n)c_\mathrm{d}(n)\\
&= \overline{\varphi}R(k).
\end{split}
\end{equation}

\textcolor{black}{Then FS, \(\hat{k}\), and the PSs, \(\hat{p}\) are jointly evaluated using coherent detector} as:
\begin{equation}\label{eqr12}
\hat{k}, \hat{p} = \mathrm{arg} \max_{k,p}~\Re\left\{\overline{\varphi}R(k)\right\}
\end{equation}

\textcolor{black}{The process of coherent detection commences with the de-chirping of the received symbol, \(y(n)\), using the down-chirp symbol, \(c_\mathrm{d}(n)\), resulting in \(r(n)\). The next step involves the calculation of the DFT of \(r(n)\), resulting in \(R(k)\). Subsequently, the detection problem, as described in eq. (\ref{eqr12}), is solved by multiplying \(R(k)\) with the conjugates of all the possible PSs. This operation yields the estimated values of both \(\hat{k}\) and \(\hat{p}\). The final step entails the conversion of the estimated \(\hat{k}\) and \(\hat{p}\) from decimal to binary representation, resulting in the bit sequences, $\hat{\lambda}$ and $\hat{\lambda}_\mathrm{PS}$, respectively. }

\begin{figure}[tb]\centering
\includegraphics[trim={0 0 0 0},clip,scale=1]{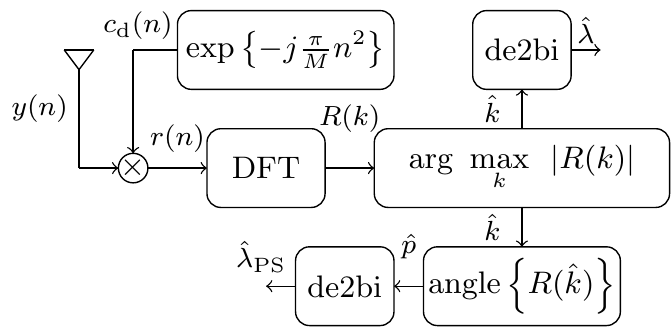}
  \caption{Semi-Coherent detector architecture for PSK-LoRa.}
\label{fig7rx}
\end{figure}
\textcolor{black}{Fig. \ref{fig7rx} illustrates the semi-coherent detector for PSK-LoRa. In semi-coherent detection, upon the evaluation of \(R(k)\), an estimation of $\hat{k}$ is made through a non-coherent approach as:}
\begin{equation}\label{eqr13}
\hat{k} = \mathrm{arg} \max_{k}~\left\vert R(k)\right\vert
\end{equation}
which \textcolor{black}{then determines} the bit sequence , \(\hat{\lambda}\) \textcolor{black}{after de2bi conversion.}

\textcolor{black}{It is imperative to recognize that the PSs cannot be determined via the non-coherent detection process described in eq. (\ref{eqr13}). Nonetheless, it is also important to acknowledge that the activated FS contains the PS information as well. Therefore, upon achieving the estimated value of $\hat{k}$, the PS can be quantified by utilizing the $\hat{k}$th  index within $R(k)$, i.e., $R(\hat{k})$ as:}
\begin{equation}\label{eqr14}
\hat{p} = \mathrm{angle} \left\{ R(\hat{k})\right\},
\end{equation}
\textcolor{black}{where $\mathrm{angle}\{\cdot\}$ constitutes the ML criterion employed for evaluating the PSs. Finally, the estimated value of the symbol PS. Finally, the estimated value of the symbol PS denoted as \(\hat{p}\) is employed to calculate the bit sequence, possessing the estimated value of $\hat{\lambda}_\mathrm{PS}$.}
\paragraph{\textcolor{black}{Takeaways}}
\textcolor{black}{PSK-LoRa is the first SC constant envelop scheme that confers an increase of over one transmitted bit per symbol over LoRa by deploying PSK alphabets. The increase in the bit depends on the cardinality of the PSK alphabets; the higher the cardinality, the greater the number of transmitted bits. Nonetheless, this augmentation necessitates a trade-off in increased detection complexity that scales proportionally to the PSK alphabets' cardinality. Given that the waveform design features PSs, their detection is feasible only through ML criteria. The coherent detection procedure employs the ML criterion to identify the FS and the PS. On the other hand, non-coherent detection cannot be utilized due to the loss of phase information. Consequently, semi-coherent detection must be employed, in which the FS is detected in a non-coherent manner, whereas PS detection relies on ML criteria. Additionally, the deployment of phase information increases the waveform design's vulnerability to frequency and phase offsets.}
\subsubsection{Slope-Shift Keying (SSK)-LoRa}
\textcolor{black}{SSK-LoRa implements varying CRs to augment the symbol cardinality compared to the conventional LoRa approach. LoRa exclusively employs an up-chirp to spread the un-chirped symbol. In contrast, the SSK-LoRa utilizes both up-chirp and down-chirp, resulting in the ability to transmit an additional bit per symbol in addition to $\lambda$.}
\begin{figure}[tb]\centering
\includegraphics[trim={0 0 0 0},clip,scale=1]{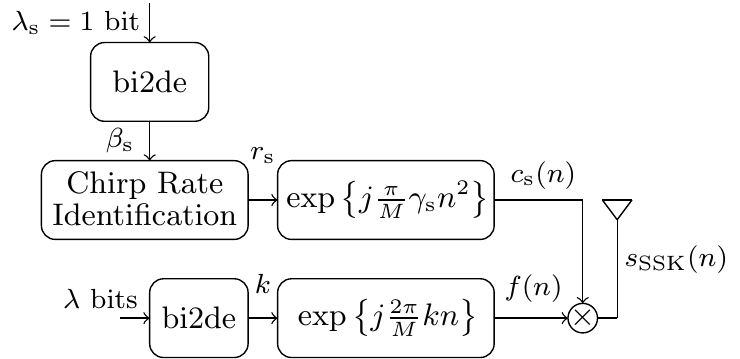}
  \caption{SSK-LoRa transmitter architecture. }
\label{fig6tx}
\end{figure}
\paragraph{Transmission}
\textcolor{black}{The transmitter architecture of the SSK-LoRa is depicted in Fig. \ref{fig6tx}. Like LoRa, $\lambda$ bits are utilized to ascertain the FS, $k$, which yields the corresponding un-chirped symbol, $f(n)$. Furthermore, an additional bit, $\lambda_{\mathrm{s}}= 1$ determines the chirping rate index, $\beta_\mathrm{s}= \llbracket 0,2^{\lambda_{\mathrm{s}}}-1\rrbracket = \left\{0,1\right\}$. This index, in turn, identifies the CR, $\gamma_\mathrm{s}$, with $\mathrm{s} =\llbracket 0, 1\rrbracket$ as:}
\begin{equation}\label{eq19}
\gamma_\mathrm{s}= \begin{dcases}
   1 & \quad \beta_\mathrm{s} = 0\\
 -1& \quad \beta_\mathrm{s} = 1
\end{dcases}.
\end{equation}

\textcolor{black}{Subsequently,} based on \(\gamma_\mathrm{s}\), the spreading symbol \(c_{\mathrm{s}}(n)\) is determined as:
\begin{equation}\label{eq20}
c_{\mathrm{s}}(n)= \begin{dcases}
    c_{\mathrm{u}}(n) & \quad \gamma_\mathrm{s} = 1\\
 c_{\mathrm{d}}(n)& \quad \gamma_\mathrm{s} = -1
\end{dcases}.
\end{equation}

\textcolor{black}{$f(n)$ undergoes a process of multiplication with the spreading symbol, $c_{\mathrm{s}}(n)$, thereby yielding the discrete-time representation of the chirped SSK-LoRa symbol, $s_\mathrm{SSK}(n)$, which is expressed as follows:}
\begin{equation}\label{eq22}
\begin{split}
s_\mathrm{SSK}(n)&= f(n)c_{\mathrm{s}}(n)\\
& = \begin{dcases}
    \exp\left\{j\frac{\pi}{M}n(2k+n)\right\} & \gamma_\mathrm{s} = 1\\
 \exp\left\{j\frac{\pi}{M}n(2k-n)\right\}& \gamma_\mathrm{s} = -1
\end{dcases}.
\end{split}
\end{equation}

The symbol energy of SSK-LoRa symbol is \(E_\mathrm{s} = \sfrac{1}{M}\sum_{n=0}^{M-1}\vert s_\mathrm{SSK}(n) \vert^2 = 1\). 
\paragraph{Detection}
\begin{figure}[tb]\centering
\includegraphics[trim={0 0 0 0},clip,scale=1]{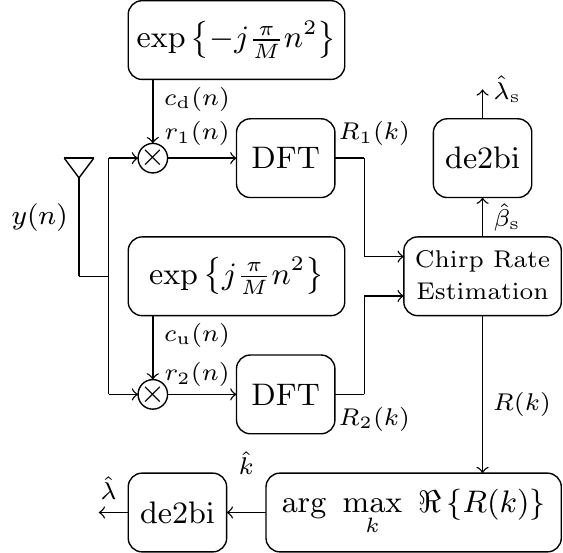}
  \caption{Coherent detector architecture for SSK-LoRa.}
\label{fig10rx}
\end{figure}
\textcolor{black}{The coherent detector for SSK-LoRa is illustrated in Fig. \ref{fig10rx}. The goal is to determine the inner product between the received symbol vector, $\boldsymbol{y}$, and the transmit SSK-LoRa symbol vector, $\boldsymbol{s}_\mathrm{SSK} = \left[{s}_\mathrm{SSK}(0), {s}_\mathrm{SSK}(1), \cdots, {s}_\mathrm{SSK}(M-1)\right]^\mathrm{T}$, i.e., to evaluate $\langle \boldsymbol{y}, \boldsymbol{s}_\mathrm{SSK}\rangle$. It is necessary to consider two distinct scenarios in this evaluation process: (i) the generation of the transmit symbol occurs through the utilization of the up-chirp, i.e., ${s}_\mathrm{SSK} = f(n)c_\mathrm{u}(n)$, and (ii) the transmit symbol is attained using the down-chirp, i.e., ${s}_\mathrm{SSK} = f(n)c_\mathrm{d}(n)$. For the first scenario, the calculation of the inner product is conducted as follows:}
\begin{equation}\label{in_pd_ssk_lora}
\begin{split}
\langle   \boldsymbol{y}, \boldsymbol{s}_\mathrm{SSK}\rangle &= \sum_{n=0}^{M-1}y(n)\overline{s}_\mathrm{SSK}(n)\\
& = \sum_{n=0}^{M-1}y(n)\overline{f}(n)c_\mathrm{d}(n)\\
& = \sum_{n=0}^{M-1}r_1(n)\overline{f}(n)\\
& = R_1(k),
\end{split}
\end{equation}
where \(r_1(n) = y(n)c_\mathrm{d}(n)\) \textcolor{black}{and $R_1(k)$ is the DFT of \(r_1(n)\)}. On the other hand, \textcolor{black}{for the second scenario}, the inner product yields:
\begin{equation}\label{in_pd_ssk_lora2}
\begin{split}
\langle   \boldsymbol{y}, \boldsymbol{s}_\mathrm{SSK}\rangle &= \sum_{n=0}^{M-1}y(n)\overline{s}_\mathrm{SSK}(n)\\
& = \sum_{n=0}^{M-1}y(n)\overline{f}(n)c_\mathrm{u}(n)\\
& = \sum_{n=0}^{M-1}r_2(n)\overline{f}(n)\\
& = R_2(k),
\end{split}
\end{equation}
where \(r_2(n) = y(n)c_\mathrm{u}(n)\) \textcolor{black}{and $R_2(k)$ is the DFT of \(r_2(n)\)}.

\textcolor{black}{Consequently, in the context of coherent detection, after calculating $R_1(k)$ and $R_2(k)$, the subsequent stage entails determining whether an up-chirp or a down-chirp was used at the transmitter. This identification is accomplished by computing the parameters  $\kappa_1^\mathrm{coh} = \max~ \Re\left\{R_1(k)\right\}$ and  $\kappa_2^\mathrm{coh} = \max~\Re\left\{R_2(k)\right\}$. Subsequently, the CR estimation module derives the determines $\hat{\beta}_\mathrm{s}$, based on these parameters as follows:}
\begin{equation}\label{eqr22}
\hat{\beta}_\mathrm{s}=\begin{dcases}
    0 & \kappa_1^\mathrm{coh}>\kappa_2^\mathrm{coh}\\
 1&  \kappa_1^\mathrm{coh}<\kappa_2^\mathrm{coh}
\end{dcases}.
\end{equation}

Once the spreading symbol \textcolor{black}{is identified}, \(\kappa_1^\mathrm{coh}\) and \(\kappa_2^\mathrm{coh}\) are then also used to determine whether the transmitted FS belongs to \(R_1(k)\) or \(R_2(k)\) as \textcolor{black}{follows}:
\begin{equation}\label{eqr23}
R(k)=\begin{dcases}
    R_1(k) & \kappa_1^\mathrm{coh}>\kappa_2^\mathrm{coh}\\
 R_2(k)&  \kappa_1^\mathrm{coh}<\kappa_2^\mathrm{coh}
\end{dcases}.
\end{equation}

\textcolor{black}{Then,} using the coherent detection, the FS is identified as:
\begin{equation}\label{eqr24}
\hat{k} = \mathrm{arg} \max_{k}~\Re\left\{R(k)\right\}.
\end{equation}

\textcolor{black}{Lastly, \(\hat{k}\) and \(\hat{\beta}_\mathrm{s}\) determine the bit sequences of lengths \(\hat{\lambda}\) and \(\hat{\lambda}_\mathrm{s}\), respectively after de2bi conversion.}  
\begin{figure}[tb]\centering
\includegraphics[trim={0 0 0 0},clip,scale=1]{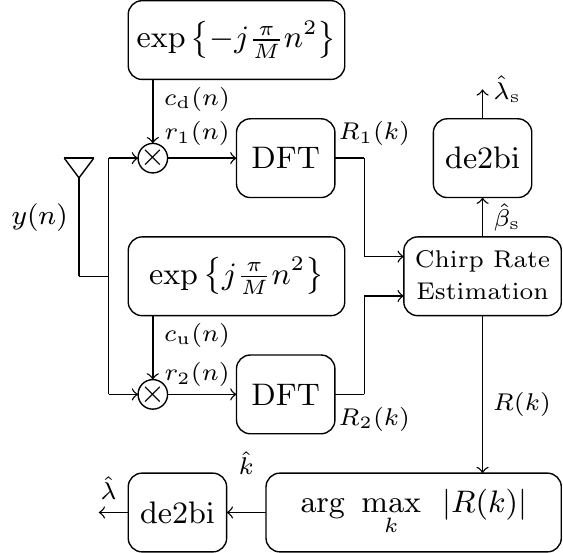}
  \caption{Non-coherent detector architecture for SSK-LoRa.}
\label{fig11rx}
\end{figure}

\textcolor{black}{The non-coherent detector configuration for SSK-LoRa is presented in Fig. \ref{fig11rx}. In contradistinction to the coherent detector, the non-coherent detector utilizes $\kappa_1^\mathrm{non-coh} = \max \left\vert R_1(k)\right\vert$ and $\kappa_2^\mathrm{non-coh} = \max \left\vert R_2(k)\right\vert$ for estimating $\hat{\beta}_\mathrm{s}$ and $R(k)$. $\hat{\beta}_\mathrm{s}$ is determined similarly to eq. (\ref{eqr22}) but using $\kappa_1^\mathrm{non-coh}$ and $\kappa_1^\mathrm{non-coh}$ instead of $\kappa_1^\mathrm{coh}$ an $\kappa_1^\mathrm{coh}$. Furthermore, the detection in the transmit FS, $\hat{k}$ is performed as follows:}
\begin{equation}\label{eqr25}
\hat{k} = \mathrm{arg} \max_{k}~\left\vert R(k)\right\vert.
\end{equation}

\textcolor{black}{The bit sequence comprising of \(\hat{\lambda}\) and \(\hat{\lambda}_\mathrm{s}\) bits are determined after de2bi conversion of \(\hat{k}\) and $\hat{\beta}_\mathrm{s}$.}
\paragraph{\textcolor{black}{Takeaways}}
\textcolor{black}{SSK-LoRa is a constant envelope SC scheme that boasts superior performance characteristics compared to other schemes, such as ICS-LoRa and E-LoRa. This scheme can transmit an extra bit per symbol relative to LoRa, which is only marginal. SSK-LoRa's waveform properties are also highly robust and provide enhanced EE. Compared to ICS-LoRa, SSK-LoRa utilizes two different CRs, which have a lesser impact on cross-correlation properties, resulting in improved BER performance. Furthermore, by not using I/Q components, SSK-LoRa increases its robustness against frequency and phase offsets. Despite these benefits, the detection complexity of the coherent and non-coherent detectors for SSK-LoRa is higher than that of LoRa. This is due to the additional computation of a DFT and related processing. Thus, while SSK-LoRa offers enhanced performance characteristics, it comes at the cost of increased computational complexity.}
\subsubsection{Discrete Chirp Rate Keying (DCRK)-LoRa}
\textcolor{black}{In the SSK-LoRa, two complementary chirp signals, i.e., an up-chirp and a down-chirp, are utilized to achieve a transmission rate of one additional bit per symbol as compared to the conventional LoRa. The extension of this framework through the use of multiple CRs (which generally exceeds two) results in a higher encoding of bits per symbol. This scheme is referred to as the DCRK-LoRa, as described in \cite{dcrk_css}, and is a generalization of the SSK-LoRa. In the DCRK-LoRa, a total of $M_\mathrm{c}$ distinct CRs can be utilized, thereby enabling the transmission of $\lambda + \log_2(M_\mathrm{c})$ bits per symbol of duration $T_\mathrm{s}$.}

\begin{figure}[tb]\centering
\includegraphics[trim={0 0 0 0},clip,scale=1]{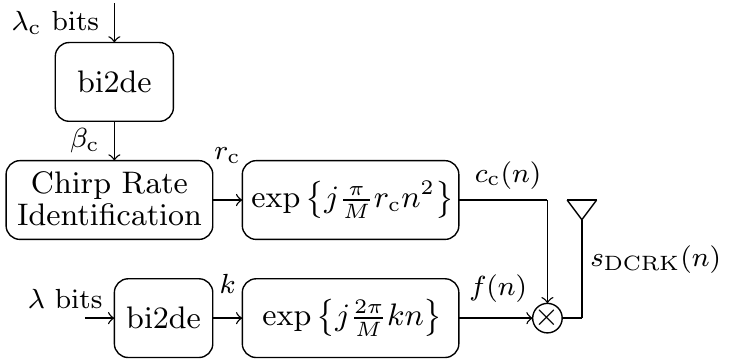}
  \caption{DCRK-LoRa transmitter architecture. }
\label{fig8tx}
\end{figure}
\paragraph{Transmission}
\textcolor{black}{The transmitter architecture of the DCRK-LoRa is depicted in Fig. \ref{fig8tx}. Within the DCRK-LoRa framework, the value of the FS parameter, $k$, is calculated through the bi2de conversion of $\lambda$ bits. Subsequently, the un-chirped symbol, $f(n)$, is derived, as expressed in eq. (\ref{sc}). In addition to the $\lambda$ bits, $\lambda_\mathrm{c}= \log_2(M_\mathrm{c})$ bits are utilized to determine the CR, $\gamma_\mathrm{c}$. The bi2de conversion of $\lambda_\mathrm{c}$ results in $\beta_\mathrm{c} = \llbracket 0, 2^{\lambda_\mathrm{c}}-1\rrbracket$ with a cardinality of $M_\mathrm{c}$, which is employed to determine $\gamma_\mathrm{c}$, which assumes $M_\mathrm{c}$ non-zero integer values through $\beta_\mathrm{c}$ as follows:}
\begin{equation}\label{eq30}
\gamma_\mathrm{c}= \begin{dcases}
    -\frac{M_\mathrm{c}}{2} & \quad \beta_\mathrm{c} = 0\\
 ~~~\cdot& \quad ~~~\cdot\\
 ~~~\cdot& \quad ~~~\cdot\\
 ~~~\cdot& \quad ~~~\cdot\\
 -1& \quad \beta_\mathrm{c} = \frac{2^{\lambda_\mathrm{c}-1}}{2}\\
 1 & \quad \beta_\mathrm{c} = \frac{2^{\lambda_\mathrm{c}-1}}{2}+1\\
~~~\cdot& \quad ~~~\cdot\\
~~~\cdot& \quad ~~~\cdot\\
~~~\cdot& \quad ~~~\cdot\\
 \frac{M_\mathrm{c}}{2}& \quad \beta_\mathrm{c} = 2^{\lambda_\mathrm{c}}-1
\end{dcases},
\end{equation}
where \(\mathrm{c}\) is the indexing variable whose values are \(\mathrm{c} = \llbracket 0, M_\mathrm{c}-1\rrbracket\). Once the CR \(\gamma_\mathrm{c}\) is determined, the spreading symbol \(c_\mathrm{c}(n)\) is \textcolor{black}{ascertained as in eq. (\ref{cr}).}%

Afterwards, \(f(n)\) is chirped using \(c_\mathrm{c}(n)\), resulting in DCRK-LoRa discrete-time symbol, \(s_\mathrm{DCRK}(n)\) as:
\begin{equation}\label{eq32}
s_\mathrm{DCRK}(n) = f(n)c_\mathrm{c}(n) =\exp\left\{j\frac{\pi}{M}n\left(2k +\gamma_\mathrm{c}n\right)\right\}. 
\end{equation}

The symbol energy of DCRK-LoRa is equal to \(E_\mathrm{s} = \sfrac{1}{M}\sum_{n=0}^{M-1}\vert s_\mathrm{DCRK}(n) \vert^2 = 1\). 
\paragraph{Detection}
\begin{figure}[tb]\centering
\includegraphics[trim={0 0 0 0},clip,scale=1]{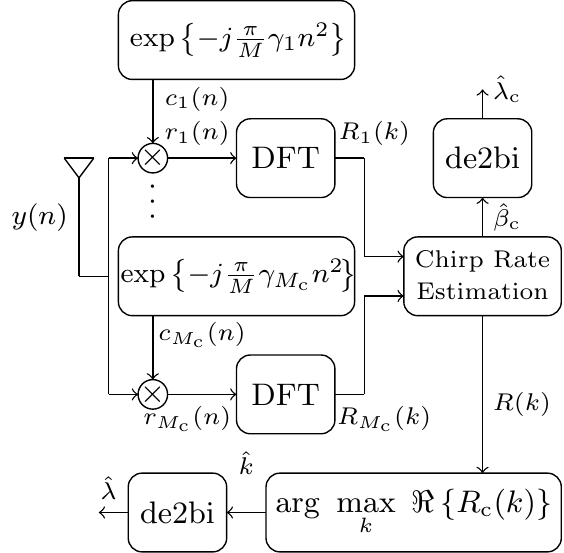}
  \caption{Coherent detector architecture for DCRK-LoRa.}
\label{fig14rx}
\end{figure}

\textcolor{black}{The coherent detector for DCRK-LoRa is depicted in Fig. \ref{fig14rx}. The coherent detection process comprises two key elements: (i) the determination of the FS of the un-chirped symbol, denoted by $\hat{k}$; and (ii) identification of the CR, represented by $\hat{\gamma}\mathrm{c}$. The proposed coherent detector for DCRK-LoRa leverages the inner product of the received signal, $\boldsymbol{y}$, and the DCRK-LoRa transmit symbol, $\boldsymbol{s}_\mathrm{DCRK}=\left[{s}_\mathrm{DCRK}(0), {s}_\mathrm{DCRK}(1), \cdots, {s}_\mathrm{DCRK}(M-1)\right]^\mathrm{T}$. The inner product is calculated as $\langle \boldsymbol{y}, \boldsymbol{s}_\mathrm{DCRK}\rangle$, yields:}
\begin{equation}\label{eqr35}
\begin{split}
\langle   \boldsymbol{y}, \boldsymbol{s}_\mathrm{DCRK}\rangle &= \sum_{n=0}^{M-1}y(n)\overline{s}_\mathrm{DCRK}(n)\\
&= \sum_{n=0}^{M-1}y(n)\overline{f}(n)\overline{c}_\mathrm{c}(n)\\
&= \sum_{n=0}^{M-1}r_\mathrm{c}(n)\overline{f}(n) \\
&= R_\mathrm{c}(k).
\end{split}
\end{equation}
\textcolor{black}{whereby, \(r_\mathrm{c}(n) = y(n)\overline{c}_\mathrm{c}(n)\). The evaluation of the CR requires the calculation of the CR considering all feasible CRs that the transmitter can utilize, as stated in eq. (\ref{eqr35}). To determine the CR, \(\kappa_\mathrm{c}^\mathrm{coh} =\max ~\Re\left\{R_\mathrm{c}(k)\right\}\) 
 is evaluated. Then, \(\hat{\beta}_\mathrm{c}\)  is attained as:}
\begin{equation}\label{eqr36}
\hat{\beta}_\mathrm{c} = \mathrm{arg} \max\limits_{\mathrm{c}}~\kappa_\mathrm{c}^\mathrm{coh}.
\end{equation}

\textcolor{black}{Then, using \(\hat{\beta}_\mathrm{c} \), the bit information in the CR, i.e., \(\hat{\lambda}_\mathrm{c}\) is attained. Moreover, the DFT outputs corresponding to \(\hat{\beta}_\mathrm{c}\), i.e. \(R_\mathrm{c}(k)\) determines the FS of the un-chirped symbol as:}
\begin{equation}\label{eqr37}
\hat{k} = \mathrm{arg} \max\limits_{k}~\Re\left\{R_\mathrm{c}(k)\right\}.
\end{equation}

The bit sequence of length \(\hat{\lambda}\) is determined after bi2de conversion of \(\hat{k}\). 
\begin{figure}[tb]\centering
\includegraphics[trim={0 0 0 0},clip,scale=1]{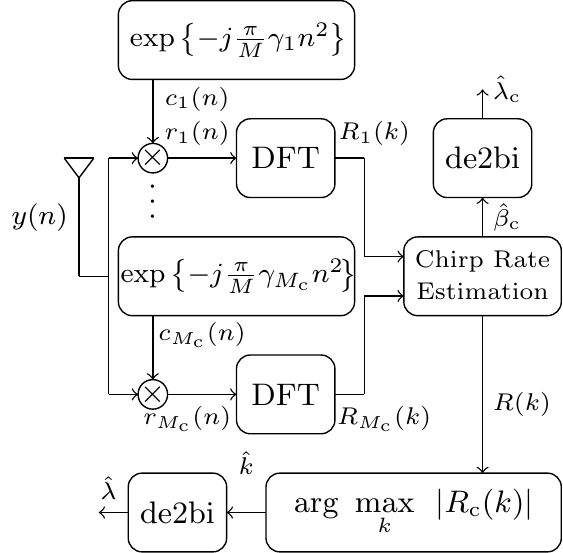}
  \caption{Non-coherent detector architecture for DCRK-LoRa.}
\label{fig15rx}
\end{figure}

\textcolor{black}{Similarly to the non-coherent detection of SSK-LoRa, the evaluation of the CR parameter, \(\hat{\beta}_\mathrm{c}\) in DCRK-LoRa non-coherent detection is achieved by computing $\kappa_\mathrm{c}^\mathrm{non-coh} = \max \left\vert R_\mathrm{c}(k)\right\vert$. The determination of the FS, \(\hat{k}\), is also performed in a non-coherent manner as:}
\begin{equation}\label{eqr38}
\hat{k} = \mathrm{arg} \max\limits_{k}~\left\vert R_\mathrm{c}(k)\right\vert.
\end{equation}

\textcolor{black}{\(\hat{\beta}_\mathrm{c}\) and \(\hat{k}\) then determine the transmitted bits of lengths \(\hat{\lambda}_\mathrm{c}\) and \(\hat{\lambda}\), respectively, after de2bi conversion.}
\paragraph{\textcolor{black}{Takeaways}}
\textcolor{black}{The waveform designs of DCRK-LoRa and SSK-LoRa demonstrate certain similarities. While SSK-LoRa utilizes only two CRs, DCRK-LoRa presents a more adaptable approach by permitting any number of CRs. In addition, DCRK-LoRa also permits transmitting a higher number of bits per symbol, which is directly proportional to the number of CRs employed. DCRK-CSS exhibits SC constant envelope characteristics, contributing to a resilient waveform design. Nonetheless, certain limitations exist within the waveform design of DCRK-LoRa. Firstly, as the number of CRs increases, the computational complexity of the detector also increases proportionally. This is a noteworthy limitation, as while it allows for a significant increase in the number of transmitted bits per symbol, the power consumption rises due to the amplified complexity. Secondly, while the correlation between chirped symbols with various CRs is low, they are not orthogonal, resulting in decreased intrinsic interference with minimal performance degradation, which can be further reduced by increasing $\lambda$.}
\subsubsection{SSK-ICS-LoRa}
\textcolor{black}{The SSK-ICS-LoRa protocol, as its name implies, integrates SSK-LoRa and ICS-LoRa \cite{ssk_ics_lora}. Furthermore, it is categorically placed within the SC taxonomy, incorporating interleaving techniques and varying CRs to facilitate the waveform design. It is pertinent to note that the SSK-LoRa enhances the number of bits per symbol by $1$ relative to LoRa, by incorporating both up-chirp and down-chirp symbols. On the other hand, the ICS-LoRa protocol also transmits $1$ additional bit per symbol relative to LoRa by employing both up-chirped symbols and their interleaved counterparts. Through the amalgamation of SSK-LoRa and ICS-LoRa, SSK-ICS-LoRa leverages the utilization of up-chirp symbols, down-chirp symbols, and interleaved counterparts, thereby elevating the number of transmitted bits per symbol by $2$ with respect to LoRa.}

\paragraph{Transmission}
\begin{figure}[tb]\centering
\includegraphics[trim={0 0 0 0},clip,scale=1]{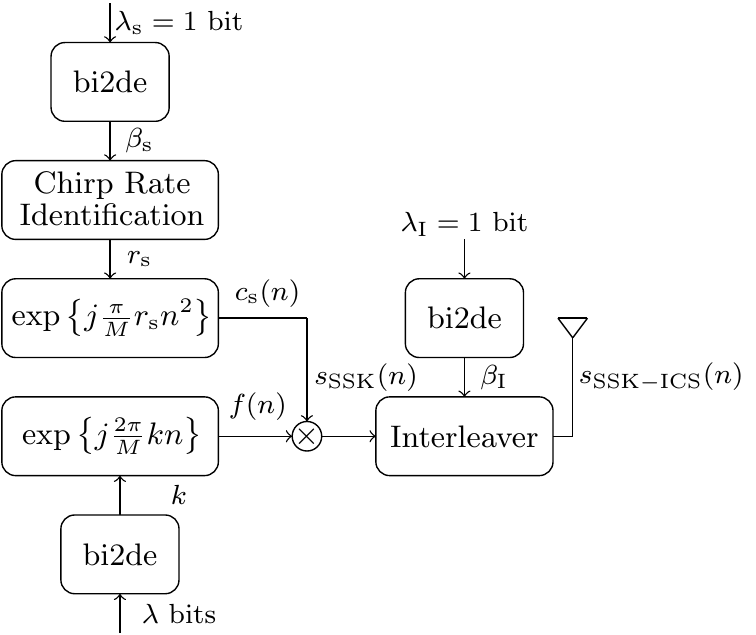}
  \caption{SSK-ICS-LoRa transmitter architecture. }
\label{fig11tx}
\end{figure}
\textcolor{black}{The transmitter architecture of SSK-ICS-LoRa is depicted in Fig. \ref{fig11tx}. Like LoRa, the number of bits, denoted by $\lambda$, determines the FS, $k$, which is then utilized to obtain the un-chirped symbol, $f(n)$. Further, the conversion of $\lambda_\mathrm{s}= 1$ bit via bi2de results in $\beta_\mathrm{s}= \left\{0,1\right\}$, which serves to identify the CR, $\gamma_\mathrm{s} = \left\{-1,1\right\}$. It is noteworthy that SSK-ICS-LoRa utilizes only two distinct CRs, similar to SSK-LoRa. Once $\gamma_\mathrm{s}$ is established, the chirp symbol, $c_\mathrm{s}(n)$ is obtained (cf. eq. \ref{cr}). The chirped symbol, $s_{\mathrm{SSK}}(n)$ is then computed by the multiplication of $f(n)$ and $c_\mathrm{s}(n)$ (cf. eq. \ref{eq22}). We can observe that $s_{\mathrm{SSK}}(n)$ is identical to the SSK-LoRa symbol.Upon determination of $s_{\mathrm{SSK}}(n)$, the value of $\lambda_\mathrm{I}$ serves to identify whether $s_{\mathrm{SSK}}(n)$ is to be interleaved or not. After bi2de conversion of $\lambda_\mathrm{I} = 1$ bit, another parameter, $\beta_\mathrm{I}$, is determined, which indicates whether the transmitter will transmit the SSK-LoRa symbol, $s_{\mathrm{SSK}}(n)$ or its interleaved version, represented as $\Pi \left[s_{\mathrm{SSK}}(n)\right]$. The discrete-time SSK-ICS-LoRa transmit symbol is defined as:}
\begin{equation}\label{eq43}
{s}_\mathrm{SSK-ICS}(n)= \begin{dcases}
    {s}_\mathrm{SSK}(n) & \quad \beta_\mathrm{I}  = 0\\
   \Pi\left[{s}_\mathrm{SSK}(n)\right] & \quad \beta_\mathrm{I}  = 1
   \end{dcases}.
\end{equation}

The symbol energy of SSK-ICS-LoRa is \(E_\mathrm{s} = \sfrac{1}{M}\sum_{n=0}^{M-1}\vert {s}_\mathrm{SSK-ICS}(n) \vert^2 = 1\).
\paragraph{Detection}
\begin{figure}[tb]\centering
\includegraphics[trim={0 0 0 0},clip,scale=1]{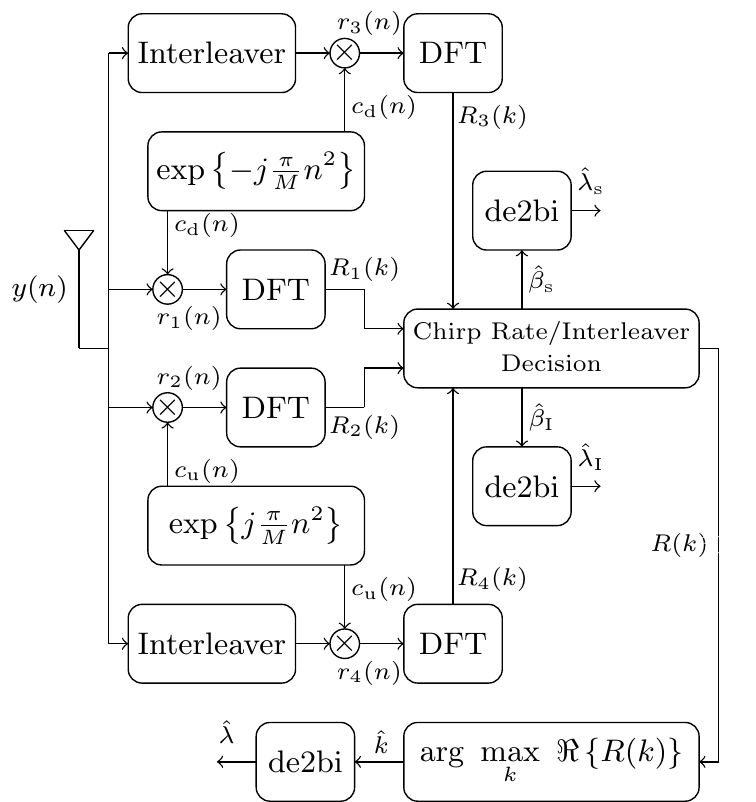}
  \caption{Coherent detector architecture for SSK-ICS-LoRa.}
\label{fig19rx}
\end{figure}

\textcolor{black}{The coherent detection mechanism employed in SSK-ICS-LoRa, as depicted in Fig. \ref{fig19rx}, encompasses four distinct possibilities concerning the transmission symbol. These possibilities can be classified as (i) utilization of the up-chirp symbol for spreading without interleaving, indicated by the binary value of the parameters, ${\beta_{\mathrm{s}},\beta_{\mathrm{I}}} = \{0,0\}$; (ii) deployment of the down-chirp symbol for spreading without interleaving, indicated by the binary value of the parameters, ${\beta_{\mathrm{s}},\beta_{\mathrm{I}}} = \{1,0\}$; (iii) utilization of the up-chirp symbol for spreading with interleaving, indicated by the binary value of the parameters, ${\beta_{\mathrm{s}},\beta_{\mathrm{I}}} = \{0,1\}$; (iv) deployment of the down-chirp symbol for spreading with interleaving, indicated by the binary value of the parameters, ${\beta_{\mathrm{s}},\beta_{\mathrm{I}}} = \{1,1\}$. In all cases, the inner product, $\langle \boldsymbol{y}, \boldsymbol{s}_{\mathrm{SSK-ICS}} \rangle$, between the received symbol, $\boldsymbol{y}$, and the SSK-ICS-LoRa transmit symbol, $\boldsymbol{s}_{\mathrm{SSK-ICS}} = \left[s_{\mathrm{SSK-ICS}}(0),s_{\mathrm{SSK-ICS}}(1),\cdots,s_{\mathrm{SSK-ICS}}(M-1)\right]^{T}$, are distinctive. For the scenario of ${\beta_{\mathrm{s}},\beta_{\mathrm{I}}} = {0,0}$, the evaluation of $\langle \boldsymbol{y}, \boldsymbol{s}_{\mathrm{SSK-ICS}} \rangle$ is as follows:}
\begin{equation}\label{in_pd_ssk_ics1}
\begin{split}
\langle   \boldsymbol{y}, \boldsymbol{s}_\mathrm{SSK-ICS}\rangle &= \sum_{n=0}^{M-1}y(n)\overline{s}_\mathrm{SSK-ICS}(n)\\
& = \sum_{n=0}^{M-1}y(n)\overline{f}(n)c_\mathrm{d}(n)\\
& = \sum_{n=0}^{M-1}r_1(n)\overline{f}(n)\\
&= R_1(k),
\end{split}
\end{equation}
where \(R_1(k)\) is the DFT of \(r_1(n)= y(n)c_\mathrm{d}(n)\). \textcolor{black}{For} \(\{\beta_\mathrm{s},\beta_\mathrm{I}\}= \{1,0\}\), \(\langle   \boldsymbol{y}, \boldsymbol{s}_\mathrm{SSK-ICS}\rangle\) evaluates to:
\begin{equation}\label{in_pd_ssk_ics2}
\begin{split}
\langle   \boldsymbol{y}, \boldsymbol{s}_\mathrm{SSK-ICS}\rangle &= \sum_{n=0}^{M-1}y(n)\overline{s}_\mathrm{SSK-ICS}(n)\\
& = \sum_{n=0}^{M-1}y(n)\overline{f}(n)c_\mathrm{u}(n)\\
& = \sum_{n=0}^{M-1}r_2(n)\overline{f}(n)\\
&= R_2(k),
\end{split}
\end{equation}
where \(R_2(k)\) is the DFT of \(r_2(n)= y(n)c_\mathrm{u}(n)\). Considering up-chirp symbol for spreading with interleaving, i.e., \(\{\beta_\mathrm{s},\beta_\mathrm{I}\}= \{0,1\}\), then \(\langle   \boldsymbol{y}, \boldsymbol{s}_\mathrm{SSK-ICS}\rangle\) results in:
\begin{equation}\label{in_pd_ssk_ics3}
\begin{split}
\langle   \boldsymbol{y}, \boldsymbol{s}_\mathrm{SSK-ICS}\rangle &= \sum_{n=0}^{M-1}y(n)\overline{s}_\mathrm{SSK-ICS}(n)\\
& = \sum_{n=0}^{M-1}\Pi\left[y(n)\right]\overline{f}(n)c_\mathrm{d}(n)\\
& = \sum_{n=0}^{M-1}r_3(n)\overline{f}(n)\\
&= R_3(k),
\end{split}
\end{equation}
where \(R_3(k)\) is the DFT of \(r_3(n)= \Pi[y(n)]c_\mathrm{d}(n)\). Lastly, considering  \(\{\beta_\mathrm{s},\beta_\mathrm{I}\}= \{1,1\}\), the inner product yields:
\begin{equation}\label{in_pd_ssk_ics4}
\begin{split}
\langle   \boldsymbol{y}, \boldsymbol{s}_\mathrm{SSK-ICS}\rangle &= \sum_{n=0}^{M-1}y(n)\overline{s}_\mathrm{SSK-ICS}(n)\\
& = \sum_{n=0}^{M-1}\Pi\left[y(n)\right]\overline{f}(n)c_\mathrm{u}(n)\\
& = \sum_{n=0}^{M-1}r_4(n)\overline{f}(n)\\
&= R_4(k),
\end{split}
\end{equation}
where \(R_4(k)\) is the DFT of \(r_4(n)= \Pi[y(n)]c_\mathrm{u}(n)\). \textcolor{black}{The subsequent procedure entails utilizing the parameters $R_1(k)$, $R_2(k)$, $R_3(k)$, and $R_4(k)$ as specified in equations (\ref{in_pd_ssk_ics1}), (\ref{in_pd_ssk_ics2}), (\ref{in_pd_ssk_ics3}) and (\ref{in_pd_ssk_ics4}), respectively to determine whether the transmitter employed an up-chirp or down-chirp and the presence of interleaving in the transmission of the chirped symbol. In this regard, the parameters, \(\kappa_2^\mathrm{coh} = \max~\Re\left \{R_2(k)\right\}\), \(\kappa_3^\mathrm{coh} = \max~\Re\left \{R_3(k)\right\}\), and \(\kappa_4^\mathrm{coh} = \max~\Re\left \{R_4(k)\right\}\) are calculated. Subsequently, based on these values, the spreading and interleaving parameters, $\hat{\beta}_\mathrm{s}$ and $\hat{\beta}_\mathrm{I} $ are ascertained as:}
\begin{equation}\label{eqr45}
\hat{\beta}_\mathrm{s} = \begin{dcases}
   0 & \quad \max\{\kappa_1^\mathrm{coh},\kappa_3^\mathrm{coh}\}> \max\{\kappa_2^\mathrm{coh},\kappa_4^\mathrm{coh}\}\\
     1 & \quad \max\{\kappa_2^\mathrm{coh},\kappa_4^\mathrm{coh}\}> \max\{\kappa_1^\mathrm{coh},\kappa_3^\mathrm{coh}\}
   \end{dcases},
\end{equation}
and 
\begin{equation}\label{eqr46}
\hat{\beta}_\mathrm{I} = \begin{dcases}
   0 & \quad \max\{\kappa_1^\mathrm{coh},\kappa_2^\mathrm{coh}\}> \max\{\kappa_3^\mathrm{coh},\kappa_4^\mathrm{coh}\}\\
     1 & \quad \max\{\kappa_3^\mathrm{coh},\kappa_4^\mathrm{coh}\}> \max\{\kappa_1^\mathrm{coh},\kappa_2^\mathrm{coh}\}
   \end{dcases},
\end{equation}
respectively. Subsequently, \(\hat{\beta}_\mathrm{s}\) and \(\hat{\beta}_\mathrm{I}\) \textcolor{black}{determine \(\hat{\lambda}_\mathrm{s}\) and \(\hat{\lambda}_\mathrm{I}\), respectively. After the evaluation of \(\hat{\beta}_\mathrm{s}\) and \(\hat{\beta}_\mathrm{I}\), the FS, \(\hat{k}\) is identified} coherently as:
\begin{equation}\label{eqr47}
\hat{k} = \mathrm{arg}\max_k~\Re\left\{R(k)\right\},
\end{equation}
where 
\begin{equation}\label{eqr48}
R(k) = \begin{dcases}
   R_1(k)& \quad \{\hat{\beta}_\mathrm{s}, \hat{\beta}_\mathrm{I}\} = \left\{0,0\right\}\\
     R_2(k) & \quad \{\hat{\beta}_\mathrm{s}, \hat{\beta}_\mathrm{I}\} = \left\{1,0\right\}\\
     R_3(k) & \quad \{\hat{\beta}_\mathrm{s}, \hat{\beta}_\mathrm{I}\} = \left\{0,1\right\}\\
      R_4(k) & \quad \{\hat{\beta}_\mathrm{s}, \hat{\beta}_\mathrm{I}\} = \left\{1,1\right\}\\
   \end{dcases},
\end{equation}

The transmit bit sequence consisting of \(\hat{\lambda}\) bits is determined after de2bi conversion of \(\hat{k}\). 
\begin{figure}[tb]\centering
\includegraphics[trim={0 0 0 0},clip,scale=1]{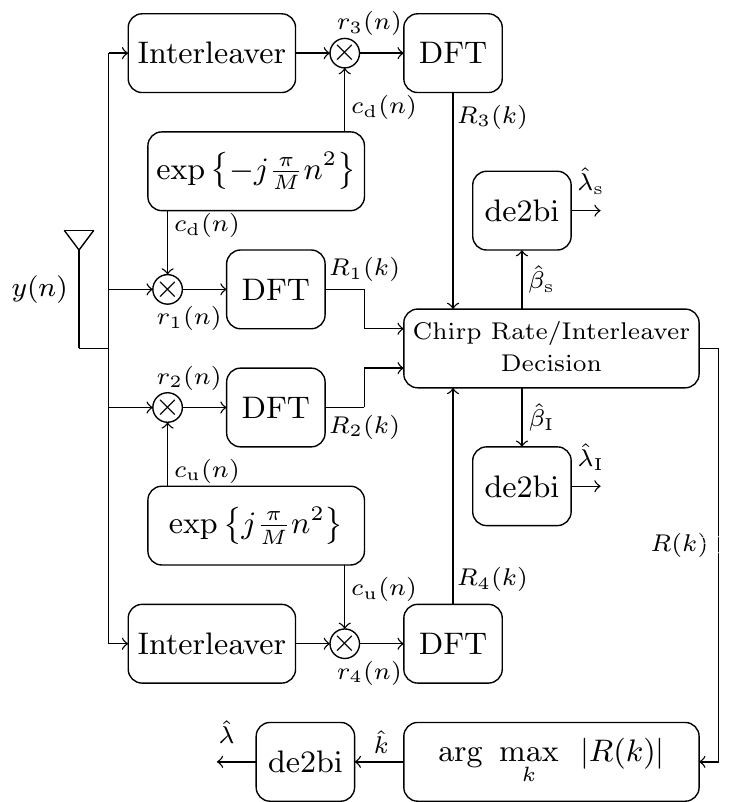}
  \caption{Non-coherent detector architecture for SSK-ICS-LoRa.}
\label{fig20rx}
\end{figure}

\textcolor{black}{The non-coherent detector for SSK-ICS-LoRa is illustrated in Fig. \ref{fig20rx}. Upon evaluating $R_1(k)$, $R_2(k)$, $R_3(k)$, and $R_4(k)$, the following differences can be observed between the coherent and non-coherent detection mechanisms. Firstly, the interleaving and CR decision on the estimations of \(\hat{\beta}_\mathrm{s}\) and \(\hat{\beta}_\mathrm{I}\) is executed based on the parameters calculated via the computation of $\kappa_1^\mathrm{non-coh} = \max\left \vert R_1(k)\right\vert$, $\kappa_2^\mathrm{non-coh} = \max\left \vert R_2(k)\right\vert$, $\kappa_3^\mathrm{non-coh} = \max\left \vert R_3(k)\right\vert$, and $\kappa_4^\mathrm{non-coh} = \max\left \vert R_4(k)\right\vert$. Secondly, the non-coherent determination of the FS, $\hat{k}$, is performed in a non-coherent manner as:}
\begin{equation}\label{eqr49}
\hat{k} = \mathrm{arg}\max_k~\left\vert R(k)\right\vert,
\end{equation}
where \(R(k)\) is same as in eq. (\ref{eqr48}). \textcolor{black}{$\hat{k}$ is then used to determine \(\hat{\lambda}\) after de2bi conversion.}
\paragraph{\textcolor{black}{Takeaways}}
\textcolor{black}{SSK-ICS-LoRa features a constant envelope and is capable of transmitting two additional bits per symbol in comparison to LoRa. The proficient waveform design properties of SSK-ICS-LoRa, including constant envelope characteristics, among others, results in the development of robustness against various channel conditions, offsets, and an improvement in energy efficiency. Nevertheless, detection complexity is notably high as the detector is compelled to identify and discriminate between up-chirp symbols, down-chirp symbols, and their interleaved counterparts. Consequently, the SSK-ICS-LoRa receiver must undertake a minimum of four DFT computations, in contrast to one DFT computation for LoRa or two DFT computations for SSK-LoRa or ICS-LoRa. Additionally, the high cross-correlation between the chirped symbol and its interleaved version in ICS-LoRa results in the degradation of the bit error rate and loss of orthogonality. Similarly, the utilization of two diverse coding rates also results in a loss of orthogonality for SSK-LoRa. Consequently, the discrete domain in SSK-ICS-LoRa's different symbols is also not orthogonal, leading to a few performance limitations.}
\subsection{\textcolor{black}{Multiple Carrier CSS Schemes}}
\subsubsection{Dual-Orthogonal Chirp Spread Spectrum (DO-CSS)}
\textcolor{black}{One of the pioneering CSS schemes belonging to MC taxonomy for LPWANs is DO-CSS  conceptualized by Vangelista, and Cattapan \cite{do_css}. The waveform design approach adopted by DO-CSS mirrors that of the LoRa in that it relies on activating only the FSs. However, unlike LoRa, which activates a single FS, DO-CSS involves the activation of two FSs, with one FS representing an even frequency and the other denoting an odd frequency. Given that each FS transmits $\lambda -1$ bits, the activation of both FSs results in the transmission of a total of $2\lambda-2$ bits per DO-CSS symbol of duration \(T_\mathrm{s}\). This simultaneous activation of two FSs corresponds to the multiplexing of two chirp symbols that exhibit different cyclic time shifts.}
\begin{figure}[tb]\centering
\includegraphics[trim={0 0 0 0},clip,scale=1]{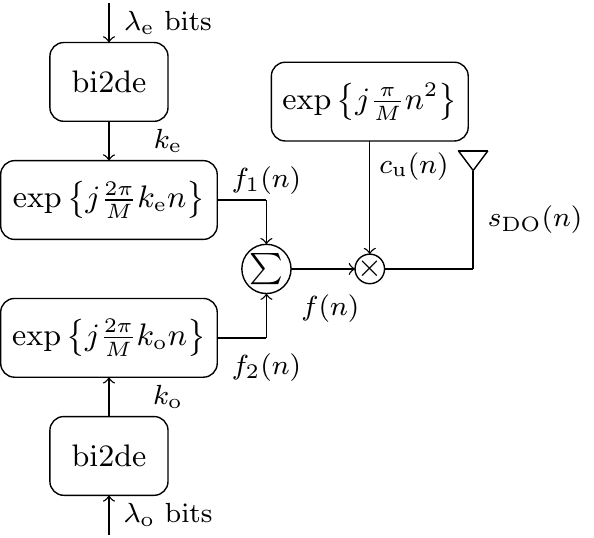}
  \caption{DO-CSS transmitter architecture. }
\label{fig5tx}
\end{figure}
\paragraph{Transmission}
\textcolor{black}{The transmitter configuration of the DO-CSS is depicted in Fig. \ref{fig5tx}. In DO-CSS, the available $M$ FSs within bandwidth $B$ are split into two distinct subsets, i.e., an even subset and an odd subset, each consisting of \(\sfrac{M}{2}\) FSs. Then, the respective even and odd FSs, \(k_\mathrm{e}\) and \(k_\mathrm{o}\), as specified by \(\lambda_\mathrm{e} = \log_2(\sfrac{M}{2})= \lambda -1\) bits for the even subset and \(\lambda_\mathrm{o} = \log_2(\sfrac{M}{2})=  \lambda -1\) bits for the odd subset are attained. Finally, the un-chirped symbols corresponding to the specified even and odd FSs, \(k_\mathrm{e}\) and \(k_\mathrm{o}\), are computed as:}
\begin{equation}\label{eq14}
f_{1}(n) = \exp\left\{j\frac{2\pi}{M}k_\mathrm{e}n\right\},
\end{equation}
and 
\begin{equation}\label{eq15}
f_{2}(n) = \exp\left\{j\frac{2\pi}{M}k_\mathrm{o}n\right\},
\end{equation}
respectively. Subsequently, \(f_{1}(n)\) and \(f_{2}(n)\) are added resulting in \textcolor{black}{\(f(n)\), that is given as:}
\begin{equation}\label{eq16}
\begin{split}
\textcolor{black}{f(n) = f_{1}(n)+f_{2}(n) = \exp\left\{j\frac{2\pi}{M}k_\mathrm{e}n\right\}+\exp\left\{j\frac{2\pi}{M}k_\mathrm{o}n\right\}.}
\end{split}
\end{equation}

\textcolor{black}{Subsequently,} \(f(n)\)  is multiplied with an up-chirp \textcolor{black}{symbol}, \(c_\mathrm{u}(n)\) resulting in discrete time chirped DO-CSS symbol \(s_\mathrm{DO}(n)\), that is given as:
\begin{equation}\label{eq17}
\begin{split}
s_\mathrm{DO}(n) &= f(n)c_\mathrm{u}(n)\\
&=\exp\left\{j\frac{\pi}{M}n\left(2k_\mathrm{e}+n\right)\right\}+\exp\left\{j\frac{\pi}{M}n\left(2k_\mathrm{o}+n\right)\right\}.
\end{split}
\end{equation}

The symbol energy of DO-CSS symbol is \(E_\mathrm{s} = \sfrac{1}{M}\sum_{n=0}^{M-1}\vert s_\mathrm{DO}(n) \vert^2 = 2\). 
\paragraph{Detection}
\textcolor{black}{In \cite{do_css}, the authors propose implementing a non-coherent detection for DO-CSS. However, in the present study, a more comprehensive examination is conducted, including the non-coherent detection mechanism and a study of the coherent detection applied to the DO-CSS framework.}
\begin{figure}[tb]\centering
\includegraphics[trim={0 0 0 0},clip,scale=1]{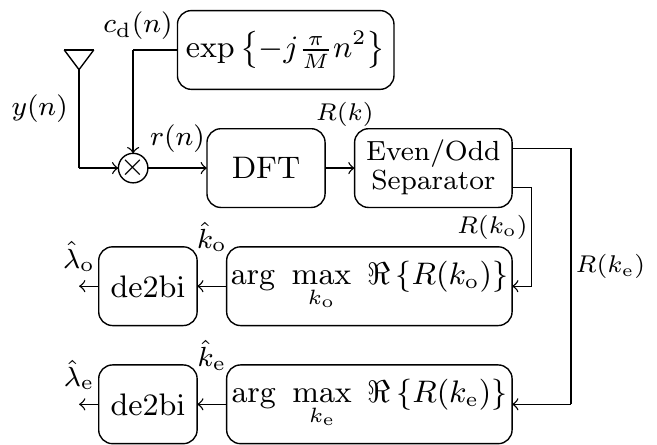}
  \caption{Coherent detector architecture for DO-CSS.}
\label{fig8rx}
\end{figure}

The inner product of the received symbol, \(\boldsymbol{y}\) and the DO-CSS transmit symbol, \(\boldsymbol{s}_\mathrm{DO} = \left[{s}_\mathrm{DO}(0), {s}_\mathrm{DO}(1), \cdots, {s}_\mathrm{DO}(M-1)\right]^\mathrm{T}\), i.e., \(\langle   \boldsymbol{y}, \boldsymbol{s}_\mathrm{DO}\rangle\) is given as:
\begin{equation}\label{eqr15}
\begin{split}
\langle   \boldsymbol{y}, \boldsymbol{s}_\mathrm{DO}\rangle & = \sum_{n=0}^{M-1}y(n)\overline{s}_\mathrm{DO}(n)\\
& = \sum_{n=0}^{M-1}y(n)\overline{f}(n)c_\mathrm{d}(n)\\
& = \sum_{n=0}^{M-1}r(n)\left\{\overline{f}_1(n) +\overline{f}_2(n)\right\}\\
& = R(k_\mathrm{e}) + R(k_\mathrm{o}).
\end{split}
\end{equation}

Then, employing eq. (\ref{eqr15}), the even and odd activated FS \textcolor{black}{are} evaluated as:
\begin{equation}\label{eqr16}
\hat{k}_\mathrm{e}, \hat{k}_\mathrm{o} = \mathrm{arg}\max_{k_\mathrm{e}, k_\mathrm{o}} ~\Re \left\{R(k_\mathrm{e}) + R(k_\mathrm{o})\right\}.
\end{equation}

\textcolor{black}{The DFT outputs, $R(k_\mathrm{e})$ and $R(k_\mathrm{o})$, respectively correspond to the DFT results obtained at the even and odd frequency indexes. The evaluation of both the estimated parameters, namely, \(\hat{k}_\mathrm{e}\) and \(\hat{k}_\mathrm{o}\), are jointly carried out following eq. (\ref{eqr16}). However, the coherent detection problem encapsulated within eq. (\ref{eqr16}) can be solved such that the separate identification of both the parameters, i.e., \(\hat{k}_\mathrm{e}\) and \(\hat{k}_\mathrm{o}\), can be achieved as follows:}
\begin{equation}\label{eqr17}
\hat{k}_\mathrm{e} = \mathrm{arg}\max_{k_\mathrm{e}} ~\Re \left\{R(k_\mathrm{e}) \right\},
\end{equation}
and
\begin{equation}\label{eqr18}
\hat{k}_\mathrm{o} = \mathrm{arg}\max_{k_\mathrm{o}}~ \Re \left\{R(k_\mathrm{o})\right\},
\end{equation}
respectively. \textcolor{black}{The coherence-based detection architecture depicted in Fig. \ref{fig8rx} elucidates the capacity for disjoint identification of \(\hat{k}_\mathrm{e}\) and \(\hat{k}_\mathrm{o}\). The coherent detection entails the de-chirping of the received symbol, \(y(n)\), using a down-chirp, \(c_\mathrm{d}(n)\), which results in \(r(n)\). After that, the DFT of \(r(n)\), i.e., \(R(k)\), is computed. The even/odd separator function then isolates the even and odd frequency components of \(R(k)\), resulting in  \(R(k_\mathrm{e})\) and \(R(k_\mathrm{o})\), respectively. Utilizing the decision criteria specified in eqs. (\ref{eqr17}) and (\ref{eqr18}), the even and odd activated frequency indexes, \(\hat{k}_\mathrm{e}\) and \(\hat{k}_\mathrm{o}\), are separately identified. Finally, the bit sequences, comprised of the corresponding \(\hat{\lambda}_\mathrm{e}\) and \(\hat{\lambda}_\mathrm{o}\) bits, are determined, utilizing \(\hat{k}_\mathrm{e}\) and \(\hat{k}_\mathrm{o}\) as inputs.}
\begin{figure}[tb]\centering
\includegraphics[trim={0 0 0 0},clip,scale=1]{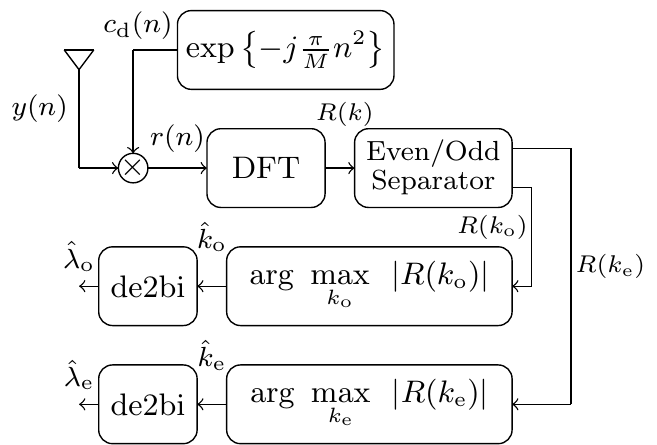}
  \caption{Non-coherent detector architecture for DO-CSS.}
\label{fig9rx}
\end{figure}
\textcolor{black}{The non-coherent detector architecture utilized for DO-CSS is depicted in reference Fig. \ref{fig9rx}. With regards to the coherent detector, the distinctive aspect lies in the fact that post-acquisition of \(R(k_\mathrm{e})\) and \(R(k_\mathrm{o})\), the activation of even and odd FSs is determined as:}
\begin{equation}\label{eqr19}
\hat{k}_\mathrm{e} = \mathrm{arg}\max_{k_\mathrm{e}}  \left\vert R(k_\mathrm{e}) \right\vert
\end{equation}
and
\begin{equation}\label{eqr20}
\hat{k}_\mathrm{o} = \mathrm{arg}\max_{k_\mathrm{o}}  \left\vert R(k_\mathrm{o})\right\vert,
\end{equation}

Afterwards, using \(\hat{k}_\mathrm{e}\) and \(\hat{k}_\mathrm{o}\), the bit sequences having respective lengths of \(\hat{\lambda}_\mathrm{e}\) and \(\hat{\lambda}_\mathrm{o}\) are ascertained. 
\paragraph{\textcolor{black}{Takeaways}}
\textcolor{black}{DO-CSS is a MC CSS scheme that enhances LoRa, a scheme from SC taxonomy, by activating two FSs instead of one. While multiple approaches can be used to activate its two FSs, DO-CSS involves the activation of an even FS and an odd FS. DO-CSS features the simplest waveform design of any MC scheme. Its principal benefit over LoRa is an almost twofold increase in the number of transmitted bits, which is a substantial gain. Nonetheless, the DO-CSS waveform design analysis reveals that its discrete-time symbol lacks a constant envelope and may exhibit a high PAPR, mainly depending on the two activated FSs. This factor could result in implementation issues since low-cost components utilized in LPWANs may have a limited linear range of operation. Although only non-coherent detection is suggested in the seminal work, both coherent and non-coherent detection of the DO-CSS symbol is possible. Moreover, if both the even and odd FSs are detected in a disjointed manner, the computation complexity of detection is similar to that of LoRa.}
\subsubsection{In-Phase and Quadrature (IQ)-CSS}
\textcolor{black}{IQ-CSS scheme is a MC approach that leverages the utilization of both I/Q components to encode the information bits onto the chirp signal \cite{iqcss}. IQ-CSS  takes advantage of the orthogonality between sine and cosine waveforms, thereby enabling the concurrent transmission of two data symbols. In contrast to E-LoRa, where the information is modulated onto either the in-phase or the quadrature component, IQ-CSS implements simultaneous encoding onto both the in-phase and quadrature components. In IQ-CSS, one FS generates the in-phase component of the un-chirped symbol. In contrast, another FS generates the quadrature component of the un-chirped symbol. The capability of transmitting information through both the in-phase and quadrature components, each having a capacity of transmitting $\lambda$ bits, results in the total number of bits transmitted per IQ-CSS symbol being double that of conventional LoRa, i.e., $2\lambda$.}
\begin{figure}[tb]\centering
\includegraphics[trim={0 0 0 0},clip,scale=1]{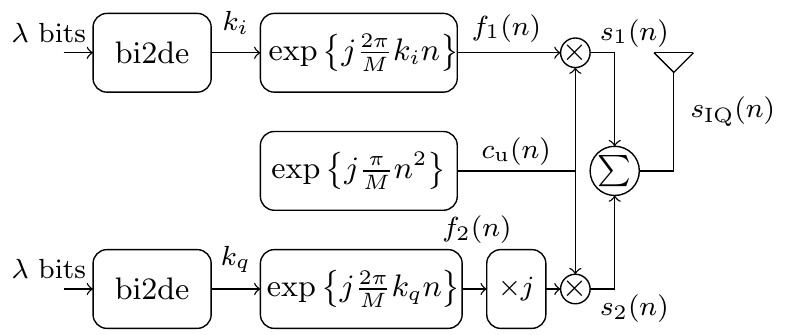}
  \caption{IQ-CSS transmitter architecture. }
\label{fig7tx}
\end{figure}
\paragraph{Transmission}
\textcolor{black}{The graphical representation of the transmitter configuration in IQ-CSS is presented in Fig. \ref{fig7tx}. Within the framework of IQ-CSS, the transmission of a total of  \(2\lambda\) bits takes place, wherein half of the bits are allocated towards determining \(k_i\), and the remaining half is earmarked to determine \(k_q\). Subsequently, the generation of in-phase and quadrature un-chirped symbols, \(f_1(n)\) and \(f_2(n)\), respectively, transpires through the utilization of \(k_i\) and \(k_q\) as follows: }

\begin{equation}\label{eq24}
f_{1}(n) = \exp\left\{j\frac{2\pi}{M}k_i n\right\},
\end{equation}
and 
\begin{equation}\label{eq25}
f_{2}(n) = \exp\left\{j\frac{2\pi}{M}k_q n\right\}.
\end{equation}

\textcolor{black}{Subsequently, the un-chirped in-phase symbol \(f_1(n)\) undergoes a multiplication with the up-chirp symbol, \(c_\mathrm{u}(n)\), resulting in the chirped in-phase symbol \(s_1(n)\). On the other hand, the un-chirped quadrature symbol \(f_2(n)\) is subjected to a phase rotation of \(\sfrac{\pi}{2}\) before its multiplication with the \(c_\mathrm{u}(n)\); thus yielding the chirped quadrature symbol \(s_2(n)\), where the chirped in-phase and quadrature symbols are given as:}
\begin{equation}\label{eq26}
s_{1}(n) = f_{1}(n)c_\mathrm{u}=  \exp\left\{j\frac{\pi}{M}n\left(2k_i +n\right)\right\},
\end{equation}
and 
\begin{equation}\label{eq27}
s_{2}(n) = jf_{2}(n)c_\mathrm{u}= j\exp\left\{j\frac{\pi}{M}n\left(2k_q+n\right)\right\}.
\end{equation}

Lastly, both \(s_{1}(n)\) and \(s_{2}(n)\) are added to attain the discrete-time IQ-CSS symbol, \textcolor{black}{which is given} as:
\begin{equation}\label{eq28}
\begin{split}
&s_\mathrm{IQ}(n) = s_{1}(n) + s_{2}(n)\\
& ~~=\exp\left\{j\frac{\pi}{M}n\left(2k_i +n\right)\right\} + j\exp\left\{j\frac{\pi}{M}n\left((2k_q+n\right)\right\}
\end{split}
\end{equation}

IQ-CSS symbol energy is equal to \(E_\mathrm{s} = \sfrac{1}{N}\sum_{n=0}^{M-1}\vert s_\mathrm{IQ}(n)  \vert^2 = 2\). 
\paragraph{Detection}
\begin{figure}[tb]\centering
\includegraphics[trim={0 0 0 0},clip,scale=1]{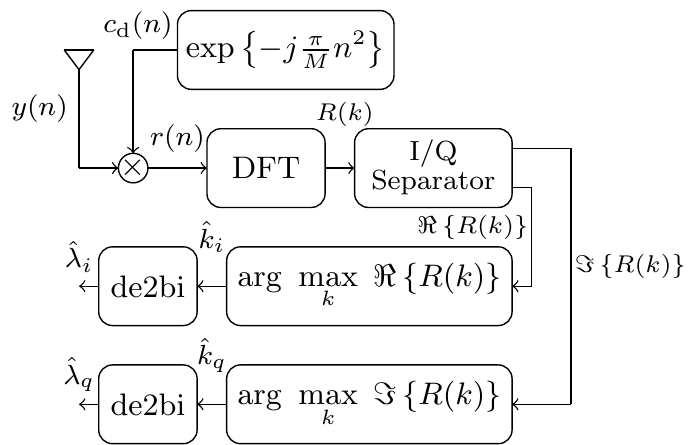}
  \caption{Coherent detector architecture for IQ-CSS.}
\label{fig12rx}
\end{figure}

In \cite{iqcss}, the authors proposed a coherent detector for IQ-CSS, which is illustrated in Fig. \ref{fig12rx}. For the coherent detection, the inner product of the received symbol, \(\boldsymbol{y}\) and the IQ-CSS transmit symbol \(\boldsymbol{s}_\mathrm{IQ} = \left[{s}_\mathrm{IQ}(0), {s}_\mathrm{IQ}(1), \cdots, {s}_\mathrm{IQ}(M-1)\right]^\mathrm{T}\), \(\langle   \boldsymbol{y}, \boldsymbol{s}_\mathrm{IQ}\rangle\) yields:
\begin{equation}\label{eqr26}
\begin{split}
\langle   \boldsymbol{y}, \boldsymbol{s}_\mathrm{IQ}\rangle  &= \sum_{n=0}^{M-1}y(n)\overline{s}_\mathrm{IQ}(n)\\
&=\sum_{n=0}^{M-1}r(n)\overline{f}_1(n)-j\sum_{n=0}^{M-1}r(n)\overline{f}_2(n)\\
&= R(k_i)-jR(k_q),
\end{split}
\end{equation}
where \(r(n)=y(n)c_\mathrm{d}(n)\). It may be noticed that when the coherent detection (cf. \ref{cd4}) is applied, the detection problem in eq. (\ref{eqr26}) can be disjointed, which leads to the following detection of the activated in-phase and quadrature FSs:
\begin{equation}\label{eqr27}
\hat{k}_i = \mathrm{arg} \max_k~\Re\left\{R(k)\right\},
\end{equation}
and
\begin{equation}\label{eqr28}
\hat{k}_q = \mathrm{arg} \max_k~\Im\left\{R(k)\right\},
\end{equation}
where \(\Re\left\{R(k_i)\right\}= \Re\left\{R(k)\right\}\) and \(\Im\left\{R(k)\right\}=\Re\left\{-jR(k_q)\right\}\).
\textcolor{black}{The coherent detector mechanism, as depicted in Fig. \ref{fig12rx}, initiates with the de-chirping of the received symbol, \(y(n)\) via application of the down-chirp, \(c_\mathrm{d}(n)\), thereby yielding \(r(n)\). Subsequently, utilization of the DFT of \(r(n)\), i.e., \(R(k)\) leads to the separation of the in-phase and quadrature components via the I/Q separation block. Finally, the in-phase and quadrature FSs are identified using eqs. (\ref{eqr27}) and (\ref{eqr28}), respectively. \(\hat{k}_i\) and \(\hat{k}_q\) are then used to determine the bits of lengths \(\hat{\lambda}_i\) and \(\hat{\lambda}_q\), respectively after de2bi conversion.}

\begin{figure}[tb]\centering
\includegraphics[trim={0 0 0 0},clip,scale=1]{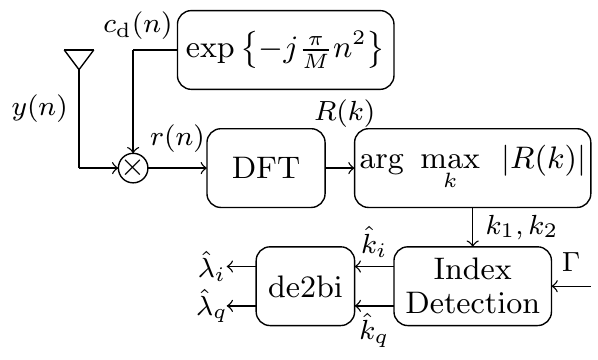}
  \caption{Non-coherent detector architecture for IQ-CSS.}
\label{fig13rx}
\end{figure}
\textcolor{black}{Recently, in \cite{iqcim}, the authors proposed a non-coherent detection mechanism for IQ-CSS, as depicted in Fig. \ref{fig13rx}. In this non-coherent detection, after performing the DFT on the received symbol, \(R(k)\), the two most prominent peaks, \(k_1\) and \(k_2\) are extracted according to the non-coherent detection principle, as specified in eq. (\ref{ncd1}) as follows: }
\begin{equation}\label{eqr29}
k_1,k_2 = \mathrm{arg} \max_k~\left\vert R(k)\right\vert,
\end{equation}
where \(\vert R(k_1)\vert > \vert R(k_2) \vert\). Subsequently, \(k_1\) and \(k_2\) are used to ascertain the FS of the in-phase and the quadrature components, \(\hat{k}_i\) and \(\hat{k}_q\). There can be two possible scenarios: (i) either \(\hat{k}_i\) and \(\hat{k}_q\) are the same, or (ii) \(\hat{k}_i\) and \(\hat{k}_q\) are different. If \(\hat{k}_i\) and \(\hat{k}_q\) are same, this implies that \(R(k_1) \gg R(k_2)\). On the other hand, for different \(\hat{k}_i\) and \(\hat{k}_q\), we are also faced with a problem of matching \(k_1\) and \(k_2\) with \(\hat{k}_i\) and \(\hat{k}_q\). 

The following criterion will decide whether \(\hat{k}_i\) and \(\hat{k}_q\)  are the same, namely:
\begin{equation}\label{eqr30}
1 < \Gamma \leq \frac{\vert R(k_1)\vert }{\vert R(k_2)\vert },
\end{equation}
where \(\Gamma > 1\) is a predefined threshold. This means that the largest peak is significantly greater than the second largest one, which implies that both FSs align as \(\hat{k}_i=\hat{k}_q=k_1\).

Now, on the other hand,  when \(\hat{k}_i\) and \(\hat{k}_q\) are distinguishable as:
\begin{equation}\label{eqr31}
1 \leq \frac{\vert R(k_1)\vert }{\vert R(k_2)\vert } < \Gamma,
\end{equation}
suggesting that both peaks have relatively similar amplitudes. In this scenario, the index detection block also calculates, the phase difference between \(R(k_1)\) and \(R(k_2)\) as:
\begin{equation}\label{eqr32}
\theta = \mathrm{phase}\left\{\overline{R}(k_1),R(k_2)\right\},
\end{equation}
\textcolor{black}{where \(\mathrm{phase}\left\{x,y\right\}\) evaluates the phase difference between \(x\) and \(y\).}

Then using \(\theta\), the decision on \(\hat{k}_i\) and \(\hat{k}_q\) is made in the index detection block as:
\begin{equation}\label{eqr33}
\hat{k}_i=\begin{dcases}
   k_1& 0 \leq \theta <\pi\\
 k_2&  -\pi \leq \theta < 0
\end{dcases},
\end{equation}
and 
\begin{equation}\label{eqr34}
\hat{k}_q=\begin{dcases}
   k_2& 0 \leq \theta <\pi\\
 k_1&  -\pi \leq \theta < 0
\end{dcases},
\end{equation}
respectively. \textcolor{black}{\(\hat{\lambda}_i\) and \(\hat{\lambda}_q\) are ascertained after de2bi conversion of \(\hat{k}_i\) and \(\hat{k}_q\), respectively.}
\paragraph{\textcolor{black}{Takeaways}}
\textcolor{black}{IQ-CSS symbol activates one FS from the in-phase component and another FS from the quadrature component, leading to a MC scheme. E-LoRa and IQ-CSS employ I/Q components for their waveform design. However, unlike E-LoRa, IQ-CSS employs both the in-phase and quadrature components simultaneously for waveform design. The number of bits transmitted per IQ-CSS symbol is twice as much as LoRa. Nonetheless, in general, the I/Q design methodology is not robust due to its inability to support non-coherent detection, and the resultant waveform is highly sensitive to phase and frequency offsets. Additionally, detecting the IQ-CSS symbol requires separating the in-phase and quadrature components, which poses severe design constraints.}
\subsubsection{Enhanced PSK-CSS (ePSK-CSS)}
\textcolor{black}{\cite{epsk_lora} proposes an extension to the PSK-LoRa technique, known as ePSK-CSS, to enhance the detection of FSs at the receiver. This is achieved by incorporating a redundant strategy that divides the bandwidth into sub-bands and activating the FF for the first sub-band and its harmonics in each subsequent sub-band. To each sub-band, a unique PS is introduced, resulting in the generation of an un-chirped symbol. As a result of using the multiple FSs activated simultaneously, the resulting symbol is the ePSK-CSS symbol. The SE of ePSK-CSS can be tailored based on the number of sub-bands and the PSs introduced in each sub-band, leading to multiple variants of ePSK-CSS. The optimal variant, consisting of two sub-bands and the utilization of quaternary PSs, can transmit \(3\) additional bits as compared to the LoRa modulation.}
\begin{figure}[tb]\centering
\includegraphics[trim={0 0 0 0},clip,scale=1]{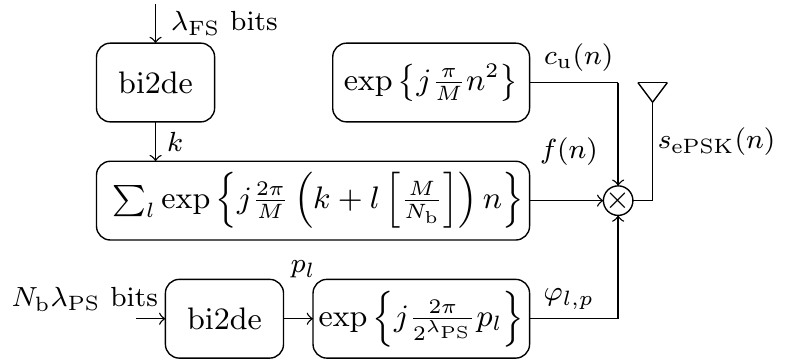}
  \caption{ePSK-CSS transmitter architecture. }
\label{fig12tx}
\end{figure}
\paragraph{Transmission}
\textcolor{black}{The transmitter architecture of ePSK-CSS, as depicted in Fig. \ref{fig12tx}. The transmitter operates under the premise of utilizing \(M\) FSs present in the bandwidth \(B\), which is segmented into \(N_\mathrm{b}\) sub-bands, each comprising of a specific number of distinct frequencies, i.e., \(\alpha = \sfrac{M}{N_\mathrm{b}}\).  The design employs a FF, \(k \in \llbracket 0, \alpha -1\rrbracket\) , within the first sub-band, which encodes a total of \(\lambda_\mathrm{FS} = \log_2(\alpha)\) bits. The remaining \(N_\mathrm{b}-1\) sub-bands are utilized to add redundancy by activating the harmonics of the aforementioned FF, \(k\), with the sequential activation of the harmonics defined as \(k+ \alpha, k+ 2\alpha, \cdots, k+(N_\mathrm{b}-1) \alpha\). This configuration enables the detection of the FF, \(k\), with a total of \(N_\mathrm{b}\) copies, thereby ensuring a robust mechanism for detection. Then,} the un-chirped symbol representing the activation of the FF and its harmonics is given as:
\begin{equation}\label{eq45}
f(n)=\exp\left\{j\frac{2\pi}{M}n\left(k+l\left[\frac{M}{N_\mathrm{b}}\right]\right)\right\},
\end{equation}
where \(l= \llbracket 1, N_\mathrm{b}\rrbracket\). Furthermore, a PS is introduced for the FF and \(l\)th harmonics, \textcolor{black}{as follows:}
\begin{equation}\label{eq46}
\varphi_{l,p} = \exp\left\{j\frac{2\pi}{2^{\lambda_\mathrm{PS}}}p_l\right\},
\end{equation}
where  \(p_l\in \llbracket 0, 2^{\lambda_\mathrm{PS}}-1\rrbracket\) with \(\lambda_\mathrm{PS}\) being the number of bits encoded in the PSs. Then, the discrete time ePSK-CSS symbol obtained by multiplying \(f(n)\),  \(\varphi_{l,p}\) and the up-chirp, \(c_\mathrm{u}\) is given as:
\begin{equation}\label{eq47}
\begin{split}
s_\mathrm{ePSK}(n) &= \sum_{l=0}^{N_\mathrm{b}-1} f(n) \varphi_{l,p}  c_\mathrm{u}(n)\\
&= \sum_{l=0}^{N_\mathrm{b}-1}\exp\left\{j\pi \left(\frac{2kn}{M} + \frac{2ln}{N_\mathrm{b}}+\frac{2p_l}{2^{\eta_\mathrm{PS}}} + \frac{n^2}{M}\right) \right\}.
\end{split}
\end{equation}

Note that the information-carrying elements are \(k\) and \(p_l\). The symbol energy of ePSK-CSS is \(E_\mathrm{s}= \sfrac{1}{M}\sum_{n=0}^{N-1}\vert s_\mathrm{ePSK}(n) \vert^2 = N_\mathrm{b}\).
\paragraph{Detection}
\begin{figure}[tb]\centering
\includegraphics[trim={0 0 0 0},clip,scale=1]{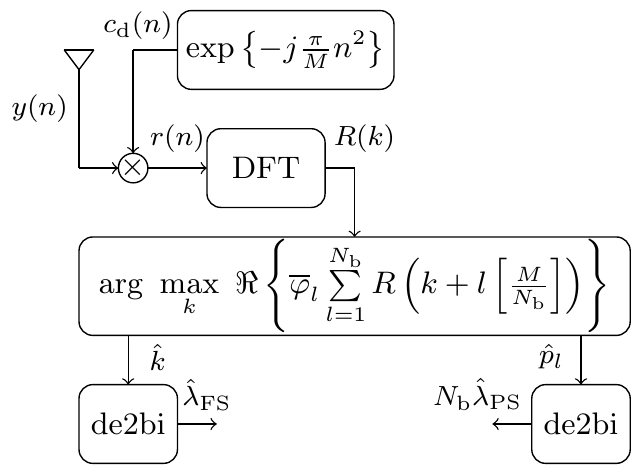}
  \caption{Coherent detector architecture for ePSK-CSS.}
\label{fig21rx}
\end{figure}
\textcolor{black}{Fig. \ref{fig21rx} portrays the coherent detection mechanism for the ePSK-CSS signal. The evaluation of the inner product between the received symbol, \( \boldsymbol{y}\), and the ePSK-CSS transmission symbol, represented by the vector,\(\boldsymbol{s}_\mathrm{ePSK} = \left[ s_\mathrm{ePSK}(0),  s_\mathrm{ePSK}(1), \cdots,  s_\mathrm{ePSK}(M-1)\right]^\mathrm{T}\), is performed as follows:}
\begin{equation}\label{eqr50}
\begin{split}
\langle \boldsymbol{y},\boldsymbol{s}_\mathrm{ePSK}\rangle&=\sum_{n=0}^{M-1}y(n) \overline{s}_\mathrm{ePSK}(n)\\
&=\sum_{n=0}^{M-1}\sum_{l=0}^{N_\mathrm{b}-1}y(n) \overline{f}(n) \overline{\varphi}_{l,p} c_\mathrm{d}(n)\\
&=\sum_{n=0}^{M-1}\sum_{l=0}^{N_\mathrm{b}-1}r(n) \overline{f}(n) \overline{\varphi}_{l,p} \\
&=\overline{\varphi}_{l,p}\sum_{l=0}^{N_\mathrm{b}-1} R\left(k+l\left[\frac{M}{N_\mathrm{b}}\right]\right)  ,
\end{split}
\end{equation}
where \(r(n) = y(n)c_\mathrm{d}(n)\), and \(R(k+l[\sfrac{M}{N_\mathrm{b}}]) \) is the DFT of \(r(n)\) evaluated at \(k+l(\sfrac{M}{N_\mathrm{b}})\) index. \textcolor{black}{Accordingly,} using coherent detection, the activated FF and the PS for the \(N_\mathrm{b}\) sub-bands can be jointly estimated as:
\begin{equation}\label{eqr51}
\hat{k},\hat{p}_l =\mathrm{arg}\max_{k,p}~\Re\left\{\overline{\varphi}_{l,p}\sum_{l=0}^{N_\mathrm{b}-1} R\left(k+l\left[\frac{N}{N_\mathrm{b}}\right]\right) \right\}.
\end{equation}

\textcolor{black}{As seen from Fig. \ref{fig21rx} and eq. (\ref{eqr51}), the received signal is first correlated with the down-chirp, \(c_\mathrm{d}(n)\), yielding, \(r(n)\), for which, DFT is evaluated resulting in \(R(k)\), before multiplying it with conjugate of all the possible PSs, \(\overline{\varphi}_{l,p}\). Finally the sum over \(N_\mathrm{b}\) is evaluated to determine \(\hat{k}\) and \(\hat{p}_l\) for all the sub-bands. \(\hat{k}\) and \(\hat{p}_l\) are then used to determine the bit sequences of lengths \(\hat{\lambda}_\mathrm{FS}\) and \(\hat{\lambda}_\mathrm{PS}\) after de2bi conversion.}
\begin{figure}[tb]\centering
\includegraphics[trim={0 0 0 0},clip,scale=1]{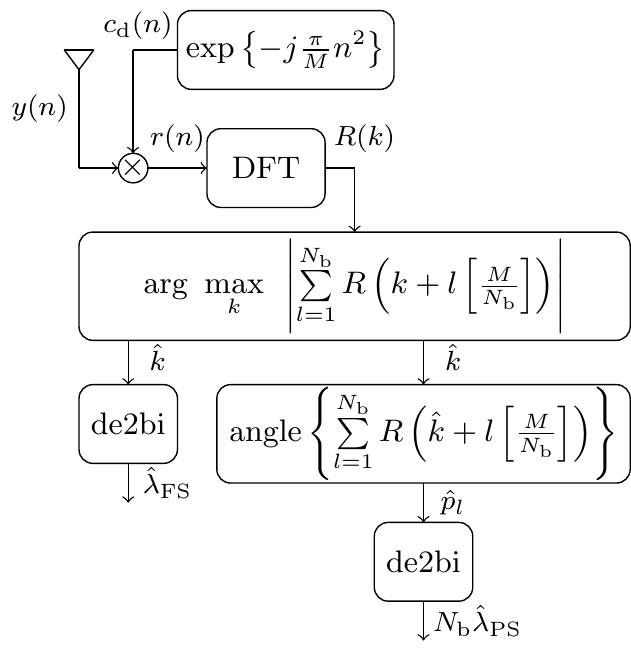}
  \caption{Semi-Coherent detector architecture for ePSK-CSS.}
\label{fig22rx}
\end{figure}

\textcolor{black}{The semi-coherent detector for ePSK-CSS is depicted in Figure \ref{fig22rx}. The semi-coherent detector for ePSK-CSS operates by performing a non-coherent evaluation of the FF and coherent determination of the PSs. The identification of the FF from the received ePSK-CSS symbol is carried out as follows:}
\begin{equation}\label{eqr52}
\hat{k}=  \mathrm{arg}\max_{k} ~\left\vert \sum_{l=0}^{N_\mathrm{b}-1} R\left(k+l\left[\frac{M}{N_\mathrm{b}}\right]\right) \right\vert.
\end{equation}

\textcolor{black}{It is important to observe that equation (\ref{eqr52}) only estimates the FF but does not determine the PSs for each sub-band. After obtaining the FF, the PS for each sub-band can be calculated by subjecting the value\(R(\hat{k}+l[\sfrac{M}{N_\mathrm{b}}])\) to decoding within the phase discriminator, as follows:}
\begin{equation}\label{eqr53}
\begin{split}
\hat{p}_l& =  \mathrm{angle}\left\{  R\left(\hat{k}+l\left[\frac{M}{N_\mathrm{b}}\right]\right)\right\},
\end{split}
\end{equation}
for \(l = 0, 1,\cdots, N_\mathrm{b}-1\), where \(\mathrm{angle}\{\cdot\}\) represents the phase discriminator.  The function of the phase discriminator is to identify the PS on \(R(\hat{k}+l[\sfrac{N}{N_\mathrm{b}}])\). It may be noticed that unlike ML detection, where the FS and PSs are detected jointly, in the semi-coherent detection, the identification of FS and PSs is disjoint. \textcolor{black}{As depicted in Figure \ref{fig22rx}, after evaluating  \(R(k)\), the first step is to identify the FF using equation (\ref{eqr52}), which computes the average of \(R(\hat{k}+l[\sfrac{N}{N_\mathrm{b}}])\) over \(N_\mathrm{b}\) sub-blocks to leverage the benefits of redundancy and evaluate the energy of the averaged symbol. Once the FF has been determined, the PSs on different sub-bands are then determined using equation (\ref{eqr53}). \(\hat{k}\) and \(\hat{p}_l\) are then used to determine the bit sequences of lengths \(\hat{\lambda}_\mathrm{FS}\) and \(\hat{\lambda}_\mathrm{PS}\) after de2bi conversion.}
\paragraph{\textcolor{black}{Takeaways}}
\textcolor{black}{The ePSK-CSS waveform design transforms the SC scheme into a MC scheme, which improves the robustness of the waveform design relative to PSK-LoRa.  ePSK-CSS activates one FF and its harmonics, thereby improving the detection of the FF, as multiple versions are available in the form of its harmonics. This improvement in the detection mechanism is likely to enhance the EE of the ePSK-CSS system. However, activating the FF and its harmonics and using different PSs may result in lower robustness in different offsets, particularly in the FO, where the FO alters each symbol independently. Additionally, like PSK-LoRa, ePSK-CSS is limited to using only coherent and semi-coherent detection, which may constrain the overall applicability of the scheme. The coherent detection process is also considerably complex, as it requires consideration of all possible combinations of the FSs and PSs to determine the activated FF and its harmonics and the used PSs. Moreover, to achieve maximum efficacy, it is crucial to undertake a comprehensive evaluation of the tradeoffs involved to ascertain the optimal number of harmonics of the FF and the PSs.}
\subsubsection{Group-based CSS (GCSS)}
\textcolor{black}{The GCSS scheme, as put forth in \cite{gcss}, bears a similarity with the ePSK-CSS in that the available FSs are partitioned into multiple groups, referred to as sub-bands in ePSK-CSS. In GCSS, one FS is selected from each group to be activated for the un-chirped symbol. Moreover, in GCSS, the trade-off between EE and SE can be altered using a configurable parameter, referred to as the group number, that determines the number of groups in which the available FSs are partitioned. A higher group number leads to an increase in SE, resulting in a decline of EE as a larger number of FSs are activated. Conversely, a lower group number results in a higher EE and lower SE.}
\begin{figure}[tb]\centering
\includegraphics[trim={0 0 0 0},clip,scale=1]{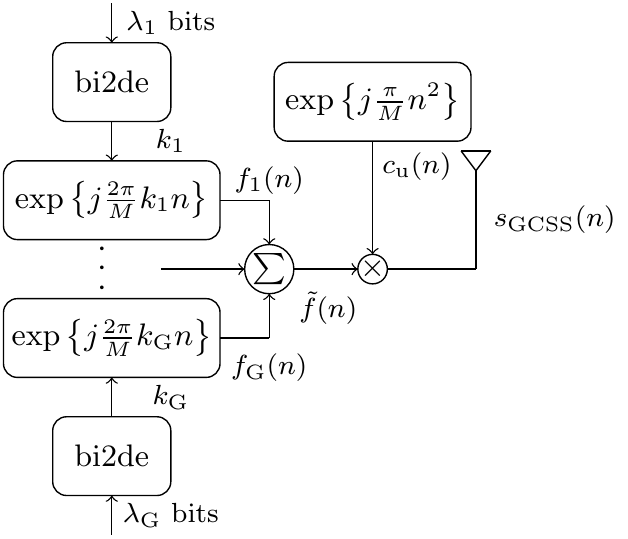}
  \caption{GCSS transmitter architecture. }
\label{fig13tx}
\end{figure}
\paragraph{Transmission}
\textcolor{black}{The transmitter of GCSS is illustrated in Fig. \ref{fig13tx}. The bandwidth \(B\) of \(M\) available FSs is subdivided into a set of \(\mathrm{G}\) sub-groups, so each sub-group consists of precisely \(\sfrac{M}{\mathrm{G}}\) FSs. The FSs within the gth sub-group can be identified via the indexes \(k_g \), which are expressed as \(k_g = \llbracket (g-1)(\sfrac{M}{\mathrm{G}}),g(\sfrac{M}{\mathrm{G}})-1 \rrbracket\), where the parameter \(g\) is defined over the range \(g = \llbracket 1, \mathrm{G}\rrbracket\). Additionally, the activation of the FS within the \(g\)th sub-group requires the utilization of \(\lambda_g = \log_2(\sfrac{M}{\mathrm{G}})\) bits of information. The activation of the FS within the first sub-group is performed by converting \(\lambda_1\)  bits into decimal representation, then using these decimal values to determine the specific FS index  \(k_1\) from the range \(k_1 = \llbracket 0, \sfrac{M}{\mathrm{G}}-1\rrbracket\). This activation process is similarly performed for the  \(\mathrm{G}\)th group with \(\lambda_\mathrm{G}\) bits, determining the FS index \(k_\mathrm{G}\) from the range \(k_\mathrm{G} = \llbracket (\mathrm{G}-1)(\sfrac{M}{G}), M-1\rrbracket\). Once all the FSs within the GG clusters have been determined, the corresponding un-chirped symbols \(f_{1}(n),f_{2}(n), \cdots, f_\mathrm{G}(n)\) can be obtained. The un-chirped symbol for the \(g\)th group is obtained as:}
\begin{equation}\label{eq49}
f_{g}(n) = \exp\left\{j \frac{2\pi}{M}k_g n\right\}.
\end{equation}

Subsequently, the un-chirped symbols from all the \(\mathrm{G}\) groups are added together resulting in:
\begin{equation}\label{eq50}
\tilde{f}(n) =\sum_{i=1}^\mathrm{G} f_i(n) =  \sum_{i=1}^\mathrm{G}\exp\left\{j \frac{2\pi}{M}k_i n\right\}.
\end{equation}

\(\tilde{f}(n) \) is then chirped using an up-chirp, \(c_\mathrm{u}(n)\) yielding a discrete-time GCSS symbol, \(s_\mathrm{GCSS}(n)\), that is given as:
\begin{equation}\label{eq51}
s_\mathrm{GCSS}(n) = \tilde{f}(n)c_\mathrm{u}(n)= \sum_{i=1}^\mathrm{G}\exp\left\{j \frac{\pi}{M}n\left(2k_i + n\right)\right\}.
\end{equation}

The symbol energy of GCSS is \(E_\mathrm{s}= \sfrac{1}{M}\sum_{n=0}^{M-1}\vert s_\mathrm{GCSS}(n) \vert^2 = \mathrm{G}\).
\paragraph{Detection}
\begin{figure}[tb]\centering
\includegraphics[trim={0 0 0 0},clip,scale=1]{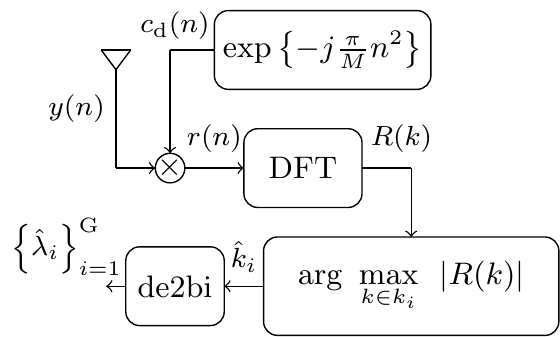}
  \caption{Non-Coherent detector architecture for GCSS.}
\label{fig23rx}
\end{figure}
The authors in \cite{gcss} only proposed a coherent detector for GCSS that is shown in Fig. \ref{fig23rx}. In the non-coherent detector, the FSs from all the groups are determined non-coherently as:
\begin{equation}\label{eqr54}
\hat{k}_i =  \mathrm{arg}\max_{k_i \in k}~ \left\vert R(k) \right\vert,
\end{equation}
where \(k_i = \llbracket (i-1)(\sfrac{M}{\mathrm{G}}),\cdots,i(\sfrac{M}{\mathrm{G}})-1\rrbracket\) for \(i = \llbracket 1, \mathrm{G}\rrbracket\). Eq. (\ref{eqr54}) implies that \(\mathrm{G}\) peaks having the highest amplitudes in the index range for each group identified by \(k_i\) needs to be evaluated. \textcolor{black}{Once \(\hat{k}_i\) are obtained for all the groups, we can also determine the respective bits of lengths \(\hat{\lambda}_i\) after de2bi conversion.}
\paragraph{\textcolor{black}{Takeaways}}
\textcolor{black}{GCSS presents a simple waveform design strategy by partitioning the available FSs into discrete groupings and then selectively activating a single FS from each group, which indicates the utilization of multiple FS activation techniques in its waveform design. This versatile technique facilitates various spectral and energy efficiencies based on the selected groupings, which positions GCSS as a fitting solution for many applications. It must be emphasized that only non-coherent detection has been analyzed in this study and the seminal GCSS work; notwithstanding, both coherent and non-coherent detection mechanisms for the GCSS scheme are feasible. Nevertheless, the coherent detection modus operandi is accompanied by a substantial degree of intricacy, which increases proportionally with an increase in the number of groups used. Therefore, thoughtful contemplation is required to identify the optimal number of groups to be implemented. On the other hand, non-coherent detection is straightforward, involving the calculation of only one DFT.}
\subsubsection{Time Domain Multiplex-CSS (TDM-CSS) }
\textcolor{black}{TDM-CSS, a member of the MC taxonomic classification of CSS schemes, leverages the concept of time domain multiplexing as a waveform design. The fundamental notion behind TDM-CSS is to concurrently embody up-chirp and down-chirp symbols, thereby yielding a TDM symbol \cite{tdm_lora}. Unlike SSK-LoRa, which utilizes different CRs to encode informational bits, TDM-CSS instead opts for multiplexing two LoRa waveforms with different CRs, i.e., one LoRa symbol using an up-chirp chirp and the other using a down-chirp symbol. It is noteworthy to emphasize that while SSK-LoRa exclusively employs a single FS, TDM-CSS, on the other hand, employs two distinct FSs as it multiplexes two LoRa symbols with different CRs. Subsequently, the resultant of the summation of the two chirped symbols produces the TDM-CSS symbol. The number of bits conveyed within the TDM-CSS symbol is double that of LoRa and is the same as that of IQ-CSS.}
\begin{figure}[tb]\centering
\includegraphics[trim={0 0 0 0},clip,scale=1]{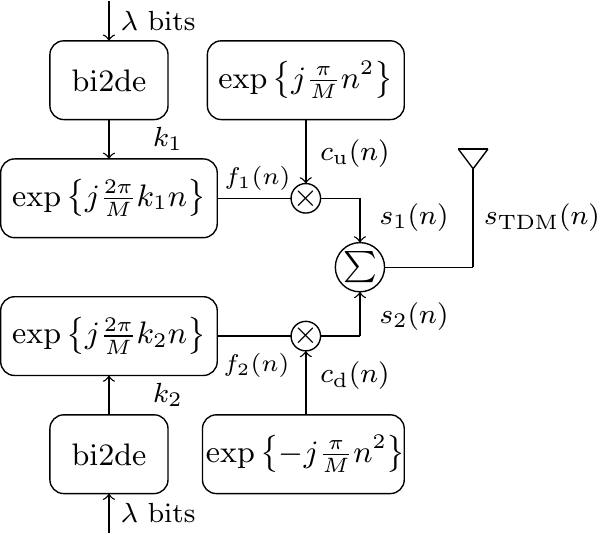}
  \caption{TDM-CSS transmitter architecture. }
\label{fig14tx}
\end{figure}
\paragraph{Transmission}
\textcolor{black}{The TDM-CSS transmitter configuration is depicted in Fig. \ref{fig14tx}. It encompasses two separate bit sequences, each comprised of a total of \(\lambda= \log_2(M)\) bits, that serve to determine two distinct FSs, \(k_1\) and \(k_2\). Upon determination of these FSs, the two un-chirped symbols, correspondingly aligned with the said FSs, are obtained as follows:}
\begin{equation}\label{eq53}
f_{1}(n)= \exp\left\{j\frac{2\pi}{M}k_1 n\right\},
\end{equation}
and 
\begin{equation}\label{eq54}
f_{2}(n)= \exp\left\{j\frac{2\pi}{M}k_2 n\right\},
\end{equation}
respectively. Subsequently, \(f_{1}(n)\) is multiplied with an up-chirp, \(c_\mathrm{u}(n)\), \(f_{2}(n)\) is multiplied with a down-chirp, \(c_\mathrm{d}(n)\) resulting in two distinct chirped symbols, \(s_{1}(n)\) and \(s_{2}(n)\) which in discrete-time are given as:
\begin{equation}\label{eq55}
s_{1}(n) = f_{1}(n)c_\mathrm{u}(n)= \exp\left\{j\frac{\pi}{M}n\left(2k_1+n\right)\right\},
\end{equation}
and 
\begin{equation}\label{eq56}
s_{2}(n) = f_{2}(n)c_\mathrm{d}(n)= \exp\left\{j\frac{\pi}{M}n\left(2k_2-n\right)\right\}.
\end{equation}

Lastly, the two chirped symbols, \(s_{1}(n)\) and \(s_{2}(n)\)  are added producing the TDM-CSS symbol, \(s_\mathrm{TDM}(n)\) that is given as:
\begin{equation}\label{eq57}
\begin{split}
s_\mathrm{TDM}(n)&= s_{1}(n)+s_{2}(n)\\
& = \exp\left\{j\frac{\pi}{M}n\left(2k_2-n\right)\right\} \!+\!\exp\left\{j\frac{\pi}{M}n\left(2k_2-n\right)\right\}.
\end{split}
\end{equation}

The symbol energy of TDM-CSS symbol is \(E_\mathrm{s}= \sfrac{1}{M}\sum_{n=0}^{M-1}\vert s_\mathrm{TDM}(n)\vert^2 = 2\).
\paragraph{Detection}
\textcolor{black}{In \cite{tdm_lora}, the authors have proposed the non-coherent detection for TDM-CSS; however, it is also feasible to implement the coherent detection mechanism. In light of this, we will thoroughly examine both the coherent and non-coherent detection modalities for TDM-CSS.}
\begin{figure}[tb]\centering
\includegraphics[trim={0 0 0 0},clip,scale=1]{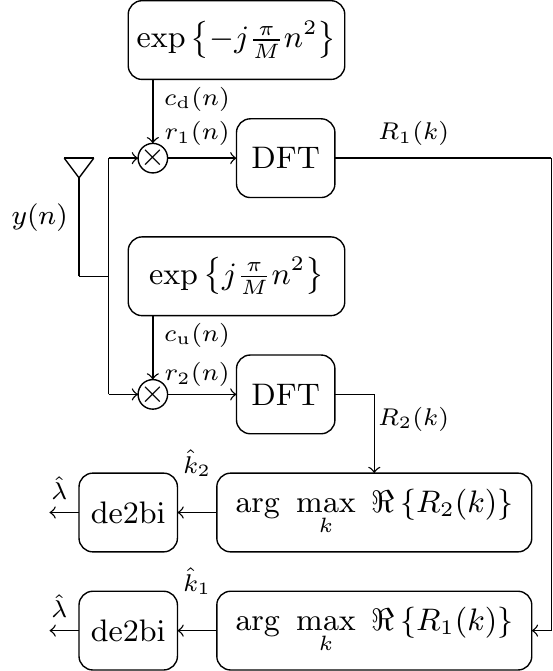}
  \caption{Coherent detector architecture for TDM-CSS.}
\label{fig24rx}
\end{figure}

The coherent detector architecture for TDM-CSS is shown in Fig. \ref{fig24rx}. The inner product of the received symbol, \(\boldsymbol{y}\) and the transmit symbol,  \(\boldsymbol{s}_\mathrm{TDM} = \left[ s_\mathrm{TDM}(0),  s_\mathrm{TDM}(1), \cdots,  s_\mathrm{TDM}(M-1)\right]^\mathrm{T}\), i.e., \(\langle \boldsymbol{y},\boldsymbol{s}_\mathrm{TDM} \rangle\) results in:
\begin{equation}\label{eqr55}
\begin{split}
\langle \boldsymbol{y},\boldsymbol{s}_\mathrm{TDM} \rangle & = \sum_{n=0}^{M-1}y(n)\overline{s}_\mathrm{TDM}(n)\\
&= \sum_{n=0}^{M-1}y(n)\left\{\overline{f}_1(n)c_\mathrm{d}(n) + \overline{f}_2(n)c_\mathrm{u}(n)\right\}\\
& = \sum_{n=0}^{M-1}r_1(n)\overline{f}_1(n) +\sum_{n=0}^{M-1}r_2(n)\overline{f}_2(n)  \\
&= R_1(k_1) + R_2(k_2),
\end{split}
\end{equation}
\textcolor{black}{where \(r_1(n) = y(n)c_\mathrm{d}(n)\) and \(r_2(n) = y(n)c_\mathrm{u}(n)\).}

\textcolor{black}{Given that \(k_1\) and \(k_2\) are elements belonging to \(k= \llbracket 0, M-1\rrbracket\), it follows that \(R_1(k_1)\) and \(R_2(k_2)\) can be re-expressed as \(R_1(k)\) and \(R_2(k)\), respectively. The next step involves the coherent identification of the FSs in the up-chirp stream and the down-chirp stream in a disjoint manner as follows:}
\begin{equation}\label{eqr56}
\hat{k}_1 = \mathrm{arg}\max\limits_{k_1\in k} ~\Re\left\{R_1(k)\right\},
\end{equation}
and
\begin{equation}\label{eqr57}
\hat{k}_2 = \mathrm{arg}\max\limits_{k_2\in k}~ \Re\left\{R_2(k)\right\},
\end{equation}
respectively. It may be noticed from Fig. \ref{fig24rx} that firstly the received symbol, \(y(n)\) is multiplied with both the down-chirp and the up-chirp, resulting in \(r_1(n)\) and \(r_2(n)\), respectively. Then DFT is evaluated for both \(r_1(n)\) and \(r_2(n)\) that yields \(R_1(k)\) and \(R_2(k)\), respectively. Subsequently, the FSs in the up-chirp stream and the down-chirp stream are identified using eqs. (\ref{eqr56}) and (\ref{eqr57}), respectively. \textcolor{black}{\(\hat{k}_1\) and \(\hat{k}_2\) then yield \(\hat{\lambda}_1\) and \(\hat{\lambda}_2\) after de2bi conversion.}

\begin{figure}[tb]\centering
\includegraphics[trim={0 0 0 0},clip,scale=1]{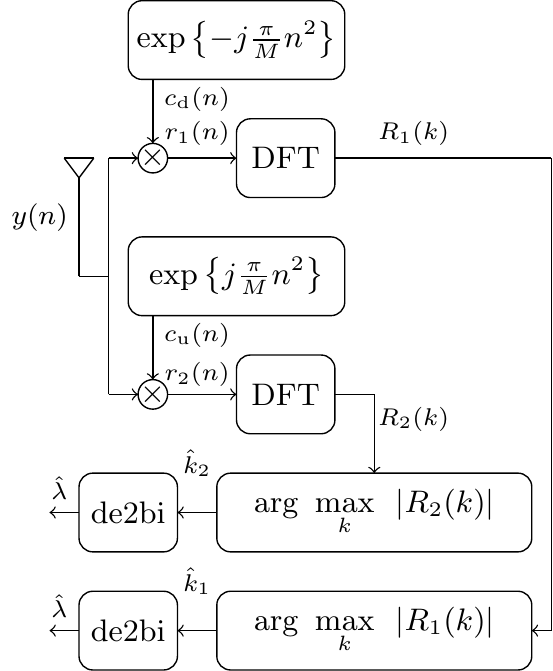}
  \caption{Non-Coherent detector architecture for TDM-CSS.}
\label{fig25rx}
\end{figure}

The non-coherent detector configuration for TDM-CSS is presented in Fig. \ref{fig25rx}. In the non-coherent detection, both the FSs in the up-chirp stream and the down-chirp stream are disjointly estimated as:
\begin{equation}\label{eqr58}
\hat{k}_1 = \mathrm{arg}\max\limits_{k_1\in k} ~\left\vert R_1(k)\right\vert,
\end{equation}
and
\begin{equation}\label{eqr59}
\hat{k}_2 = \mathrm{arg}\max\limits_{k_2\in k}~ \left\vert R_2(k)\right\vert,
\end{equation}
respectively. \textcolor{black}{The de2bi conversion of \(\hat{k}_1\) and \(\hat{k}_2\) result in \(\hat{\lambda}_1\) and \(\hat{\lambda}_2\), respectively.}
\paragraph{\textcolor{black}{Takeaways}}
\textcolor{black}{TDM presents a remarkably flexible design methodology for MC CSS schemes. It provides the capacity to multiplex several chirped symbols, which vary in rate, to generate highly efficient schemes. The TDM-CSS scheme is the fundamental application of the TDM design methodology, whereby two chirped symbols, each activated by a single FS, are multiplexed. Relative to LoRa, TDM-CSS doubles the number of bits transmitted per symbol and permits coherent and non-coherent detection of the received symbol. However, the complexity of both detection techniques surpasses that of LoRa, as two DFT computations and corresponding processing measures are necessary to identify the FSs in each of the multiplexed chirped symbols.}
\subsubsection{In-phase and Quadrature TDM-CSS}
\textcolor{black}{IQ-CSS operates by modulating the I/Q components of the un-chirped symbol using an upward chirp. In contrast, IQ-TDM-CSS implements the multiplexing of two IQ-CSS symbols, each with a differing CR. One of the IQ-CSS symbols is subjected to an up-chirp, while the other is subjected to a down-chirp. It is worth noting that the multiplexed symbols still utilize the I/Q components, which are identical to those utilized in the IQ-CSS symbol. As a result, the SE in IQ-TDM-CSS is double that of IQ-CSS and four times that of LoRa. However, the overall complexity of IQ-TDM-CSS is significantly greater than that of both LoRa and IQ-CSS.}
\begin{figure}[tb]\centering
\includegraphics[trim={0 0 0 0},clip,scale=1]{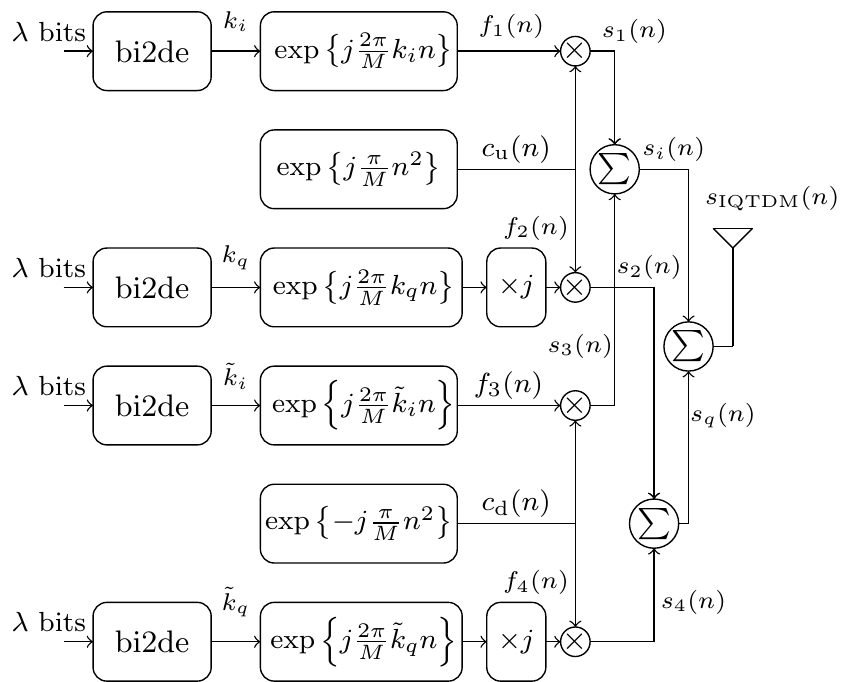}
  \caption{IQ-TDM-CSS transmitter architecture. }
\label{fig15tx}
\end{figure}
\paragraph{Transmission}
\textcolor{black}{The transmitter configuration of the IQ-TDM-CSS scheme is depicted in Fig. \ref{fig15tx}. An examination of the architecture reveals the utilization of four distinct bit sequences, each containing $\lambda$ bits. These sequences determine four FSs, which are denoted as $k_i$, $k_q$, $\tilde{k}_i$, and $\tilde{k}_q$. Of particular significance is the recognition that $k_i$ and $\tilde{k}_i$ serve as the FS for the determination of the in-phase un-chirped symbols, $f_1(n)$ and $f_3(n)$, following bi2de conversion. Conversely, $k_q$ and $\tilde{k}_q$ are utilized to ascertain the un-chirped quadrature symbols, $f_2(n)$ and $f_4(n)$. These in-phase and quadrature un-chirped symbols are given by:}
\begin{equation}\label{eq59}
f_1(n) = \exp\left\{j\frac{2\pi}{M}k_i n\right\},
\end{equation}
\begin{equation}\label{eq60}
f_2(n) = j\exp\left\{j\frac{2\pi}{M}\tilde{k}_i n\right\},
\end{equation}
\begin{equation}\label{eq61}
f_3(n) = \exp\left\{j\frac{2\pi}{M}k_q n\right\},
\end{equation}
and
\begin{equation}\label{eq62}
f_4(n) = j\exp\left\{j\frac{2\pi}{M}\tilde{k}_q n\right\},
\end{equation}
respectively. \textcolor{black}{We may observe} that a phase rotation of \(\sfrac{\pi}{2}\) is included in the quadrature un-chirped symbols. Subsequently, the in-phase un-chirped symbols, \(f_1(n)\) and \(f_3(n)\) are multiplied with an up-chirp and down-chirp symbols, respectively. The resulting symbols, \(s_1(n)\) and \(s_3(n)\) are given as:
\begin{equation}\label{eq63}
s_1(n) = \exp\left\{j\frac{\pi}{M}n\left(2k_i+n\right)\right\},
\end{equation}
and
\begin{equation}\label{eq64}
s_3(n) = \exp\left\{j\frac{\pi}{M}n\left(2\tilde{k}_i-n\right)\right\}.
\end{equation}

Similarly, the quadrature un-chirped symbols, \(f_2(n)\) and \(f_4(n)\) are \textcolor{black}{respectively} multiplied with up-chirp and down-chirp symbols, yielding \(s_2(n)\) and \(s_4(n)\) that are given as:
\begin{equation}\label{eq65}
s_2(n) = j\exp\left\{j\frac{\pi}{M}n\left(2k_q+n\right)\right\},
\end{equation}
and
\begin{equation}\label{eq66}
s_4(n) = j\exp\left\{j\frac{\pi}{M}n\left(2\tilde{k}_q-n\right)\right\}.
\end{equation}

Next, the in-phase up-chirped symbol and in-phase down-chirped symbols are added together to attain a consolidated in-phase chirped symbol, \(s_i(n)\), i.e.,: 
\begin{equation}\label{eq67}
\begin{split}
s_i(n) &= s_1(n) + s_3(n)\\
&= \exp\left\{j\frac{\pi}{M}n\left(2k_i+n\right)\right\} + \exp\left\{j\frac{\pi}{M}n\left(2\tilde{k}_i-n\right)\right\}.
\end{split}
\end{equation}

Similarly, the up-chirped and down-chirped quadrature symbols are added together to attain quadrature chirped symbol, \(s_q(n)\), \textcolor{black}{which} is given as: 
\begin{equation}\label{eq68}
\begin{split}
s_q(n) &= s_2(n) + s_4(n)\\
&= j\left[\exp\left\{j\frac{\pi}{M}n\left(2k_q+n\right)\right\} + \exp\left\{j\frac{\pi}{M}n\left(2\tilde{k}_q-n\right)\right\}\right].
\end{split}
\end{equation}

Lastly, the in-phase and the quadrature symbols, \(s_i(n) \) and \(s_q(n) \) are added together to attain the IQ-TDM-CSS symbol, \(s_\mathrm{IQTDM}(n)\), i.e., :
\begin{equation}\label{eq69}
\begin{split}
&s_\mathrm{IQTDM}(n) = s_i(n) + s_q(n)\\
&~~~= \exp\left\{j\frac{\pi}{M}n\left(2k_i+n\right)\right\} + j\exp\left\{j\frac{\pi}{M}n\left(2k_q+n\right)\right\}\\
&~~~~+\exp\left\{j\frac{\pi}{M}n\left(2\tilde{k}_i+n\right)\right\}+j\exp\left\{j\frac{\pi}{M}n\left(2\tilde{k}_q+n\right)\right\}.
\end{split}
\end{equation}

The symbol energy of IQ-TDM-CSS symbol is \(E_\mathrm{s}= \sfrac{1}{M}\sum_{n=0}^{M-1}\vert  s_\mathrm{IQTDM}(n) \vert^2 = 4\).
\paragraph{Detection}
\begin{figure}[tb]\centering
\includegraphics[trim={0 0 0 0},clip,scale=1]{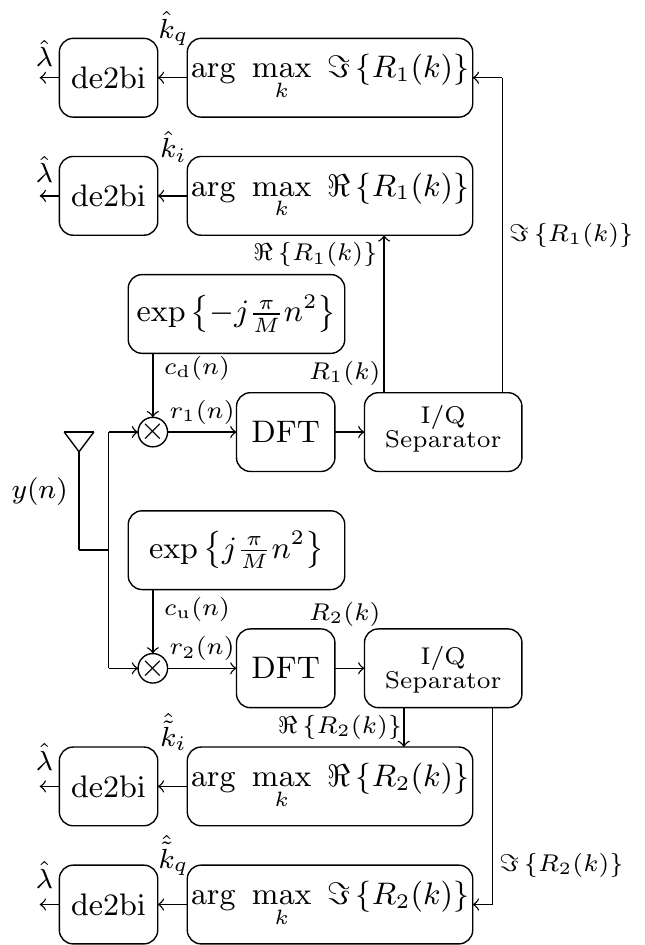}
  \caption{Coherent detector architecture for IQ-TDM-CSS.}
\label{fig26rx}
\end{figure}

The coherent detector architecture for IQ-TDM-CSS is presented in Fig.\ref{fig26rx}. The inner product of the received symbol vector, \(\boldsymbol{y}\) and the the transmit symbol vector, \(\boldsymbol{s}_\mathrm{IQTDM} = \left[s_\mathrm{IQTDM}(0),s_\mathrm{IQTDM}(1), \cdots, s_\mathrm{IQTDM}(M-1)\right]^\mathrm{T}\) is evaluated as:
\begin{equation}\label{eqr60}
\begin{split}
\langle \boldsymbol{y}, \boldsymbol{s}_\mathrm{IQTDM} \rangle & = \sum_{n=0}^{M-1}y(n)\overline{s}_\mathrm{IQTDM}(n)\\
&=  \sum_{n=0}^{M-1}y(n) \left\{\overline{f}_1(n)c_\mathrm{d}(n)+\overline{f}_2(n)c_\mathrm{d}(n)\right.\\
&\left.~~~~+ \overline{f}_3(n)c_\mathrm{u}(n) + \overline{f}_4(n)c_\mathrm{u}(n)\right\}\\
& = \sum_{n=0}^{M-1}r_1(n)\overline{f}_1(n) +  \sum_{n=0}^{M-1}r_1(n)\overline{f}_2(n)\\
&~~~~+ \sum_{n=0}^{M-1}r_2(n)\overline{f}_3(n) +  \sum_{n=0}^{M-1}r_2(n)\overline{f}_4(n)\\
& = R_1(k_i) -jR_1(k_q) +  R_2(\tilde{k}_i) -jR_2(\tilde{k}_q).
\end{split}
\end{equation}

Since \(k_i,k_q,\tilde{k}_i,\tilde{k}_q \in k\), where \(k= \llbracket 0, M-1\rrbracket\), consequently, \(R_1(k_i), R_1(k_q), R_2(\tilde{k}_i)\), and \(R_2(\tilde{k}_q)\) can be simply written as  \(R_1(k), R_1(k), R_2(k)\), and \(R_2(k)\). According to the coherent detection criterion and
according to eq. (\ref{eqr60}), the received symbol, \(y(n)\) is despreaded using both the down-chirp and the up-chirp yielding \(r_1(n)\) and \(r_2(n)\), respectively. Then, the DFT operation is performed on  \(r_1(n)\) and \(r_2(n)\) that results in \(R_1(k)\) and \(R_2(k)\), respectively. Then, the FS in the in-phase and quadrature components of the up-chirp stream and down-chirp stream is evaluated after I/Q separation performed by the I/Q separator as:
\begin{equation}\label{eqr61}
\hat{k}_i = \mathrm{arg}\max\limits_{k_i\in k}~ \Re\left\{R_1(k)\right\},
\end{equation}
\begin{equation}\label{eqr62}
\hat{k}_q = \mathrm{arg}\max\limits_{k_q\in k}~ \Im\left\{R_1(k)\right\},
\end{equation}
\begin{equation}\label{eqr63}
\hat{\tilde{k}}_i = \mathrm{arg}\max\limits_{\tilde{k}_i\in k}~ \Re\left\{R_2(k)\right\},
\end{equation}
and 
\begin{equation}\label{eqr64}
\hat{\tilde{k}}_q = \mathrm{arg}\max\limits_{\tilde{k}_q\in k}~ \Im\left\{R_2(k)\right\},
\end{equation}
respectively. Note that eqs. (\ref{eqr62}) and (\ref{eqr64}) follows from the fact that \(\Re\left\{-jx\right\} = \Im\left\{x\right\}\), where \(x\) is an arbitrary complex number. \textcolor{black}{\(\hat{k}_i\), \(\hat{k}_q\), \(\hat{\tilde{k}}_i\), and \(\hat{\tilde{k}}_q\) then determine the transmitted bit sequences.}
\paragraph{\textcolor{black}{Takeaways}}
\textcolor{black}{The IQ-TDM-CSS design methodology is a sophisticated technique that employs both time domain multiplexing and I/Q components to develop a waveform for the MC CSS scheme. The amalgamation of these design techniques has proven to yield a substantial increase in the number of bits transmitted per symbol, enhancing the system's overall efficiency. However, using I/Q components renders the scheme less robust against potential offsets that could arise from low-cost devices, constraining the scheme's scope. Additionally, the IQ-TDM-CSS design only permits coherent detection, as retrieving information from I/Q components is a complex process. The complexity of coherent detection is significantly complex, as separating I/Q components necessitates strict synchronization and the computation of two DFTs.}
\subsubsection{Dual-Mode CSS (DM-CSS)}
\textcolor{black}{DM-CSS is a MC CSS scheme that incorporates the utilization of multiple FSs, varies the chirp rate, and employs PSs to enhance the waveform design. This scheme is a refined variant of the DO-CSS, which utilizes only the even and odd FSs of the un-chirped symbol. In contrast, DM-CSS not only utilizes the even and odd subcarriers of the un-chirped symbol but also enables the chirping of the symbol through either an up-chirp or down-chirp, similar to SSK-LoRa. Furthermore, the even and odd FSs incorporate binary PS. Thus, \(3\) additional bits can be transmitted per symbol compared to the DO-CSS.}
\begin{figure}[tb]\centering
\includegraphics[trim={0 0 0 0},clip,scale=1]{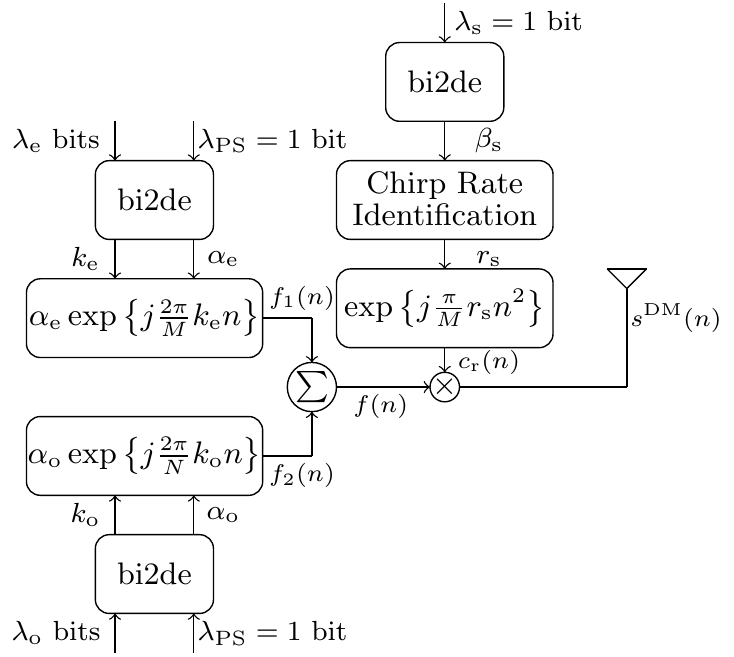}
  \caption{DM-CSS transmitter architecture. }
\label{fig16tx}
\end{figure}
\paragraph{Transmission}
\textcolor{black}{The configuration of the transmitter for DM-CSS is depicted in Fig. \ref{fig16tx}. Given that a total of \(M\) FSs are available within a bandwidth of \(B\), it is established that half of these FSs are even, and the remaining half are odd. Thus, the number of bits utilized to determine the activation of even FS, i.e., \(k_\mathrm{e}\), is \(\lambda_\mathrm{e}= \log_2(\sfrac{M}{2})\). Similarly, the number of bits required to activate the odd FS, i.e., \(k_\mathrm{o}\), is \(\lambda_\mathrm{o}= \log_2(\sfrac{M}{2})\). Additionally, the PS of the activated FSs, either even or odd, can take on values of either \(0\) or \(\pi\) radians. This is achieved by employing  \(\lambda_\mathrm{PS}=2\) bits for both even and odd FSs. The PS of the even FS is represented by \(\alpha_\mathrm{e}\), while that of the odd FS is \(\alpha_\mathrm{o}\). The PS for both even and odd FSs is mathematically represented as:}
\begin{equation}\label{eq70}
\alpha_\mathrm{e}/\alpha_\mathrm{o}= \begin{dcases}
   1 & \quad \lambda_\mathrm{PS} = 0\\
 -1& \quad \lambda_\mathrm{PS} = 1
\end{dcases}.
\end{equation}

Using \(k_\mathrm{e}\) and \(k_\mathrm{o}\), the un-chirped symbols corresponding to even and odd FSs are attained by:
\begin{equation}\label{eq71}
f_1(n) = \alpha_\mathrm{e}\exp\left\{j\frac{2\pi}{M}k_\mathrm{e}n\right\},
\end{equation}
and 
\begin{equation}\label{eq72}
f_2(n) = \alpha_\mathrm{o}\exp\left\{j\frac{2\pi}{M}k_\mathrm{o}n\right\},
\end{equation}
respectively. The consolidated un-chirped symbol, \(f(n)\) is attained by adding \(f_1(n)\) and \(f_2(n)\) as:
\begin{equation}\label{eq73}
\begin{split}
f(n) &= f_1(n) + f_2(n)\\
&= \alpha_\mathrm{e}\exp\left\{j\frac{2\pi}{N}k_\mathrm{e}n\right\}+\alpha_\mathrm{o}\exp\left\{j\frac{2\pi}{N}k_\mathrm{o}n\right\}.
\end{split}
\end{equation}

Afterward, using the same principle as in eqs. (\ref{eq19}) and (\ref{eq20}), \(\lambda_\mathrm{s} = 1\) bit determines \(\beta_\mathrm{s}\), which in turn determines whether \(f(n)\) will be chirped using an up-chirp or a down-chirp.  If \(f(n)\) is chirped with the up-chirp symbol, i.e., \(\gamma_\mathrm{s} = 1\) and \(c_\mathrm{s}(n) = c_\mathrm{u}(n)\), then, the chirped DM-CSS symbol is given as:
\begin{equation}\label{eq74}
\begin{split}
&s_\mathrm{DM}(n)=f(n)c_\mathrm{u}(n)\\
&~~~=\alpha_\mathrm{e}\exp\left\{j\frac{\pi}{N}n \left(2k_\mathrm{e}+n\right)\right\}+\alpha_\mathrm{o}\exp\left\{j\frac{\pi}{N}n \left(2k_\mathrm{o}+n\right)\right\}.
\end{split}
\end{equation}

On the contrary, \(f(n)\) will be chirped with the down-chirp symbol when \(\gamma_\mathrm{s} = -1\), i.e., \(c_\mathrm{s}(n) = c_\mathrm{d}(n)\),  then, the chirped symbol DM-CSS symbol is given as:
\begin{equation}\label{eq75}
\begin{split}
&s_\mathrm{DM}(n)=f(n)c_\mathrm{d}(n)\\
&~~~=\alpha_\mathrm{e}\exp\left\{j\frac{2\pi}{N}n \left(k_\mathrm{e}-n\right)\right\}+\alpha_\mathrm{o}\exp\left\{j\frac{2\pi}{N}n \left(k_\mathrm{o}-n\right)\right\}.
\end{split}
\end{equation}

The symbol energy, \(E_\mathrm{s}\) of \(s_\mathrm{DM}(n)\) irrespective of whether \(s(n)\) is chirped with an up-chirp or a down-chirp is \(E_\mathrm{s}=\sfrac{1}{N}\sum_{n=0}^{N-1}\vert s_\mathrm{DM}(n)\vert^2=2\).
\paragraph{Detection}
\textcolor{black}{In the seminal work presented in \cite{dm_css}, we limited our scope to the non-coherent detection mechanism for the DM-CSS. Nevertheless, it must be noted that implementing a coherent detection approach for the DM-CSS is also a viable option and shall be elucidated in the subsequent discourse.}
\begin{figure}[tb]\centering
\includegraphics[trim={0 0 0 0},clip,scale=1]{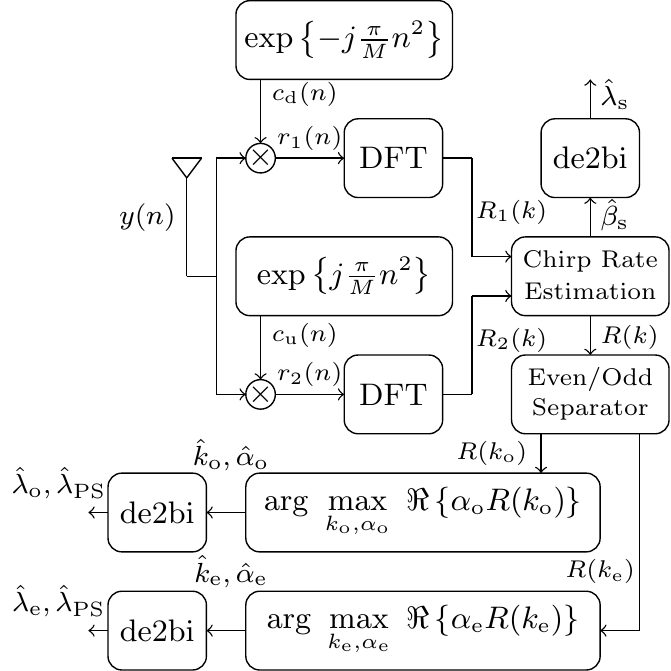}
  \caption{Coherent detector architecture for DM-CSS.}
\label{fig27rx}
\end{figure}

The coherent detector architecture for DM-CSS is illustrated in Fig. \ref{fig27rx}. The inner product of the received symbol vector, \(\boldsymbol{y}\) and the transmit symbol vector, \(\boldsymbol{s}_\mathrm{DM}= \left[{s}_\mathrm{DM}(0), {s}_\mathrm{DM}(1), \cdots, {s}_\mathrm{DM}(M-1)\right]^\mathrm{T}\) will yield two different results \textcolor{black}{given that either an up-chirp symbol or a down-chirp symbol can be employed for spreading the un-chirped symbol at the transmitter.} 

The inner product, \(\langle \boldsymbol{y},\boldsymbol{s}_\mathrm{DM}  \rangle\) when an up-chirp is used at the transmitter is given as:
\begin{equation}\label{eqr65}
\begin{split}
\langle \boldsymbol{y}, \boldsymbol{s}_\mathrm{DM} \rangle &= \sum_{n=0}^{M-1}y(n)\overline{s}_\mathrm{DM}(n)\\
&= \sum_{n=0}^{M-1}y(n)\overline{f}(n)c_\mathrm{d}(n) \\
&= \sum_{n=0}^{M-1}r_1(n)\overline{f}(n)\\
& = \alpha_\mathrm{e}R_1(k_\mathrm{e}) + \alpha_\mathrm{o}R_1(k_\mathrm{o}),
\end{split}
\end{equation}
\textcolor{black}{where \(r_1(n) = y(n)c_\mathrm{d}(n)\).}

On the other hand, when down-chirp is used at the transmitter, then, the inner product, \(\langle \boldsymbol{y},\boldsymbol{s}_\mathrm{DM}  \rangle\) yields:
\begin{equation}\label{eqr66}
\begin{split}
\langle \boldsymbol{y}, \boldsymbol{s}_\mathrm{DM} \rangle &= \sum_{n=0}^{M-1}y(n)\overline{s}_\mathrm{DM}(n)\\
&= \sum_{n=0}^{M-1}y(n)\overline{f}(n)c_\mathrm{u}(n) \\
&= \sum_{n=0}^{M-1}r_2(n)\overline{f}(n)\\
& = \alpha_\mathrm{e}R_2(k_\mathrm{e}) + \alpha_\mathrm{o}R_2(k_\mathrm{o}),
\end{split}
\end{equation}
\textcolor{black}{where \(r_2(n) = y(n)c_\mathrm{u}(n)\).}

\textcolor{black}{The mathematical relationship between the even and odd FSs can be succinctly expressed as} \(R_1(k_\mathrm{e}) + R_1(k_\mathrm{o}) = R_1(k)\) and \(R_2(k_\mathrm{e}) + R_2(k_\mathrm{o}) = R_2(k)\), where \(k= \llbracket 0, M-1\rrbracket\), \textcolor{black}{\(R_1(k)\) and \(R_2(k)\) are the DFT outputs of \(r_1(n)\) and \(r_2(n)\), respectively}. The first step is to determine whether an up-chirp or a down-chirp was used at the transmitter for spreading the un-chirped symbol. In order to do so, firstly, \(\kappa_1^\mathrm{coh} = \max\{\Re\left\{R_1(k)\right\}\}\) and \(\kappa_2^\mathrm{coh} = \max\{\Re\left\{R_2(k)\right\}\}\) are ascertained, which leads to the estimation of the CR as:
\begin{equation}\label{eqr67}
\hat{\beta}_\mathrm{s} = \begin{dcases}
   0 & \quad \kappa_1^\mathrm{coh}>\kappa_2^\mathrm{coh}\\
     1 & \quad \kappa_1^\mathrm{coh}<\kappa_2^\mathrm{coh}
   \end{dcases},
\end{equation}
using \(\hat{\beta}_\mathrm{s}\), \(\hat{\lambda}_\mathrm{s}\) is attained after de2bi conversion. 

\(\kappa_1^\mathrm{coh}\) and \(\kappa_2^\mathrm{coh}\) also help determine whether \(R_1(k)\) or \(R_2(k)\) contain the largest peak and should be used for further processing as:
\begin{equation}\label{eqr68}
R(k) = \begin{dcases}
   R_1(k)& \quad \kappa_1^\mathrm{coh}>\kappa_2^\mathrm{coh}\\
     R_2(k) & \quad \kappa_1^\mathrm{coh}<\kappa_2^\mathrm{coh}
   \end{dcases}.
\end{equation}

Subsequently, the even and the odd FSs and the PSs on the even and odd FSs are determined jointly in a coherent manner after the separation of even and odd frequency indexes using even/odd separator as:
\begin{equation}\label{eqr69}
\hat{k}_\mathrm{e},\hat{\alpha}_\mathrm{e} = \mathrm{arg}\max\limits_{k_\mathrm{e}, \alpha_\mathrm{e}}~ \Re\left\{\alpha_\mathrm{e}R(k_\mathrm{e})\right\},
\end{equation}
and
\begin{equation}\label{eqr70}
\hat{k}_\mathrm{o},\hat{\alpha}_\mathrm{o} = \mathrm{arg}\max\limits_{k_\mathrm{o}, \alpha_\mathrm{o}}~ \Re\left\{\alpha_\mathrm{o}R(k_\mathrm{o})\right\},
\end{equation}
respectively. Using \(\hat{k}_\mathrm{e}\) and \(\hat{k}_\mathrm{o}\), \(\hat{\lambda}_\mathrm{e}\) and \(\hat{\lambda}_\mathrm{o}\) are ascertained using de2bi conversion.
\begin{figure}[tb]\centering
\includegraphics[trim={0 0 0 0},clip,scale=1]{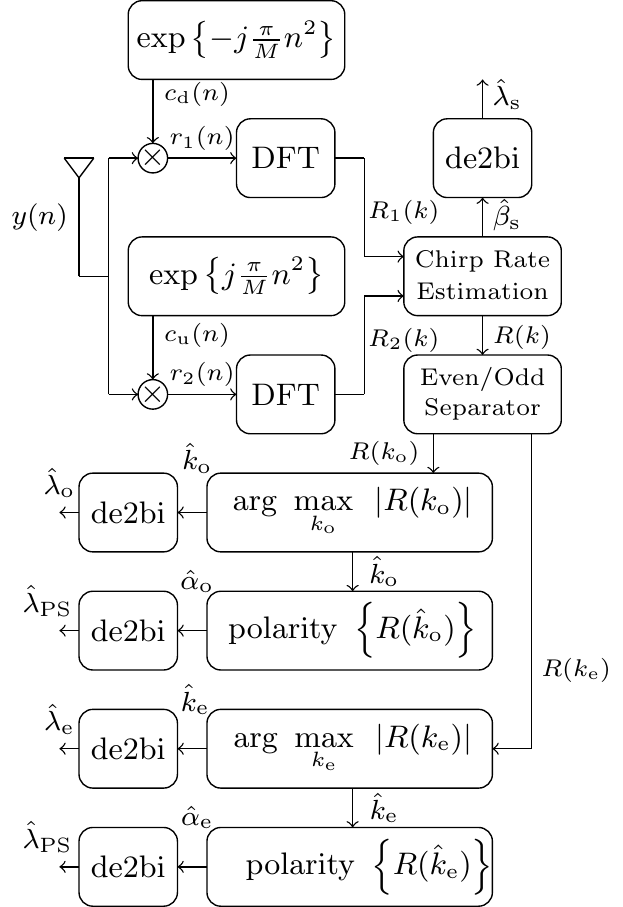}
  \caption{Non-coherent detector architecture for DM-CSS.}
\label{fig28rx}
\end{figure}

The non-coherent detector configuration for DM-CSS is provided in Fig. \ref{fig28rx}. In the non-coherent detector, the CR is identified using \(\kappa_1^\mathrm{non-ch} =\max\vert R_1(k)\vert \) and \(\kappa_2^\mathrm{non-coh} = \max \vert R_2(k)\vert\). Moreover, the activated even and odd FSs are estimated as:
\begin{equation}\label{eqa1}
\hat{k}_\mathrm{e}= \mathrm{arg}\max_{k_\mathrm{e}}~\left \vert R(k_\mathrm{e}) \right\vert,
\end{equation}
and 
\begin{equation}\label{eqa2}
\hat{k}_\mathrm{o}= \mathrm{arg}\max_{k_\mathrm{o}}~~\left \vert R(k_\mathrm{o}) \right\vert,
\end{equation}
respectively. Subsequently, the PSs of the even and odd frequencies are respectively evaluated by determining the polarity \(R(\hat{k}_\mathrm{e})\) and \(R(\hat{k}_\mathrm{o})\) as:
\begin{equation}\label{eqa3}
\hat{\alpha}_\mathrm{e}= \mathrm{polarity}\left\{R(\hat{k}_\mathrm{e})\right\},
\end{equation}
and 
\begin{equation}\label{eqa4}
\hat{\alpha}_\mathrm{o}= \mathrm{polarity}\left\{R(\hat{k}_\mathrm{o})\right\},
\end{equation}
where \(\mathrm{polarity}\{\cdot\}\) is the ML criterion to determine the polarity of the input frequency index. \textcolor{black}{Note that \(\mathrm{polarity}\{\cdot\}\) can only be used for binary PSs.} If the output of \(\mathrm{polarity}\{\cdot\} > 0\), then \(\hat{\alpha}_\mathrm{e}/\hat{\alpha}_\mathrm{o} = 1\), conversely, \(\hat{\alpha}_\mathrm{e}/\hat{\alpha}_\mathrm{o} = -1\). It is accentuated that if the channel's phase rotation is \(\geq \sfrac{\pi}{2}\), then, only coherent detection would be viable because the \(\mathrm{polarity}\{\cdot\}\) function will no longer be applicable.
\paragraph{\textcolor{black}{Takeaways}}
\textcolor{black}{DM-CSS is a MC CSS scheme that employs numerous FSs and PSs to enhance the number of transmitted bits per symbol. To achieve this, multiple FSs are activated, following a similar principle to that used in DO-CSS. In addition, DM-CSS solely employs binary PSs, obviating the need for ML detection, which can be attained simply by determining the polarity of the activated FS index. Thus, correctly identifying the activated FS is paramount, as an incorrect determination could cascade into a series of errors. The scheme can utilize coherent and semi-coherent detection mechanisms, the latter term employed due to assessing the binary PS. However, this differs from the semi-coherent detections used for PSK-LoRa or ePSK-LoRa, as the ML angle operation is not required. Instead, a polarity operation suffices, which is significantly less complex and straightforward compared to the former procedure. Additionally, only one DFT operation is necessary to determine the FS. The discrete domain time DM-CSS symbols are also orthogonal. Nonetheless, utilizing multiple FSs may increase the PAPR based on the activated FSs, potentially posing implementation issues like any other MC scheme.}
\subsection{\textcolor{black}{Multiple Carrier with Index Modulation CSS Schemes}}

\subsubsection{FSCSS with Index Modulation (FSCSS-IM)}
\textcolor{black}{The FSCSS-IM approach was proposed by Hanif and Nguyen in a seminal work \cite{fscss_im}. This scheme represents the pioneering effort in integrating IM with the CSS technique. In FSCSS-IM, multiple FSs are concurrently activated for the un-chirped symbol based on a predetermined pattern, referred to as FAP in this study. It is emphasized that in the FSCSS-IM approach, the information bits are transmitted in the FAP instead of the activated frequency subcarriers of the un-chirped symbol or the cyclic-time shift of the chirped symbol, as observed in the LoRa scheme. It is worth mentioning that the CIM approach presented in \cite{iqcim} is equivalent to the FSCSS-IM method.
FSCSS-IM has the potential to transmit a higher number of bits per symbol, owing to the immense number of possible FAPs. However, it is noted that two to three FSs in the FAP are typically activated, as activating a more significant number of FSs may degrade the energy efficiency and significantly increase the system complexity. The CSS approach in which all the chirps are multiplexed is called OCDM, which has been investigated in \cite{ocdm}. Conversely, the amalgamation of IM with OCDM results in OCDM-IM, described in \cite{ocdm_im}.}

\begin{figure}[tb]\centering
\includegraphics[trim={0 0 0 0},clip,scale=1]{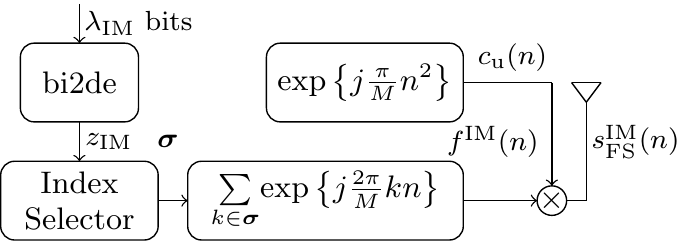}
  \caption{FSCSS-IM transmitter architecture. }
\label{fig9tx}
\end{figure}
\paragraph{Transmission}
\textcolor{black}{The transmitter configuration FSCSS-IM is depicted in Figure \ref{fig9tx}.  Notably, a maximum of \(M\) FSs can be selectively enabled. The set of these available FSs is denoted as \(\boldsymbol{\Omega} = \left\{0,1,\cdots, M-1\right\}\). Suppose that a total of \(\varsigma\) FSs are to be activated. Then, the bi2de conversion of\(\lambda_\mathrm{IM} = \lfloor \log_2 \binom{M}{\varsigma}\rfloor\) bits results in \(z_\mathrm{IM} = \llbracket 0, 2^{\lambda_\mathrm{IM}}-1 \rrbracket\), which determines the set of FSs to be enabled, i.e., the FAP, \(\boldsymbol{\sigma}= \{\sigma_1, \sigma_2,\cdots, \sigma_\varsigma\}\in \boldsymbol{\Omega}\).  The mapping of  \(z_\mathrm{IM}\) to a unique \(\boldsymbol{\sigma}\) is accomplished through the utilization of different combinatorial algorithms, as detailed in \cite{mapping}.  The set \(\boldsymbol{\sigma}\)comprises all the FSs that are to be enabled, i.e., \(\boldsymbol{\sigma} = \left\{k_1, k_2,\cdots, k_\varsigma \right\}\). Subsequently, the un-chirped symbol is obtained as:}
\begin{equation}\label{eq34}
f^\mathrm{IM}(n) = \sum_{k\in \boldsymbol{\sigma}}\exp\left\{j\frac{2\pi}{M}k n\right\}.
\end{equation}

\textcolor{black}{\(f^\mathrm{IM}(n)\) comprising of multiple activated FSs according to a selected FAP is subsequently  chirped} using an up-chirp symbol as:
\begin{equation}\label{eq35}
s_\mathrm{FS}^\mathrm{IM}(n) = f^\mathrm{IM}(n)c_\mathrm{u}(n)=\sum_{k\in \boldsymbol{\sigma}}\exp\left\{j\frac{2\pi}{M}n(k+n)\right\}.
\end{equation}
The symbol energy of FSCSS-IM symbol is equal to \(E_\mathrm{s} = \sfrac{1}{M}\sum_{n=0}^{M-1}\vert s_\mathrm{FS}^\mathrm{IM}(n) \vert^2 = \varsigma\). 
\paragraph{Detection}
\begin{figure}[tb]\centering
\includegraphics[trim={0 0 0 0},clip,scale=1]{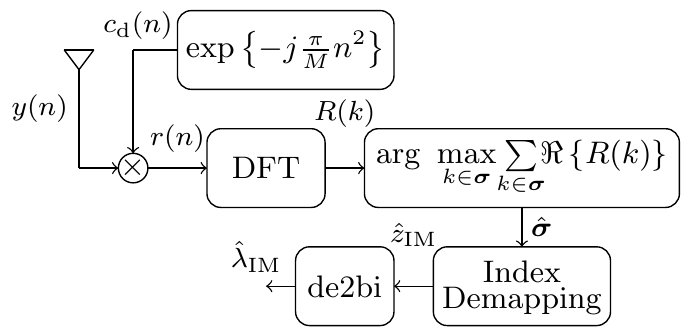}
  \caption{Coherent detector architecture for FSCSS-IM.}
\label{fig16rx}
\end{figure}
The coherent detector for FSCSS-IM is presented in Fig. \ref{fig16rx}. The inner product of \(\boldsymbol{y}\) and the FSCSS-IM transmit symbol \(\boldsymbol{s}_\mathrm{FS}^\mathrm{IM} = \left[{s}_\mathrm{FS}^\mathrm{IM}(0), {s}_\mathrm{FS}^\mathrm{IM}(1), \cdots, {s}_\mathrm{FS}^\mathrm{IM}(M-1)\right]^\mathrm{T}\), i.e., \(\langle   \boldsymbol{y}, \boldsymbol{s}_\mathrm{FS}^\mathrm{IM}\rangle\) results in:
\begin{equation}\label{eqr39}
\begin{split}
\langle   \boldsymbol{y}, \boldsymbol{s}_\mathrm{FS}^\mathrm{IM}\rangle &= \sum_{n=0}^{M-1}y(n)s_\mathrm{FS}^\mathrm{IM}\\
&=\sum_{n=0}^{M-1}\sum_{k\in \boldsymbol{\sigma}}y(n)\overline{f}^\mathrm{IM}(n)c_\mathrm{d}(n)\\
&= \sum_{n=0}^{M-1}\sum_{k\in \boldsymbol{\sigma}}r(n)\overline{f}^\mathrm{IM}(n) \\
&= \sum_{k\in \boldsymbol{\sigma}} R(k).
\end{split}
\end{equation}

Thus, using eq. (\ref{eqr39}) and the coherent detection principle (cf. eq. \ref{cd3}), the \(\varsigma\) FSs, i.e., FAP is determined considering the indexes of \(\varsigma\) highest peaks in the real components of \( R(k)\) as:
\begin{equation}\label{eqr40}
\hat{\boldsymbol{\sigma}} = \mathrm{arg} \max\limits_{k \in \boldsymbol{\sigma}} ~\Re\left\{\sum_{k\in \boldsymbol{\sigma}} R(k)\right\}.
\end{equation}

The coherent detection configuration shows that after de-chirping the received symbol \(y(n)\) using a down-chirp, \(c_\mathrm{d}(n)\), \(r(n)\) is obtained. The latter is then fed to the DFT, which yields \(R(k)\). Subsequently, \(\varsigma\) highest peaks in the real part of \(R(k)\) are determined using the decision criterion as in eq. (\ref{eqr32}) which results in \(\hat{\boldsymbol{\sigma}} \), using which the transmitted bits, \(\hat{\lambda}_\mathrm{IM}\) are determined after de2bi conversion of output of index demapping algorithm, i.e., \(\hat{z}_\mathrm{IM}\), \textcolor{black}{which in turn will yield \(\hat{\lambda}_\mathrm{IM}\) bits.}
\begin{figure}[tb]\centering
\includegraphics[trim={0 0 0 0},clip,scale=1]{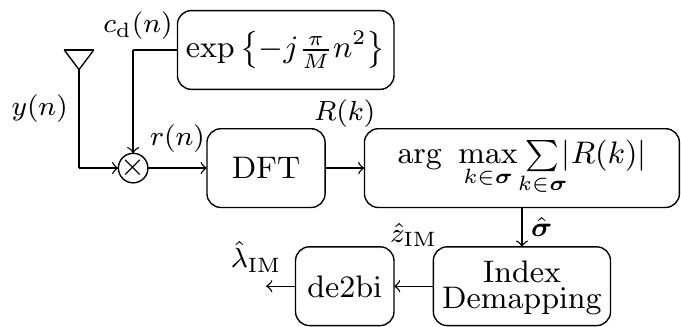}
  \caption{Non-coherent detector architecture for FSCSS-IM.}
\label{fig17rx}
\end{figure}

The non-coherent detector for FSCSS-IM is illustrated in Fig. \ref{fig17rx}. \(\hat{\boldsymbol{\sigma}}\) for FSCSS-IM is evaluated non-coherently as:
\begin{equation}\label{eqr41}
\hat{\boldsymbol{\sigma}} = \mathrm{arg} \max\limits_{k \in \boldsymbol{\sigma}} \left\vert \sum_{k\in \boldsymbol{\sigma}} R(k)\right\vert,
\end{equation}
which implies the the \(\varsigma\) peaks are determined considering both the real and imaginary components of \(R(k)\). Subsequently, the integer corresponding to the \(\hat{\boldsymbol{\sigma}}\), \(\hat{z}_\mathrm{IM}\) is determined that leads to the determination of the bit sequence having \(\hat{\lambda}_\mathrm{IM}\) bits.
\paragraph{\textcolor{black}{Takeaways}}
\textcolor{black}{The FSCSS-IM scheme is a pioneering proposal for IM-based MC CSS designed specifically for LPWANs. This innovative scheme effectively harnesses multiple FS through the IM percept, enabling a multitude of FAPs to be acquired using various combinatorial algorithms. The resultant versatility makes it ideal not only for LPWANs but also for applications requiring high data rates. The number of bits transmitted per symbol is determined by the available FAPs, which, in turn, depend on the number of activated FSs and the total number of FSs to be activated. However, it should be noted that activating multiple FSs, like any other MC scheme, amplifies the PAPR, causing some implementation problems. FSCSS-IM also allows for both coherent and non-coherent detection. Coherent detection, however, can be very complicated, depending on the number of available FAPs. Non-coherent detection, on the other hand, is more straightforward as it does not need to consider the possible FAPs. It is vital to keep the number of FSs in the FAP low for low-power applications in LPWANs because activating a higher number of FSs in the FAP would increase the number of bits transmitted per symbol while negatively impacting the EE.}
\subsubsection{In-phase and Quadrature Chirp Index Modulation (IQ-CIM)}
\textcolor{black}{The FSCSS-IM system, described in \cite{fscss_im}, integrates IM with CSS and imparts flexibility to the scheme. Another low-power approach that leverages IM is IQ-CIM, which is presented in \cite{iqcim}. IQ-CIM constitutes an integration of IQ-CSS and IM, offering more flexible modulation than FSCSS-IM at the cost of increased computational demands. Compared to FSCSS-IM, which uses IM only for in-phase un-chirped symbols, IQ-CSS implements IM for both in-phase and quadrature components of the un-chirped symbols. Essentially, the principle behind IQ-CIM involves employing two separate FSCSS-IM branches, one for each of the in-phase and quadrature components. This necessitates two unique FAPs, which can transmit information. As a result, when the number of activated FSs utilized in both FAPs is equal, IQ-CIM can transmit double the number of bits relative to FSCSS-IM.}
\paragraph{Transmission} 
\begin{figure}[tb]\centering
\includegraphics[trim={0 0 0 0},clip,scale=1]{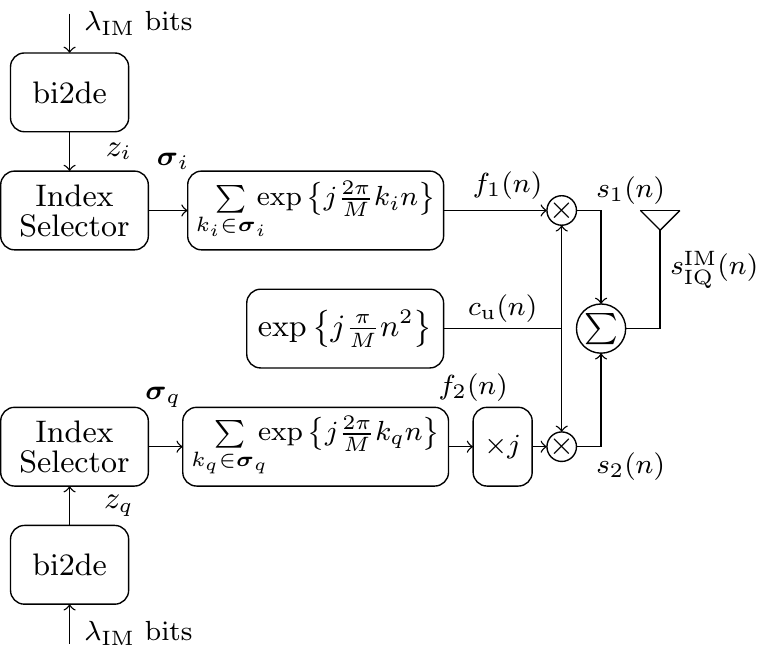}
  \caption{IQ-CIM transmitter architecture. }
\label{fig10tx}
\end{figure}

\textcolor{black}{The architecture of the IQ-CIM transmitter is depicted in Fig. \ref{fig10tx}. Following the bi2de conversion of  \(\lambda_\mathrm{IM} = \lfloor \log_2 \binom{M}{\varsigma_i} \rfloor\) and \(\lambda_\mathrm{IM}= \lfloor \log_2 \binom{M}{\varsigma_q} \rfloor\) bits, the corresponding integers of\(z_i\) and \(z_q\)  are computed,  where \(\varsigma_i\) and \(\varsigma_q\) represent the number of activated FSs for the in-phase and quadrature un-chirped symbols, respectively. The index selector block inputs these values to obtain the distinct FAPs, represented by  \(\boldsymbol{\sigma}_i \in \boldsymbol{\Omega}\) and \(\boldsymbol{\sigma}_q\in \boldsymbol{\Omega}\), where  \(\boldsymbol{\Omega} = \{0,1,\cdots, M-1\}\) is the set of all possible FSs.
Assuming that \(\varsigma_i = \varsigma_q = \varsigma \leq \sfrac{M}{2}\) FSs are activated for both the in-phase and quadrature components, then \(\boldsymbol{\sigma}_i = \left\{k_{i,1}, k_{i,2}, \cdots k_{i, \varsigma}\right\}\) and \(\boldsymbol{\sigma}_q = \left\{k_{q,1}, k_{q,2}, \cdots k_{q, \varsigma}\right\}\). Activating all FSs in   \(\boldsymbol{\sigma}_i\) and \(\boldsymbol{\sigma}_q \) subsequently yields the following un-chirped in-phase and quadrature symbols:}
\begin{equation}\label{eq37}
f_{1}(n) = \sum_{k_i\in \boldsymbol{\sigma}_i}\exp\left\{j\frac{2\pi}{M}k_i n\right\},
\end{equation}
and 
\begin{equation}\label{eq38}
f_{2}(n) = \sum_{k_q\in \boldsymbol{\sigma}_q}\exp\left\{j\frac{2\pi}{M}k_q n\right\},
\end{equation}
respectively. \(f_{1}(n)\) and \(f_{2}(n)\) are \textcolor{black}{then} chirped using an up-chirp symbol to attain the chirped in-phase and quadrature symbols as \textcolor{black}{follows}:
\begin{equation}\label{eq39}
s_{1}(n) =f_1(n)c_\mathrm{u}(n)= \sum_{k_i\in \boldsymbol{\sigma}_i}\exp\left\{j\frac{2\pi}{M}n\left(k_i + n\right) \right\},
\end{equation}
and 
\begin{equation}\label{eq40}
s_{2}(n) =f_2(n)c_\mathrm{u}(n)= \sum_{k_q\in \boldsymbol{\sigma}_q}\exp\left\{j\frac{2\pi}{M}n\left(k_q+ n\right) \right\},
\end{equation}
respectively. Subsequently, the chirped in-phase and quadrature chirped symbols are combined to attain IQ-CIM symbol as:
\begin{equation}\label{eq41}
\begin{split}
&s_\mathrm{IQ}^{\mathrm{IM}}(n) = s_{1}(n) + js_{2}(n)\\
&~~= \sum_{k_i\in \boldsymbol{\sigma}_i}\!\!\!\exp\left\{\!j\frac{2\pi}{M}n\left(k_i + n\right)\! \right\} \!+\!\! \sum_{k_q\in \boldsymbol{\sigma}_q}\!\!\!\exp\left\{\!j\frac{2\pi}{M}n\left(k_q+ n\right)\! \right\}.
\end{split}
\end{equation}

Considering that \(\varsigma_i = \varsigma_q = \varsigma\) FSs are activated for both the in-phase and the quadrature components, then, the symbol energy of IQ-CIM symbol is \(E_\mathrm{s} = \sfrac{1}{M}\sum_{n=0}^{M-1}\vert s_\mathrm{IQ}^{\mathrm{IM}}(n)  \vert^2 = 2\varsigma\). 
\paragraph{Detection}
\begin{figure}[tb]\centering
\includegraphics[trim={0 0 0 0},clip,scale=1]{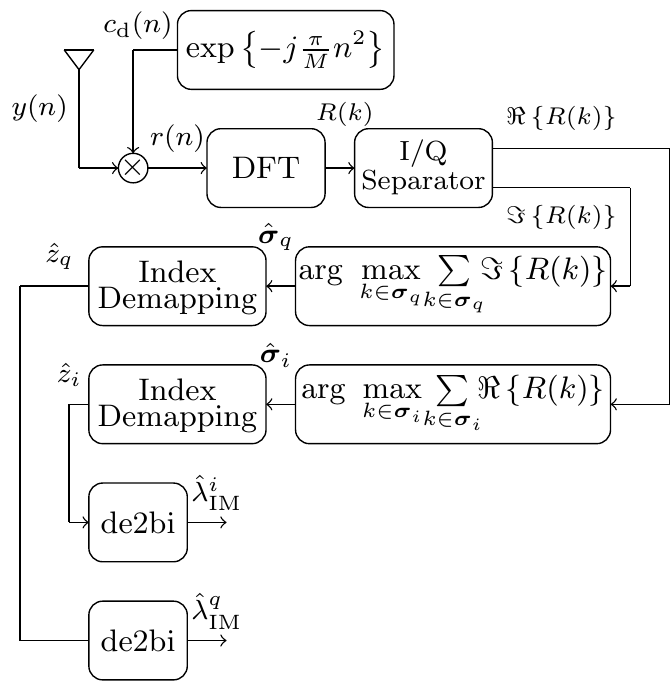}
  \caption{Coherent detector architecture for IQ-CIM.}
\label{fig18rx}
\end{figure}

\textcolor{black}{The transmission of data within the IQ-CIM framework utilizes both the in-phase and quadrature components, thereby necessitating the utilization of coherent detection. Notably, the utilization of non-coherent detection, as exhibited in IQ-CSS, is not a viable option within the IQ-CIM framework. The configuration for the implementation of a coherent detector within the IQ-CIM is depicted in Fig. \ref{fig18rx}.} The inner product of \(\boldsymbol{y}\) and the IQ-CIM transmit symbol \(\boldsymbol{s}_\mathrm{IQ}^\mathrm{IM} = \left[{s}_\mathrm{IQ}^\mathrm{IM}(0), {s}_\mathrm{IQ}^\mathrm{IM}(1), \cdots, {s}_\mathrm{IQ}^\mathrm{IM}(M-1)\right]^\mathrm{T}\), i.e., \(\langle   \boldsymbol{y}, \boldsymbol{s}_\mathrm{IQ}^\mathrm{IM}\rangle\) results in:
\begin{equation}\label{eqr42}
\begin{split}
&\langle   \boldsymbol{y}, \boldsymbol{s}_\mathrm{IQ}^\mathrm{IM}\rangle = \sum_{n=0}^{M-1}y(n)s_\mathrm{IQ}^\mathrm{IM}\\
&~= \sum_{n=0}^{M-1}\left\{\sum_{k_i\in \boldsymbol{\sigma}_i}y(n)\overline{f}_1(n) c_\mathrm{d}(n) -j\sum_{k_q\in \boldsymbol{\sigma}_q}y(n)\overline{f}_2(n)c_\mathrm{d}(n)\right\}\\
&~= \sum_{n=0}^{M-1}\sum_{k_i\in \boldsymbol{\sigma}_i}r(n)\overline{f}_1(n) -j\sum_{n=0}^{M-1}\sum_{k_q\in \boldsymbol{\sigma}_q}r(n)\overline{f}_2(n)\\
&~= \sum_{k_i\in \boldsymbol{\sigma}_i}R(k_i) -j\sum_{k_q\in \boldsymbol{\sigma}_q}R(k_q).
\end{split}
\end{equation}

Using eq. (\ref{eqr42}), the activated in-phase and quadrature FAPs, \(\hat{\boldsymbol{\sigma}}_i \) and \(\hat{\boldsymbol{\sigma}}_q\) using coherent detection \textcolor{black}{are} evaluated as:
\begin{equation}\label{eqr43}
\hat{\boldsymbol{\sigma}}_i = \mathrm{arg} \max\limits_{k_i \in \boldsymbol{\sigma}_i} \Re\left\{ \sum_{k_i\in \boldsymbol{\sigma}_i} R(k_i)\right\},
\end{equation}
and 
\begin{equation}\label{eqr44}
\hat{\boldsymbol{\sigma}}_q = \mathrm{arg} \max\limits_{k_q \in \boldsymbol{\sigma}_q} \Im\left\{ \sum_{k_q\in \boldsymbol{\sigma}_q} R(k_q)\right\},
\end{equation}
respectively, where \(\Re\left\{j\sum_{k_q\in \boldsymbol{\sigma}_q} R(k_q)\right\} = \Im\left\{ \sum_{k_q\in \boldsymbol{\sigma}_q} R(k_q)\right\}\). \(\hat{\boldsymbol{\sigma}}_i \) and \(\hat{\boldsymbol{\sigma}}_q\) are then used for index demapping that yields \(\hat{z}_i\) and \(\hat{z}_q\). Finally, both indices are precessed to determine the transmitted bit sequences having \(\hat{\lambda}_\mathrm{IM}^i\) and \(\hat{\lambda}_\mathrm{IM}^q\) bits after de2bi conversion. 
\paragraph{\textcolor{black}{Takeaways}}
\textcolor{black}{The IQ-CIM methodology involves utilizing the I/Q components in tandem with IM, in contrast to the FSCSS-IM methodology, which solely uses the IM within the in-phase component. Due to the employment of I/Q components, IQ-CIM necessitates the utilization of two distinct FAPs, resulting in a transmission of twice the number of bits per IQ-CIM symbol relative to FSCSS-IM. Implementing I/Q components within IQ-CIM makes it susceptible to the different offsets of low-cost devices in the LPWANs. Moreover, the detection mechanism for IQ-CIM is limited to coherent detection. The coherent detection process within IQ-CIM is more complicated than that of FSCSS-IM, as the FAPs must be identified within both the in-phase and quadrature domains, potentially leading to an increase in power consumption. Therefore, a comprehensive investigation is imperative to ascertain the optimal number of activated FS within the FAP to achieve compatibility for different LPWAN applications.}
\section{Performance Analysis}\label{sec4}
\textcolor{black}{Upon completing the discussion regarding the waveform design, it is necessary to analyze the performance of the previously detailed CSS schemes. In light of this, the current section conducts a comparison between the CSS schemes regarding the attainable SE, the SNR per bit required for achieving a specific SE (also known as EE), and the BER performance in the AWGN channel, along with BER performance when considering various offsets that may arise in LPWANs.}

\textcolor{black}{The performance metrics elucidated above are fundamental in evaluating the performance of a modulation scheme based on a particular design methodology. For example, the study of achievable SE holds the utmost importance in designing and optimizing a system that delivers optimal performance, capacity, and reliability. Analyzing EE in a modulation scheme is crucial in designing and optimizing a system to improve battery life, sustainability, cost, and network performance. BER performance is vital in assessing the QoS, designing optimal system configurations, evaluating channel estimation techniques, and analyzing interference effects within a given system.
Furthermore, it is worth mentioning that cost-effective LPWAN devices may incur PO and FO. These aforementioned offsets can considerably impact the quality of a communication system. If the PO exceeds a certain threshold, it may result in errors in the received signal, leading to a higher BER and a decrease in QoS. Similarly, if the FO exceeds a specific limit, it may cause inter-symbol interference (ISI) and inter-carrier interference (ICI), leading to a higher BER and a reduction in QoS.}
\subsection{Spectral Efficiency Analysis}
\textcolor{black}{To analyze the SE of different CSS schemes, we shall assume that the bandwidth occupied by the symbols for any CSS scheme is \(B= M\Delta f\). Additionally, our analysis shall consider the symbol duration for all the CSS schemes as \(T_\mathrm{s}\).}

\textcolor{black}{In general, it is observed that the CSS schemes that belong to the SC CSS taxonomy exhibit lower SE in comparison to those that belong to the MC or MC-IM categories. It is noteworthy to emphasize that the primary goal of various CSS schemes proposed in the literature is to enhance the SE relative to the LoRa. This is because LoRa is capable of achieving exceedingly low SE.}

\textcolor{black}{In the sequel, we shall provide a detailed analysis of the achievable SE for each CSS scheme individually.}
\subsubsection{\textcolor{black}{Single Carrier CSS Schemes}}
\begin{itemize}
\item \textcolor{black}{In LoRa, the fundamental unit of information transmission is the \(M\) feasible FSs, resulting in \(\lambda= \log_2(M)\) bits transmitted per LoRa symbol. Therefore, the SE of LoRa is \(\eta_\mathrm{LoRa} =\sfrac{\lambda}{M}\) bits/s/Hz. Notably, the total number of possible symbols we can employ in LoRa is \(M_\mathrm{LoRa} = M = 2^\lambda\). As the number of bits to be transmitted increases, the cardinality of the set of admissible symbols for LoRa also increases accordingly.}
\item \textcolor{black}{In the ICS-LoRa system, \(\lambda\) bits are encoded through the FS. Furthermore, the system transmits \(\lambda_\mathrm{I} = 1\) bit for every symbol by considering the transmission of either the LoRa symbol or its interleaved counterpart. Thus, the total number of bits transmitted per ICS-LoRa symbol is \(\lambda+1\), resulting in \(M_\mathrm{ICS} = 2^{\lambda+1}=2M\) possible symbols for ICS-LoRa. The resulting SE for ICS-LoRa is \(\eta_\mathrm{ICS} = \sfrac{(\lambda + 1)}{M}\) bits/s/Hz.}
\item \textcolor{black}{In the E-LoRa, the FS conveys \(\lambda\) information bits. Furthermore, an additional \(\lambda_\mathrm{IQ}=1\) bit determines whether we should use the in-phase or quadrature component of the FS. Consequently, the total number of bits transmitted in an E-LoRa symbol is \(\lambda + 1\), making the possible E-LoRa symbols that can be transmitted is equal to \(M_\mathrm{E-LoRa} = 2^{\lambda+1}=2M\). The resulting SE of E-LoRa is \(\eta_\mathrm{E-LoRa} = \sfrac{(\lambda + 1)}{M}\) bits/s/Hz.}
\item \textcolor{black}{In the PSK-LoRa, the FSs and PSs of the un-chirped symbols serve as the information-bearing elements that transmit \(\lambda\) and \(\lambda_\mathrm{PS}= \log_2(M_\varphi)\) bits, respectively, resulting in a total of \(\lambda + \lambda_\mathrm{PS}\) bits per symbol. The resulting SE and the number of possible symbols for PSK-LoRa are \(\eta_\mathrm{PSK} = \sfrac{(\lambda + \lambda_\mathrm{PS})}{M}\) bits/s/Hz and \(M_\mathrm{PSK}=2^{\lambda+\lambda_\mathrm{PS}}\), respectively. If binary PSs, i.e., \(M_\varphi =2\) is considered, then \(\lambda_\mathrm{PS}=1\), whereas for quaternary PSs, \(M_\varphi =4\), \(\lambda_\mathrm{PS}=2\). In \cite{psk_lora}, it has been demonstrated that the optimal number of PSs that result in the highest EE is\( M_\varphi =4\). Therefore, the optimal variant of PSK-LoRa, referred to as QPSK-LoRa, transmits \(\lambda +2\)  bits per symbol and has a SE of \(\eta_\mathrm{PSK} = \sfrac{(\lambda+2)}{M}\) bits/s/Hz, with the total number of possible symbols being \(M_\mathrm{PSK}=2^{\lambda +2} = 4M\).}
\item \textcolor{black}{In SSK-LoRa, the information conveyed through the transmission is bifurcated into two separate components: the FS of the un-chirped symbol and the slope-shift of the chirp symbol. The FS encodes \(\lambda\) bits, while the slope-shift encodes \(\lambda_\mathrm{s} = 1\) bit. Therefore, the number of bits transmitted in a SSK-LoRa symbol is \(\lambda +1\). Consequently, the SE of SSK-LoRa is calculated as \(\eta_\mathrm{SSK} = \sfrac{(\lambda +1)}{M}\) bits/s/Hz. Additionally, the number of distinct symbols that can be transmitted using SSK-LoRa is calculated as \(M_\mathrm{SSK} = 2^{\lambda +1}= 2M\).}
\item In DCRK-LoRa, in addition to \(\lambda\) bits that are transmitted in the FS of the un-chirped symbol, \(\lambda_\mathrm{c}= \log_2(M_\mathrm{c})\) bits can be transmitted in the discrete CR, where \(M_\mathrm{c}\) is the total number of CRs. So, a total of \(\lambda + \lambda_\mathrm{c}\)  bits are transmitted per DCRK-LoRa symbol yielding a SE of \(\eta_\mathrm{DCRK} = \sfrac{(\lambda + \lambda_\mathrm{c}}{M})\) bits/s/Hz. Therefore, the total number of possible symbols in DCRK-LoRa is equal to \(M_\mathrm{DCRK} = 2^{\lambda + \lambda_\mathrm{c}}\).
\item SSK-ICS-LoRa combines ICS-LoRa and SSK-LoRa. In addition to the \(\lambda\) bits transmitted in the FS of the un-chirped symbol, \(\lambda_\mathrm{s} = 1\) bit decides whether an up-chirp or a down-chirp shall be used at the transmitter, and \(\lambda_\mathrm{I}= 1\) bit determines if the chirped symbol will be interleaved or not. Thus, a total of \(\lambda + \lambda_\mathrm{s}+\lambda_\mathrm{I} = \lambda +2\) bits can be transmitted per SSK-ICS-LoRa symbol. Moreover, the number of possible symbols for SSK-ICS-LoRa is \(M_\mathrm{SSK-ICS} = 2^{\lambda+2}= 4M\). As a result, the SE of SSK-ICS-LoRa is \(\eta_\mathrm{SSK-ICS} = \sfrac{(\lambda+2)}{M}\) bits/s/Hz.
\end{itemize}
\paragraph{\textcolor{black}{Takeaways}}
\textcolor{black}{Based on the SE analysis, it is apparent that various SC CSS schemes can only marginally enhance the number of bits conveyed per symbol compared to LoRa. The increase is either one bit per symbol or, at most, three bits per symbol. Accordingly, the cardinality of the possible symbols for SC CSS schemes increases between two and eight times. The number of possible symbols is enumerated for look-up-table (LUT) implementations. It is highlighted that that higher SE requires more LUT memory and will also increase the coherent detection complexity. The overall SE improvement relative to LoRa is also inadequate, as depicted in Table \ref{se_sc}. This lack of substantial SE increase can be attributed to the fact that SC CSS schemes uphold the integrity of the waveform design, thereby retaining its robustness. Despite this, the lower SE may still pose a predicament for these schemes due to the lower achievable SE problem. }
\renewcommand{\arraystretch}{1.3}
\begin{table*}[h]
  \caption{\textcolor{black}{SE Characteristics of SC CSS Schemes}}
   \label{se_sc}
  \centering
  \color{black}\begin{tabular}{*{4}{c}}
    \hline
    \hline
    \bfseries{CSS Scheme}    & \bfseries {SE} & \bfseries {SE increase w.r.t. LoRa} & \bfseries {Number of Possible Symbols}  \\
    \hline
    \hline
    LoRa & \(\frac{\lambda}{M}\) & \(-\)& \(2^\lambda =M\)  \\
     E-LoRa & \(\frac{\lambda +1}{M}\) & \(\frac{1}{M}\)& \(2^{\lambda+1}= 2M\)  \\
     ICS-LoRa & \(\frac{\lambda +1}{M}\) & \(\frac{1}{M}\)& \(2^{\lambda+1}= 2M\)  \\
     PSK-LoRa & \(\frac{\lambda +\lambda_\mathrm{PS}}{M}\)  & \(\frac{\lambda_\mathrm{PS}}{M}\) & \(2^{\lambda+\lambda_\mathrm{PS}}\)\\
     SSK-LoRa & \(\frac{\lambda +1}{M}\) & \(\frac{1}{M}\)& \(2^{\lambda+1}= 2M\)\\
DCRK-LoRa & \(\frac{\lambda +\log_2(M_\mathrm{c})}{M}\)& \(\frac{\log_2(M_\mathrm{c})}{M}\) & \(2^{\lambda+\log_2(M_\mathrm{c})}\) \\
SSK-ICS-LoRa & \(\frac{\lambda +2}{M}\) & \(\frac{2}{M}\)& \(2^{\lambda+2}= 4M\) \\
    \hline
    \hline
  \end{tabular}
\end{table*}
\subsubsection{\textcolor{black}{Multiple Carrier CSS Schemes}}
\begin{itemize}
\item In DO-CSS, the information is transmitted via the even and the odd FSs, i.e., \(k_\mathrm{e}\) and \(k_\mathrm{o}\). Since there are \(\sfrac{M}{2}\) even and \(\sfrac{M}{2}\) odd FSs, then, the number of bits that can be transmitted by either the even or the odd FS is \(\lambda -1\). This results in a total of \(2\lambda -2\) bits are transmitted per DO-CSS symbol and a total of \(M_\mathrm{DO} = 2^{2\lambda-2}= M^2 - 4\) possible symbols. Moreover, the SE of DO-CSS is \(\eta_\mathrm{DO} = \sfrac{(2\lambda -2) }{M}\) bits/s/Hz.
\item In IQ-CSS, a total of \(2\lambda\) bits are transmitted per IQ-CSS symbol, where the first half \textcolor{black}{is} embedded in the in-phase part and the second half \textcolor{black}{in} the quadrature part. Therefore, the total number of possible IQ-CSS symbols are \(M_\mathrm{IQ}= 2^{2\lambda}= M^2\) and the SE is \(\eta_\mathrm{IQ} = \sfrac{2\lambda}{M}\) bits/s/Hz.%
\item In ePSK-CSS, the FF is ascertained using \(\lambda_\mathrm{FS} = \log_2(\sfrac{M}{N_\mathrm{b}})\) bits. Moreover, the FF and its \(N_\mathrm{b}-1\) harmonics each carry \(\lambda_\mathrm{PS}\) bits in the PS. Thus, the total number of bits transmitted per ePSK-CSS symbol is \(\lambda_\mathrm{FS} +N_\mathrm{b}\lambda_\mathrm{PS}\). The resulting SE and \textcolor{black}{the} total number of possible symbols for ePSK-CSS is  \(\eta_\mathrm{ePSK} = \sfrac{(\lambda_\mathrm{FS} +N_\mathrm{b}\lambda_\mathrm{PS})}{M}\) bits/s/Hz, and \(M_\mathrm{ePSK} = 2^{\lambda_\mathrm{FS} +N_\mathrm{b}\lambda_\mathrm{PS}}\). It had been demonstrated in \cite{epsk_lora} that in the optimal variant of ePSK-CSS, there are quaternary PSs and \(N_\mathrm{b}=2\). Accordingly, the number of bits transmitted by the optimal variant of ePSK-CSS is \(\lambda +3\), and the total number of possible symbols is \(M_\mathrm{ePSK}= 8M\). 
\item The The number of bits transmitted in GCSS depends on the number of groups, \(\mathrm{G}\). The higher the number of groups, the higher the number of transmitted bits. This is because the number of bits transmitted by the \(g\)th group is \(\lambda_g = \log_2(\sfrac{M}{\mathrm{G}})\). \textcolor{black}{The number of bits transmitted in each group may be the same.} Thus, when there are \(\mathrm{G}\) groups, the total number of bits transmitted per GCSS symbol is \(\mathrm{G}\log_2(\sfrac{M}{\mathrm{G}})\). This results in a SE of  \(\eta_\mathrm{GCSS} = \sfrac{(\mathrm{G}\log_2(\sfrac{M}{\mathrm{G}}))}{M}\) bits/s/Hz. Moreover, the number of possible symbols in GCSS having \(\mathrm{G}\) groups is \(M_\mathrm{GCSS}= 2^{\mathrm{G}\log_2(\sfrac{M}{\mathrm{G}})}\).
\item In TDM-CSS, \(\lambda\) bits are to be transmitted in the up-chirped symbol, and another \(\lambda\) bits can be transmitted in the down-chirped symbol. Thus, a total of \(2\lambda\) bits can be transmitted per TDM-CSS symbol. A total of \(M_\mathrm{TDM}= 2^{2\lambda} = M^2\) symbols are possible for TDM-CSS, and its SE is \(\eta_\mathrm{TDM} = \sfrac{2\lambda}{M}\) bits/s/Hz, which is the same as that of IQ-CSS and twice as much when compared to LoRa.
\item IQ-TDM-CSS consists of \textcolor{black}{two in-phase and quadrature} symbols, each of which \textcolor{black}{transmits} \(\lambda\) bits. To be more precise, two TDM-CSS schemes are simultaneously transmitted, one on the in-phase and the other on the quadrature components. Thus, a total of \(4\lambda\) bits can be transmitted per IQ-TDM-CSS symbol. Accordingly, the SE of IQ-TDM-CSS is \(\eta_\mathrm{IQTDM} = \sfrac{4\lambda}{M}\) bits/s/Hz, which is the double of \(\eta_\mathrm{IQ-CSS}\). Moreover, the number of possible symbols for IQ-TDM-CSS is \(M_\mathrm{IQTDM}= M^4\).
\item In \textcolor{black}{the} DM-CSS symbol, \(\lambda_\mathrm{e} = \log_2(\sfrac{N}{2})\) and \(\eta_\mathrm{o} = \log_2(\sfrac{N}{2})\) bits are encoded in the activated even and odd FSs, i.e., \(k_\mathrm{e}\) and \(k_\mathrm{o}\). In addition, \(2\lambda_\mathrm{PS}=2\log_2(2)=2\) bits are encoded in the PSs of activated even and odd FSs, i.e., \(\alpha_\mathrm{e}\) and \(\alpha_\mathrm{o}\). Lastly, \(\lambda_\mathrm{s}=\log_2(2)=1\) bit is encoded in the slope of the chirp. Therefore, the total number of bits that can be transmitted per DM-CSS symbol is \( \lambda_\mathrm{e} + \lambda_\mathrm{o} + 2\lambda_\mathrm{PS} + \lambda_\mathrm{s} = 2\lambda + 1\). Thus, the SE of DM-CSS is \( \eta_\mathrm{DM} = \sfrac{(2\lambda + 1)}{M}\) bits/s/Hz. In DM-CSS, a total of \(M_\mathrm{DM}= 2^{2\lambda+1}= M^2 + 2\) distinct symbols is possible.
\end{itemize}
\paragraph{\textcolor{black}{Takeaways}}
\textcolor{black}{Diverging from SC CSS schemes, the MC CSS schemes substantially increase the number of bits conveyed per symbol, generating a notably higher SE than LoRa and its SC CSS equivalents (as evidenced in Table \ref{se_mc}). IQ-TDM-CSS offers the most considerable SE increase among the MC CSS techniques, delivering a quadruple SE relative to LoRa. Additionally, other methods increase the SE relative to LoRa by twofold or threefold, with ePSK-LoRa exhibiting the most minimal increase, increasing the number of transmitted bits by only two. The inferior increase in SE for ePSK-LoRa is due to its extra robustness in determining the FS and the PS, which hinders its SE increase. While MC CSS schemes might be more versatile in achieving diverse and higher SEs than their SC counterparts, they come at the cost of a less robust waveform design, which gives rise to some implementation complications and could lead to diminished performance in the presence of different offsets. The number of potential waveforms also surges considerably for MC CSS schemes, necessitating a higher LUT memory. Furthermore, the complexity of the coherent detection would also rise unless dis-joint detection is employed.}
\renewcommand{\arraystretch}{1.3}
\begin{table*}[h]
  \caption{\textcolor{black}{SE Characteristics of MC CSS Schemes}}
   \label{se_mc}
  \centering
  \color{black}\begin{tabular}{*{4}{c}}
    \hline
    \hline
    \bfseries{CSS Scheme}    & \bfseries {SE} & \bfseries {SE increase w.r.t. LoRa} & \bfseries {Number of Possible Symbols}  \\
    \hline
    \hline
    
      IQ-CSS & \(\frac{2\lambda}{M}\)  & \(\frac{\lambda}{M}\) & \(2^{2\lambda}= M^2\)\\
ePSK-CSS & \(\frac{\log_2(\sfrac{M}{N_\mathrm{b}}) + N_\mathrm{b}\lambda_\mathrm{PS}}{M}\)&  \(\frac{2}{M}\) (for \(\{N_\mathrm{b},M_\varphi\}=\{4,2\}\)) & \(2^{\lambda_\mathrm{FS} +N_\mathrm{b}\lambda_\mathrm{PS}}\) \\
          GCSS & \(\frac{\mathrm{G}\log_2(\sfrac{M}{\mathrm{G}})}{M}\) & \(\frac{2\lambda -2}{M}\) (for \(\mathrm{G}=2\)) & \(2^{\mathrm{G}\log_2(\sfrac{M}{\mathrm{G}})}\) \\
 TDM-CSS & \(\frac{2\lambda}{M}\) & \(\frac{\lambda}{M}\)& \(2^{2\lambda}= M^2\) \\
IQ-TDM-CSS & \(\frac{4\lambda}{M}\) & \(\frac{3\lambda}{M}\)& \(2^{4\lambda}= M^4\) \\
DM-CSS & \(\frac{2\lambda +1}{M}\) & \(\frac{\lambda + 1}{M}\)& \(2^{2\lambda+1}= M^2 +2\)  \\
    \hline
    \hline
  \end{tabular}
\end{table*}
\subsubsection{\textcolor{black}{Multiple Carrier with Index Modulation CSS Schemes}}
\begin{itemize}
\item In FSCSS-IM, a total of \(\lambda_\mathrm{IM} = \lfloor \log_2 \binom{M}{\varsigma}\rfloor\) bits are transmitted per symbol. It may be noticed that the total number of distinct FAPs used for transmission in FSCSS-IM may significantly exceed \(M\). Consequently, the number of bits transmitted in FSCSS-IM is significantly higher than that of LoRa. For example, for \(M= 128\) and \(\varsigma = 2\), the number of bits that can be transmitted per symbol is \(12\) bits, whereas, for the same occupied bandwidth, LoRa is capable of transmitting only \(7\) bits. It is also highlighted that the maximum number of bits that can be transmitted per symbol would be maximum when \(\varsigma \approx \sfrac{M}{2}\), however, in this case, the EE of would be considerably poor relative to LoRa because an increase in \(\varsigma\) diminishes the EE. Moreover, FSCSS-IM is extremely flexible in achieving different spectral efficiencies by changing \(\varsigma\). The SE of FSCSS-IM is \(\eta_\mathrm{FS}^\mathrm{IM} = \sfrac{\lambda_\mathrm{IM}}{M}\) bits/s/Hz, and the number of possible symbols that can be transmitted is \(M_\mathrm{FS}^\mathrm{IM}= 2^{\lambda_\mathrm{IM}}\).
\item In IQ-CIM, both the in-phase and the quadrature un-chirped symbols employ IM\textcolor{black}{; therefore,} a total of \(2\lambda_\mathrm{IM}\) bits can be transmitted per symbol provided the same number of FSs are used for both the in-phase and the quadrature un-chirped symbols. This implies that the number of bits transmitted per IQ-CIM symbol is twice relative to that of the FSCSS-IM  symbol of the same duration. The total number of possible IQ-CIM symbols are \(M_\mathrm{IQ}^\mathrm{IM}= 2^{2\lambda_\mathrm{IM}}\). Moreover, the SE of IQ-CIM is \(\eta_\mathrm{IQ}^\mathrm{IM} = \sfrac{2\lambda_\mathrm{IM}}{M}\) bits/s/Hz implying that \(\eta_\mathrm{IQ}^\mathrm{IM}  = 2\eta_\mathrm{FS}^\mathrm{IM}\).  
\end{itemize}
\paragraph{\textcolor{black}{Takeaways}}
\textcolor{black}{Similarly to MC CSS schemes, MC-IM CSS schemes also significantly boost the SE relative to SC CSS schemes, as indicated in Table \ref{se_mc_im}. By employing two active FSs, FSCSS-IM nearly doubles the number of bits conveyed per symbol, whereas activating two in-phase and quadrature FSs, increases the number of bits by almost threefold for IQ-CIM. Activating a higher number of FSs could further raise the SE but at the expense of intensified MC transmission effects such as higher PAPR, reduced robustness to offsets, and frequency-selective fading, among others. MC-IM schemes represent the most pliable taxonomy for furnishing diverse SEs, as activating a higher number of FSs increases the SE. However, detection complexity typically increases alongside the number of activated FSs, potentially acting as a restricting factor for these approaches.}
\renewcommand{\arraystretch}{1.5}
\begin{table*}[h]
  \caption{\textcolor{black}{SE Characteristics of MC-IM CSS Schemes}}
   \label{se_mc_im}
  \centering
  \color{black}\begin{tabular}{*{4}{c}}
    \hline
    \hline
    \bfseries{CSS Scheme}    & \bfseries {SE} & \bfseries {SE increase w.r.t. LoRa} & \bfseries {Number of Possible Symbols}  \\
    \hline
    \hline
    FSCSS-IM & \(\frac{\lfloor \log_2 \binom{M}{\varsigma}\rfloor}{M}\) & \(\frac{\lambda-2}{M}\) {(2 active indices)} &\(2^{\lambda_\mathrm{IM}}\)\\
     IQ-CIM & \(\frac{\lfloor \log_2 \binom{M}{\varsigma_i}\rfloor + \lfloor \log_2 \binom{M}{\varsigma_q}\rfloor}{M}\) & \(\frac{2\lambda-4}{M}\) (2 active indices) & \(2^{2\lambda_\mathrm{IM}}\)\\
    \hline
    \hline
  \end{tabular}
\end{table*}
\subsubsection{\textcolor{black}{Gropings based on Achievable SE}}
\textcolor{black}{Due to varying spectral efficiencies exhibited by different CSS schemes, as evidenced by Tables \ref{se_sc}, \ref{se_mc}, and \ref{se_mc_im}, it is imperative to conduct a just and equitable evaluation by comparing only those schemes which possess comparable spectral efficiencies. Hence, we have partitioned the methods into six groups predicated on their corresponding achievable spectral efficiencies, as tabulated in Table \ref{tab3}. The specifics of these groups are enumerated below:}
\begin{itemize}
\item \textcolor{black}{Group 1 comprises SC CSS schemes that exhibit a SE falling within the confines of \([\sfrac{\lambda}{M},\sfrac{(\lambda+1)}{M}]\) bits/s/Hz. An astute observation discloses that solely the schemes included within this grouping, specifically LoRa, ICS-LoRa, E-LoRa, and SSK-LoRa, possess the capability of transmitting an additional bit in comparison to the classical LoRa.}
\item \textcolor{black}{Group 2 comprises SC CSS methodologies that can confer a SE of \(\sfrac{(\lambda+2)}{M}\) bits/s/Hz, namely PSK-LoRa, and SSK-LoRa. It is noteworthy that we consider PSK-LoRa, which employs quaternary PSs, and is thus capable of attaining a SE of \(\sfrac{(\lambda+2)}{M}\) bits/s/Hz. It is discernible that the strategies encompassed within this cluster have the potential to amplify the number of transmitted bits per symbol by two.}
\item \textcolor{black}{Group 3 comprises schemes that achieve a SE of \(\sfrac{(\lambda+3)}{M}\) bits/s/Hz. This category encompasses methodologies that can augment the number of transmitted bits by three compared to the LoRa. Consequently, this group includes one SC CSS scheme and one MC CSS scheme, namely DCRK-LoRa with having \(M_\mathrm{c}=8\) CRs and the optimal variant of ePSK-CSS, utilizing quaternary PSs and \(N_\mathrm{b}=2\).}
\item \textcolor{black}{Group 4 is comprised of DO-CSS and GCSS methodologies that transmit two bits fewer than twice the number of bits transmitted by LoRa. Consequently, the SE considered for this category is \(\sfrac{(2\lambda-2)}{M}\) bits/s/Hz. Notably, both schemes incorporated within this group are MC CSS schemes.}
\item \textcolor{black}{Group 5 pertains to the set of MC CSS schemes that exhibit a SE greater than \(\sfrac{2\lambda}{M}\) bits/s/Hz. It is recalled that the peak SE achieved by the methodologies elucidated in this examination is \(\sfrac{4\lambda}{M}\) bits/s/Hz. Henceforth, the maximum SE encompassed by this group is delimited to \(\sfrac{4\lambda}{M}\) bits/s/Hz. The trio of IQ-CSS, TDM-CSS, and DM-CSS fall under the ambit of this group.}
\item \textcolor{black}{Group 6 comprises MC-IM CSS schemes, that are, FSCSS-IM and IQ-CIM.}
\end{itemize}

In the sequel, we retain this grouping of the CSS schemes when different performance metrics shall be elucidated.
\begingroup
\setlength{\tabcolsep}{6pt} 
\renewcommand{\arraystretch}{1.2} 
\begin{table}[tbh]
\caption{Partitioning of CSS schemes into groups based on achievable spectral efficiencies.}
\centering
\begin{tabular}{c||c||c}
\hline
\hline
\textbf{Group} & \textbf{SE  (bits/s/Hz)} &  \textbf{CSS schemes} \\
\hline
\hline
Group 1& \(\left[\frac{\lambda}{M},\frac{\lambda+1}{M}\right]\) & LoRa, ICS-LoRa,\\
{} & {}& E-LoRa, SSK-LoRa\\
\hline
Group 2&\(\frac{\lambda+2}{M}\) & PSK-LoRa, SSK-ICS-LoRa\\
\hline
Group 3&\(\frac{\lambda+3}{M}\) & ePSK-CSS, DCRK-LoRa\\
\hline
Group 4&\(\frac{2\lambda-2}{M}\) & DO-CSS, GCSS\\
\hline
Group 5&\(\left[\frac{2\lambda}{M},\frac{4\lambda}{M}\right]\) & IQ-CSS, TDM-CSS,\\
{}& {}& IQ-TD-CSS, DM-CSS\\
\hline
Group 6&IM Approaches & FSCSS-IM, IQ-CIM\\
\hline 
\hline
\end{tabular}
\label{tab3}
\end{table}
\endgroup
\subsection{Spectral Efficiency vs Energy Efficiency Performance}\label{ee_perf}
\textcolor{black}{In this section, we evaluate the SE versus EE performance of various schemes that have already been elucidated in detail in the preceding section regarding their waveform design. The attainable SE has already been examined previously. Conversely, the EE is established by appraising \(E_\mathrm{b}/N_0=\sfrac{(E_\mathrm{s}T_\mathrm{s})}{(\eta N_0)}\) that is obligatory to achieve a BER of \(10^{-3}\) for a specified SE. Furthermore, the SE is altered by modifying \(\lambda = \llbracket 6,12\rrbracket\). It is noteworthy that the \(\lambda\) is characterized for LoRa. However, the SE of all the schemes is also a function of \(\lambda\). Additionally, we maintain the categorization of various CSS schemes, as established in the previous section. The SE versus EE for all the groups is evaluated in an AWGN channel. The performance of coherent and non-coherent detection mechanisms is evaluated separately. Some modulations, such as E-LoRa, IQ-CIM, and IQ-TDM-CSS, do not possess non-coherent detection mechanisms; thus, their performance is not represented. For other modulations, including GCSS, coherent detection mechanisms are not present in the literature. Finally, LoRa is regarded as a benchmarking scheme because it is the extensively implemented CSS scheme for LPWANs, and its performance is compared to that of the schemes in each group.}
\subsubsection{Group 1}
\begin{figure}[tb]\centering
\includegraphics[trim={16 0 0 0},clip,scale=0.69]{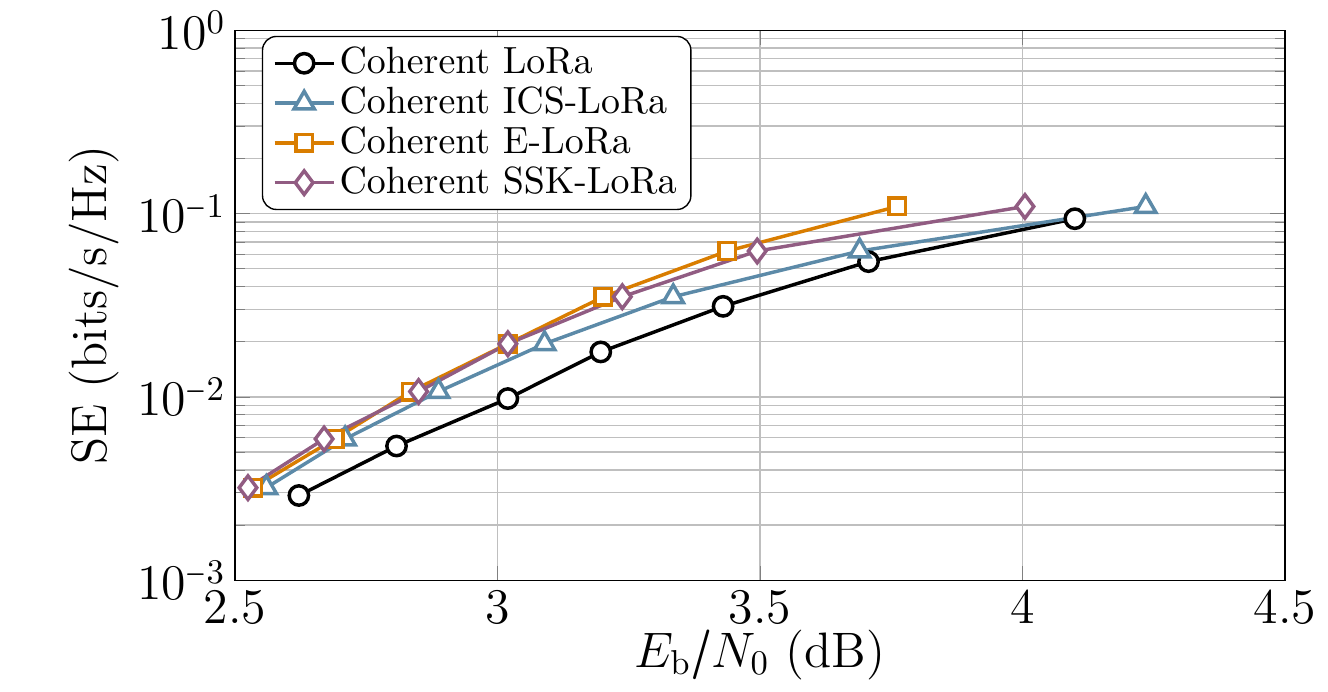}
  \caption{SE versus required SNR per bit for group 1 schemes considering coherent detection and target BER of \(10^{-3}\) in AWGN channel.}
\label{fig1_ee}
\end{figure}
\textcolor{black}{Fig. \ref{fig1_ee} illustrates a comparative performance analysis of LoRa, ICS-LoRa, E-LoRa, and SSK-LoRa using coherent detection. The observation reveals that E-LoRa and SSK-LoRa exhibit similar EE performance versus \(E_\mathrm{b}/N_0\), whereas the performance of ICS-LoRa has shown a slight deterioration. This is because the distinct interleaved chirp symbols present a relatively higher cross-correlation compared to the chirp symbols in E-LoRa and SSK-LoRa. This results in a loss of approximately \(0.2\) dB relative to E-LoRa and SSK-LoRa when \(\lambda = 12\)  (i.e., when SE \(\simeq 0.0029\) bits/s/Hz) and a loss of about \(0.35\) dB relative to E-LoRa when \(\lambda = 6\) (i.e., when SE \(\simeq 0.09375\) bits/s/Hz). However, all the aforementioned schemes perform better than the coherently detected LoRa, implying that the required energy per bit for correct detection is reduced for the schemes in group 1.}
\begin{figure}[tb]\centering
\includegraphics[trim={16 0 0 0},clip,scale=0.69]{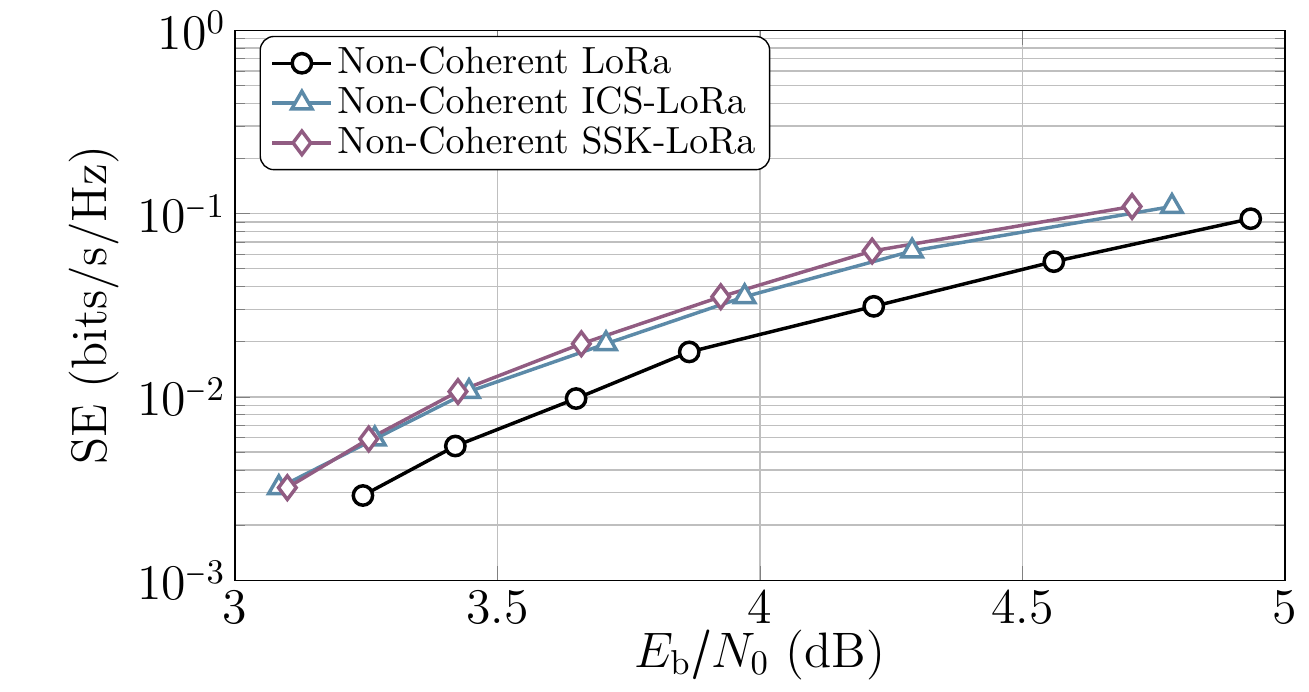}
  \caption{SE versus required SNR per bit for group 1 schemes considering non-coherent detection and target BER of \(10^{-3}\) in AWGN channel.}
\label{fig2_ee}
\end{figure}

\textcolor{black}{On the contrary, we can make a distinct observation from Fig. \ref{fig2_ee}, which illustrates the EE performance versus \(E_\mathrm{b}/N_0\) using a non-coherent detection mechanism. We can note that the performance of SSK-LoRa exhibits a marginal improvement compared to ICS-LoRa. This result is attributed to the lower cross-correlation between the chirped symbols for SSK-LoRa, which significantly increases the minimum Euclidean distance between the distinct chirp symbols. However, the performance of both ICS-LoRa and SSK-LoRa still surpasses that of non-coherently detected LoRa. It should be noted that the coherent E-LoRa is not depicted, as the information is lost in the phase-shifted un-chirped symbol.}
\subsubsection{Group 2}
\begin{figure}[tb]\centering
\includegraphics[trim={16 0 0 0},clip,scale=0.69]{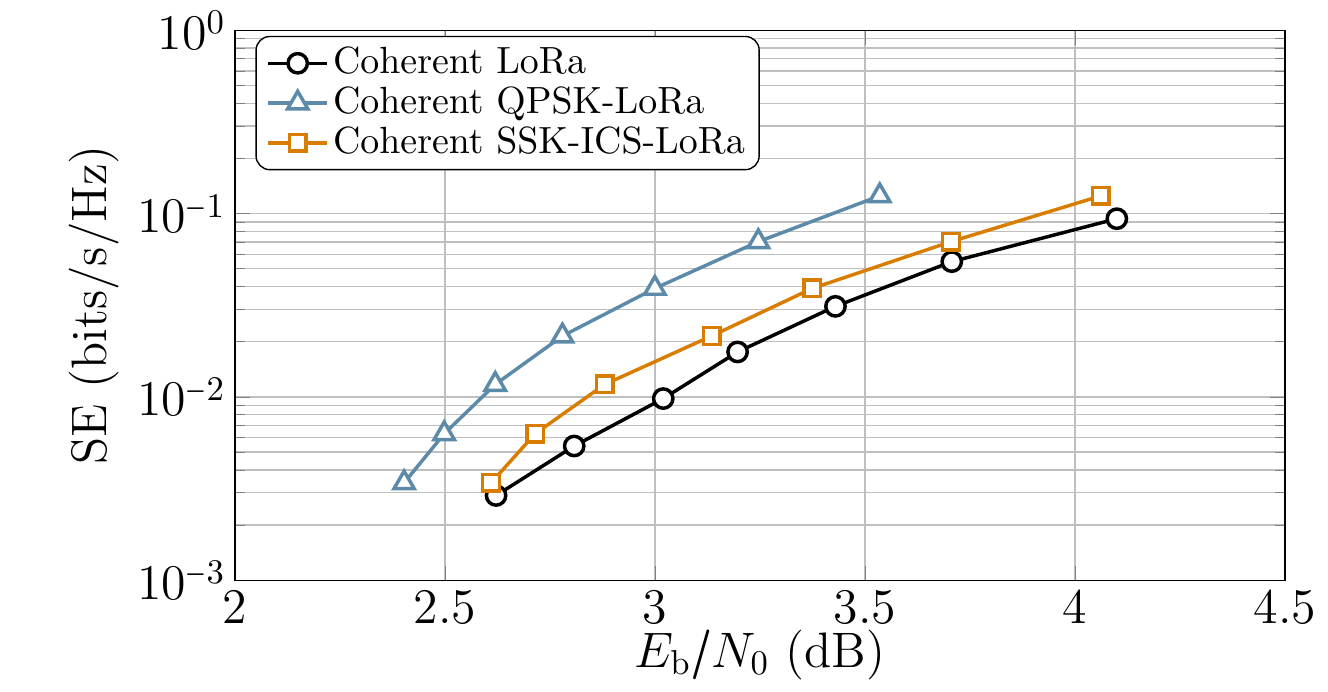}
  \caption{SE versus required SNR per bit for group 2 schemes considering coherent detection and target BER of \(10^{-3}\) in AWGN channel.}
\label{fig3_ee}
\end{figure}

\textcolor{black}{Fig. \ref{fig3_ee} compares PSK-LoRa with quaternary PSs (QPSK-LoRa) and SSK-ICS-LoRa using coherent detection and the coherent detection performance of LoRa. Firstly, we can observe that the performance of LoRa is significantly deteriorated compared to QPSK-LoRa and SSK-ICS-LoRa. Secondly, QPSK-LoRa exhibits better performance than SSK-ICS-LoRa. This is because PS does not affect the orthogonality of symbols. In contrast, the interleaved versions of either up-chirped or down-chirped symbols cause a high cross-correlation, affecting orthogonality.}
\begin{figure}[tb]\centering
\includegraphics[trim={16 0 0 0},clip,scale=0.69]{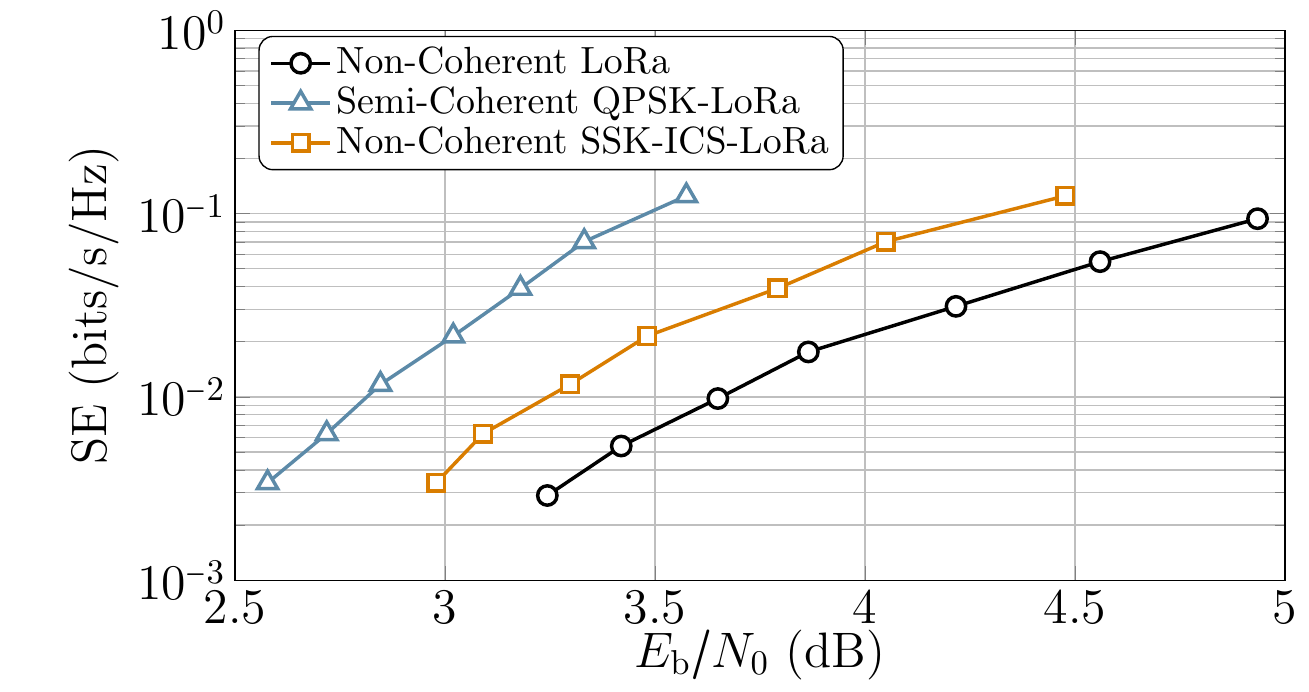}
  \caption{SE versus required SNR per bit for group 2 schemes considering non-coherent detection and target BER of \(10^{-3}\) in AWGN channel.}
\label{fig4_ee}
\end{figure}

\textcolor{black}{Fig. \ref{fig4_ee} illustrates the performance of coherent LoRa and SSK-ICS-LoRa. Notably, non-coherent detection for QPSK-LoRa is impossible due to the presence of the PS in the un-chirped symbol. Therefore, we introduce the implementation of semi-coherent detection for QPSK-LoRa, where the FS is non-coherently detected. In contrast, the PS is coherently detected using the ML criterion. This approach is deemed necessary because it would be impossible to determine the PS using non-coherent detection in the presence of channel phase rotation. In terms of performance, SSK-ICS-LoRa outperforms non-coherent LoRa, but it is inferior to semi-coherent QPSK-LoRa. However, comparing non-coherent SSK-ICS-LoRa to semi-coherent QPSK-LoRa would be unfair, as there is a significant difference in computational complexity involved in the detection process.}
\subsubsection{Group 3}
\begin{figure}[tb]\centering
\includegraphics[trim={16 0 0 0},clip,scale=0.69]{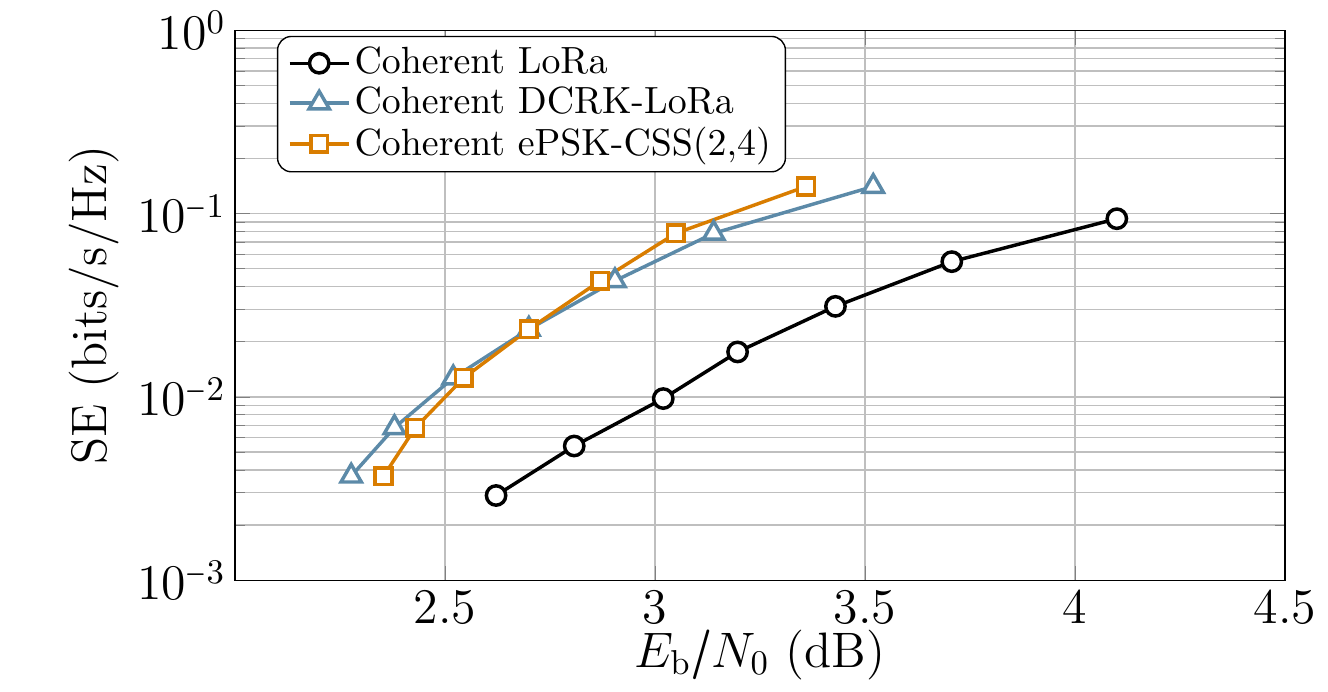}
  \caption{SE versus required SNR per bit for group 3 schemes considering coherent detection and target BER of \(10^{-3}\) in AWGN channel.}
\label{fig5_ee}
\end{figure}
\textcolor{black}{It is pertinent to note that ePSK-CSS\((2,4)\) can transmit \(\lambda+3\) bits per symbol. In contrast, DCRK-LoRa is an exceptionally versatile scheme, as it can transmit more than \(\lambda+3\) bits per symbol by augmenting the number of discrete CRs, \(M_\mathrm{c}\) with the caveat of heightened complexity in the transceiver.}

\textcolor{black}{Fig. \ref{fig5_ee} elucidates the coherent detection performance of DCRK-LoRa and ePSK-CSS(\(2,4\)). In this context, we consider \(M_\mathrm{c} = 8\) that yields a SE of \(\sfrac{(\lambda+3)}{M}\)  bits/s/Hz. The coherent detection performance of ePSK-CSS(\(2,4\)) is marginally superior to DCRK-LoRa for the range of \(\lambda = \llbracket 6,8 \rrbracket\). In contrast, for \(\lambda = \llbracket 10,12 \rrbracket\), the performance of DCRK-LoRa outperforms that of ePSK-CSS(\(2,4\)). It is worth mentioning that both DCRK-LoRa and ePSK-CSS(\(2,4\)) exhibit superior performance compared to coherent LoRa.}

\begin{figure}[h]\centering
\includegraphics[trim={16 0 0 0},clip,scale=0.69]{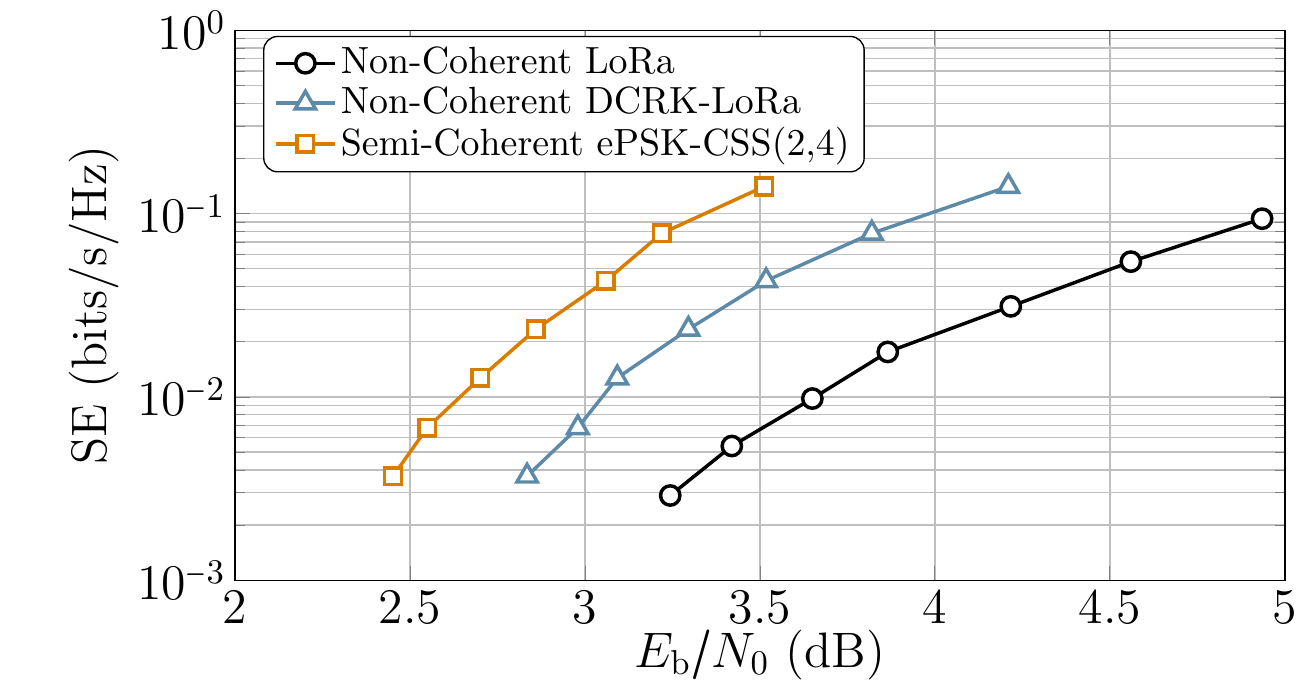}
  \caption{SE versus required SNR per bit for group 3 schemes considering non-coherent detection and target BER of \(10^{-3}\) in AWGN channel.}
\label{fig6_ee}
\end{figure}

\textcolor{black}{Fig. \ref{fig6_ee} portrays the comparative performance of non-coherent LoRa, DCRK-LoRa, and semi-coherent ePSK-CSS(\(2,4\)). It is noteworthy that the non-coherent DCRK-LoRa outperforms non-coherent LoRa. However, the semi-coherent ePSK-CSS(\(2,4\)) outperforms both DCRK-LoRa and LoRa. It should be emphasized that the implementation of semi-coherent detection for ePSK-CSS(\(2,4\)) necessitates channel equalization, which may diminish the practicality of this approach.}
\subsubsection{Group 4}
\textcolor{black}{Group 4 constitutes DO-CSS and GCSS, specifically  GCSS with \(\mathrm{G}=2\) due to its superior EE. It is worth noting that DO-CSS and GCSS with \(\mathrm{G}=2\) attain the same SE of \(\sfrac{(2\lambda-2)}{M}\)  bits/s/Hz.}
\begin{figure}[t]\centering
\includegraphics[trim={16 0 0 0},clip,scale=0.69]{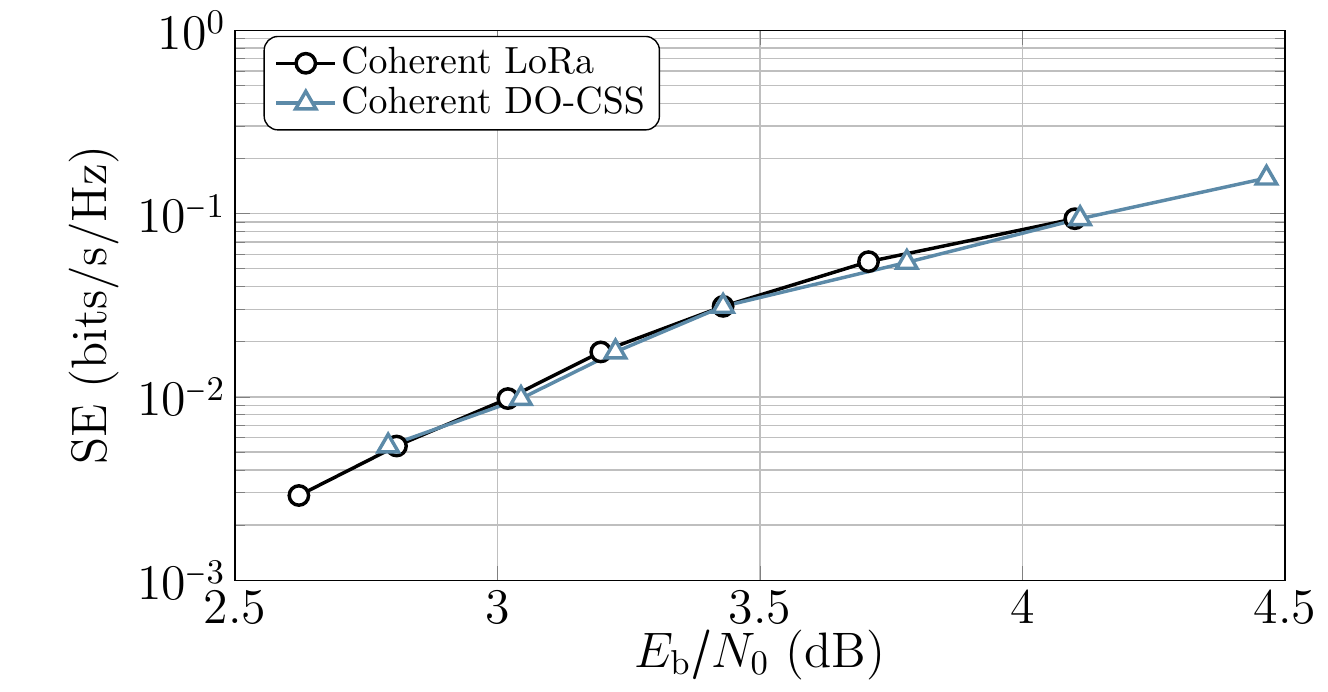}
  \caption{SE versus required SNR per bit for group 4 schemes considering coherent detection and target BER of \(10^{-3}\) in AWGN channel.}
\label{fig7_ee}
\end{figure}

\textcolor{black}{Fig. \ref{fig7_ee} in this study depicts the EE performance as a function of the required \(E_\mathrm{b}/N_0\) for group 4 of modulation schemes, comprising DO-CSS and GCSS. The coherent detector for GCSS is unavailable; hence only the performance of DO-CSS is demonstrated and compared with that of coherently detected LoRa. Notably, DO-CSS has an overall higher SE compared to LoRa, but the performance is indistinguishable at a fixed SE.}
\begin{figure}[t]\centering
\includegraphics[trim={16 0 0 0},clip,scale=0.69]{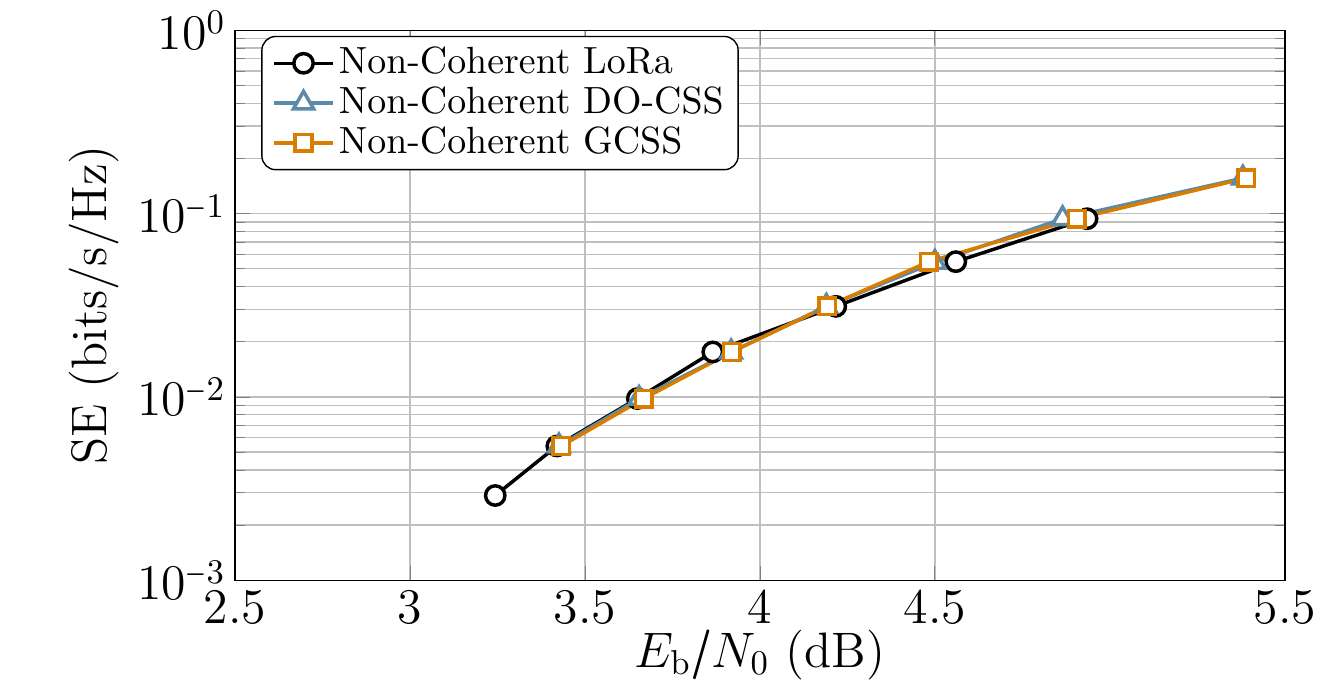}
  \caption{SE versus required SNR per bit for group 4 schemes considering non-coherent detection and target BER of \(10^{-3}\) in AWGN channel.}
\label{fig8_ee}
\end{figure}

Fig. \ref{fig8_ee} shows the performance of non-coherently detected DO-CSS, GCSS, and LoRa. We can observe that all the schemes exhibit similar performances. 
\subsubsection{Group 5}
\begin{figure}[t]\centering
\includegraphics[trim={16 0 0 0},clip,scale=0.69]{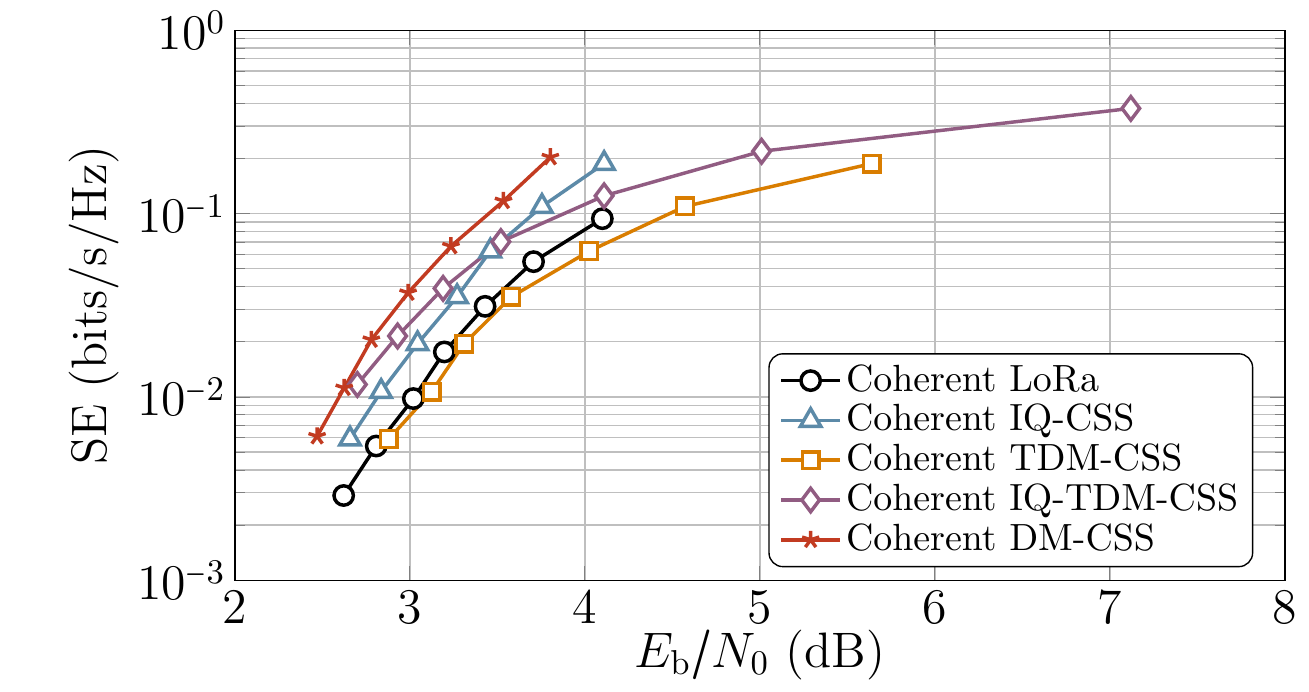}
  \caption{SE versus required SNR per bit for group 5 schemes considering coherent detection and target BER of \(10^{-3}\) in AWGN channel.}
\label{fig9_ee}
\end{figure}
\textcolor{black}{Fig. \ref{fig9_ee} depicts the coherent detection performance of IQ-CSS, TDM-CSS, IQ-TDM-CSS, DM-CSS, and LoRa. Notably, the coherently detected IQ-CSS, TDM-CSS, IQ-TDM-CSS, and DM-CSS exhibit higher spectral efficiencies and superior performance than LoRa. Among the remaining approaches in this group, DM-CSS achieves the best performance. IQ-TDM-CSS outperforms IQ-CSS, TDM-CSS, and LoRa when \(\lambda > 8\). However, for \(\lambda = \{6,7\}\) , IQ-TDM-CSS's performance is inadequate relative to other methods. Herefore, IQ-TDM-CSS's benefit, i.e., higher achievable spectral efficiencies, may not be advantageous. Furthermore, while TDM-CSS can achieve higher achievable SE, its EE performance is worse than that of LoRa. IQ-CSS performs better than LoRa and TDM-CSS but worse than IQ-TDM-CSS for \(\lambda > 8\) and DM-CSS.}

\begin{figure}[tb]\centering
\includegraphics[trim={16 0 0 0},clip,scale=0.69]{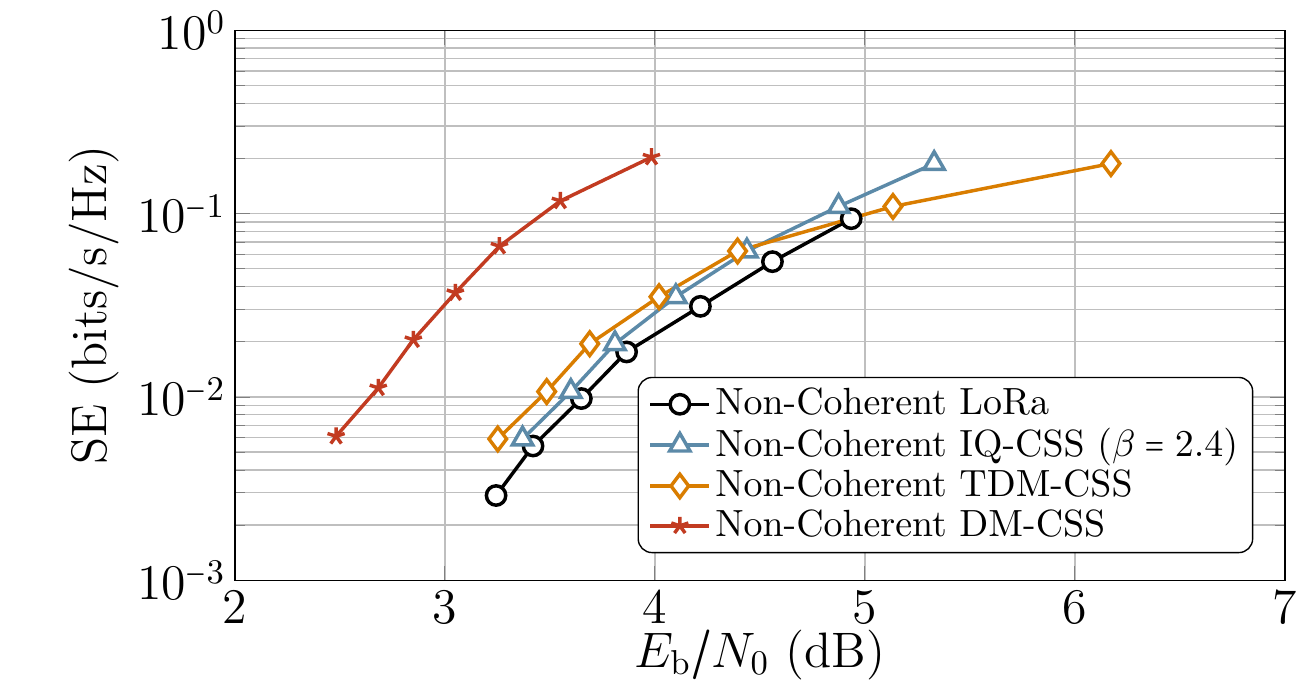}
  \caption{SE versus required SNR per bit for group 5 schemes considering non-coherent detection and target BER of \(10^{-3}\) in AWGN channel.}
\label{fig10_ee}
\end{figure}

\textcolor{black}{Fig. \ref{fig10_ee} depicts the non-coherent detection performance of the modulation techniques: IQ-CSS with \(\beta =2.4\), TDM-CSS, and DM-CSS. It is important to note that the non-coherent detector for IQ-TDM-CSS is unavailable. DM-CSS stands out as the best performer among this category's modulation techniques. In particular, at\(\lambda = 6\)), DM-CSS exhibits a performance gain of approximately  \(1.2\) dB compared to non-coherent IQ-CSS with \(\beta =2.4\) and \(2.1\) dB  concerning non-coherent TDM-CSS. Furthermore, DM-CSS indicates a performance improvement of roughly \(0.8\)  dB compared to the other modulation techniques at \(\lambda = 12\). It is worth noting that TDM-CSS performs even worse than LoRa for \(\lambda < 7\).}
\subsubsection{Group 6}
\begin{figure}[tb]\centering
\includegraphics[trim={16 0 0 0},clip,scale=0.69]{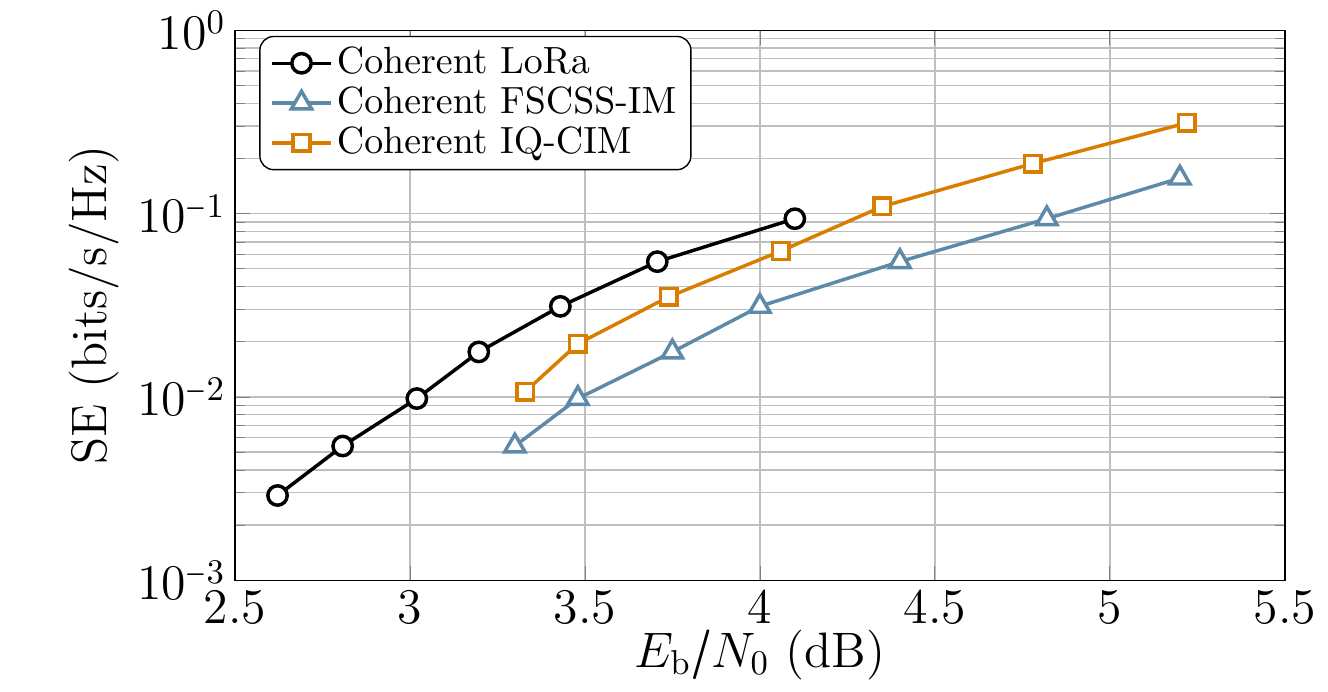}
  \caption{SE versus required SNR per bit for group 6 schemes considering coherent detection and target BER of \(10^{-3}\) in AWGN channel.}
\label{fig11_ee}
\end{figure}

\textcolor{black}{Fig. \ref{fig11_ee} depicts the EE performance of coherent detection for two IM approaches, namely FSCSS-IM and IQ-CIM. We assume two active indexes for both methods, i.e., \(\varsigma= \varsigma_i=\varsigma_q = 2\). Despite these schemes' flexibility in terms of achievable spectral efficiencies (if more indexes are activated), their EE performance for coherent detection is found to be even worse than that of coherently detected LoRa. Furthermore, IQ-CIM has the potential to attain high SE compared to FSCSS-IM and is more energy-efficient than the latter.}
\begin{figure}[tb]\centering
\includegraphics[trim={16 0 0 0},clip,scale=0.69]{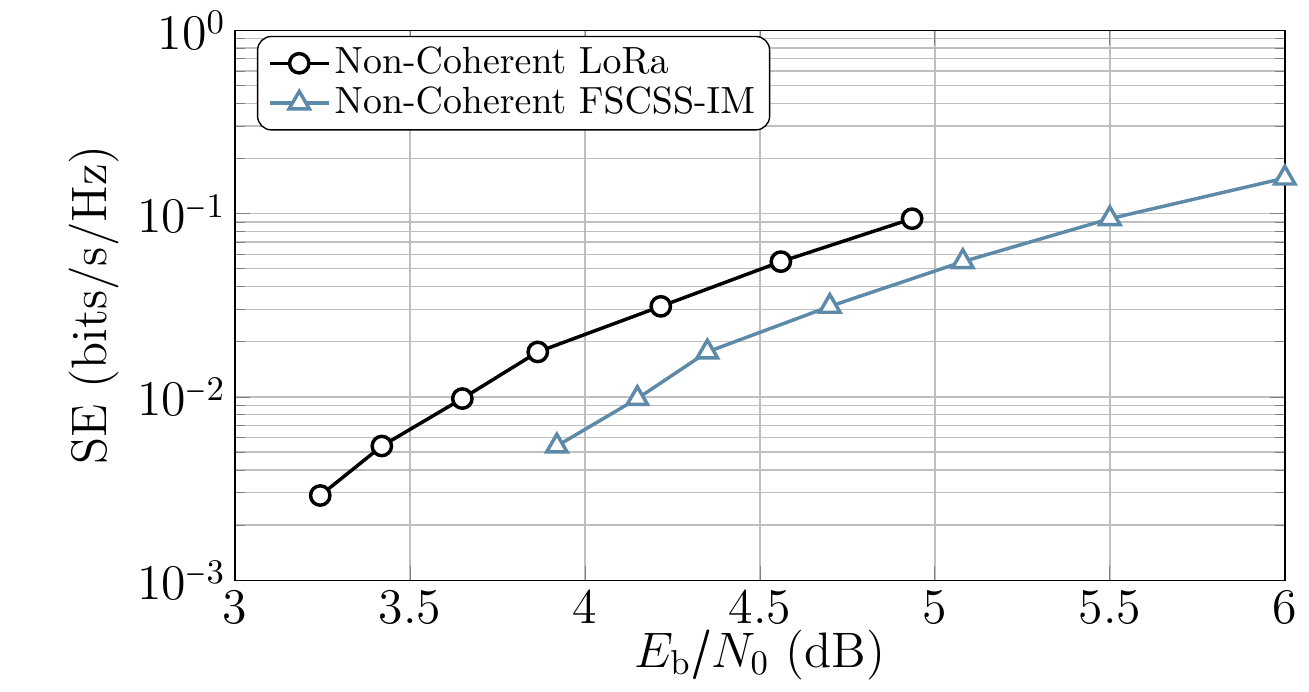}
  \caption{SE versus required SNR per bit for group 6 schemes considering non-coherent detection and target BER of \(10^{-3}\) in AWGN channel.}
\label{fig12_ee}
\end{figure}

\textcolor{black}{In Fig. \ref{fig12_ee}, we demonstrate the non-coherent detection performance of FSCSS-IM and LoRa. Notably, a non-coherent detector for IQ-CIM is impossible as it employs active indexes for both the in-phase and quadrature components of the un-chirped symbols. The results indicate that FSCSS-IM can achieve higher spectral efficiency with two active indexes. However, it performs worse than non-coherently detected LoRa regarding EE.}
\subsubsection{\textcolor{black}{Takeaways}}
\textcolor{black}{The most energy-efficient versions considering the SE versus EE performance are listed in Table  \ref{eff_se_ee}. Employing the SE versus EE metric provides a valuable means of distinguishing between schemes in distinct groups that possess optimal performance in coherent and non-coherent detection modes within an AWGN channel. In particular, this metric offers a rudimentary indication of the most resilient waveform design in a simplified scenario. However, results may diverge if a channel with impairments or a dispersive channel is considered, requiring an examination of various impairments, which is done in the sequel of this study. Within group 1, SSK-LoRa outperforms other schemes in the same group for both coherent and non-coherent detection mechanisms, indicating that using different slope rates represents the most robust waveform design for attaining a higher SE with higher EE. Similarly, QPSK-LoRa performs optimally for coherent and semi-coherent detection within group 2. In group 3, DCRK-LoRa demonstrates the best performance for coherent detection, while ePSK-CSS(\(2,4\)) exhibits the best performance concerning semi-coherent detection. The most robust waveform design for achieving a higher SE and EE with SC CSS schemes is DCRK-LoRa, which utilizes multiple chirp rates. However, the complexity of DCRK-LoRa is comparatively high when contrasted with other SC CSS schemes. }

\textcolor{black}{Within group 4, DO-CSS exhibits the best performance for coherent detection. At the same time, both DO-CSS and GCSS yield identical outcomes for non-coherent detection, as their frame structures are relatively comparable with \(\mathrm{G}=2\). In group 5, DM-CSS proves optimal for both coherent and non-coherent detection modes, indicating that DM-CSS represents the most resilient waveform design in MC schemes. It can achieve the highest SE of any MC scheme with a lower SNR per bit requirement. Regarding IM schemes, IQ-CIM yields the best results for coherent detection, while FSCSS-IM is the superior option for non-coherent detection.}

\renewcommand{\arraystretch}{1}
\begin{table*}[h]
  \caption{\textcolor{black}{Most energy-efficient schemes from each group for different spectral efficiencies.}}
   \label{eff_se_ee}
  \centering
  \color{black}\begin{tabular}{*{3}{c}}
    \hline
    \hline
    \bfseries{Group}    & \bfseries {Coherent Detection} & \bfseries {Non-Coherent Detection} \\
    \hline
    \hline
    
   Group 1 & SSK-LoRa & SSK-LoRa\\
   Group 2 & QPSK-LoRa & QPSK-LoRa (Semi-Coherent)\\
   Group 3 & DCRK-LoRa & ePSK-CSS(\(2,4\)) (Semi-Coherent)\\
   Group 4 & DO-CSS & Same Performance for both DO-CSS and GCSS\\
   Group 5 & DM-CSS & DM-CSS\\
   Group 6 & IQ-CIM & FSCSS-IM\\
    \hline
    \hline
  \end{tabular}
\end{table*}
\subsection{Bit-Error Rate Performance}
\textcolor{black}{In this section, we delve into the BER performance of the discussed schemes with the assumption of \(\lambda =8\)  and AWGN channel. Furthermore, we use the BER performance of LoRa, both coherent and non-coherent detection mechanisms, as a baseline for comparison. To provide a comprehensive analysis, we categorize the schemes into different groups, as specified in Table \ref{tab3}, and present the performance of each group individually in subsequent subsections.}
\subsubsection{Group 1}
\begin{figure}[tb]\centering
\includegraphics[trim={18 0 0 0},clip,scale=0.86]{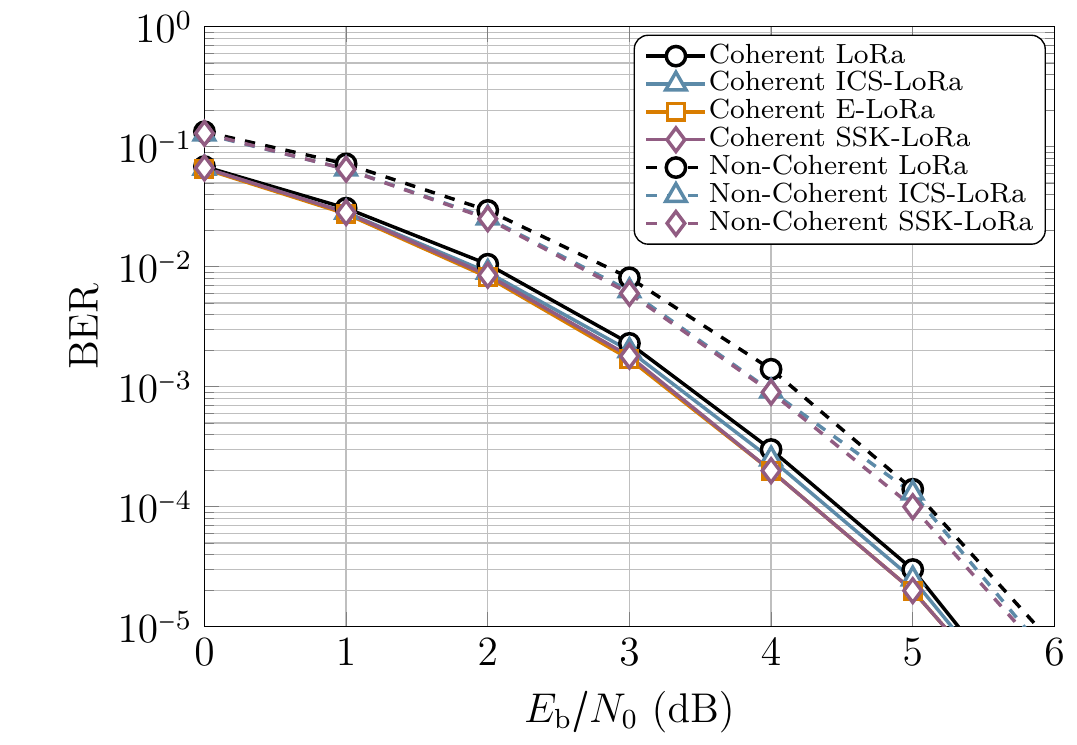}
  \caption{BER performance for group 1 schemes considering coherent/non-coherent detection in AWGN channel for \(\lambda = 8\).}
\label{fig1_ber}
\end{figure}

\textcolor{black}{Per Table \ref{tab3}, the first group encompasses LoRa, ICS-LoRa, E-LoRa, and SSK-LoRa. Fig. \ref{fig1_ber} delineates the BER performance of these approaches, wherein coherent SSK-LoRa and E-LoRa evince superior performance compared to coherent LoRa and ICS-LoRa. While ICS-LoRa with coherent detection demonstrates marginal improvement over coherently detected LoRa, SSK-LoRa, and ICS-LoRa with non-coherent detection, outperform non-coherently detected LoRa. It is pertinent to note that non-coherent detection of E-LoRa is not feasible. The superiority of ICS-LoRa, E-LoRa, and SSK-LoRa over LoRa stems from the lower required SNR/bit essential for correct bit detection.}
\subsubsection{Group 2}
\begin{figure}[tb]\centering
\includegraphics[trim={18 0 0 0},clip,scale=0.86]{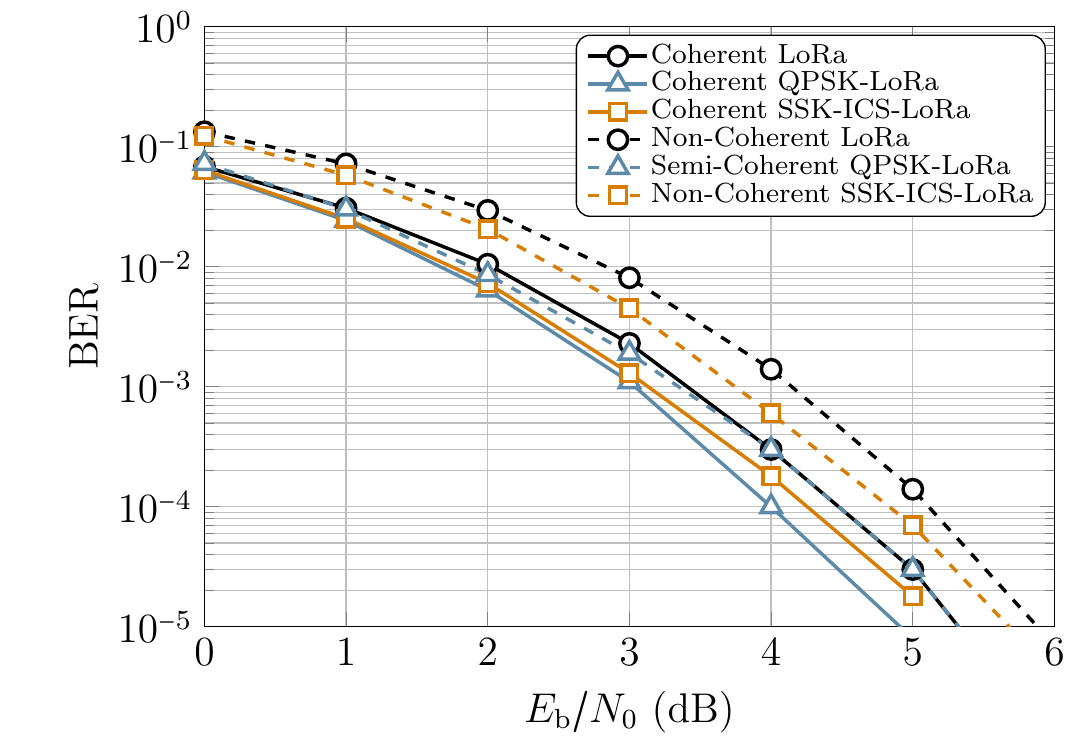}
  \caption{BER performance for group 2 schemes considering coherent/non-coherent detection in AWGN channel for \(\lambda = 8\).}
\label{fig2_ber}
\end{figure}

\textcolor{black}{The BER performances of the schemes in group 2 are delineated in Fig. \ref{fig2_ber}. This section elucidates PSK-LoRa with quaternary PSs, i.e., QPSK-LoRa. It is apparent that coherently detected QPSK-LoRa surpasses detected SSK-ICS-LoRa and LoRa. Nonetheless, coherently and non-coherently detected SSK-ICS-LoRa surpass coherently and non-coherently detected LoRa. Note that non-coherent detection for QPSK-LoRa is unattainable; nevertheless, it can perform semi-coherent detection. The performance of semi-coherent QPSK-LoRa is identical to coherent LoRa. However, it is recalled that the semi-coherent detection of QPSK-LoRa necessitates synchronization and equalization at the receiver. }
\subsubsection{Group 3}
\begin{figure}[tb]\centering
\includegraphics[trim={18 0 0 0},clip,scale=0.86]{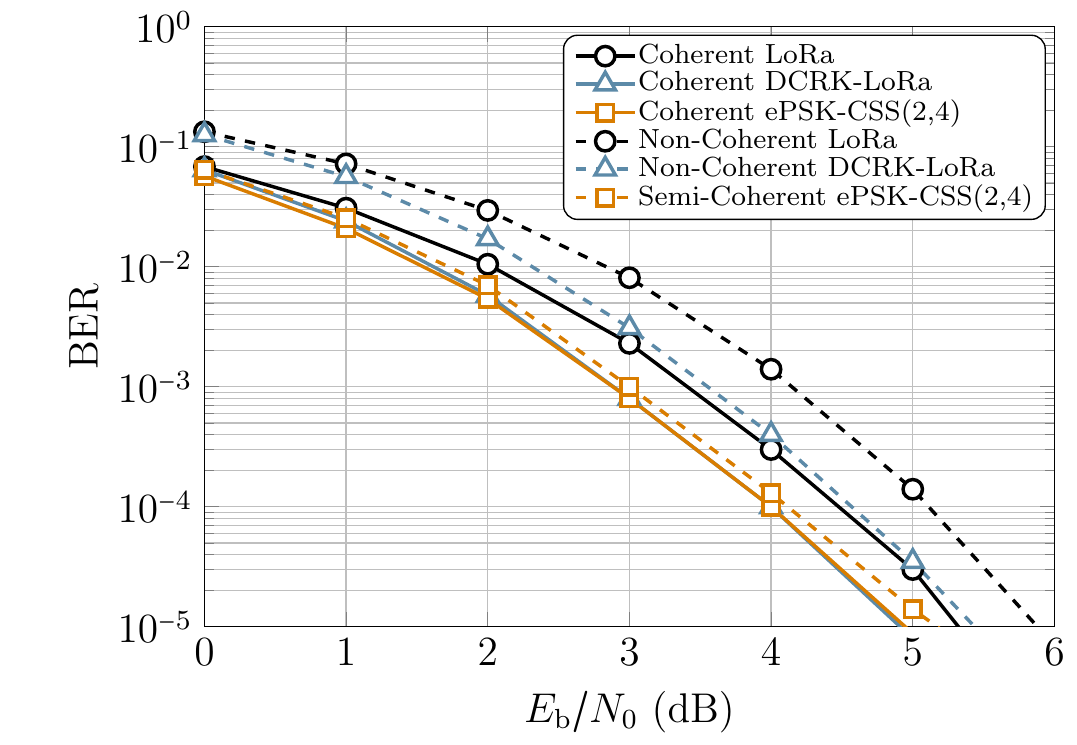}
  \caption{BER performance for group 3 schemes considering coherent/non-coherent detection in AWGN channel for \(\lambda = 8\).}
\label{fig3_ber}
\end{figure}

\textcolor{black}{Fig. \ref{fig3_ber} showcases the bit error rate (BER) performances of group 3 schemes, namely DCRK-LoRa and ePSK-CSS(\(2,4\)), in comparison with LoRa. We can observe that the coherent detection performance of both DCRK-LoRa and ePSK-CSS(\(2,4\)) is similar, and both outperform the coherently detected LoRa. We should note that non-coherent detection is infeasible for ePSK-CSS(\(2,4\)); hence, we employ a semi-coherent detection mechanism. On the other hand, the non-coherently detected DCRK-LoRa performs better than the non-coherent detection of LoRa. However, the semi-coherently detected ePSK-CSS(\(2,4\)) exhibits slightly inferior performance compablack to the coherent detection of ePSK-CSS(\(2,4\). It should be emphasized that the receiver's complexity for DCRK-LoRa, regardless of the detection method, is considerably higher than that of ePSK-CSS(\(2,4\)).}
\subsubsection{Group 4}
\begin{figure}[tb]\centering
\includegraphics[trim={18 0 0 0},clip,scale=0.86]{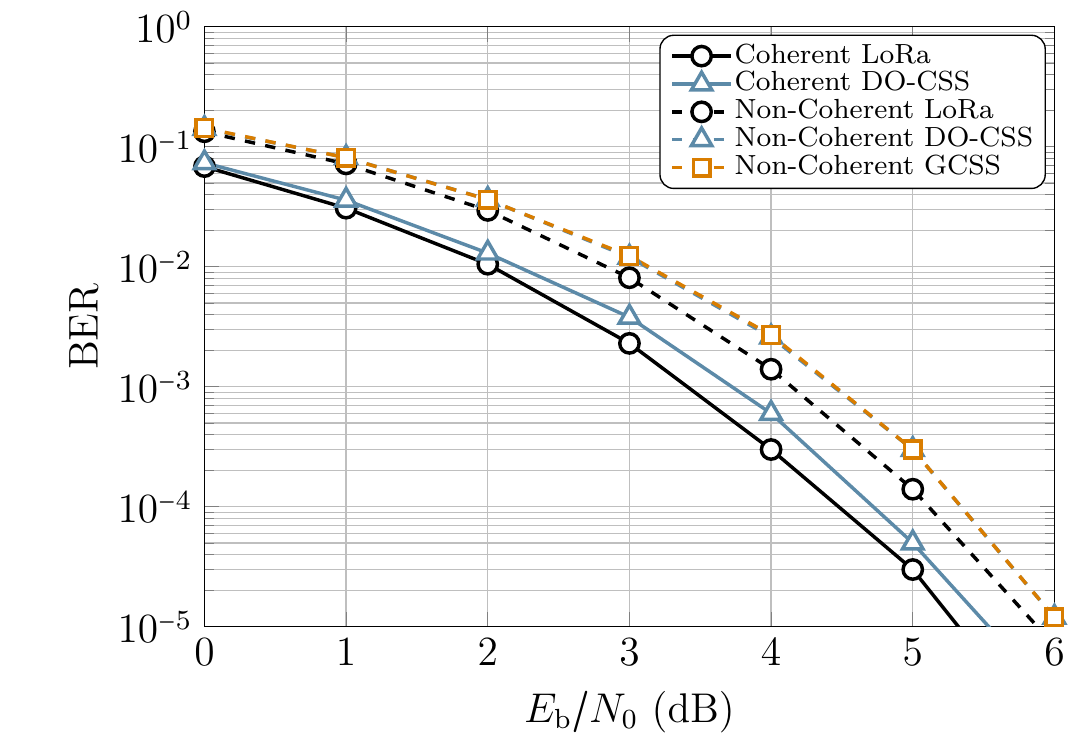}
  \caption{BER performance for group 4 schemes considering coherent/non-coherent detection in AWGN channel for \(\lambda = 8\).}
\label{fig4_ber}
\end{figure}

\textcolor{black}{In this section, we shall present the BER performance of schemes in Group 4, which comprises DO-CSS and GCSS. Note that no coherent detection mechanism has been proposed in the literature for GCSS; thus, we shall only showcase the performance of its non-coherent receiver. The BER performance of coherently detected DO-CSS demonstrates only a marginal degradation compared to detected LoRa, as illustrated in Fig. \ref{fig4_ber}. It is important to recall that the achievable SE for DO-CSS is \(\sfrac{\lambda-2}{M}\) bits/s/Hz higher relative to LoRa. Furthermore, both DO-CSS and GCSS exhibit similar performances concerning non-coherent detection, which is somewhat worse than non-coherently detected LoRa.}
\subsubsection{Group 5}
\begin{figure}[tb]\centering
\includegraphics[trim={18 0 0 0},clip,scale=0.86]{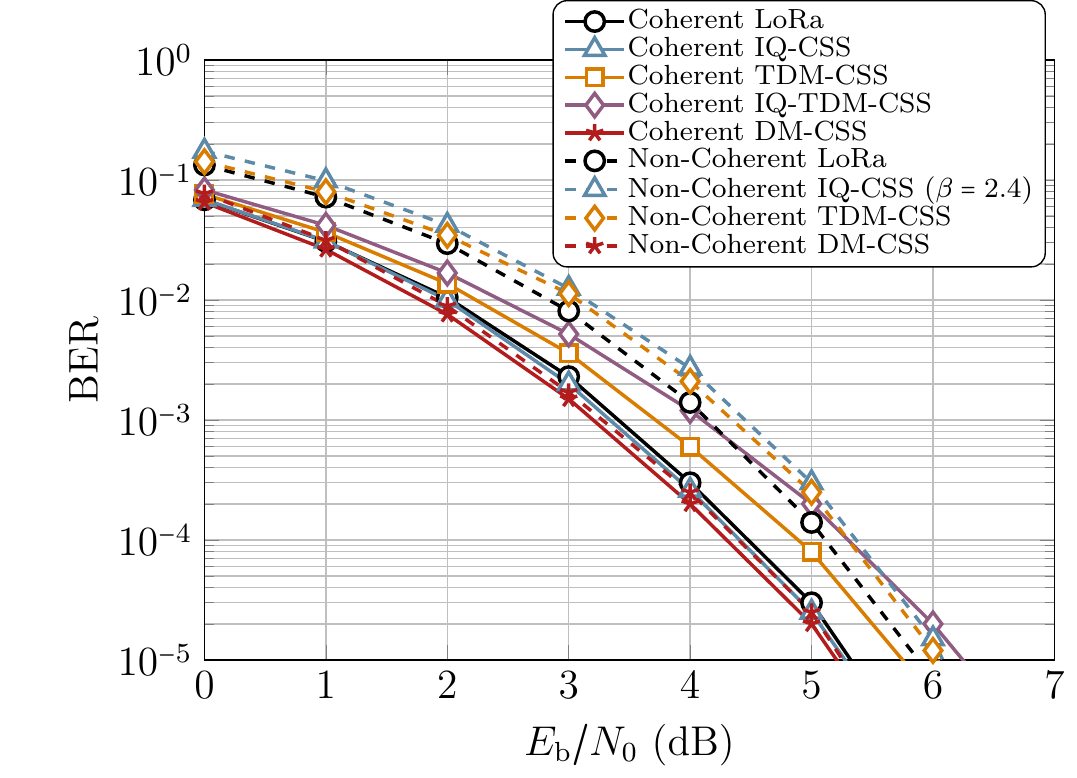}
  \caption{BER performance for group 5 schemes considering coherent/non-coherent detection in AWGN channel for \(\lambda = 8\).}
\label{fig5_ber}
\end{figure}

\textcolor{black}{Group 5 comprises several schemes, namely, IQ-CSS, TDM-CSS, IQ-TDM-CSS, and DM-CSS, whose BER performances under coherent and non-coherent detection are compared in Fig. \ref{fig5_ber}. It is pertinent to note that IQ-CSS, TDM-CSS, IQ-TDM-CSS, and DM-CSS all offer an increase in SE of \(\sfrac{\lambda}{M}\), \(\sfrac{\lambda}{M}\), \(\sfrac{3\lambda}{M}\), and \(\sfrac{(\lambda+1)}{M}\), respectively, relative to LoRa. As regards the BER performance, we can observe that the coherently detected IQ-CSS outperforms coherently detected TDM-CSS, IQ-TDM-CSS, and LoRa. Conversely, the non-coherently detected IQ-CSS with \(\beta = 2.4\) exhibits a worse BER performance than that of non-coherently detected LoRa, TDM-CSS, and DM-CSS. It is worth mentioning that non-coherent detection is not feasible for IQ-TDM-CSS. Although the BER performance of coherently detected TDM-CSS is better than coherently detected IQ-TDM-CSS, the latter offers the highest SE increase of \(\sfrac{3\lambda}{M}\) bits/s/Hz over TDM-LoRa. Nevertheless, the BER performance of coherently detected IQ-TDM-CSS is worse than that of all other schemes under coherent detection. Finally, we can infer that DM-CSS outperforms all other schemes in the group, both under coherent and non-coherent detection, regarding BER performance.}
\subsubsection{Group 6}
\begin{figure}[tb]\centering
\includegraphics[trim={18 0 0 0},clip,scale=0.86]{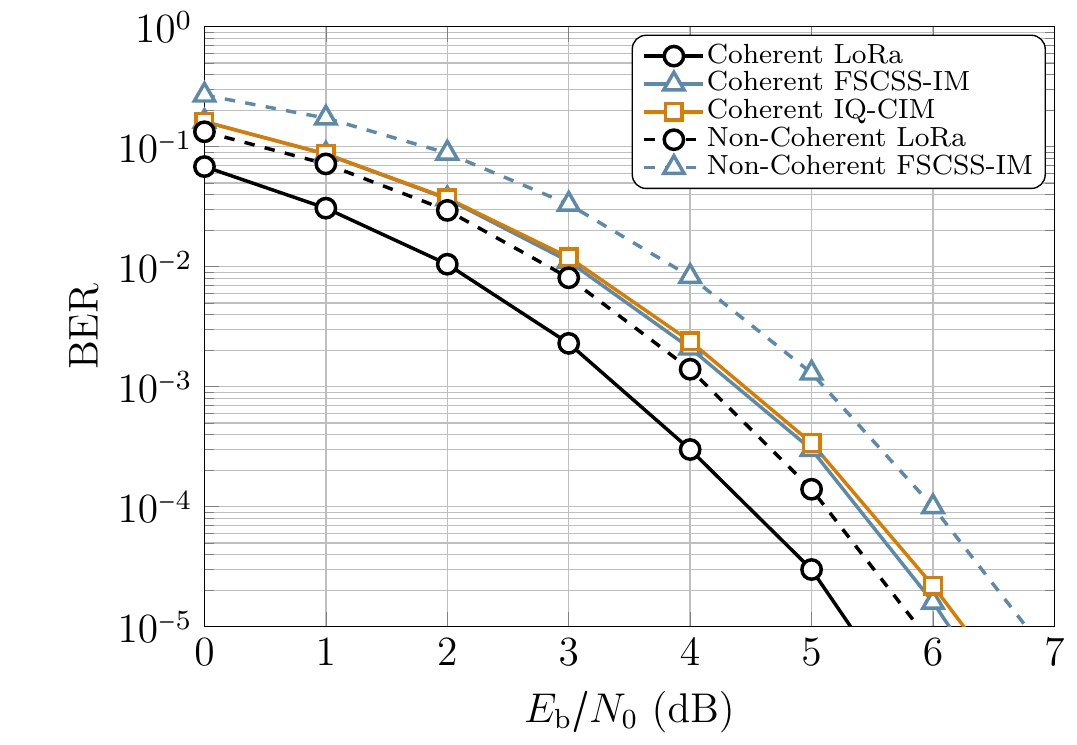}
  \caption{BER performance for group 6 schemes considering coherent/non-coherent detection in AWGN channel for \(\lambda = 8\).}
\label{fig6_ber}
\end{figure}

\textcolor{black}{In the sixth group, we focus on the IM schemes, namely FSCSS-IM and IQ-CIM, where FSCSS-IM employs \(\varsigma = 2\) and IQ-CIM employs \(\varsigma_i = \varsigma_q =2\). The BER performances of both schemes are depicted in Fig. \ref{fig6_ber}. We can observe that the BER performance of FSCSS-IM using both coherent and non-coherent detection is inferior to that of LoRa; however, its SE is higher. Moreover, the BER performance of coherently detected IQ-CIM is nearly comparable to that of coherently detected FSCSS-IM. We should note that the achievable SE of IQ-CIM is twice that of FSCSS-IM and is \(\sfrac{2\lambda-4}{M}\) bits/s/Hz higher than LoRa when \(\varsigma_i = \varsigma_q =2\).}
\subsubsection{\textcolor{black}{Takeaways}}
\textcolor{black}{The BER performance curves exhibit a consistent trend with the SE versus EE analysis, as they both utilize the same channel model. Consequently, the insights gleaned from the SE versus EE analysis apply to the BER analysis in the AWGN channel. Notably, the SSK-LoRa waveform, when detected coherently or non-coherently, is optimal for group 1. In contrast, the QPSK-LoRa waveform detected coherently or semi-coherently delivered the best performance for group 2. The DCRK-LoRa waveform outperforms the other designs in group 3, whereas the semi-coherently detected ePSK-CSS(2,4) waveform is the most effective. Once again, DCRK-LoRa represents the most resilient waveform design for a CSS scheme.}

\textcolor{black}{For groups other than group 3, we employ the MC CSS schemes. In group 4, DO-CSS is the most effective option for adopting coherent detection. Conversely, when we use non-coherent detection, DO-CSS and GCSS yield equivalent performance levels. In group 5, DM-CSS maintains its superiority for both coherent and non-coherent detection mechanisms. Given that the CSS schemes in group 5 boast better SE than those in group 4, DM-CSS's waveform design is the most robust among the MC CSS schemes.}

\textcolor{black}{Regarding the MC-IM scheme, coherently detected IQ-CIM outperforms FSCSS-IM, whereas the non-coherent detection mechanism does not provide enough information to discern FSCSS-IM's performance.}
\subsection{Bit-Error Performance considering Phase Offset}
\textcolor{black}{This section will examine the efficacy of all the prescribed methods considering PO, a well-documented phenomenon prevalent in low-cost devices. For this purpose, we shall scrutinize the received symbol that both PO and AWGN corrupt, and it can be expressed as follows:}
\begin{equation}\label{po_eq}
y(n) = \exp\{j\psi\} s(n) + w(n),
\end{equation}
\textcolor{black}{where \(\psi\) symbolizes the PO. We shall conduct a thorough analysis of the efficacy of each prescribed approach while accounting for the varied values of PO, specifically \(\psi = \sfrac{\pi}{8}\) radians, \(\psi = \sfrac{\pi}{4}\) radians, and \(\lambda = 8\). Additionally, we shall adhere to the same grouping classification criteria of the schemes enumerated in Table \ref{tab3}.}
\subsubsection{Group 1}
\begin{figure}[tb]\centering
\includegraphics[trim={18 0 0 0},clip,scale=0.86]{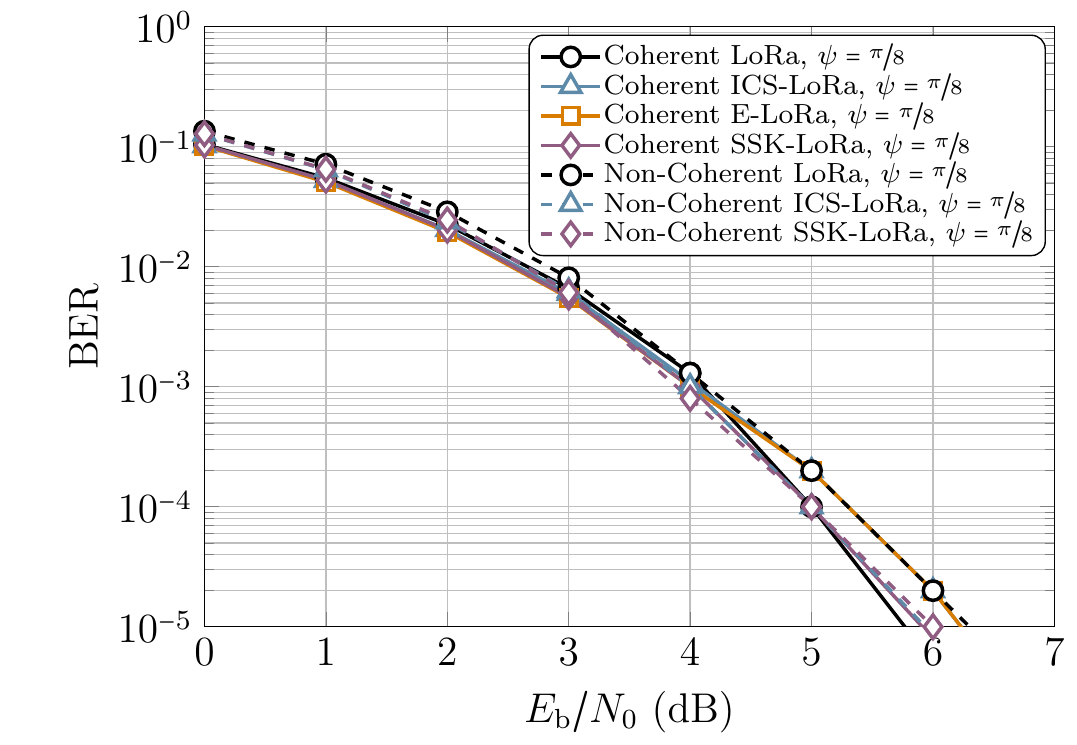}
  \caption{BER performance for group 1 schemes considering coherent/non-coherent detection, \(\psi=\sfrac{\pi}{8}\), AWGN channel, and \(\lambda = 8\).}
\label{fig1_ber_po}
\end{figure}

\textcolor{black}{Fig. \ref{fig1_ber_po} provide insights into the BER performances of schemes that have been classified within group 1 while taking into consideration the impact of PO, specifically at \(\psi = \sfrac{\pi}{8}\) radians. We can observe that the employment of coherent detection in LoRa negatively impacts BER performance in the presence of PO. In contrast, the BER performance of non-coherently detected LoRa remains unaltered. Moreover, it is noteworthy that the utilization of coherent detection engenders a more significant impact on BER performance than when non-coherent detection is employed. In particular, for \(\psi = \sfrac{\pi}{8}\) radians, the BER performances of ICS-LoRa, and SSK-LoRa, while utilizing both coherent detection and non-coherent detection are almost identical to that of coherently detected LoRa. Conversely, the BER performance of E-LoRa is most adversely affected in the presence of PO and is akin to that of non-coherently detected LoRa.}

\begin{figure}[tb]\centering
\includegraphics[trim={18 0 0 0},clip,scale=0.86]{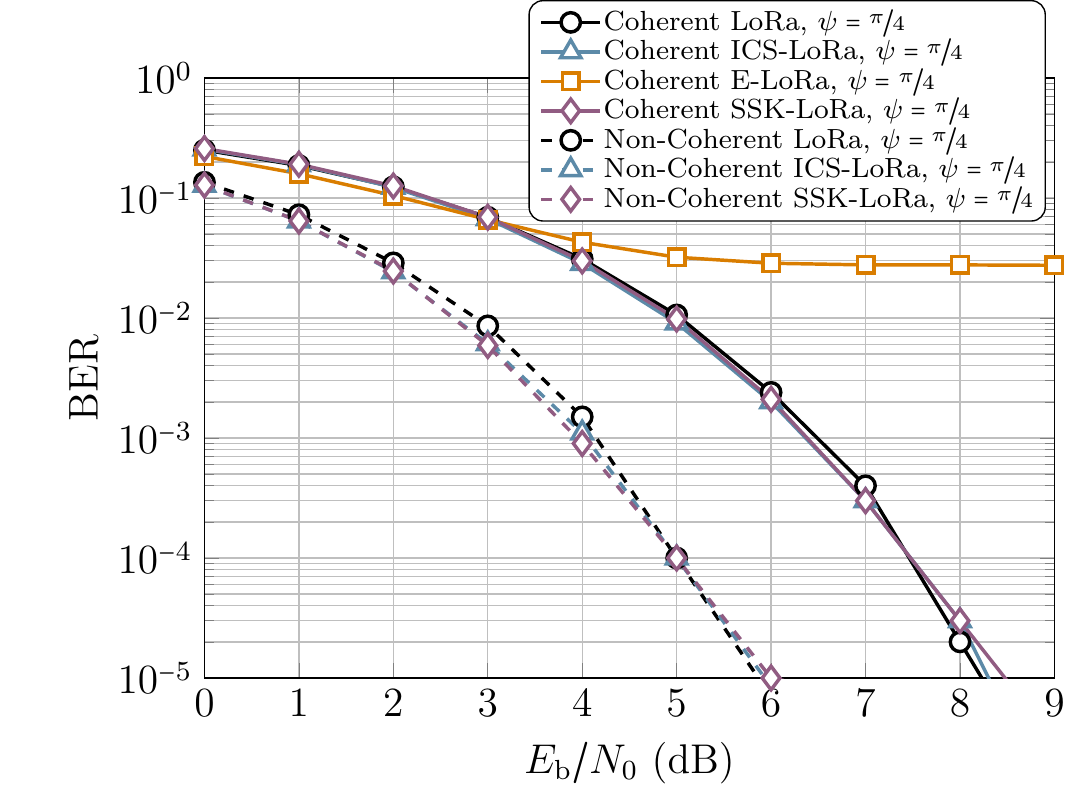}
  \caption{BER performance for group 1 schemes considering coherent/non-coherent detection, \(\psi=\sfrac{\pi}{4}\), AWGN channel, and \(\lambda = 8\).}
\label{fig2_ber_po}
\end{figure}

\textcolor{black}{Fig. \ref{fig2_ber_po} illustrates the performance of schemes that have been classified within group 1 while considering \(\psi = \sfrac{\pi}{4}\) radians. Notably, it is evident from Fig. \ref{fig2_ber_po} that the BER performance of all the schemes that have employed coherent detection undergoes a severe impact, which results in a degradation of over \(2\) dB in the \(\sfrac{E_{\mathrm{b}}}{N_0}\) metric. Given that the information is transmitted via the phase of the un-chirped symbols, coherently detected E-LoRa suffers from inferior BER performance. On the other hand, the BER performance of LoRa, ICS-LoRa, and SSK-LoRa, which have employed non-coherent detection, remains oblivious to the increase in PO, thereby attesting to the supremacy of non-coherent detection over coherent detection.}
\subsubsection{Group 2}
\begin{figure}[tb]\centering
\includegraphics[trim={18 0 0 0},clip,scale=0.86]{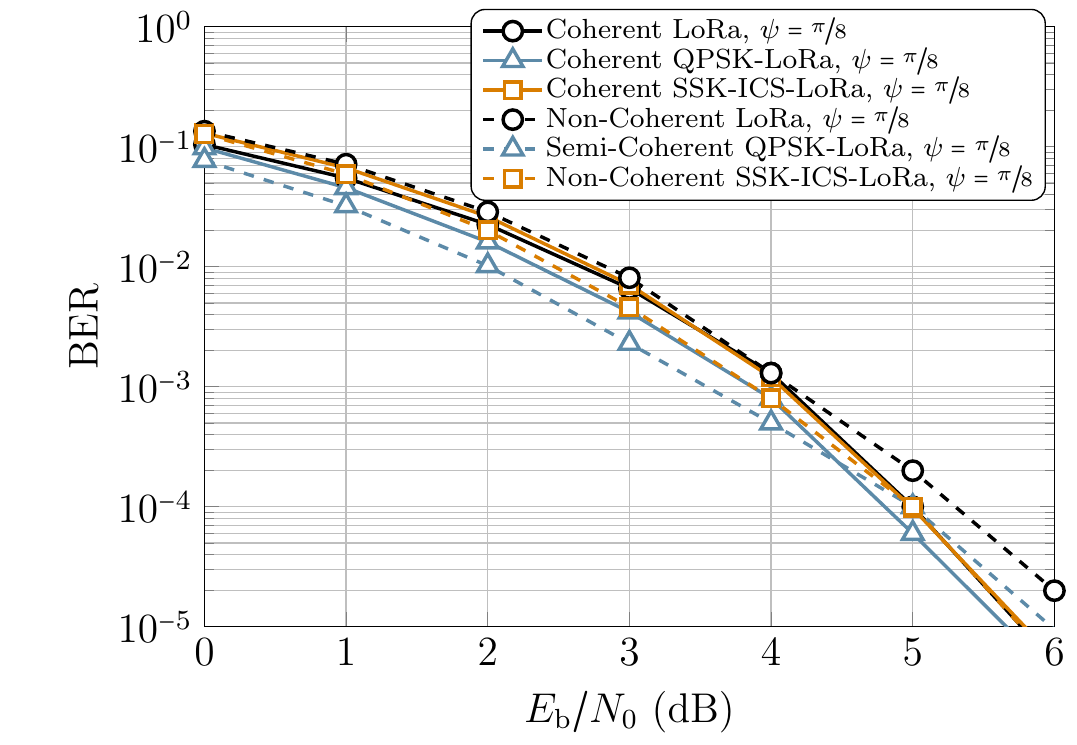}
  \caption{BER performance for group 2 schemes considering coherent/non-coherent detection, \(\psi=\sfrac{\pi}{8}\), AWGN channel, and \(\lambda = 8\).}
\label{fig3_ber_po}
\end{figure}

\textcolor{black}{In Fig. \ref{fig3_ber_po}, the BER performances of QPSK-LoRa and SSK-ICS-LoRa in the presence of PO at \(\psi =\sfrac{\pi}{8}\) radians are presented. It is worth noting that the BER performance of QPSK-LoRa, which relies on PS modulation, experiences a deterioration of approximately \(0.5\) dB for \(\psi =\sfrac{\pi}{8}\) radians, compared to when \(\psi =0\) radians when both coherent detection and semi-coherent detection are employed. This is attributable to the negative influence of PO on PS modulation. As for SSK-ICS-LoRa, the BER performance using non-coherent detection remains unaffected, while the BER performance using coherent detection undergoes a degradation of approximately \(1\) dB when \(\psi =\sfrac{\pi}{8}\) radians relative to when \(\psi =0\) radians. It is also worth noting that the BER performance concerning non-coherent detection for SSK-ICS-LoRa is marginally better when compared to the use of coherent detection for low \(\sfrac{E_\mathrm{b}}{N_0}\) values. Finally, the BER performance of coherently detected LoRa and SSK-ICS-LoRa is indistinguishable, implying that SSK-ICS-LoRa suffers a higher degradation in the presence of PO due to the high cross-correlation between different symbols.}

\begin{figure}[tb]\centering
\includegraphics[trim={18 0 0 0},clip,scale=0.86]{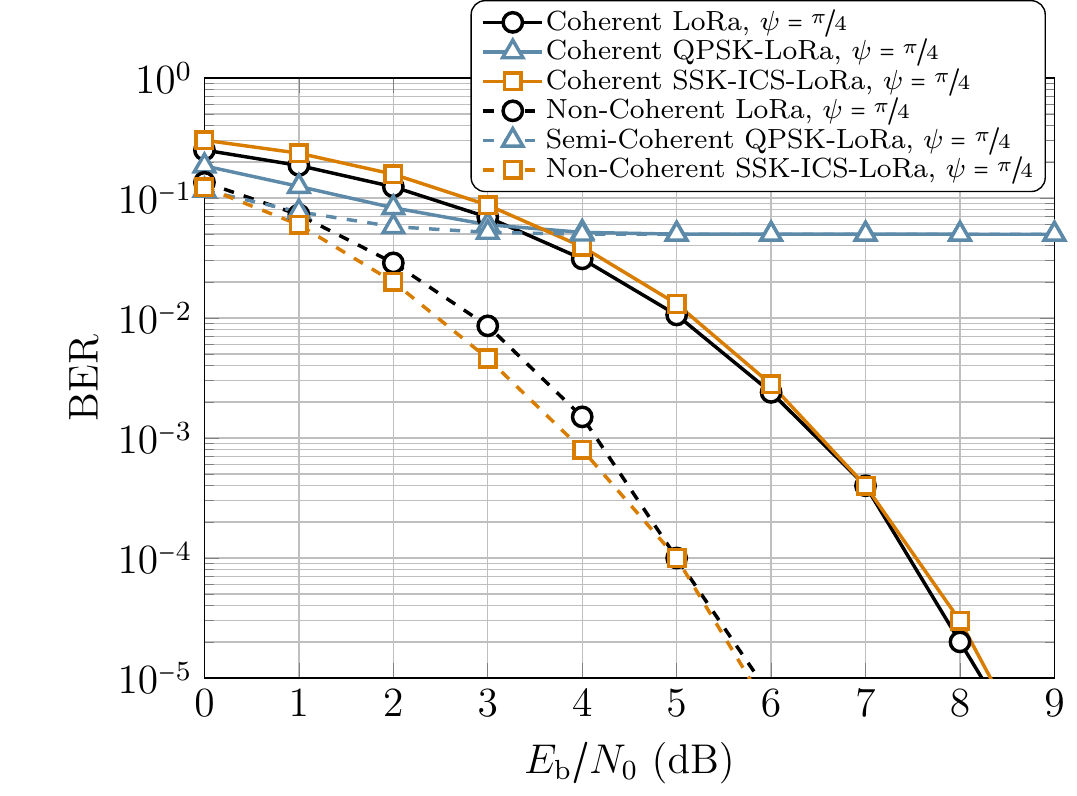}
  \caption{BER performance for group 2 schemes considering coherent/non-coherent detection, \(\psi=\sfrac{\pi}{4}\), AWGN channel, and \(\lambda = 8\).}
\label{fig4_ber_po}
\end{figure}

\textcolor{black}{The BER performance of group 2 schemes in the presence of PO is illustrated in Fig. \ref{fig4_ber_po}, where \(\psi =\sfrac{\pi}{4}\) radians. As depicted by the BER curves, an error floor of around \(10^{-1}\) is observed for both coherent and semi-coherent detection mechanisms at \(\psi =\sfrac{\pi}{4}\) radians due to the influence of high PO on the PSs that are integrated into FS. Notably, the BER performance of coherently and non-coherently detected SSK-ICS-LoRa is almost the same as coherently and non-coherently detected LoRa, respectively. Furthermore, the BER performance of converging schemes exhibits degradation in coherent detection compared to non-coherent detection.}
\subsubsection{Group 3}
\begin{figure}[tb]\centering
\includegraphics[trim={18 0 0 0},clip,scale=0.86]{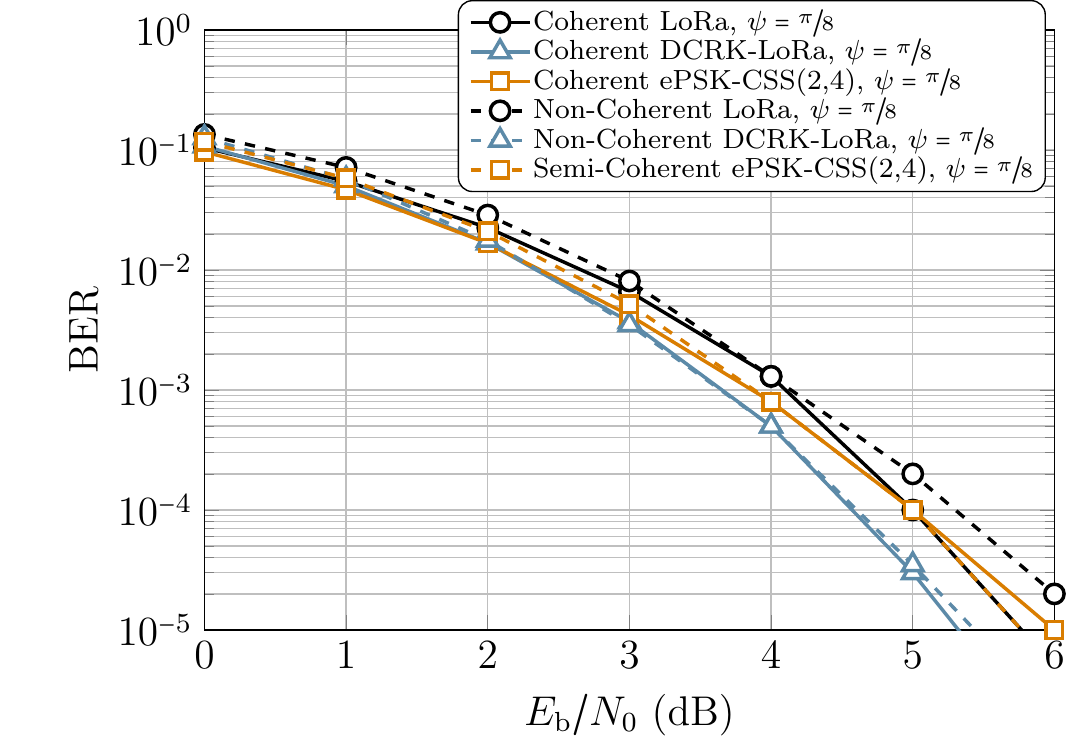}
  \caption{BER performance for group 3 schemes considering coherent/non-coherent detection, \(\psi=\sfrac{\pi}{8}\), AWGN channel, and \(\lambda = 8\).}
\label{fig5_ber_po}
\end{figure}

\textcolor{black}{Fig. \ref{fig5_ber_po} showcases the BER performance of DCRK-LoRa and ePSK-CSS(\(2,4\)) in the presence of PO of \(\psi =\sfrac{\pi}{8}\) radians. It is evident from the results that the BER performance of coherently detected DCRK-LoRa degrades by \(0.5\) dB when PO of \(\psi =\sfrac{\pi}{8}\) radians is introduced compared to when PO is absent. In contrast, the BER performance of non-coherently detected DCRK-LoRa remains unaffected under the same conditions. Additionally, the BER performance of coherently and non-coherently detected DCRK-LoRa is identical when \(\psi =\sfrac{\pi}{8}\) radians. On the other hand, the BER performance of ePSK-CSS(\(2,4\)) using coherent and semi-coherent detection methods suffers degradation of about \(1\) dB when PO of \(\psi =\sfrac{\pi}{8}\) radians is introduced. Moreover, the BER performance of DCRK-LoRa using either coherent or non-coherent detection outperforms that of LoRa and ePSK-CSS(\(2,4\)) using coherent or semi-coherent/non-coherent detection schemes.}

\begin{figure}[tb]\centering
\includegraphics[trim={18 0 0 0},clip,scale=0.86]{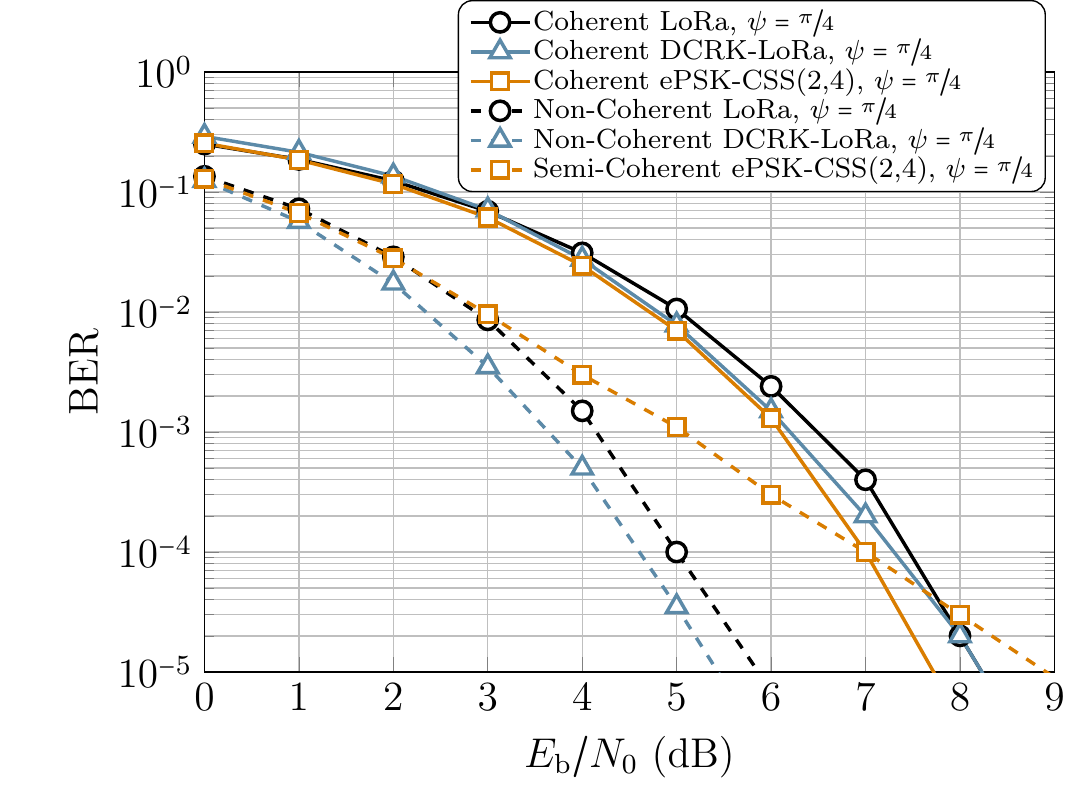}
  \caption{BER performance for group 3 schemes considering coherent/non-coherent detection, \(\psi=\sfrac{\pi}{4}\), AWGN channel, and \(\lambda = 8\).}
\label{fig6_ber_po}
\end{figure}

\textcolor{black}{Fig. \ref{fig6_ber_po} displays the BER performance of group 3 schemes for a PO of \(\psi =\sfrac{\pi}{4}\) radians. The BER performance of non-coherently detected DCRK-LoRa remains impervious to the introduction of PO from \(\psi =\sfrac{\pi}{8}\) radians to \(\psi =\sfrac{\pi}{4}\) radians. However, the BER performance of DCRK-LoRa using coherent detection undergoes a substantial degradation of \(2.5\) dB as PO rises from \(\psi= 0\) radians to \(\psi =\sfrac{\pi}{4}\) radians. The semi-coherently detected PSK-LoRa displays a less favorable BER performance compared to semi-coherently detected ePSK-CSS(\(2,4\)), but both experience significant performance degradation. The coherently detected ePSK-CSS(\(2,4\)) also suffers a BER degradation of \(2.5\) dB as PO is raised from \(\psi =0\) radians to \(\psi =\sfrac{\pi}{4}\) radians. Remarkably, coherently detected ePSK-CSS(\(2,4\)) exhibits slightly better BER performance than DCRK-LoRa for high \(\sfrac{E_\mathrm{b}}{N_0}\) values.}

\subsubsection{Group 4}
\begin{figure}[tb]\centering
\includegraphics[trim={18 0 0 0},clip,scale=0.86]{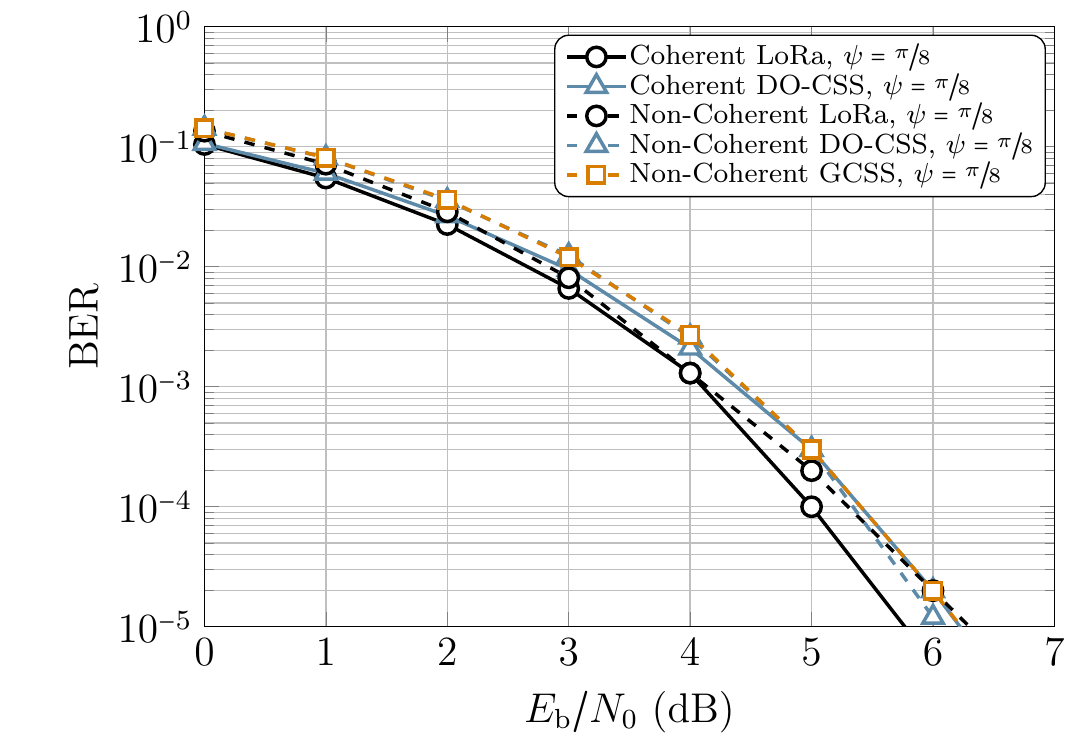}
  \caption{BER performance for group 4 schemes considering coherent/non-coherent detection, \(\psi=\sfrac{\pi}{8}\), AWGN channel, and \(\lambda = 8\).}
\label{fig7_ber_po}
\end{figure}

\textcolor{black}{The BER performance of DO-CSS and GCSS in the presence of PO of \(\psi =\sfrac{\pi}{8}\) radians using coherently and non-coherently detected schemes is depicted in Fig. \ref{fig7_ber_po}. It is discernible that the coherently detected DO-CSS exhibits a BER degradation of approximately \(0.5\) dB as the PO increases from \(\psi =0\) radians to \(\psi =\sfrac{\pi}{8}\) radians. In contrast, the non-coherently detected DO-CSS does not demonstrate any PO degradation. Furthermore, the BER performances of coherently detected and non-coherently detected DO-CSS are almost indistinguishable. Regarding non-coherently detected GCSS, its BER performance remains unaltered, especially in the presence of PO. The BER performances of coherently and non-coherently detected LoRa remain superior to those of coherently and non-coherently detected DO-CSS and non-coherently detected GCSS, as observed from Fig. \ref{fig7_ber_po}.} 

\begin{figure}[tb]\centering
\includegraphics[trim={18 0 0 0},clip,scale=0.86]{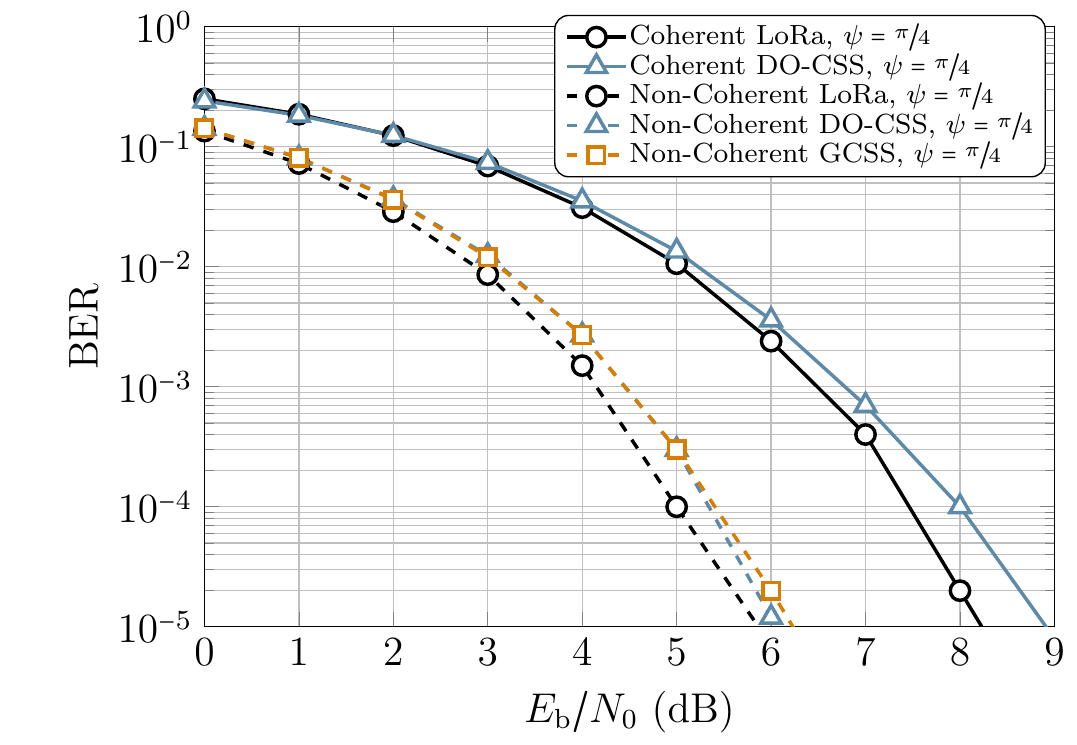}
  \caption{BER performance for group 4 schemes considering coherent/non-coherent detection, \(\psi=\sfrac{\pi}{4}\), AWGN channel, and \(\lambda = 8\).}
\label{fig8_ber_po}
\end{figure}

\textcolor{black}{Fig. \ref{fig8_ber_po} depicts a substantial increase of approximately \(3\) dB in BER for coherently detected DO-CSS when PO equals \(\psi =\sfrac{\pi}{4}\) radians. However, non-coherently detected DO-CSS and GCSS exhibit no deterioration in BER, even when PO is elevated from \(\psi =0\) radians to \(\psi =\sfrac{\pi}{4}\) radians.} 
\subsubsection{Group 5}
\begin{figure}[tb]\centering
\includegraphics[trim={18 0 0 0},clip,scale=0.86]{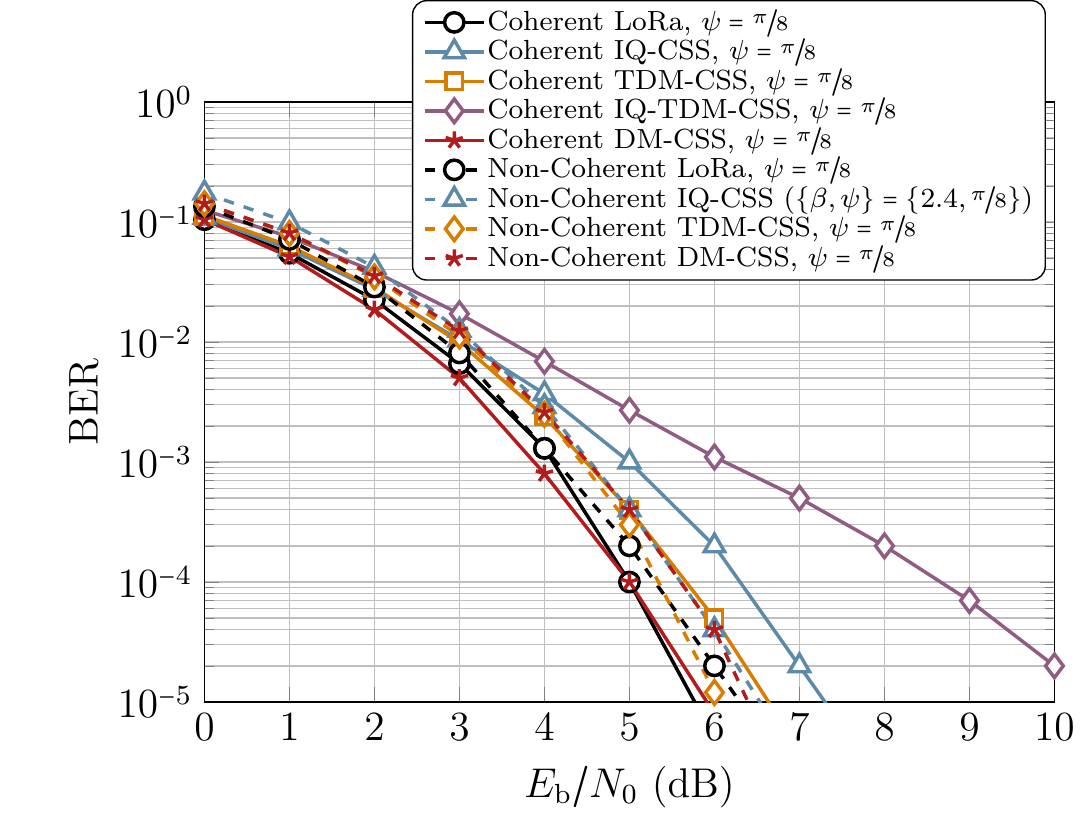}
  \caption{BER performance for group 5 schemes considering coherent/non-coherent detection, \(\psi=\sfrac{\pi}{8}\), AWGN channel, and \(\lambda = 8\).}
\label{fig9_ber_po}
\end{figure}

\textcolor{black}{Figure \ref{fig9_ber_po} demonstrates the BER performances of group 5 schemes in the presence of PO with \(\psi =\sfrac{\pi}{8}\) radians. It is evident from the figure that the coherently detected IQ-CSS and TDM-CSS incur BER degradations of approximately \(1.5\) dB and \(0.5\) dB, respectively, compared to their performance in the absence of PO. However, non-coherently detected IQ-CSS and TDM-CSS exhibit PO-agnostic BER performance. Coherently detected IQ-TDM-CSS suffers a severe degradation in BER performance, as indicated in Fig. \ref{fig9_ber_po}. Furthermore, coherently and non-coherently detected DM-CSS experience BER penalties of about \(1\) dB and \(1.5\) dB, respectively, when PO increases from \(\psi =0\) radians to \(\psi =\sfrac{\pi}{8}\) radians. Among coherently detected approaches, DM-CSS performs the best in terms of BER performance, whereas, among non-coherently detected approaches, TDM-CSS, apart from LoRa, provides the best performance.}

\begin{figure}[tb]\centering
\includegraphics[trim={18 0 0 0},clip,scale=0.86]{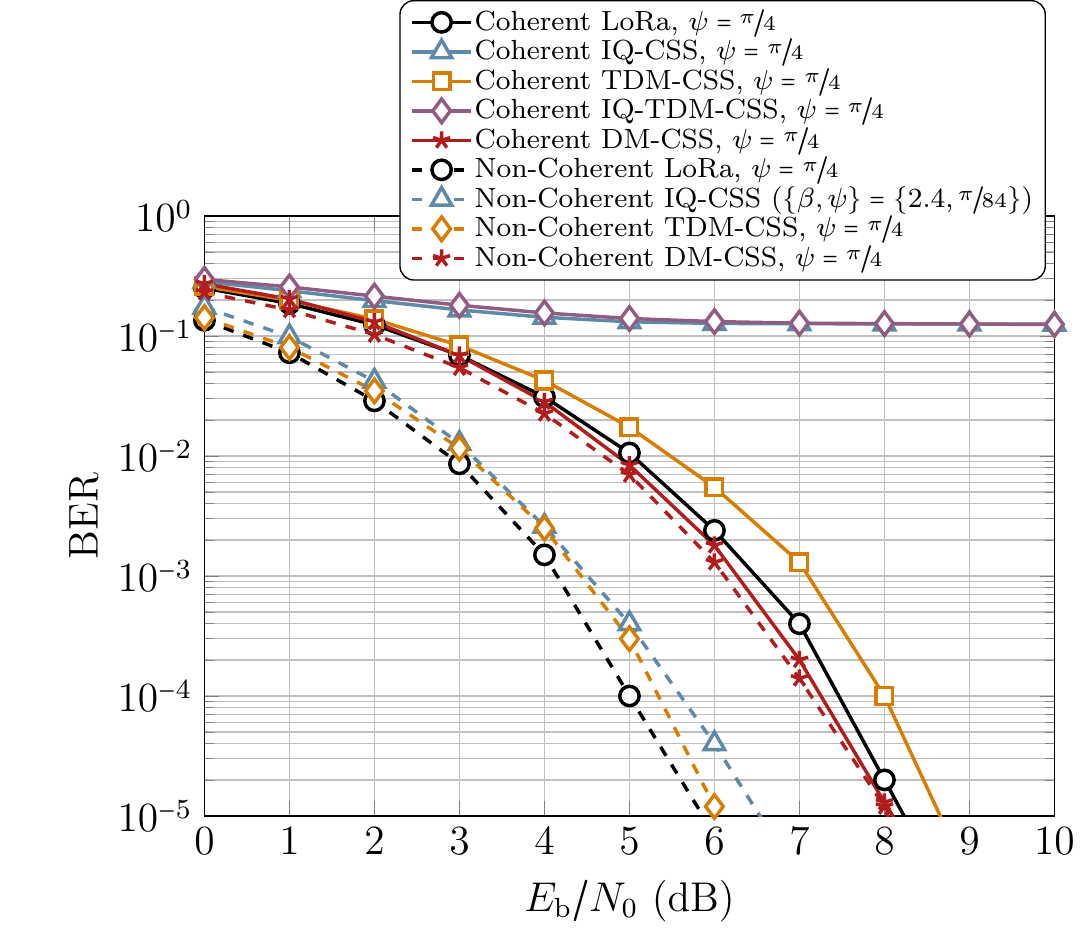}
  \caption{BER performance for group 5 schemes considering coherent/non-coherent detection, \(\psi=\sfrac{\pi}{4}\), AWGN channel, and \(\lambda = 8\).}
\label{fig10_ber_po}
\end{figure}

\textcolor{black}{The BER performances of the fifth group of schemes are illustrated in Fig. \ref{fig10_ber_po} for PO equal to \(\psi =\sfrac{\pi}{4}\) radians. It can be observed that coherently and non-coherently detected IQ-CSS and coherently detected IQ-TDM-CSS experience a debilitating error floor at BER of \(10^{-1}\) for \(\psi =\sfrac{\pi}{4}\) radians. This indicates a significant decline in BER performance. In contrast, the BER performance of non-coherently detected TDM-CSS remains unchanged. For coherently detected TDM-CSS, the BER performance experiences a decline of approximately \(3.2\) dB when PO is \(\psi =\sfrac{\pi}{4}\) radians compared to when no PO is present. Furthermore, the BER performance of coherently and non-coherently detected DM-CSS suffers a penalty of \(2.8\) dB when PO increases from \(\psi =0\) radians to \(\psi =\sfrac{\pi}{4}\) radians due to the encoding of binary PSs in the signal structure. Overall, non-coherently detected TDM-CSS provides the most optimal BER performance.}
\subsubsection{Group 6}
\begin{figure}[tb]\centering
\includegraphics[trim={18 0 0 0},clip,scale=0.86]{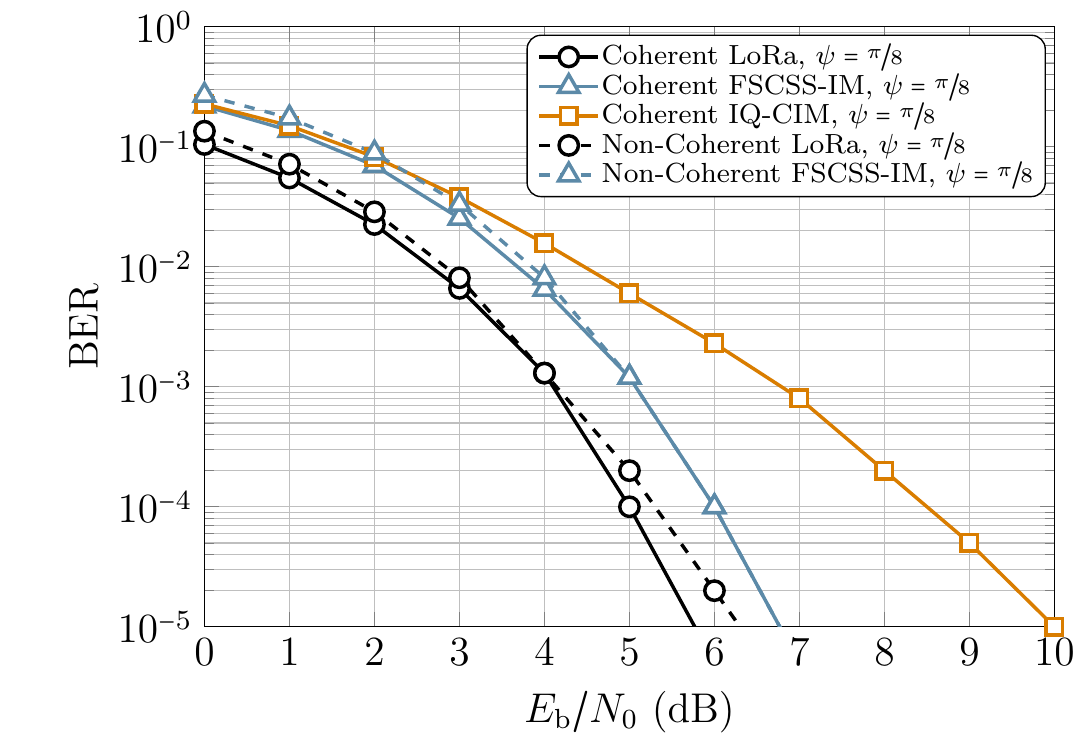}
  \caption{BER performance for group 6 schemes considering coherent/non-coherent detection, \(\psi=\sfrac{\pi}{8}\), AWGN channel, and \(\lambda = 8\).}
\label{fig11_ber_po}
\end{figure}

\textcolor{black}{Fig. \ref{fig11_ber_po} showcases the BER performances of IM techniques when the PO is \(\psi =\sfrac{\pi}{8}\) radians. In this context, we set the values of \(\varsigma\), \(\varsigma_i\) and \(\varsigma_q\) to \(2\) for FSCSS-IM and IQ-CIM. Evidently, coherently detected FSCSS-IM incurs a BER penalty of approximately \(1\) dB when the PO is increased to \(\psi =\sfrac{\pi}{8}\) radians in comparison to the scenario with no PO. Meanwhile, non-coherently detected FSCSS-IM presents a stable BER performance despite the PO increase from \(\psi =0\) radians to \(\psi =\sfrac{\pi}{8}\) radians. Conversely, coherently detected IQ-CIM significantly degrades BER performance as PO increases. }

\begin{figure}[tb]\centering
\includegraphics[trim={18 0 0 0},clip,scale=0.86]{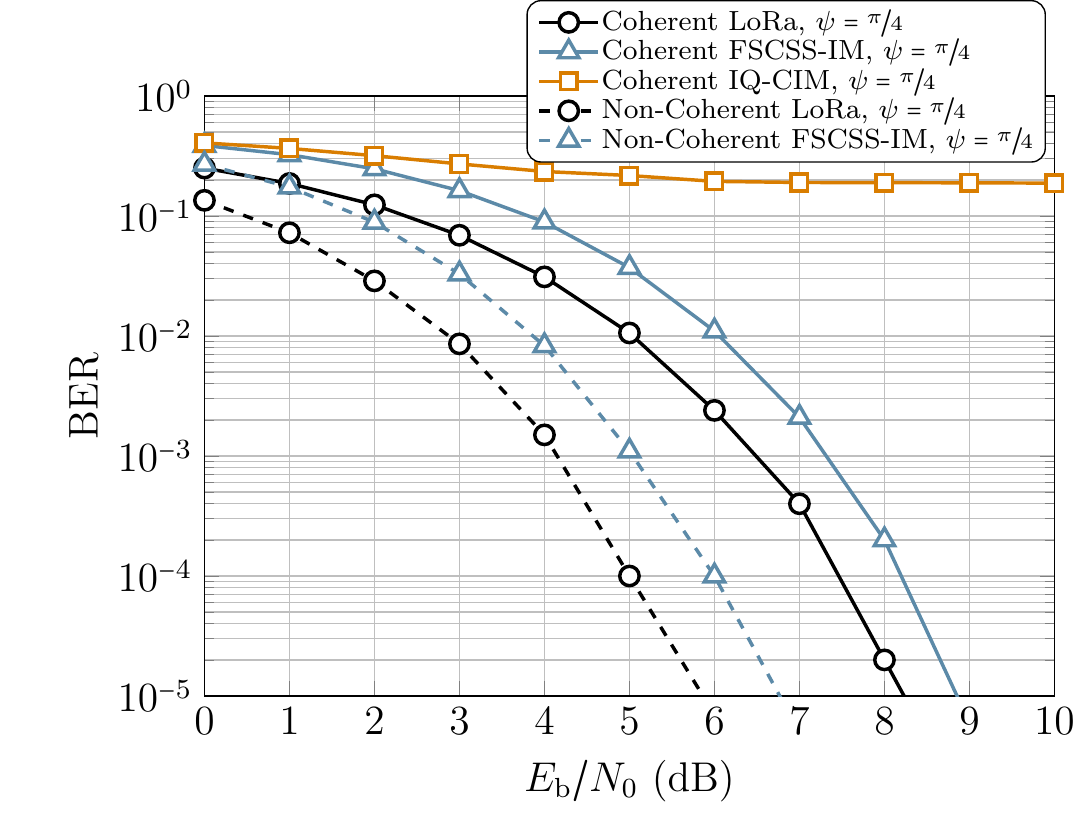}
  \caption{BER performance for group 6 schemes considering coherent/non-coherent detection, \(\psi=\sfrac{\pi}{4}\), AWGN channel, and \(\lambda = 8\).}
\label{fig12_ber_po}
\end{figure}

\textcolor{black}{Upon close examination of the BER performances of FSCSS-IM and IQ-CIM, depicted in Fig. \ref{fig12_ber_po}, it is discernible that coherently detected FSCSS-IM undergoes a severe degradation of nearly \(3\) dB in BER as PO escalates from \(\psi =0\) radians to \(\psi =\sfrac{\pi}{4}\) radians. In contrast, the BER performance of coherently detected IQ-CIM experiences a high error floor in the presence of a PO of \(\psi =\sfrac{\pi}{4}\) radians, thereby rendering its performance suboptimal. Consequently, FSCSS-IM outperforms IQ-CIM in terms of BER performance under PO.}
\subsubsection{\textcolor{black}{Takeaways}}
\textcolor{black}{The most robust scheme against PO in each group is listed in Table  \ref{eff_po}. It is of utmost importance to emphasize that the BER performance undergoes significant changes relative to the AWGN channel when the presence of PO is considered, particularly in LPWANs where cost-effective devices are utilized. In the first group of schemes, coherent detection demonstrates similar performance for all the schemes except for E-LoRa, which exhibits a high level of sensitivity to PO. In the case of non-coherent detection, all the schemes show identical performances except for E-LoRa, which lacks a non-coherent detector. The performance of SSK-ICS-LoRa is optimal for both coherent and non-coherent detection in group 2, whereas QPSK-LoRa exhibits the best performance for the AWGN channel. However, considering the impact of PO, the performance of QPSK-LoRa deteriorates due to the significant effect of PO on the detection of the PSs incorporated in the waveform design. In group 3, DCRK-LoRa exhibits the best coherent and non-coherent detection performance. Although in the AWGN case, the semi-coherently detected ePSK-CSS(2,4) scheme outperforms non-coherently detected DCRK-LoRa, this changes with the incorporation of PO, as PO has a degrading impact on the PSs incorporated in the ePSK-CSS(2,4) waveform design. The DCRK-LoRa scheme represents the best waveform design for SC CSS schemes.}

\textcolor{black}{DO-CSS exhibits the best performance for coherent detection for the group 4 schemes based on MC CSS. It is important to note that no coherent detector exists or has been proposed for GCSS in this study. In contrast, both schemes exhibit similar performance when non-coherent detection is applied. The optimal performance in group 5 is observed in the TDM-CSS scheme. This is contrary to the performance of the AWGN channel. The change in performance in DM-CSS could be attributed to the use of binary PSs in the waveform design or the interference caused by multiple chirp rates. Finally, in group 6, the best performance is exhibited by the FSCSS-IM scheme, as the I/Q components employed in IQ-CIM are susceptible to POs.}

\textcolor{black}{The observed trend in the performance of the schemes indicates that waveform designs incorporating PSs or using I/Q components are generally susceptible to POs and exhibit poorer performance compared to those schemes that do not employ these design methodologies.}
\renewcommand{\arraystretch}{1}
\begin{table*}[h]
  \caption{\textcolor{black}{PO robust schemes from different groups.}}
   \label{eff_po}
  \centering
  \color{black}\begin{tabular}{*{3}{c}}
    \hline
    \hline
    \bfseries{Group}    & \bfseries {Coherent Detection} & \bfseries {Non-Coherent Detection} \\
    \hline
    \hline
    
   Group 1 & Same Performance apart from E-LoRa & Same Performance\\
   Group 2 & SSK-ICS-LoRa & SSK-ICS-LoRa\\
   Group 3 & DCRK-LoRa & DCRK-LoRa\\
   Group 4 & DO-CSS & Same Performance\\
   Group 5 & TDM-CSS & TDM-CSS\\
   Group 6 & FSCSS-IM & FSCSS-IM\\
    \hline
    \hline
  \end{tabular}
\end{table*}
\subsection{Bit-Error Performance considering Frequency Offset}
\textcolor{black}{In this section, we analyze the BER behavior of various CSS approaches in the presence of FO. The carrier FO results in the linear accumulation of phase rotations from one symbol to the next. Consequently, the received symbol that encompasses the effects of FO can be mathematically expressed as:}
\begin{equation}
y(n) = \exp\left\{\frac{j2\pi \Delta f n}{M}\right\} s(n) + w(n), 
\end{equation}
where \(\Delta f\) is the FO in Hertz (Hz)\footnote{\(\Delta f\) could also be seen as residual FO. This is reasonable as IoT modems (for example, blacktooth) normally implement a carrier frequency offset compensator before demodulation. Thus, after compensation, \(\Delta f\) will reflect the residual FO.}. To evaluate the BER performance, we consider FOs of \(\Delta f = 0.1\) Hz, \(\Delta f = 0.2\) Hz, \(\lambda = 8\), and AWGN channel for all the CSS schemes.
\subsubsection{Group 1}
\begin{figure}[tb]\centering
\includegraphics[trim={18 0 0 0},clip,scale=0.86]{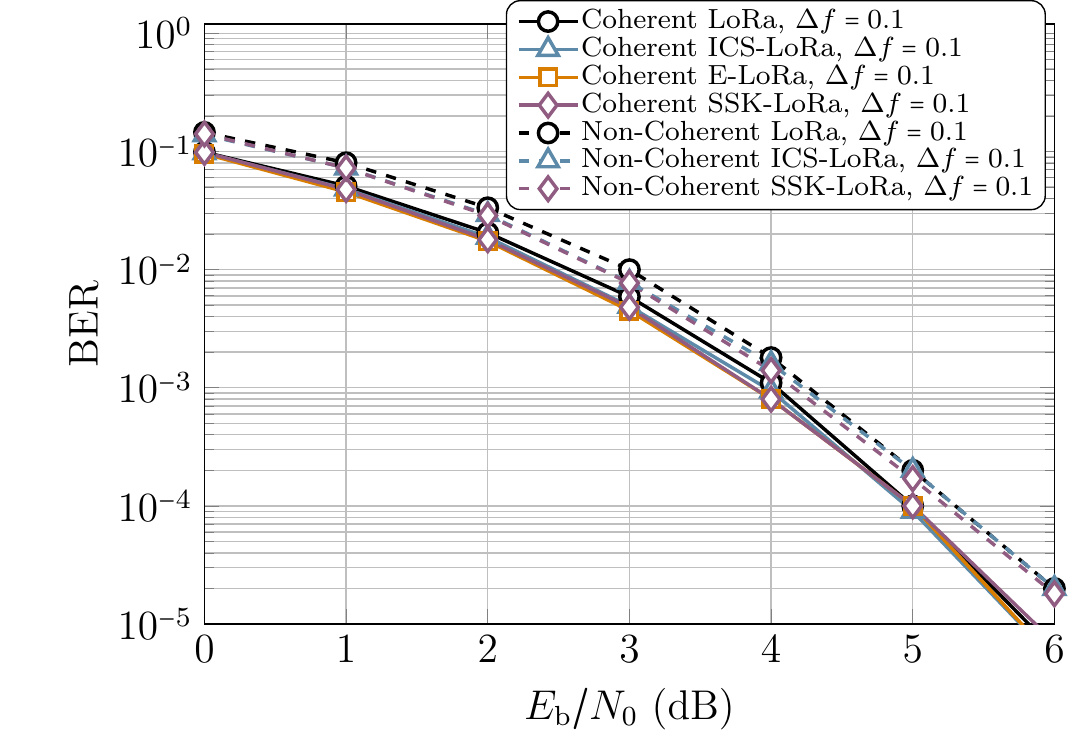}
  \caption{BER performance for group 1 schemes considering coherent/non-coherent detection, \(\Delta f=0.1\), AWGN channel, and \(\lambda = 8\).}
\label{fig1_ber_fo}
\end{figure}

\textcolor{black}{The BER performances of CSS schemes belonging to group 1 under FO of \(\Delta f = 0.1\) Hz are depicted in Fig. \ref{fig1_ber_fo}. Notably, all group 1 schemes exhibit no BER deterioration under non-coherent detection when confronted with a FO of \(\Delta f = 0.1\) Hz. On the contrary, when non-coherent detection is employed, all the schemes demonstrate a BER degradation of roughly \(0.6\) dB upon the advent of \(\Delta f = 0.1\) Hz, relative to the case of \(\Delta f = 0\) Hz. The BER performances of SSK-LoRa and ICS-LoRa emerge as the most exceptional in both scenarios of coherent and non-coherent detections.}

\begin{figure}[tb]\centering
\includegraphics[trim={18 0 0 0},clip,scale=0.86]{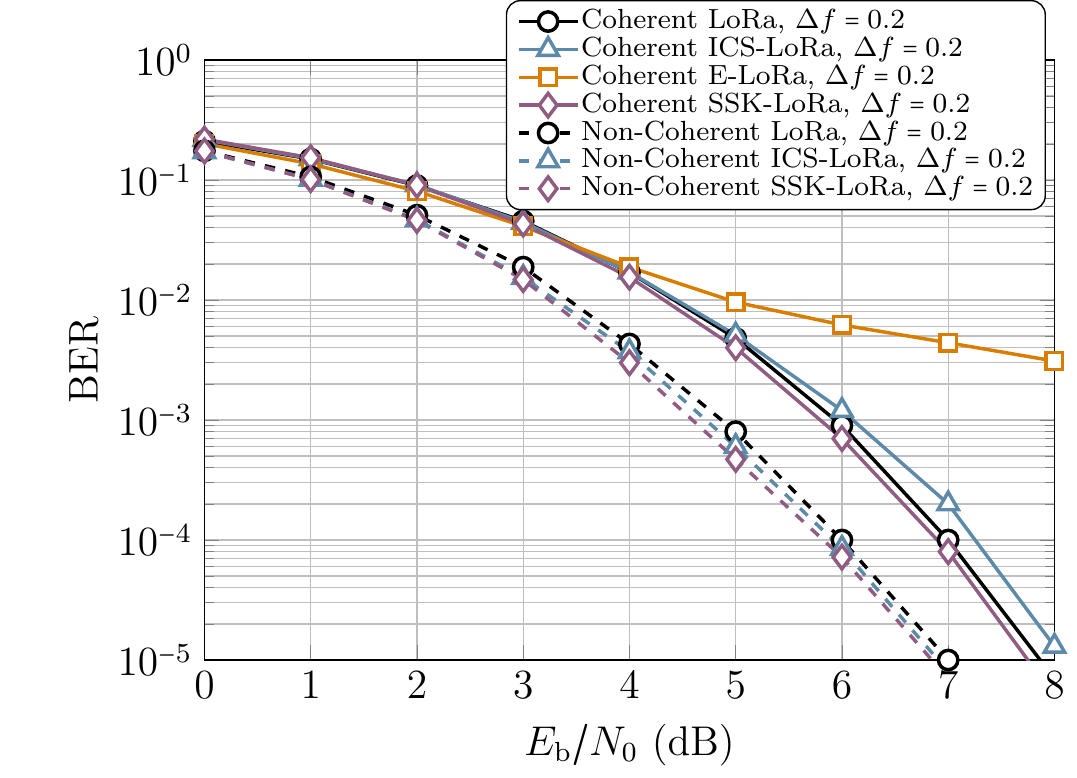}
  \caption{BER performance for group 1 schemes considering coherent/non-coherent detection, \(\Delta f=0.2\), AWGN channel, and \(\lambda = 8\).}
\label{fig2_ber_fo}
\end{figure}

\textcolor{black}{The BER performances of the first group of schemes, subject to a FO of \(\Delta f = 0.2\) Hz, are depicted in Fig. \ref{fig2_ber_fo}. The outcomes of the coherent detection for LoRa, ICS-LoRa, and SSK-LoRa indicate a degradation of approximately \(3\) dB. However, in the same FO scenario, non-coherent detection of these schemes does not manifest any BER degradation, as demonstrated in the figure. Meanwhile, the performance of E-LoRa experiences a severe impact, as shown in Fig. \ref{fig2_ber_fo}. Notably, the BER performances of SSK-LoRa for coherently and non-coherently detected schemes exhibit the best results among the other schemes utilizing coherent and non-coherent detection techniques, respectively.}
\subsubsection{Group 2}
\begin{figure}[tb]\centering
\includegraphics[trim={18 0 0 0},clip,scale=0.86]{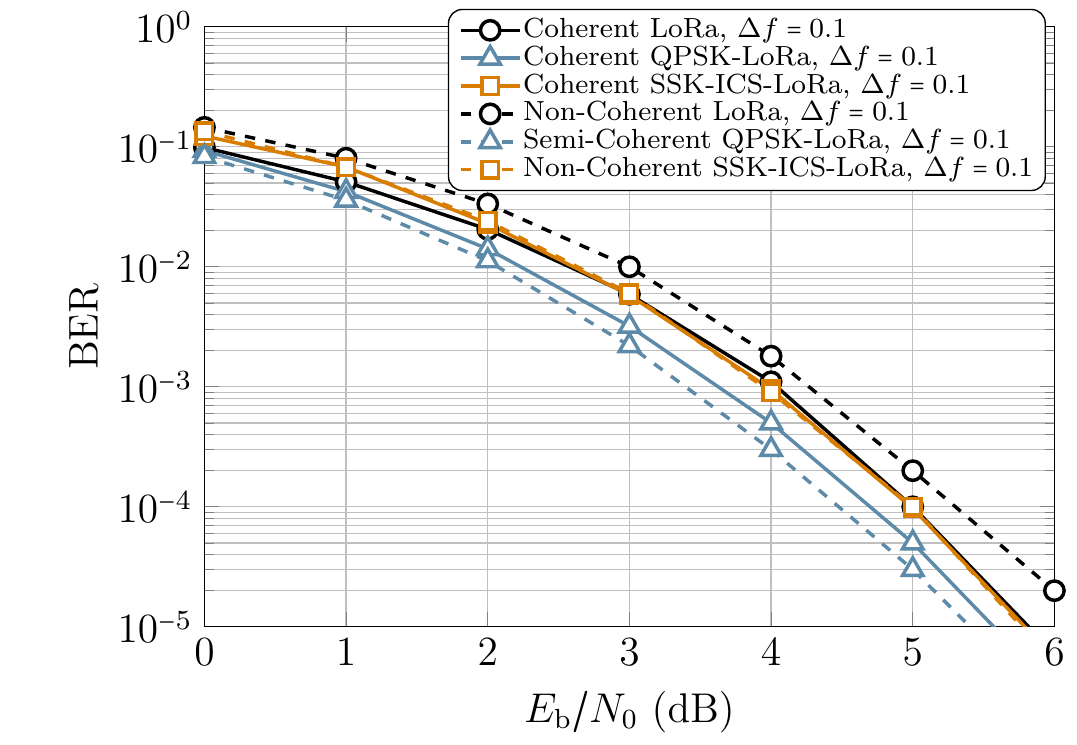}
  \caption{BER performance for group 2 schemes considering coherent/non-coherent detection, \(\Delta f=0.1\), AWGN channel, and \(\lambda = 8\).}
\label{fig3_ber_fo}
\end{figure}

\textcolor{black}{Fig. \ref{fig3_ber_fo} presents a comparative analysis of the BER performances of group 2 schemes, namely QPSK-LoRa and SSK-ICS-LoRa, under the influence of FO at \(\Delta f = 0.1\) Hz. The results indicate that coherent detection for the aforementioned schemes leads to considerable degradation of nearly \(0.6\) dB when FO is set to \(\Delta f = 0.1\) Hz. Conversely, the BER performance of QPSK-LoRa utilizing semi-coherent detection remains unaffected by the FO at the same frequency offset. However, non-coherently detected SSK-ICS-LoRa experiences a decline in BER performance of around \(0.2\) dB under the influence of the same FO. Consequently, the best performance is obtained by the semi-coherently detected QPSK-LoRa.}

\begin{figure}[tb]\centering
\includegraphics[trim={18 0 0 0},clip,scale=0.86]{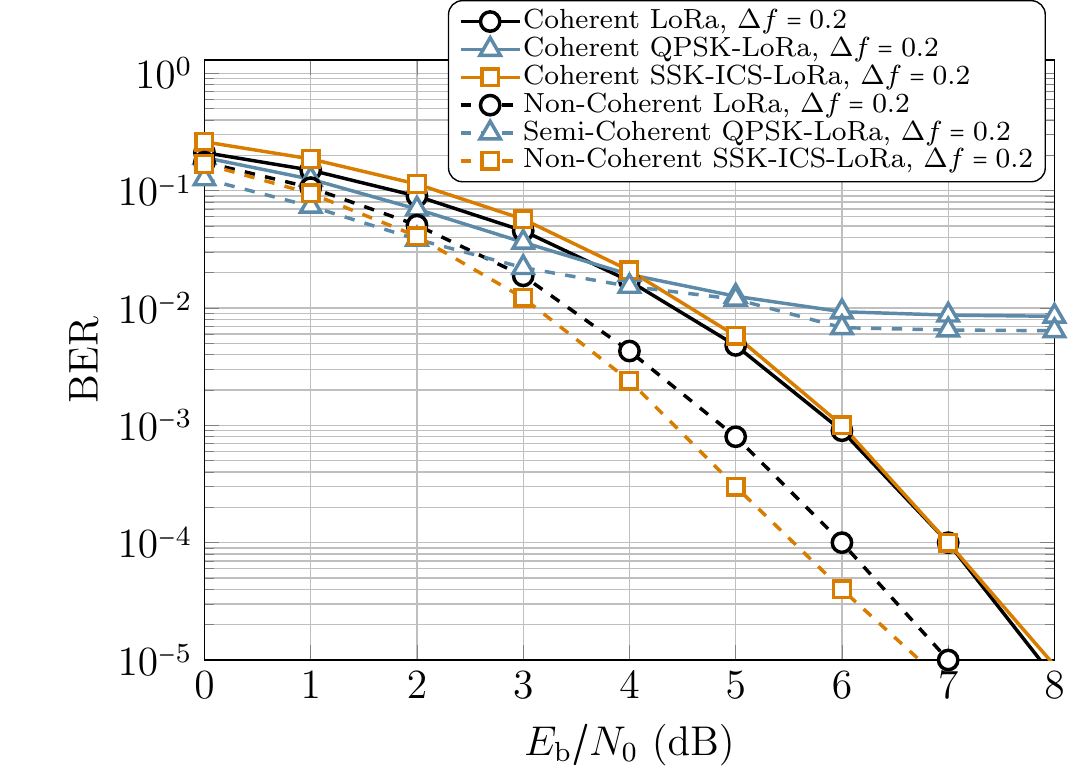}
  \caption{BER performance for group 2 schemes considering coherent/non-coherent detection, \(\Delta f=0.2\), AWGN channel, and \(\lambda = 8\).}
\label{fig4_ber_fo}
\end{figure}

\textcolor{black}{The BER performances of QPSK-LoRa and SSK-ICS-LoRa are depicted in Fig. \ref{fig4_ber_fo} at a FO of \(\Delta f = 0.2\) Hz. The results illustrate that although the coherently and semi-coherently detected QPSK-LoRa outperformed SSK-ICS-LoRa at \(\Delta f = 0.1\) Hz, FO increased by \(0.1\) Hz, i.e., \(\Delta f = 0.2\) Hz severely impacts the BER performances of the coherently and non-coherently detected QPSK-LoRa. The coherent detection for QPSK-LoRa experiences a significant degradation in performance, suffering a loss of approximately \(2.5\) dB when FO increases from \(\Delta f = 0\) Hz to \(\Delta f = 0.2\) Hz. In contrast, the BER performances of SSK-ICS-LoRa coherently and non-coherently detected remain the most optimal under such conditions.}
\subsubsection{Group 3}
\begin{figure}[tb]\centering
\includegraphics[trim={18 0 0 0},clip,scale=0.86]{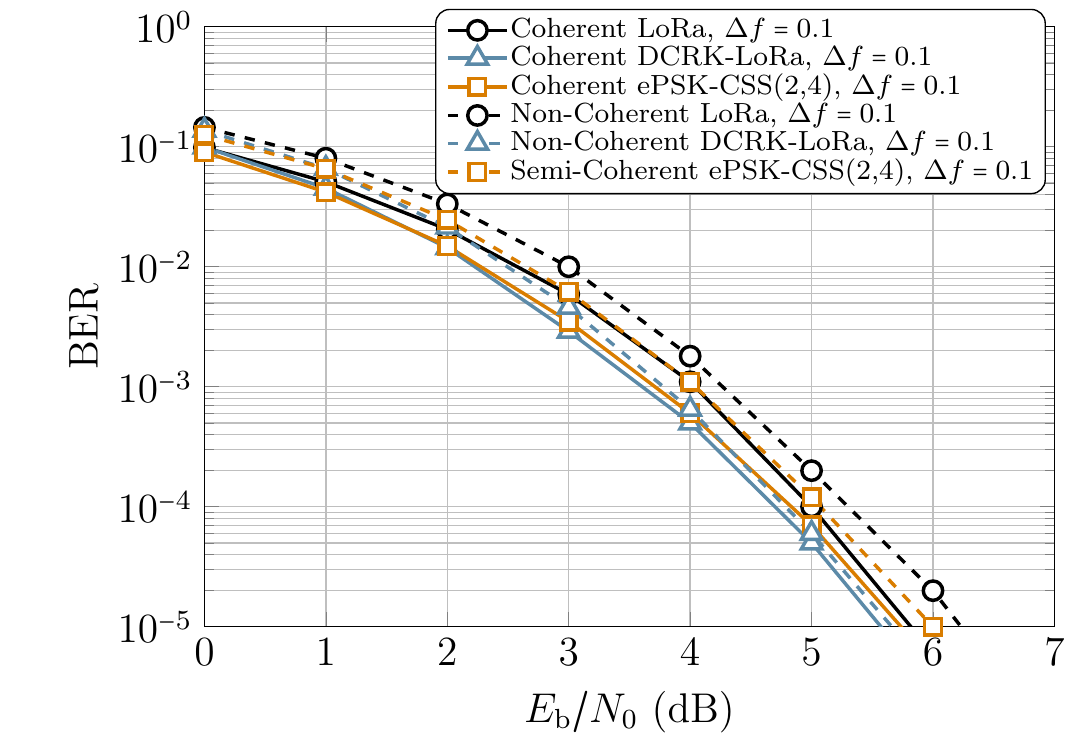}
  \caption{BER performance for group 3 schemes considering coherent/non-coherent detection, \(\Delta f=0.1\), AWGN channel, and \(\lambda = 8\).}
\label{fig5_ber_fo}
\end{figure}

\textcolor{black}{Fig. \ref{fig5_ber_fo} displays the BER performances of group 3 schemes at a FO of \(\Delta f = 0.1\) Hz. The results indicate that both coherently detected DCRK-LoRa and ePSK-CSS(\(2,4\)) suffer degradation of roughly \(0.6\) dB in BER performance. Moreover, the BER performances of non-coherently detected DCRK-LoRa and semi-coherently detected ePSK-CSS(\(2,4\)) decline by approximately \(0.3\) dB and \(1\) dB, respectively. Compared to coherently and semi-coherently detected ePSK-CSS(\(2,4\)), coherently and non-coherently detected DCRK-LoRa exhibits better BER performance.}

\begin{figure}[tb]\centering
\includegraphics[trim={18 0 0 0},clip,scale=0.86]{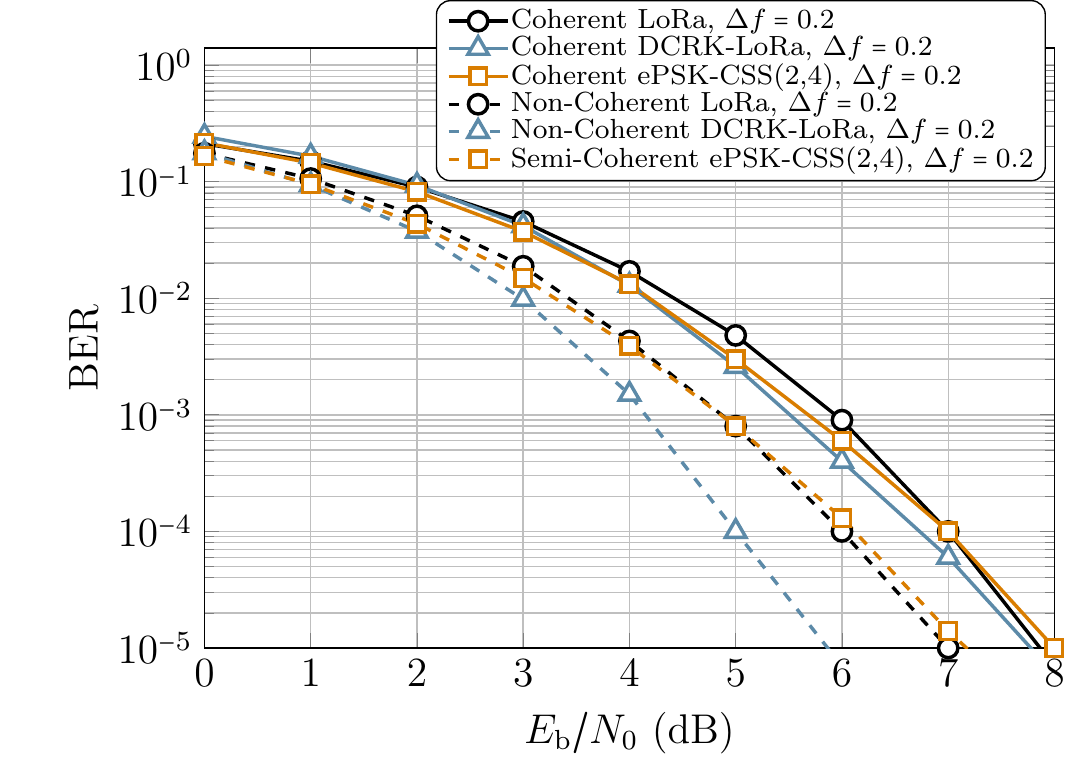}
  \caption{BER performance for group 3 schemes considering coherent/non-coherent detection, \(\Delta f=0.2\), AWGN channel, and \(\lambda = 8\).}
\label{fig6_ber_fo}
\end{figure}

\textcolor{black}{Fig. \ref{fig6_ber_fo} demonstrates the BER performances of DCRK-LoRa and ePSK-CSS(\(2,4\)) at a higher FO of \(\Delta f = 0.2\) Hz. The results reveal that both coherently detected DCRK-LoRa and ePSK-CSS(\(2,4\)) experience a significant degradation of approximately \(2.5\) dB and \(2.7\) dB, respectively, in terms of BER performance. Regarding the BER performance achieved using non-coherent detection, we observe that DCRK-LoRa suffers a loss of \(0.6\) dB, whereas semi-coherently detected ePSK-CSS(\(2,4\)) exhibits a loss of around \(2\) dB as FO increases from \(\Delta f = 0\) Hz to \(\Delta f = 0.2\) Hz. Overall, the performance of non-coherently detected DCRK-LoRa outperforms other detection schemes in the group.}
\subsubsection{Group 4}
\begin{figure}[tb]\centering
\includegraphics[trim={18 0 0 0},clip,scale=0.86]{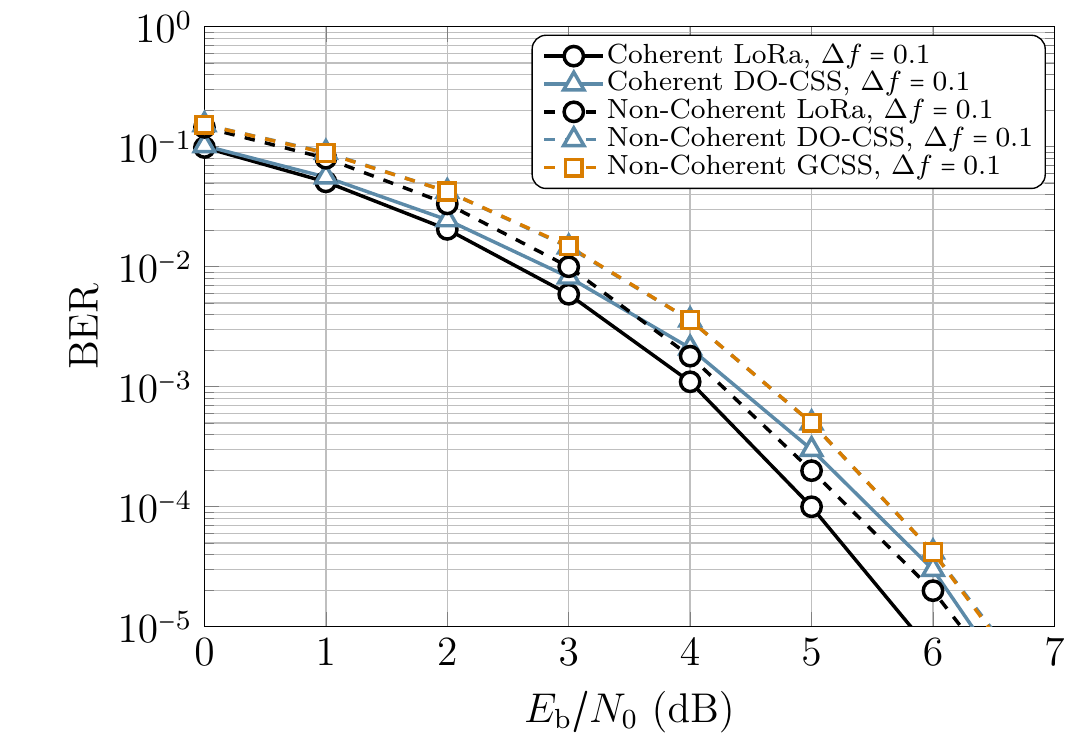}
  \caption{BER performance for group 4 schemes considering coherent/non-coherent detection, \(\Delta f=0.1\), AWGN channel, and \(\lambda = 8\).}
\label{fig7_ber_fo}
\end{figure}

\textcolor{black}{Fig. \ref{fig7_ber_fo} illustrates the BER characteristics of the group 5 modulation schemes under consideration, with \(\Delta f\) of \(0.1\) Hz. It is evident that the deployment of coherent detection for the DO-CSS scheme leads to a BER degradation of approximately \(0.5\) dB with respect to the FO-free scenario. Conversely, deploying non-coherent detection for DO-CSS and GCSS schemes results in a BER deterioration of \(0.1\) dB when the FO is increased from \(0\) Hz to \(0.1\) Hz.}

\begin{figure}[tb]\centering
\includegraphics[trim={18 0 0 0},clip,scale=0.86]{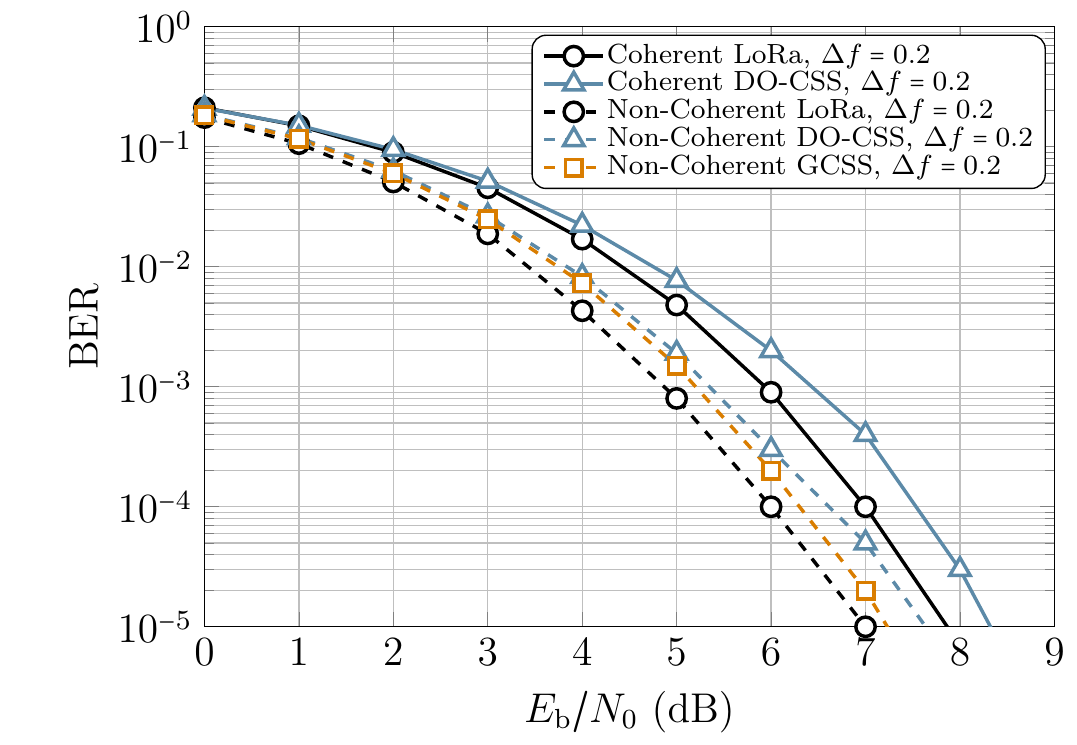}
  \caption{BER performance for group 4 schemes considering coherent/non-coherent detection, \(\Delta f=0.2\), AWGN channel, and \(\lambda = 8\).}
\label{fig8_ber_fo}
\end{figure}

\textcolor{black}{The BER behavior of DO-CSS and GCSS schemes is depicted in Fig. \ref{fig8_ber_fo}, considering a FO of \(0.2\) Hz. Upon increasing the FO from \(0\) Hz to \(0.2\) Hz, using non-coherent detection for DO-CSS and GCSS schemes leads to a BER degradation of \(0.7\) dB and \(0.6\) dB, respectively. Meanwhile, the coherent detection of DO-CSS in the same scenario results in a BER deterioration of \(2.7\) dB. Among the aforementioned modulation schemes, GCSS presents superior or, at the very least, equivalent performance compared to DO-CSS.}
\subsubsection{Group 5}
\begin{figure}[tb]\centering
\includegraphics[trim={18 0 0 0},clip,scale=0.86]{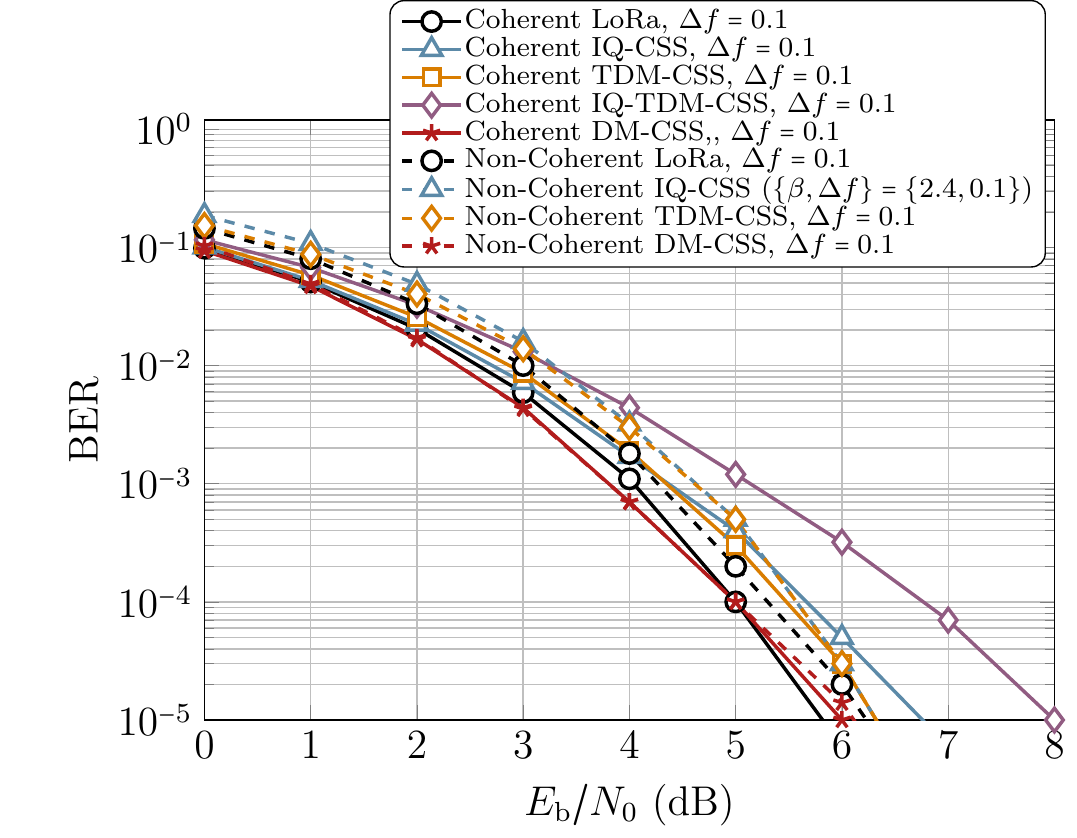}
  \caption{BER performance for group 5 schemes considering coherent/non-coherent detection, \(\Delta f=0.1\), AWGN channel, and \(\lambda = 8\).}
\label{fig9_ber_fo}
\end{figure}

\textcolor{black}{Fig. \ref{fig9_ber_fo} displays the BER performances of the modulation schemes belonging to group 5, with a FO of \(0.1\) Hz. Upon increasing the FO from 0 Hz to \(0.1\) Hz, a degradation of \(1\) dB and \(0.1\) dB can be observed in the BER performance of coherently and non-coherently detected IQ-CSS, respectively. The same FO increment results in a BER deterioration of \(0.4\) dB in both coherently and non-coherently detected TDM-CSS. However, coherently detected IQ-TDM-CSS experiences severe losses in the same scenario. Finally, the DM-CSS exhibits a BER loss of \(0.7\) dB for both coherent and non-coherent detection. We can observe that the BER performance of DM-CSS, whether coherently or non-coherently detected, remains superior to that of the other schemes when subjected to coherent or non-coherent detection.}

\begin{figure}[tb]\centering
\includegraphics[trim={18 0 0 0},clip,scale=0.86]{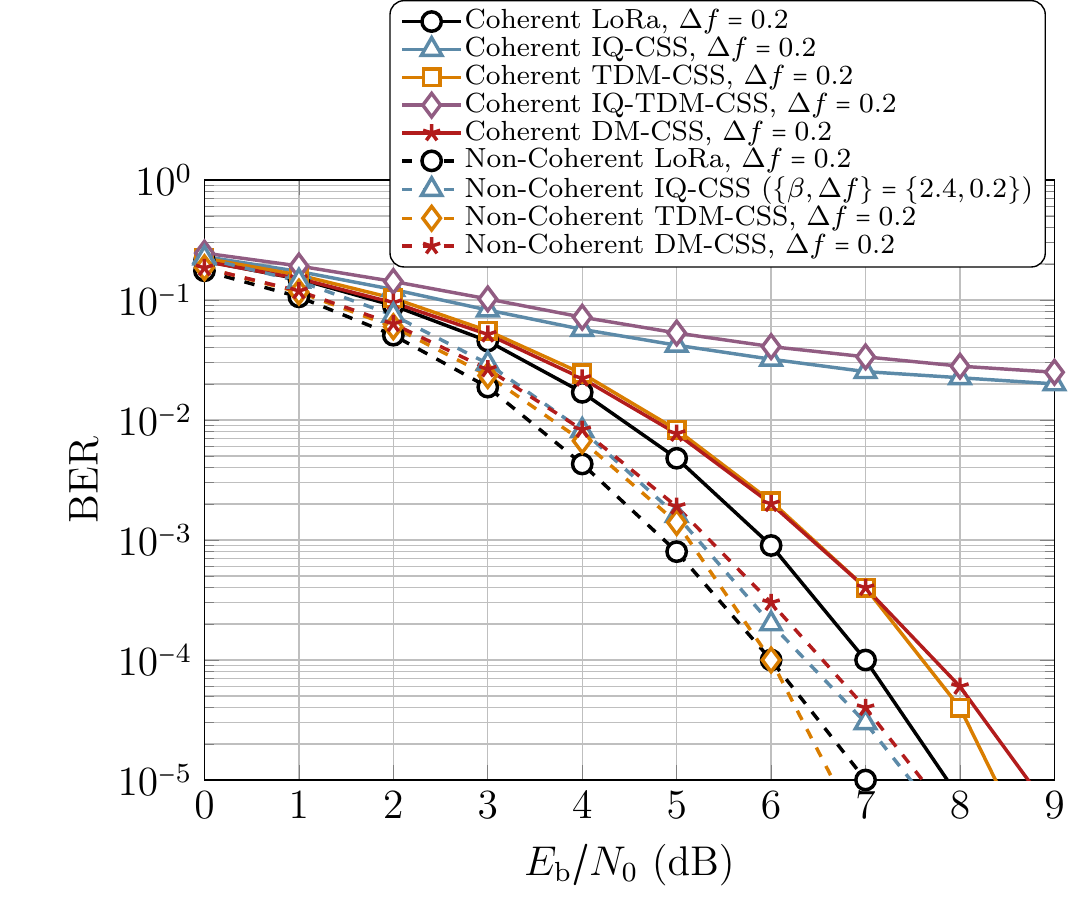}
  \caption{BER performance for group 5 schemes considering coherent/non-coherent detection, \(\Delta f=0.2\), AWGN channel, and \(\lambda = 8\).}
\label{fig10_ber_fo}
\end{figure}

\textcolor{black}{In Fig. \ref{fig10_ber_fo}, the BER performance of group 5 modulation schemes is depicted upon increasing the FO from \(0\) Hz to \(0.2\) Hz. Firstly, it is noteworthy that coherently detected IQ-CSS and IQ-TDM-CSS achieve an error floor of approximately \(10^{-1}\) in BER. Additionally, the BER performance of non-coherently detected IQ-CSS, TDM-CSS, and DM-CSS experiences a BER degradation of \(0.6\) dB, \(0.7\) dB, and \(2\) dB, respectively, under the same FO conditions. Meanwhile, the BER outcomes of coherently detected TDM-CSS and DM-CSS deteriorate by \(2.7\) dB and \(3.4\) dB, respectively. Overall, non-coherently detected TDM-CSS presents the best BER performance when subjected to a frequency offset of \(\Delta f = 0.2\) Hz.}
\subsubsection{Group 6}
\begin{figure}[tb]\centering
\includegraphics[trim={18 0 0 0},clip,scale=0.86]{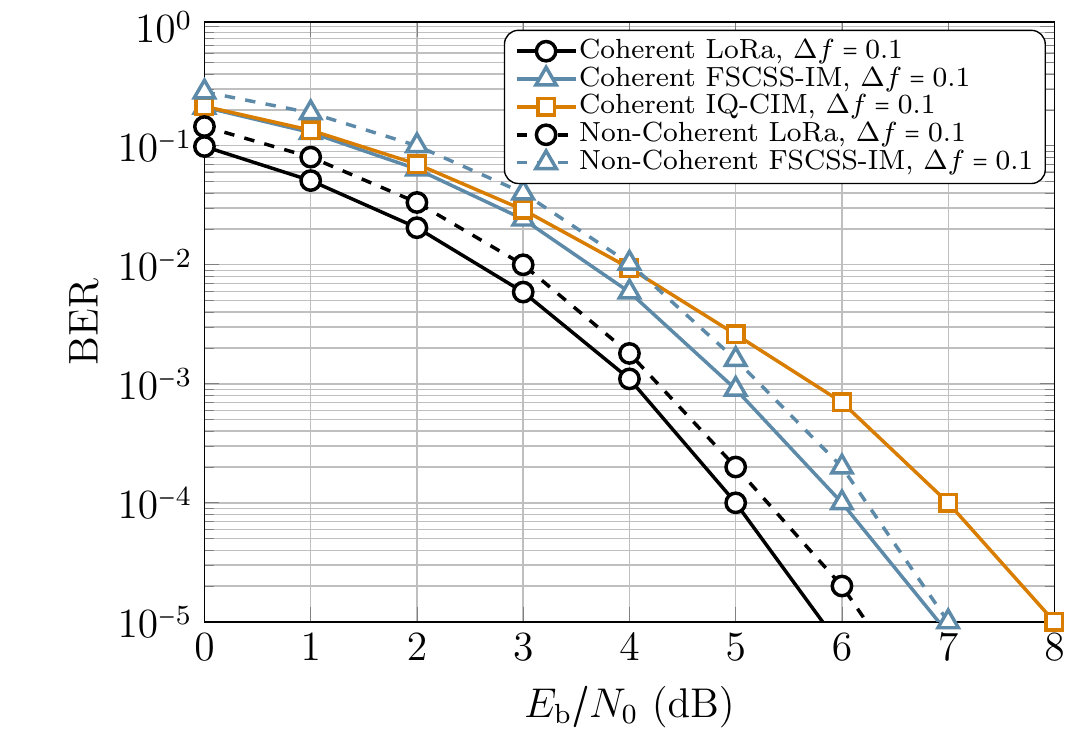}
  \caption{BER performance for group 6 schemes considering coherent/non-coherent detection, \(\Delta f=0.1\), AWGN channel, and \(\lambda = 8\).}
\label{fig11_ber_fo}
\end{figure}

\textcolor{black}{In Fig. \ref{fig11_ber_fo}, the BER performances of the FSCSS-IM and IQ-CIM schemes are depicted for a FO of \(\Delta f = 0.1\) Hz. For both methods, we set the values of \(\varsigma\), \(\varsigma_i\), and \(\varsigma_q\) to \(2\). Notably, coherently detected FSCSS-IM and IQ-CIM incur a BER degradation of approximately \(0.7\) dB and \(1.3\) dB, respectively, while non-coherently detected FSCSS-IM merely experiences a minimal loss of \(0.1\) dB under the same FO conditions. Among the investigated IM schemes, FSCSS-IM yields superior BER performance.}

\begin{figure}[tb]\centering
\includegraphics[trim={18 0 0 0},clip,scale=0.86]{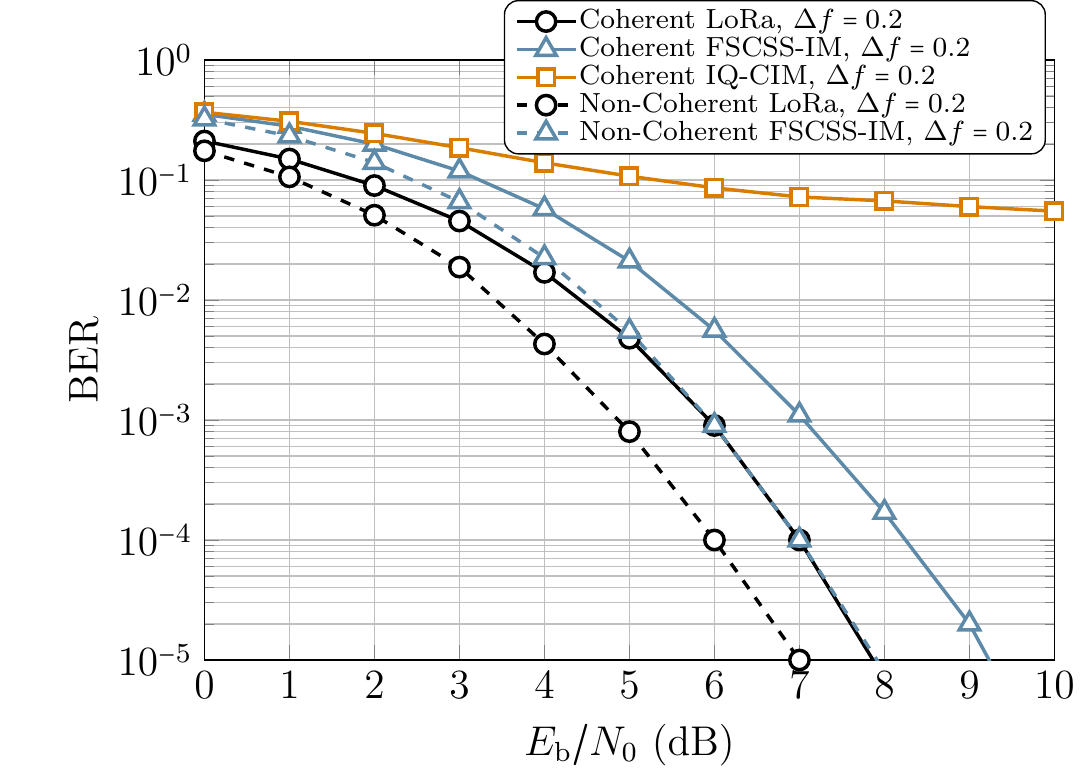}
  \caption{BER performance for group 6 schemes considering coherent/non-coherent detection, \(\Delta f=0.2\), AWGN channel, and \(\lambda = 8\).}
\label{fig12_ber_fo}
\end{figure}

\textcolor{black}{Upon analyzing the BER performances of the IM approaches, namely FSCSS-IM and IQ-CIM, depicted in Fig. \ref{fig12_ber_fo}, it is observed that the coherent detection of IQ-CIM experiences a significant degradation when corrupted with FO of \(\Delta f = 0.2\) Hz. However, in the case of FSCSS-IM, when the FO is increased from \(\Delta f = 0\) Hz to \(\Delta f = 0.2\) Hz, both coherent and non-coherent detections exhibit deteriorations of approximately \(2.8\) dB and \(1\) dB, respectively.}
\subsubsection{\textcolor{black}{Takeaways}}
\textcolor{black}{Table \ref{eff_fo} presents the optimal schemes for achieving robustness against FO impairments in each group. Group 1 exhibits the most robust waveform design against FO, with SSK-LoRa being the preferred scheme for both coherent and non-coherent detection mechanisms. Meanwhile, in group 2, SSK-ICS-LoRa is the most robust scheme for coherent and non-coherent detection mechanisms. In group 3, DCRK-LoRa demonstrates optimal coherent and non-coherent detection performance. The results indicate that DCRK-LoRa is the most efficient scheme among all the schemes evaluated for its superior performance and robust waveform design against FO in the SC CSS scheme.}

\textcolor{black}{In group 4, the DO-CSS exhibits the best performance for coherent detection, while the GCSS shows the best results for non-coherent detection. The performance of GCSS in non-coherent detection may be attributed to the separation of the FSs in un-chirped symbols, which provides sufficient space between them. However, in DO-CSS, two consecutive FSs may be activated, leading to incorrect detection of the activated FS. Therefore, it is essential to analyze the interference that MC CSS may cause. In group 5, TDM-CSS is the optimal scheme for achieving higher SE and optimal robustness against FO. This is because the waveform design of other counterparts either incorporates PSs or uses I/Q components, which results in a less robust waveform design.}

\textcolor{black}{Regarding the MC-IM schemes, the FSCSS-IM demonstrates the best performance. The IQ-CIM scheme is vulnerable to deterioration caused by FO due to its reliance on I/Q components. Thus, based on the findings, the optimal scheme for achieving higher SE and robustness against FO in the MC CSS scheme is TDM-CSS, while the FSCSS-IM is the preferred choice for MC-IM schemes due to its superior performance.}
\renewcommand{\arraystretch}{1}
\begin{table*}[h]
  \caption{\textcolor{black}{FO robust schemes from different groups.}}
   \label{eff_fo}
  \centering
  \color{black}\begin{tabular}{*{3}{c}}
    \hline
    \hline
    \bfseries{Group}    & \bfseries {Coherent Detection} & \bfseries {Non-Coherent Detection} \\
    \hline
    \hline
    
   Group 1 & SSK-LoRa & SSK-LoRa\\
   Group 2 & SSK-ICS-LoRa & SSK-ICS-LoRa\\
   Group 3 & DCRK-LoRa & DCRK-LoRa\\
   Group 4 & DO-CSS & GCSS\\
   Group 5 & TDM-CSS & TDM-CSS\\
   Group 6 & FSCSS-IM & FSCSS-IM\\
    \hline
    \hline
  \end{tabular}
\end{table*}
\subsection{\textcolor{black}{General Summary}}
\textcolor{black}{The outcomes derived from the SE versus EE and BER performance analysis in the AWGN channel merely indicate the system's operation under ideal circumstances. In contrast, the results acquired from assessing the influence of PO and FO consider certain distortions that may arise in low-cost LPWAN systems. Although we may obtain BER performance results by considering the Rayleigh channel, the expectation is that the outcomes will follow the same trend as in the AWGN channel.}

\textcolor{black}{Based on the results, it is evident that the AWGN channel's performance may not always provide a precise representation of scheme performance in realistic environments. However, it does provide a solid foundation to ascertain which scheme can achieve higher SE and EE. The results obtained from the AWGN channel illustrate the attainability of SE performance without accounting for the detrimental impact of diverse distortions on the waveform design. Consequently, the performance of different schemes differs when other distortions are considered.}

\textcolor{black}{For instance, the performance of QPSK-LoRa appears ideal in the AWGN channel. Still, as soon as PO and FO are considered, the performance of QPSK-LoRa declines significantly, rendering it unsuitable for deployment in the circumstances with significant PO or FO. Therefore, a scheme's performance must be evaluated extensively for various scenarios to analyze waveform design thoroughly. In the subsequent section, we shall examine the outcomes derived from this analysis in the context of different waveform designs.}
\section{\textcolor{black}{Merits and Limitations of Different Waveform Designs}}\label{sec5}
\textcolor{black}{Within this section, we shall undertake a comprehensive assessment of the advantages and constraints inherent in various methodologies based on the outcomes procured from the preceding segment. 
\subsection{\textcolor{black}{Important Characteristics for Waveform Design}}
While there may exist a plethora of attributes that we may utilize to ascertain the merits and limitations of a CSS approach, for an exploratory inquiry, it may be reasonable to commence with an assessment of the following characteristics:}
\begin{enumerate}
\item \textit{Constant envelope property:} \textcolor{black}{If a waveform has a constant envelope, the resulting PAPR would be low, facilitating straightforward design and implementation of systems utilizing low-cost equipment with a limited linear range of operation. Conversely, a high PAPR could have a negative impact on the EE. This is due to the Input Back Off at which the power amplifier operates. A waveform with high PAPR is more prone to clipping, which effectively nullifies the anticipated enhancement in BER that would ordinarily be expected with an increase in \(\sfrac{E_\mathrm{b}}{N_0}\). Notably, only CSS schemes falling within the SC taxonomy exhibit constant envelope properties, while MC CSS schemes do not manifest this attribute.}
\item \textit{EE improvement with respect to LoRa:} \textcolor{black}{EE analysis of a CSS scheme gives insights into power consumption. If an approach can demonstrate superior EE in contrast to LoRa, the battery terminals utilized in LPWAN nodes would consume less power, resulting in an extended battery life while maintaining equivalent or better BER. In other words, a more energy-efficient scheme than LoRa would significantly benefit LPWAN nodes by reducing the frequency of battery replacement or recharge requirements while simultaneously offering a similar level of BER accuracy. EE analysis can offer valuable insights into the robustness of a waveform design. Generally, when only one FS is activated for an un-chirped symbol, we can expect an improvement in EE. Conversely, activating multiple FSs tends to decrease EE, highlighting the need to select the number of FSs in MC CSS schemes carefully. It is important to note that increasing the number of FSs could lead to higher SE, but it would invariably lead to a reduction in EE. Therefore, a balance must be struck between SE and EE when selecting the number of FSs to activate in MC CSS schemes.}
\item \textit{SE improvement with respect to LoRa:} \textcolor{black}{If a scheme demonstrates superior SE compared to LoRa, it will transmit information over the air in less time, reducing latency between transmission and reception. The SE of a scheme is determined by the number of activated FSs for un-chirped symbols. Fewer activated FSs lead to a lower SE, whereas increasing the number of activated FSs would result in a higher SE. Therefore, MC CSS schemes exhibit a higher SE than SC CSS schemes. It is worth noting that the choice of waveform design methodology significantly impacts SE. Selecting a design that increases SE while retaining robustness against various channel impairments is crucial. Some designs may significantly increase SE compared to other methods; however, certain designs may trade off robustness against impairments that may occur in LPWANs to achieve higher SE. Therefore, careful consideration must be given to the choice of design that balances SE and robustness against channel impairments.}
\item \textit{Possibility of coherent detection:} \textcolor{black}{Coherent detection is a feature of waveform design considered secondary in importance. However, in certain situations, it is necessary to incorporate a coherent detection mechanism, such as when the received signal is affected by PSs or changes in its carrier frequency. Despite its importance in some scenarios, the possibility of implementing coherent detection can significantly increase the cost and complexity of a system. On the one hand, the cost of the system will rise as additional processing is needed for functions like equalization and synchronization. On the other hand, the increased complexity may also directly affect the system's power requirements.}
\item \textit{Possibility of non-coherent detection:} \textcolor{black}{Similar to coherent detection, incorporating a non-coherent detection mechanism is also considered to be of secondary importance for a waveform design. Non-coherent detection is typically utilized when the received signal is not subject to PSs or changes in its carrier frequency, which can happen when the transmitter and receiver are stationary, and the channel is relatively stable. Having the option of a non-coherent detection mechanism is advantageous since it allows for the system to be implemented with lower costs and power requirements. As a result, non-coherent detection is often used in low-cost consumer electronics where cost and power efficiency are critical considerations. Overall, non-coherent detection is a beneficial alternative when accurate signal recovery is not vital, and a coherent detection mechanism's cost and power requirements outweigh its benefits.}
\item \textit{Robustness against PO:} \textcolor{black}{In low-cost LPWAN devices, it is expected that PO would occur due to a variety of factors, such as frequency drifts, doppler shifts, multipath fading, etc. If the receiver cannot compensate for the PO, the recovered symbols may be distorted or lost entirely, leading to a degradation in the quality of the communication link. This can result in decreased data rates, increased error rates, and reduced range and reliability of the system. Overall, robustness against phase offset is critical for maintaining the quality and reliability of a communication system, ensuring that we can achieve accurate symbol recovery even in the presence of PSs and carrier frequency changes.}
\item \textit{Robustness against FO:} \textcolor{black}{FO is another common issue in low-cost LPWAN systems, similar to PO. FO occurs when the transmitted signal's frequency differs from that of the receiver's local oscillator. This can be caused by differences in the crystal oscillator frequency, doppler shift, and temperature fluctuations. FO can lead to signal distortion, inter-symbol interference, and reduced overall system performance. If the waveform design is not robust to FO, the receiver may fail to demodulate the signal correctly, resulting in lost or corrupted data. Therefore, designing communication systems with FO robustness is crucial to ensure reliable and robust communication. Designing a waveform resilient to FO makes it possible to achieve accurate and dependable data transmission while keeping costs low. The robustness of the waveform design to FO is critical in achieving better system performance overall.}
\item \textit{\textcolor{black}{Complexity:}} \textcolor{black}{The complexity of a waveform transceiver is a critical aspect of its design. The ideal transceiver should have minimal complexity, directly impacting its cost, power consumption, reliability, and ease of integration. A complex transceiver typically requires more components, increasing manufacturing and maintenance costs. By minimizing the complexity of the transceiver, we can significantly reduce production and maintenance costs. Additionally, a complex transceiver architecture may require more power to operate than a simpler one, leading to higher energy consumption and shorter battery life in portable devices. Keeping the transceiver's complexity low can minimize power consumption, which is especially crucial in battery-powered devices. A complex transceiver may also be more prone to failures due to the increased number of components and their interactions, leading to higher maintainability costs. A simpler transceiver is less likely to experience such issues, improving reliability. Furthermore, a simpler transceiver is easier to integrate into larger systems, such as IoT devices. Complex transceivers may require more specialized interfaces or software drivers, increasing the time and effort needed to incorporate them into a more extensive system. Overall, reducing the complexity of a transceiver can lead to significant benefits, making it an attractive choice for many applications in terms of cost, power consumption, reliability, and ease of integration.}
\end{enumerate}
\subsection{\textcolor{black}{Analysis of Characteristics for Different Waveform Design}}
\subsubsection{\textcolor{black}{Activation of Frequency Shift(s)}}
\textcolor{black}{There are three ways to use this design methodology: activating a single FS, activating multiple FSs, and activating multiple FSs using IM precept. Activating a single FS for the un-chirped symbol leads to LoRa, which has a constant envelope and low PAPR. As a result, the implementation is straightforward, with no apparent design limitations, and the possibility of both coherent and non-coherent detection. Moreover, LoRa is a robust scheme against FO and PO with low implementation complexity.On the other hand, activating multiple FSs either normally or using IM precept results in schemes known as DO-CSS, GCSS, and FSCSS-IM, which do not have a constant envelope and have a high PAPR. Therefore, it is more challenging to implement practically, and the problems associated with high PAPR may arise. However, activating multiple FSs increases the SE compared to LoRa, making DO-CSS, GCSS, and FSCSS-IM more energy-efficient for a given SE in most cases. Coherent and non-coherent detection is possible with reasonable complexity, but the robustness against FO and PO is affected. However, the impact on performance due to these offsets is unavoidable if multiple FSs are activated.}
\subsubsection{\textcolor{black}{Interleaving the Chirped Symbol}}
\textcolor{black}{If only one FS is activated for the un-chirped symbol and the chirped symbol is interleaved, the resulting scheme is known as ICS-LoRa. Interleaving the chirped symbol does not affect the constant envelope characteristics, and the higher cardinality of possible symbols increases the resulting SE compared to LoRa. Simulation results indicate that ICS-LoRa is also energy-efficient relative to LoRa under different channel conditions, as it is robust against PO and FO. Additionally, while the seminal work only presented the non-coherent detector, this study analyzes both coherent and non-coherent detection mechanisms. Lastly, it is important to note that the ICS-LoRa transceiver's complexity is higher than that of LoRa.}
\subsubsection{\textcolor{black}{Using Phase Shift(s)}}
\textcolor{black}{PSs are versatile tools that can be utilized in both SC and MC CSS schemes. In SC schemes, PSK-LoRa leverages PS to encode extra bits in the transmitted symbol, while ePSK-CSS employs PS along with some robustness to develop the MC CSS scheme. While PS can significantly enhance SE, detecting PS is not a simple task requiring a ML detector. In PSK-LoRa and ePSK-CSS, the FSs can be detected in both a coherent and non-coherent manner; however, detecting the PS necessitates coherent ML mechanisms, increasing the detection's complexity and limiting its applicability to only coherent detection. In this work, a detector that utilizes a semi-coherent approach to detect the FS and an ML detector to estimate the PS is referred to as a semi-coherent detector, which can be applied to both PSK-LoRa and ePSK-CSS. However, the practicality of a semi-coherent detector is questionable. Additionally, the EE of the schemes utilizing this design methodology is superior to LoRa in more straightforward scenarios. However, the robustness of these schemes against PO and FO is either non-existent or severely limited. Specifically, PSK-LoRa is not robust against PO and FO, whereas ePSK-CSS is slightly more robust due to the diversity used in waveform design. The transceiver's complexity in schemes utilizing this design methodology is high because determining the PS using ML criteria is necessary. However, we can reduce the complexity of these schemes by employing only two PSs. In this case, determining the polarity of the activated FS would provide information about the activated PS, obviating the need for ML detection.}
\subsubsection{\textcolor{black}{Chirp Rates Variation}}
\textcolor{black}{Generally, a specific CR is typically used to modulate an un-modulated symbol. However, it is possible to utilize multiple CRs instead of just one. When two CRs are used, this approach is called SSK-LoRa, while the more generalized scheme allowing for multiple CRs is known as DCRK-LoRa. Both of these schemes fall within the SC CSS taxonomy. Using multiple CRs offers a straightforward means of increasing SE, with the SE being directly proportional to the number of CRs utilized. As a result, these design methodologies typically boast higher SE values than classical LoRa schemes. In turn, this increased SE translates to higher EE for a given SE. Moreover, these schemes are relatively simple and can accommodate both coherent and non-coherent detection mechanisms: Moreover, the schemes employing this design methodology offer good robustness against PO and FO. However, it is important to note that using multiple CRs can also cause interference, degrading system performance. Lastly, it is worth noting that the complexity of these schemes increases linearly with the number of CRs used. Therefore, it is crucial to consider the constraints associated with system complexity when designing the waveform.}
\subsubsection{\textcolor{black}{Chirp Rates Variation and Interleaving the Chirped Symbol}}
\textcolor{black}{The SSK-ICS-LoRa scheme belongs to the SC CSS taxonomy, utilizing varying CRs and interleaving the design methodology. This approach only uses two different CRs, and retains the constant envelope characteristics, resulting in a low PAPR. This design methodology makes the implementation of the scheme straightforward. Compared to LoRa, the CSS scheme resulting from this design methodology offers a higher SE and is more energy-efficient, as demonstrated in the previous section. Additionally, the design methodology allows for the implementation of coherent and non-coherent detection mechanisms without stringent limitations. However, the complexity of the receiver is higher relative to LoRa since it needs to detect whether the transmitted symbols are interleaved or not, as well as different CRs. Despite this limitation, the scheme resulting from this design methodology is robust against PO and FO issues commonly encountered in LPWANs.}
\subsubsection{\textcolor{black}{Using Phase Shift(s) and Varying Chirp Rates}}
\textcolor{black}{In this design methodology, the PSs are incorporated in the un-chirped symbols, while varying CRs are used for spreading the un-chirped symbol. The DM-CSS scheme employs this methodology and utilizes two distinct PSs, simplifying detection by obviating the ML detection requirement to ascertain the PS, unlike the other schemes, which use more than two PSs. Instead, the identified FSs' polarity specifies the PS. Furthermore, it only employs two varying CRs, which reduces the interference caused by using multiple CRs. As a MC CSS scheme, the ensuing waveform lacks a constant envelope. However, this procedure can also generate a SC waveform with a constant envelope. This waveform design facilitates both coherent and non-coherent detection mechanisms. Nevertheless, determining the polarity of the identified FSs is indispensable to deciding on the binary PSs encompassed in the un-chirped symbols, culminating in an additional complexity overhead. Additionally, akin to other techniques that utilize varying CRs, the complexity of the transceiver increases linearly with an increase in CRs. Hence, the complexity of this waveform design surpasses that of LoRa. Besides, the MC characteristics of this waveform design have an adverse effect on its robustness against offsets. Although it possesses marginal resilience against these offsets, it is not significantly profound.}
\subsubsection{\textcolor{black}{Time Domain Multiplexing of Chirped Symbols}}
\textcolor{black}{The TDM design methodology utilizes multiplexing of chirped symbols with different CRs, thereby forming MC CSS schemes. In the existing literature, the schemes that employ this methodology are classified as TDM-CSS and IQ-TDM-CSS. We categorize IQ-TDM-CSS as a scheme that follows the I/Q components design methodology, which will be elaborated upon subsequently. As the resulting scheme is a member of the MC taxonomy, it does not possess a constant envelope. Despite this limitation, this methodology provides an ingenious waveform design that yields schemes with high SE and improved EE compared to LoRa. Furthermore, this design can implement coherent and non-coherent detection mechanisms if only the time-domain symbols are multiplexed without I/Q components. The results from the previous section demonstrate that TDM-CSS has adequate resilience against the PO and FO. In fact, TDM-CSS outperforms the other schemes in its category in the presence of both PO and FO. Lastly, the complexity of the schemes employing this design methodology increases with the number of multiplexed symbols. In other words, the more multiplexed symbols, the higher the complexity of the transceiver.}
\subsubsection{\textcolor{black}{Use of In-phase and Quadrature Components}}
\textcolor{black}{The design methodology under consideration employs the I/Q components of the un-chirped symbol to encode information. This design methodology can be utilized for SC, MC, and MC-IM CSS taxonomies. The state-of-the-art schemes that adopt this methodology include E-LoRa, IQ-CSS, IQ-CIM, and IQ-TDM-CSS. It is noteworthy that IQ-CIM and IQ-TDM-CSS also integrate IM principles and TDM design methodology into the waveform design. E-LoRa is the only scheme mentioned above with a constant envelope, whereas the other schemes lack a constant one. This design methodology effectively improves SE and EE. However, the EE enhancement is only evident in simple channel conditions, whereas dispersive channels, in general, severely impact the performance of the schemes employing this design methodology. Furthermore, these schemes typically do not support non-coherent detectors due to the information in I/Q components. It is important to note that a non-coherent detector is available for IQ-CSS in the literature. The design methodology is highly sensitive to offsets that may exist in LPWANs. As indicated by the results, the performance of the schemes resulting from this methodology is severely degraded in the presence of PO and FO. The complexity of these schemes is dependent on the waveform design. In other words, the complexity of the design methodology differs when IM is combined with it versus TDM. Nevertheless, separating I/Q components at the receiver may result in severe implementation issues. In conclusion, this design methodology is the least robust among the ones discussed here.}
\begin{table*}[h]
  \caption{\textcolor{black}{Merits and Limitations of different design methodologies for SC CSS schemes. '-' refers to not applicable.}}
   \label{meritslimitations_sc}
  \centering
  \color{black}\begin{tabular}{*{9}{c}}
    \hline
    \hline
   \bfseries{Design}  & \bfseries{Constant}    & \bfseries {SE} & \bfseries {EE}& \bfseries {Coherent}  & \bfseries {Non-coherent} &  \bfseries {PO} & \bfseries {FO} & \bfseries {Complexity}    \\
     \bfseries {Methodology}  & \bfseries {Envelope}    &  \bfseries{Improv.} & \bfseries {Improv.}& \bfseries {Detection}  & \bfseries {Detection} &  \bfseries{Robust} & \bfseries{Robust}  & {}  \\
    \hline
    \hline
   Activation  & \checkmark & - & - & \checkmark & \checkmark &  \checkmark & \checkmark & Low\\
   of FS(s) & &  &  &  &  &   &\\
    \hline
 Interleaving & \checkmark & \checkmark & \checkmark &\checkmark & \checkmark & \checkmark & \checkmark & Low\\
  \hline
  Use of PS(s) & \checkmark  & \checkmark & \checkmark &\checkmark & \xmark  & \xmark & \xmark & High\\
   \hline
 CRs Variations & \checkmark & \checkmark & \checkmark &\checkmark & \checkmark & \checkmark & \checkmark & Low to High\\
  \hline
  CRs Variations & \checkmark & \checkmark & \checkmark & \checkmark & \checkmark & \checkmark & \checkmark & Medium\\
and Interleaving &  &  &  & &  & & &\\
 \hline
  CRs Variations & -  & - & - & - & - & - & - & -\\
and use of PSs &  &  &  & &  &  & &\\
 \hline
  TDM & -  & - & - & - & - & - & - & -\\
   \hline
 Use of I/Q & \checkmark &\checkmark  & \checkmark & \checkmark & \xmark  & \xmark &\xmark & High\\
 Components &  &  &  & &  & &\\
    \hline
    \hline
  \end{tabular}
\end{table*}
\begin{table*}[h]
  \caption{\textcolor{black}{Merits and Limitations of different design methodologies for MC CSS schemes. '-' refers to not applicable.}}
   \label{meritslimitations_mc}
  \centering
  \color{black}\begin{tabular}{*{9}{c}}
    \hline
    \hline
   \bfseries{Design}  & \bfseries{Constant}    & \bfseries {SE} & \bfseries {EE}& \bfseries {Coherent}  & \bfseries {Non-coherent} &  \bfseries {PO} & \bfseries {FO} & \bfseries {Complexity}    \\
     \bfseries {Methodology}  & \bfseries {Envelope}    &  \bfseries{Improv.} & \bfseries {Improv.}& \bfseries {Detection}  & \bfseries {Detection} &  \bfseries{Robust} & \bfseries{Robust}  & {}  \\
    \hline
    \hline
   Activation  & \xmark & \checkmark & \checkmark & \checkmark & \checkmark &  \checkmark & \checkmark & Low\\
   of FS(s) & &  &  &  &  &   &\\
    \hline
 Interleaving  & -  & - & - & - & - & - & - & -\\
  \hline
  Use of PS(s) & \xmark  & \checkmark & \checkmark &\checkmark & \xmark  & \xmark & \checkmark & High\\
   \hline
 CRs Variations  & -  & - & - & - & - & - & - & -\\
  \hline
  CRs Variations  & -  & - & - & - & - & - & - & -\\
and Interleaving &  &  &  & &  & & &\\
 \hline
  CRs Variations  & \xmark  & \checkmark & \checkmark &\checkmark & \checkmark  & \checkmark & \checkmark & Medium\\
and use of PSs &  &  &  & &  &  & &\\
 \hline
  TDM & \xmark  & \checkmark & \checkmark &\checkmark & \checkmark  & \checkmark & \checkmark & Medium\\
   \hline
 Use of I/Q & \xmark  & \checkmark & \checkmark &\checkmark & \xmark  & \xmark & \xmark & High\\
 Components &  &  &  & &  & &\\
    \hline
    \hline
  \end{tabular}
\end{table*}
\subsection{\textcolor{black}{Takeaways}}
\textcolor{black}{Tables \ref{meritslimitations_sc} and \ref{meritslimitations_mc} present a succinct overview of the benefits and shortcomings of the various design methodologies employed in the symbol structure of SC and MC CSS schemes, respectively. '-' signifies the inapplicability of a particular design methodology for the given classification of CSS schemes. For instance, TDM does not apply to SC CSS schemes, whereas CR variations have not been studied in MC CSS schemes.}

\textcolor{black}{Three design strategies have been identified as the most effective for SC schemes: interleaving, CR variation, and CR variation coupled with interleaving. However, it is worth noting that interleaving chirped symbols creates interference on the FS adjacent to the transmitted FS. Furthermore, using different CRs leads to interference, with a conspicuous absence of studies investigating such interference. If we can ascertain the nature of such interference, we may employ various techniques for mitigating interference to circumvent it, thereby potentially improving the performance of these schemes.} 

\textcolor{black}{Additionally, it is observable that the simultaneous use of various CRs and PSs, and TDM have been evaluated as design methodologies for SC CSS schemes in the literature. Although TDM is unsuitable for SC schemes, using different CRs and PSs may yield new waveform designs compatible with LPWANs.}

\textcolor{black}{The MC CSS has been found to exhibit optimal performance when employing three design methodologies: the activation of FSs, the use of varied CR in conjunction with diverse PSs, and TDM. TDM is the most versatile scheme owing to its ability to multiplex different chirped symbols with varying CRs, although this comes at the cost of significant interference from the multiplexed chirped symbols. Despite its potential for performance improvement, the literature lacks studies investigating this interference. Moreover, the scheme utilizing CR variation and different PSs, known as DM-CSS, employs binary PSs to mitigate the transceiver's complexity while utilizing only two distinct CRs. Investigating using a higher number of CRs may result in novel waveform designs.}

\textcolor{black}{Notably, the design methodologies of interleaving, CR variation, and CR variation with interleaving have yet to be studied in the context of MC CSS schemes. Their application is expected to lead to new waveform designs compatible with LPWANs.}

\textcolor{black}{Furthermore, it has been established that utilizing I/Q components as a design methodology for both SC and MC CSS schemes exhibits the worst performance, despite the significant improvement in SE. Simulation results illustrate that, in most cases, the performance falls below par. Moreover, in the case of SC CSS schemes, using PSs also does not result in efficient performance.}
\subsection{\textcolor{black}{Analysis of Characteristics for CSS Schemes}}
\textcolor{black}{In Table \ref{tab4}, we have summarized the performances of the studied schemes for the performance metrics listed above. In Table \ref{tab4}, we have used \checkmark and \xmark \(~\)to illustrate whether a certain function is possible for a given approach. On the other hand, a different symbol represents different levels of performance, e.g., \(+\) means a slight improvement in performance or minute robustness against a given offset, \(++\) means a considerable improvement in performance, or high robustness against an offset, whereas, \(+++\) is used for high-performance improvement. }
\renewcommand{\arraystretch}{1.1}
\begin{table*}[ht] 
  \centering
   \caption{Summary of attained performances for different schemes for performance metrics (Const. Env.: Constant Envelope, SE imp.: SE improvement, EE imp.: EE improvement, Coh. Detec.: Coherent Detection, Non-Coh. Detec.: Non-Coherent Detection. }
   \resizebox{\textwidth}{!}{%
  \begin{tabular}{@{}cccccccc@{}}
    \toprule
       \toprule
  \bfseries {Modulation}    & \bfseries {Const. Env.} & \bfseries {SE imp.} & \bfseries {EE imp.} & \bfseries {Coh. Detec.} & \bfseries {Non-Coh. Detec.}  & \bfseries {Robustness to PO} & \bfseries {Robustness to FO}  \\
    \midrule
        \midrule
    LoRa & \checkmark  & \xmark & \xmark & \checkmark & \checkmark & \(++\) & \(++\)  \\
     E-LoRa & \checkmark & \(+\) & \(+\) & \checkmark & \xmark & \xmark & \xmark \\
     ICS-LoRa & \checkmark & \(+\) & \(+\)  & \checkmark & \checkmark & \(++\) & \(++\)   \\
     PSK-LoRa & \checkmark  & \(+\) & \(++\)  & \checkmark & \xmark & \xmark & \xmark\\
 DO-CSS & \xmark  & \xmark & \(++\) & \checkmark  & \checkmark & \(+\) & \(+\) \\
      SSK-LoRa & \checkmark & \(+\)& \(+\)  & \checkmark & \checkmark & \(++\) & \(++\)\\
      IQ-CSS & \xmark  & \(++\) & \(+\)& \checkmark & \checkmark  & \xmark & \xmark\\
DCRK-LoRa & \checkmark & \(+\) & \(++\)& \checkmark & \checkmark & \(++\) & \(++\)\\
FSCSS-IM & \xmark & \(++\) &\xmark & \checkmark & \checkmark  & \(+\) & \(+\)\\
     IQ-CIM & \xmark & \(++\)  & \xmark & \checkmark & \xmark & \xmark & \xmark\\
SSK-ICS-LoRa & \checkmark & \(+\)& \(++\) & \checkmark & \checkmark & \(++\) & \(++\)\\
ePSK-CSS & \xmark&  \(+\) & \(++\) & \checkmark & \xmark & \(+\) & \(+\)\\
          GCSS & \xmark & \(++\) & \xmark & \checkmark & \checkmark & \(+\) & \(+\)\\
 TDM-CSS & \xmark & \(++\)& \(+\) & \checkmark & \checkmark & \(+\) & \(+\)\\
IQ-TDM-CSS & \xmark & \(+++\)& \(++\) & \checkmark & \xmark & \xmark & \xmark\\
DM-CSS & \xmark & \(++\)& \(+++\) & \checkmark & \checkmark & \(+\) & \(+\)\\
    \bottomrule
      \bottomrule
  \end{tabular}}
   \label{tab4}
\end{table*}

\textcolor{black}{Now, we also identify different CSS waveform designs which provide the best performance in each group in terms of the characteristics mentioned above. Table \ref{tab_optimal_group} lists the best scheme from each group.}
\begin{itemize}
\item \textcolor{black}{Among the approaches in group 1, SSK-LoRa scheme has emerged as the most promising, owing to its constant envelope characteristic, yielding higher SE and EE than the LoRa scheme. It is highlighted that SSK-LoRa uses \textit{varying CRs} as a design methodology. Furthermore, SSK-LoRa can employ both coherent and non-coherent detection, thus achieving a high level of immunity against PO and FO. It is worth noting that the ICS-LoRa also shares these attributes. However, SSK-LoRa outperforms ICS-LoRa, resulting in a lower cross-correlation between the symbols.}
\item \textcolor{black}{Within the second group, the SSK-ICS-LoRa scheme exhibits superior performance across multiple performance metrics, indicating its effectiveness as a design strategy. Specifically, SSK-ICS-LoRa leverages \textit{CR variation and interleaving} technique as the design methodology. Conversely, PSK-LoRa, a competing scheme within group 2, displays inadequacy in its capacity for non-coherent detection and susceptibility to both the PO and FO, rendering it less robust compared to SSK-ICS-LoRa.}
\item \textcolor{black}{While DCRK-LoRa exhibits commendable performance across all performance metrics, the formidable complexity of its receiver is potentially a constraining factor. DCRK-LoRa leverages \textit{CR variation} as a fundamental design strategy, differentiating it from its group 3 counterpart, ePSK-CSS, which lacks a non-coherent receiver and suffers from comparatively lower robustness against the PO and FO. Notwithstanding its heightened complexity with increased CRs implementation, DCRK-LoRa emerges as the optimal alternative, affording the capacity to attain higher spectral efficiencies by employing varying CRs.}
\item \textcolor{black}{Within the fourth group, DO-CSS and GCSS with \(\mathrm{G} = 2\) demonstrate comparable performance, capitalizing on utilizing \textit{multiple FSs} as a core design methodology. Nevertheless, attaining superior SE is possible for GCSS by increasing \(\mathrm{G}\), albeit at the potential expense of a decrease in EE and an increase in vulnerability to offsets. Consequently, GCSS emerges as the most advantageous alternative, achieving an optimal trade-off between SE and ancillary parameters. Nevertheless, the absence of a coherent detector for GCSS mandates DO-CSS as the preferred option for superior performance across the aforementioned performance metrics.}
\item \textcolor{black}{Within group 5, DM-CSS emerges as the optimal performer within an ideal non-linear channel; however, in the presence of PO and FO, TDM-CSS exhibits superior performance to DM-CSS. It is worth noting that TDM-CSS transmits one less bit per symbol compared to DM-CSS. Furthermore, the non-coherent detector for DM-CSS is only functional if the phase rotation is less than \(\sfrac{\pi}{2}\). While IQ-CSS and IQ-TDM-CSS exhibit poor robustness against PO and FO, the latter scheme affords an exceptionally high bit rate compared to other counterparts within the group but is prone to severe vulnerability to offsets. Ultimately, the superior performance of TDM-CSS establishes it as the optimal approach within group 5. TDM-CSS uses \textit{time-domain multiplexing} as the design methodology. }
\item \textcolor{black}{Although the potential for higher SE with IQ-CIM within the IM approaches in group 6, its inherent limitations, such as the absence of a coherent receiver and limited resistance to PO and FO, render it a less attractive alternative than FSCSS-IM. Notably, FSCSS-IM permits coherent detection and exhibits adequate robustness against offsets at \(\varsigma =2\), although such robustness may deteriorate for higher values of \(\varsigma\). FSCSS-IM uses \textit{activation of multiple FSs} as the design strategy. }
\end{itemize}
\begingroup
\setlength{\tabcolsep}{6pt} 
\renewcommand{\arraystretch}{1.2} 
\begin{table}[tbh]
\caption{\textcolor{black}{Categorization of the schemes with optimal performance from each group.}}
\centering
\color{black}\begin{tabular}{llcc}
\hline
\hline
\textbf{Group} &  \textbf{Optimal Scheme} \\
\hline
\hline
Group 1& SSK-LoRa\\

Group 2& SK-ICS-LoRa\\

Group 3& DCRK-LoRa\\

Group 4& DO-CSS, GCSS\\

Group 5& TDM-CSS\\

Group 6& FSCSS-IM\\
\hline 
\hline
\end{tabular}
\label{tab_optimal_group}
\end{table}
\endgroup
\section{\textcolor{black}{Open Issues and Future Research Directions}}\label{framework}
\textcolor{black}{This section will expound upon several unresolved concerns and future trajectories regarding waveform design in the LPWANs. These avenues for inquiry extend from innovating waveform designs to practically implementing established counterparts of LoRa which demonstrate better performances.}
\subsection{\textcolor{black}{Use of Different Design Methodologies}}
\textcolor{black}{This study scrutinized eight design methodologies for their potential implementation in waveform design for LPWANs. It has been observed that certain design methodologies are limited to specific taxonomies; for example, TDM can only be utilized in developing an MC CSS scheme. Moreover, Tables \ref{meritslimitations_sc} and \ref{meritslimitations_mc} indicate that specific design methodologies are exclusive to one of the three taxonomies. For instance, interleaving is only studied for the SC category rather than the MC or MC-IM taxonomies. Thus, an area for future research would be exploring the potential of such design methodologies for various taxonomies. The study reveals that design methodologies such as interleaving, CRs variation, and PS usage have yet to be investigated for MC or MC-IM schemes.}

\textcolor{black}{On the other hand, the use of CR variation and PS has yet to be studied for SC schemes. However, interleaving, CRs variation, and PS utilization are the design methods that demonstrate the best results for SC schemes. In fact, ICS-LoRa, SSK-LoRa, and SSK-ICS-LoRa are among the most formidable competitors to classical LoRa systems. Additionally, the CR variation and PS use methodology is employed by DM-CSS, which is one of the best schemes for a linear channel.}

\textcolor{black}{Hence, a logical extension of future research would be to apply these design methodologies to the taxonomies where they still need to be included. Some of these methodologies can be easily incorporated into other taxonomies; for instance, CRs can be varied effortlessly for DO-CSS, leading to improved spectral efficiencies. However, implementing these methodologies can also pose certain limitations. For example, utilizing different CRs can cause interference, even in SC CSS schemes. For MC CSS schemes, this interference may be more pronounced. Nevertheless, a comprehensive study is needed to elaborate on the merits and limitations of these waveform designs.}

\textcolor{black}{Some design methodologies offer significant flexibility in waveform design, such as CRs variation and TDM. While CR variation can be readily incorporated into waveform design, it only leads to a marginal increase in SE. In contrast, TDM is an adaptable design methodology, and though the seminal work on TDM primarily focused on TDM-CSS and IQ-TDM-CSS schemes, TDM can be applied in various ways. For instance, DM-TDM-CSS employs the TDM strategy, yielding superior performance and resilience against different offsets than IQ-TDM-CSS while having almost the same achievable SE. Although DM-TDM-CSS is not included in this study due to the exclusion criteria mentioned in Table \ref{inclusion_exclusion_css}, it is worth noting that the TDM methodology is not limited to TDM-CSS, IQ-TDM-CSS, or DM-TDM-CSS but can be implemented in various ways. Therefore, a comprehensive study is necessary to investigate the potential of different waveform designs that can be developed using TDM design methodology.}
\subsection{\textcolor{black}{Development of New Design Methodology}}
\textcolor{black}{Another conceivable avenue for future exploration in waveform design is the development of a new design methodology. One promising possibility is the fusion of binary PS with some established design strategies. As expounded in the antecedent sections of this inquiry, PSs can be integrated into un-chirped symbols to enhance SE based on the cardinality of the number of PSs. However, using PSs necessitates employing an ML detector to distinguish between the various PSs at the receiver, thereby increasing the complexity of the receiver.}

\textcolor{black}{However, if two PSs, 0 and pi, are utilized, the requirement of an ML detector can be relaxed. In such cases, the polarity function would suffice to determine the binary PSs. We can integrate this feature with most design methodologies, such as activating FS(s), interleaving, CR variation, CR variation, and interleaving, TDM, and using I/Q components. Furthermore, we could apply this novel design methodology to SC, MC, and MC-IM taxonomies. It is expected that for SC schemes, this design method would not compromise crucial characteristics of the waveform design, such as constant envelope and resilience to offsets. For MC schemes, it is projected that this design method would maintain the robustness of the waveform design.}

\textcolor{black}{For a scheme belonging to the SC category, this design method would augment the number of transmitted bits by one. Conversely, for schemes from MC or MC-IM taxonomies, the increase in the number of bits depends on the number of activated FSs. For instance, in DO-CSS and TDM-CSS, this design method could amplify the number of transmitted bits by two.}

\textcolor{black}{In the previous section, we underscored the flexibility of the CR variation and TDM design methods. Combining binary PSs with CR variation and TDM could be the most efficacious SC and MC scheme design methods, respectively.}

\textcolor{black}{A comprehensive investigation must scrutinize the benefits and limitations of the schemes adopting this design method. Notwithstanding this, a preliminary analysis evinces that this method is unlikely to detrimentally impact the integrity of a waveform design.}
\subsection{\textcolor{black}{Interference and Orthogonality Analysis of CSS Schemes}}
\textcolor{black}{Empirical evidence suggests that different design methods employed to develop MC CSS schemes can result in interference. The primary cause of interference is activating multiple FSs or utilizing different CRs. In general, the magnitude of interference decreases with an increase in \(M\), the number of FSs employed. However, when \(M\) is relatively low, the interference can be substantial. Therefore, conducting an analytical evaluation of interference incurred in MC schemes and identifying the optimal design methodology that causes minimal interference is imperative. Identification of the nature of interference could facilitate the development of design-specific interference mitigation techniques, thereby improving the overall performance of the schemes. Nevertheless, comprehensive interference analysis is necessary to accomplish this objective.}

\textcolor{black}{An additional aspect we could analytically evaluate is the orthogonality of the CSS schemes. LoRa is orthogonal in the discrete-time domain; however, there is a lack of comprehensive research on the orthogonality of other CSS alternatives to LoRa, including SC or MC. Analyzing orthogonality is essential because non-orthogonal schemes cause interference between two symbols. Furthermore, the absence of orthogonality affects detection mechanisms, as the cross-correlation of the symbols is non-zero, leading to performance degradation. Therefore, a comprehensive study of the orthogonality properties of CSS schemes could aid in identifying efficient design methodologies.}
\subsection{\textcolor{black}{Development of MC-IM Schemes}}
\textcolor{black}{In this study, we conducted an assessment of three diverse taxonomies: SC, MC, and MC-IM. These taxonomies employ various design methodologies for waveform design. Among them, the most adaptable taxonomy is MC-IM, which enables different spectral and energy efficiencies contingent upon the number of activated FSs. This enhances the applicability of MC-IM schemes. Regrettably, the existing literature on MC-IM schemes is deficient and employs only suboptimal design methodologies, such as activating multiple FSs or utilizing I/Q components. These design methodologies are suboptimal, considering the diverse characteristics mentioned earlier. Therefore, developing efficient MC-IM schemes using the most optimal design methodologies is crucial. This could result in more effective and robust waveform designs that can also be adaptable to attaining different spectral and energy efficiencies, depending on the specific application requirements.}
\subsection{\textcolor{black}{Practical Implementation of LoRa Competitor Waveforms}}
\textcolor{black}{Despite the numerous alternatives to LoRa proposed in the literature that have demonstrated superior performance, practical implementations of different alternatives to LoRa have been limitedly explored. The pervasiveness of LoRa can be attributed to its practical implementation and the theoretical and empirical investigations of its performance. Hence, practically implementing the most optimal alternatives to LoRa could make a substantial contribution. Additionally, such an investigation could help identify the bottlenecks in implementing diverse waveform designs.}
\section{Conclusions}\label{sec6}
\textcolor{black}{This study offers a thorough analysis of different CSS-based waveform designs for LPWANs, which fall under various taxonomies of the CSS schemes and adopt distinct design methodologies to form the symbol structure. The prevalent CSS scheme employed in LPWANs is LoRa, but it suffers from low transmission rates. Several CSS alternatives have been proposed in the literature to address this issue, which aims to enhance SE and improve EE. However, the current state-of-the-art mainly focuses on the practical implementation and theoretical aspects of LoRa, neglecting other alternatives for IoT deployment and system evaluation.}

\textcolor{black}{To fill this research gap, this work comprehensively elucidates different taxonomies of CSS schemes and various design methods to present a comprehensive waveform design for some CSS alternatives to LoRa. Moreover, as different schemes yield different spectral efficiencies, the CSS schemes are grouped based on the achievable spectral efficiencies to analyze schemes with similar spectral efficiencies. Simulations are performed under various channel conditions to determine the optimal design methodology. The simulation results demonstrate that MC and MC-IM taxonomies schemes can achieve higher spectral efficiencies than SC schemes at the cost of less robustness against the PO and FO. Conversely, SC CSS schemes are more resilient against offsets.}

\textcolor{black}{Based on the simulation results, we identify that specific design methodologies provide the best results for SC and MC taxonomies. For SC taxonomy, the optimal results are obtained when CRs are varied or when CRs are varied, along with the interleaving of the chirped symbols. The schemes for the SC taxonomy that provide the best performance are SSK-LoRa, SSK-ICS-LoRa, and DCRK-LoRa. On the other hand, the schemes that provide the best MC taxonomy performance are activating multiple FSs and TDM. Additionally, we note that the literature lacks the implementation of different design methods for MC-IM. The least robust design methodologies for both SC and MC schemes are the use of PSs and I/Q components, as these design methodologies increase the transceiver's complexity and do not allow non-coherent detection in most cases. }

\textcolor{black}{Based on the state-of-the-art, we propose various open research issues that involve developing new design methodologies and investigating new waveform designs based on efficient design methodologies explored in the study. Moreover, we identify that only some schemes belonging to MC-IM taxonomy have been investigated in the literature; hence, this can be a potential direction for future research to study different design methodologies for MC-IM schemes. Furthermore, we recognize that a comprehensive analytical analysis is needed to explore the orthogonality and the nature of interference in the CSS schemes.}

\textcolor{black}{In conclusion, the presented waveform design in this work is expected to inspire future research in this field.}
\bibliographystyle{unsrt}
\bibliography{biblio}
\end{document}